\documentclass[12pt,preprint]{aastex}

\usepackage{epsfig}
\usepackage{subfigure}
\usepackage{amsthm,amsmath,amssymb}
\usepackage{mathrsfs}
\usepackage[displaymath]{lineno}
\usepackage{booktabs}
\usepackage{multirow}

\shorttitle{GRB Precursors}

\shortauthors{Li \& Mao}

\begin{document}

\title{Temporal Analysis of GRB Precursors in the Third $Swift$-BAT Catalog}
\author{Liande Li$^{1,2,3}$ and Jirong Mao$^{1,2,4}$}
\affil{$^{1}$Yunnan Observatories, Chinese Academy of Sciences, 650011 Kunming, Yunnan Province, People's Republic of China; jirongmao@mail.ynao.ac.cn\\
       $^{2}$Center for Astronomical Mega-Science, Chinese Academy of Sciences, 20A Datun Road, Chaoyang District, 100012 Beijing, People's Republic of China; \\
       $^{3}$University of Chinese Academy of Sciences, 100049 Beijing, People's Republic of China; \\
       $^{4}$Key Laboratory for the Structure and Evolution of Celestial Objects, Chinese Academy of Sciences, 650011 Kunming, People's Republic of China \\
       }

\begin{abstract}
We select 52 long gamma-ray bursts(GRBs) having the precursor activity in the third Swift-BAT catalog.
Each episode shown in both the precursors and the main bursts is fitted by the Norris function.
We systematically analyze the temporal properties for both the precursors and the main bursts.
We do not find any significant difference between the temporal profile of the precursor and that of the main burst.
The photon count of the precursor is related to that of the main burst.
It is indicated that the precursor and the main burst might have the same physical origin, as the precursor and the main burst follow the same $\tau_p-\omega$ relation.
However, we do not find the explicit relation between the energy release of the precursor and the quiescent time. 
Some theoretical models, such as the fallback collapsar scenario and the jet-cocoon scenario, may be helpful to
explain the GRB-precursor phenomena.
\end{abstract}

\section{INTRODUCTION}

Gamma-ray bursts (GRBs) are the most violent explosion events in the universe.
A GRB lightcurve often shows multiple emission pulses. These pulses are separated by some quiescent intervals.
One of the pulses has the largest peak intensity among all the pluses, and this pulse is usually called as the main burst.
Sometimes a weaker pulse before the main burst is shown, and this weaker pulse is identified as the precursor.
It is important to perform the precursor research to understand the physical mechanisms of the GRB central engine.
Some theoretical models can be constrained by the GRB-precursor observations.

GRB 900126, as the first GRB with a clear precursor, was detected by Ginga satellite in 1990, and the precursor spectrum can be fitted by a blackbody (BB) shape (Murakami et al. 1991).
Koshut et al. (1995) defined the GRB precursor by two conditions. First, the peak intensity of the precursor must be less than that of the main burst. Second, the quiescent time should be at least as long as the duration of the main burst. Here, the quiescent time is defined to be from the end of the precursor to the beginning of the main burst. Then,
they searched the GRB precursors in the BATSE catalog.
It was not clear that the precursor emission is harder or softer than the main-burst emission.
Lazzati (2005) also identified the GRB precursors in the BATSE catalog. The sample contained the bright GRBs with $T_{90} \geq 5$ s.
They used a single power-law (PL) model and a BB model respectively to fit the precursor spectra.
Most spectra have the good fitting by the PL model. 
It was found that the precursors are softer than the main bursts.
Burlon et al. (2008; labeled as ``B08" in Table 1) searched the precursors in the Swift-observed GRBs, and the GRBs with the precursors have known redshifts.
After they compared the spectral index of the precursor to that of the main burst, they found that there is no major difference between the precursor and the main burst.
Subsequently, Burlon et al. (2009)
identified the GRBs with the precursors in the BATSE catalog.
By the spectral analysis, they concluded that the precursor and main burst are originated from the same mechanism.
Hu et al. (2014; labeled as ``H14" in Table 1)
used the Bayesian block algorithm to search the GRB precursors observed by the Swift-BAT.
They also suggested by the spectral analysis that the origin of the precursor is consistent to that of the main burst.
Charisi et al. (2015) searched the isolated emission episodes from the Fermi-GBM, Swift-BAT and BATSE detections.
They identified the emission episode with the highest peak intensity as the main burst. The emissions
prior to the main burst are called as the pre-peak events, and the emissions after the main burst are called as the post-peak events.
By comparing some temporal parameters of the pre/post-peak event to those of the main burst,
they did not find the correlation between the pre/post-peak event and the main burst.
Lan et al. (2018; labeled as ``L18" in Table 1) used a Bayesian block algorithm to search the two-episode GRB emission in the Fermi-GBM data. 
They did not find any significant correlation between the duration of the first episode and that of the second episode.
Coppin et al. (2020; labeled as ``C20" in Table 1) searched the GRB precursors in the Fermi-GBM catalog. Some faint precursors were selected.
They found that the distribution of the quiescent time has a bimodal distribution.
It indicates that the precursors may have two progenitor types.

Some theoretical explanations have been proposed to investigate the origin of the GRB precursors.
In the fireball model, when the fireball becomes transparent, the transient radiation produced by the photosphere is released.
This radiation corresponds to the precursor emission. When the fireball becomes totally optically thin, the main burst is produced by the internal shocks (M{\'e}sz{\'a}ros \& Rees 2000; Daigne \& Mochkovitch 2002; Li 2007). Thus,
the precursor is expected to have thermal emission, 
and the quiescent time
cannot be very long. 
It is suggested in the GRB progenitor model that GRB precursor is associated with collapsar (Ramirez-Ruiz et al. 2002; Waxman \& M{\'e}sz{\'a}ros 2003; Lazzati \& Begelman 2005).
GRB jet is produced by the collapsar, and the so-called cocoon might be formed around the jet.
Strong precursor emission is released during the interaction between the jet and the cocoon.
When the jet escapes from the cocoon, the main burst is produced. Thus, the precursor is expected to have the thermal emission from the cocoon.
Furthermore, the quiescent time cannot be longer than 10 s.
Cheng \& Dai (2001) proposed a two-stage model. 
In this model, a massive star 
can collapse into a neutron star, and a preliminary jet is produced. 
Then, the neutron star reaches to the maximum mass due to accretion and collapses into a black hole. In this process, another jet is launched. 
Wang \& M{\'e}sz{\'a}ros (2007) 
further discussed the long quiescent time between the precursor and the main burst. The precursor can be produced by the first jet formed during the the initial core collapse.
The second jet is launched by the fallback accretion, and the main burst is produced.
In this fallback collapsar scenario, the quiescent time can be longer than 100 s.
Lipunova et al. (2009) suggested a similar model. 
The precursor is produced by a fast-rotating collapsar. The collapsar turns to be a spinar.
When the angular momentum of the spinar is carried away, a black hole or a neutron star is formed, and the main burst is produced.
This model also predicts a long quiescent time that can be longer than 100 s.
Some GRBs may have multiple precursors.
Lyutikov \& Usov (2000) proposed a magnetized wind model.
The magnetized wind from the GRB progenitor may produce three radiative regions:
the photosphere region, the wind instability region, and the wind interaction region to the ambient gas.
The GRB emission components are produced in these regions.
The first component is the emission from the photosphere, and a weak precursor that has a BB spectrum can be produced.
The second component is generated by the wind instability, and the third component is produced in the wind interaction with the ambient gas. The second and the third components can be identified as the two-episode main burst.
The quiescent time is 10$-$100 s.
Bernardini et al. (2013) proposed a propeller model to explain the GRB precursor. The GRB central engine is assumed to be a magnetar.
The precursor and the main burst are produced by the matter accretion along the magnetic field lines.
When the centrifugal drag halts the accretion, the system enters a propeller phase, and the GRB becomes quiescent.
If there is enough accumulated matter and the pressure can overcome the centrifugal drag, the propeller phase is finished and the accretion restarts. Thus, the multiple precursors and the very long quiescent times can be well interpreted.

The results from the data analysis suggest different physical origins of the GRB precursor. The theoretical explanations to the GRB progenitor have diversities. It is necessary to systematically investigate the GRB precursor in a certain dataset, and the results can be applied to further constrain some theoretical models. The spectral properties of the GRB precursors have been carefully examined (Koshut et al. 1995; Lazzati 2005; Burlon et al. 2008; Burlon et al. 2009; Hu et al. 2014). In this paper, we comprehensively perform the temporal analysis to the GRB precursors. The data are taken from the third Swift-BAT catalog. The temporal properties of the precursors to those of the main bursts are statistically compared.

This paper is organized as follows. In Section 2, we have the definition of the GRB precursor. The selection criteria of the GRB precursor are given. The temporal fitting process is presented. In section 3, we comprehensively analyze the temporal properties of the GRB precursors. The main results are presented. In Section 4, the results obtained in this paper are compared to those from some theoretical models. We draw the conclusions in Section 5.

\section{MOTIVATIONS, SAMPLE SELECTION, AND FITTING PROCEDURE}
\subsection{MOTIVATIONS}
We first present our scientific motivations on the GRB temporal analysis. Here, some definitions are provided. Either the prompt emission or the precursor in one GRB lightcuvre has one pulse or a few pulses. Each pulse may contain one episode or a few episodes. Each episode can be described by the fast rise and exponential decay (FRED) shape (Fenimore et al. 1996). We show the definitions in Figure 1. Furthermore, one episode is composed by many short-timescale variabilities. 
The short-time variability in one episode has the duration of $10^{-6}$-$10^{-5}$ s (Bhatt \& Bhattacharyya 2012).

It has been proposed that the short-timescale variability happens due to the internal shock collision in the fireball framework (e.g., Fenimore et al. 1996). Alternatively, turbulence has been suggested as a possible mechanism to produce GRB variabilities (Kumar \& Narayan 2009; Narayan \& Kumar 2009). 
In general, the FRED feature in a long GRB has a much longer duration than those of the short-time variabilities. 
It is interesting to investigate if the FRED feature in the precursor has the same mechanism as that in the prompt emission.
Moreover, the time interval between the precursor and the prompt emission also provides the information on both the activity timescale of the central engine and the light propagation.  
Thus, we perform the temporal analysis to find the FRED feature in both the precursor and the prompt emission of each GRB lightcurve in our sample.

\subsection{SAMPLE SELECTION AND FITTING PROCEDURE}
We select the GRBs triggered by Swift-BAT in the energy range of 15-150 keV
from the third Swift-BAT catalog (Lien et al. 2016)\footnote{https://swift.gsfc.nasa.gov/results/batgrbcat/index.html}. The catalog contains 1006 GRBs. 
We then search for the precursors in the GRB lightcurves and systematically investigate the temporal properties of the GRB precursors. 
In the catalog, GRB event data have been processed by the standard BAT software with the latest calibration database. The background is subtracted to build the GRB lightcurves.    
Each lightcurve has been built in five energy channels, which are 15-25 keV, 25-50 keV, 50-100 keV, 100-350 keV and 15-350 keV. In this paper, we use the lightcurves in the energy band of 15-350 keV.
Moreover, there are a few binning types in the catalog. The binning times of 2, 8, 16, 64 ms, and 1s are provided.
We examine the lightcurves with the different binning times in the catalog. We find that the FRED feature of not only prompt emissions but also precursors can be clearly shown in the 1 s binning lightcurves and the 64 ms binning lightcurves.
Furthermore, we consider that the 1 s binning lightcurve
may be too coarse to be used for a certain FRED-feature fitting. 
The coarse effect can be effectively reduced when we take a lightcurve with a shorter binning time. 
In this paper, we select the long GRBs ($T_{90}\ge 2$ s) to study the precursor, and we take the 64 ms binning lightcurves for the temporal analysis.
Because the FRED feature of long GRBs 
has the duration time in the order of 1 s in general, we can use the GRB lightcurves with the binning 
time of 64 ms to fit the FRED features.
The details on the
GRB-data processing have been presented by Lien et al. (2016).
We also have the cross check to the GRB samples of the short GRBs, and we are sure that the short GRBs with the extended emissions are excluded in this paper.

We perform a visual search to identify the GRBs with the precursor activity in the third Swift-BAT catalog.
The selection criteria are as follows:
(1) the one with the highest peak intensity among the emission pulses is identified as the main burst, and the emission pulses that occurred before the main burst are identified as the precursors;
(2) the photon-count rate of the precursor must fall to the background level before the start of the main burst;
(3) one pulse may contain several episodes (an episode should contain a relatively complete structure, which means that the rise, the peak, and the decay signatures are clearly shown in an episode);
(4) the GRB lightcurves with the time resolution of 64 ms are considered; 
(5) we confirm a precursor when its signature can be clearly shown in all energy-channel (15-25, 25-50, 50-100, 100-350, and 15-350 keV) lightcurves;
(6) the GRBs with the precursor activity have redshift information.
We finally identify 52 long GRBs with the precursor activity. Among the GRBs with the redshift measurement, the ratio between the GRBs with the precursors and the total GRBs in the third Swift-BAT catalog is 13.8\%.
We draw a cartoon to illustrate the episodes, the precursors, the main burst, and the quiescent time of one GRB in Figure 1.

The redshift distribution of the selected GRBs has the range of 0.54 $-$ 6.32\footnote{The redshift numbers are listed in Tables 1 and 2. They are taken from the webpage of https://www.mpe.mpg.de/$\sim$jcg/grbgen.html. Most redshift numbers are spectroscopically determined. A few exceptions are the ones with the upper-limit numbers. They are determined by the photometric measurement. We take all the redshift numbers mentioned above in the paper.}. We see the distribution in the left panel of Figure 2. 
The GRB duration is identified as $T_{90}$. $T_{90}$ values are taken from the Swift-BAT catalog. We see the $T_{90}$ distribution of the selected GRBs in the right panel of Figure 2.

A GRB episode can be simply described 
by the FRED shape (Fenimore et al. 1996). The FRED shape is widely used to perform the GRB temporal analysis (Norris et al. 1996; Lee et al. 2000).
In this paper, we use the fitting procedure provided by Norris et al. (2005).
The fitting function is $I(t)$ = $A$$\lambda$$e^{-{\tau_{1}}/{(t-t_{s})}-{(t-t_{s})}/{\tau_{2}}}$,
where $\lambda$ = $e^{2\mu}$, $\mu$ = {({$\tau_{1}$}$/${$\tau_{2}$})}$^{1/2}$, $A$ is the maximum intensity of an episode, and $t_s$ is the start time of an episode. 
The peak time is $t_p$ = $\tau$$_{p}$ + $t_s$ = {({$\tau$$_{1}$}{$\tau$$_{2}$})}$^{1/2}$ + $t_s$, where $\tau_{p}$ ={({$\tau$$_{1}$}{$\tau$$_{2}$})}$^{1/2}$. The pulse width is $\omega = \tau_d + \tau_r = \tau_{2} (1+4\mu) ^{1/2}$, and the pulse asymmetry is $\kappa=(\tau_d-\tau_r)/(\tau_d+\tau_r) = (1+4\mu) ^{-1/2}$, where the rise time is $\tau_r=\tau_2[(1+4\mu)^{1/2}-1]/2$, and the decay time is $\tau_d=\tau_2[(1+4\mu)^{1/2}+1]/2$.
We use the fitting function to fit all the episodes in both the precursors and the main bursts\footnote{In the GRB sample of Norris et al. (2005), one pulse only contains one episode. Thus, the episode can be called as {\it pulse}. In this paper, one pulse may contain several episodes.}.
We use the maximum likelihood estimation (MLE) method to fit the FRED feature.
We totally analyze 52 GRB lightcurves. Sixty-four precursor pulses are identified.
If one GRB has one precursor, we find 41 GRBs belong to this kind. We find 10 GRBs, and each GRB has two precursors. There is one GRB with three precursors in our sample.
The fitting parameters are listed in Table 1. The error of the fitting parameter has the confidence interval of 1$\sigma$. The general fitting process and the fitting curves of the episodes in each GRB are illustrated in detail in Appendix. The residual of the fitting is also shown in each panel.

\section{RESULTS}

We obtain the statistical results of the episode-shape parameters for both the precursors and the main bursts.
The distributions of $A$, $t_s$, $\tau_1$, $\tau_2$, $\tau_p$, $\omega$, $k$, $\tau_r$, and $\tau_d$ are shown in (a)-(i) of Figure 3, respectively. The distributions of the redshift-corrected $t_s$, $\tau_1$, $\tau_2$, $\tau_p$, $\omega$, $\tau_r$, and $\tau_d$ are shown in (a)-(g) of Figure 4, respectively\footnote{Some precursors were occurred before their GRB triggers, and the start times of the precursors have negative numbers. We arbitrarily add 300 s to the start time of each episode for the statistics.}.

In order to investigate the difference of the temporal property between the precursor and the main burst,
we perform the Kolmogorov-Smirnov (K-S) test to the Norris-function parameter distributions for both the precursor and the main burst.
We list the K-S test results in Table 2.
It is shown that the peak intensity of the precursor is smaller than that of the main burst, because
we identify the main burst that has the maximum peak intensity in one GRB lightcurve. It is also shown that the start time of the precursor is earlier than that of the main burst, because the precursor occurs earlier than the main burst in one GRB lightcurve.
The K-S tests to the distributions for the parameters of $\tau$$_{1}$, $\tau$$_{2}$, $\tau$$_{p}$, $\omega$, $\kappa$, $\tau_{r}$, and $\tau_{d}$ indicate that
the temporal shape of the precursor and that of the main burst are similar.

The quiescent time $\Delta t_{q,i}$ is defined as the time from the end of one pulse to the beginning of the subsequent pulse, where the order of the quiescent time in one GRB is labeled with ``i". The total quiescent time $\Delta t_q$ can be calculated by summing up all $\Delta t_{q,i}$ numbers in the GRB.
Here, the beginning of one pulse is defined by the time that the first episode in the pulse rises to reach to the $1/e$ of the episode peak intensity, and the end of one pulse is defined by the time that the last episode in the pulse decays to reach to the $1/e$ of the episode peak intensity.
The distribution of the quiescent time $\Delta t_{q,i}$ is shown in the left panel of Figure 5, and the distribution of the quiescent time $\Delta t_q$ is shown in the right panel of Figure 5.
The mean value of the quiescent time $\Delta_{q,i}$ in the rest frame is 14.6 s with the standard variation of 23.9 s.
We note that the quiescent times of some GRBs in the rest frame can be longer than 100 s.
For example, the quiescent time of GRB 110709B in the rest frame can reach up to 170 s.
The details are listed in Table 3.

Besides the quiescent time, the time interval between the peak of a pulse to the
peak of another pulse may also provide the information on the process of the GRB energy release.
We can define the time interval $\Delta t_{p,i}$ that is the time between the peak time of one pulse to the peak time of the subsequent pulse, where the order of the peak time in one GRB is labeled ``i".
The total peak-time interval $\Delta t_p$ can be calculated by summing up all $\Delta t_{p,i}$ numbers in the GRB.
Here, the peak time of one pulse is the time that corresponds to the maximum intensity in all the episodes of one pulse, if the pulse contains several episodes.
The distribution of the peak-time $\Delta t_{p,i}$ is shown in the left panel of Figure 6, and the distribution of the peak-time
$\Delta t_p$ is shown in the right panel of Figure 6.
The mean value of the peak-time $\Delta t_{p,i}$ is 21.7 s with the standard variation of 28.8 s.
The details are listed in Table 3.

In order to further examine some possible correlations between the precursor and the main burst,
we compare the episode parameters in the precursors to those in the main bursts.
The results are shown in Figure 7.
The parameters of $\tau_1$, $\tau_2$, $\tau_p$, $\omega$, $\tau_r$, and $\tau_d$ can be redshift-corrected. We compare the redshift-corrected parameters in the precursors to those in the main bursts. The results are shown in Figure 8. We perform the correlation analysis to quantitatively
examine the relation between the precursor and the main burst. The correlation coefficient and the $p$-value are given in each panel of Figures 7 and 8.
We do not find any tight correlation between the precursor and the main burst.

Sometimes one pulse contains several episodes in one GRB. We average the episode parameters of the precursors, and we also average the episode parameters of the main burst in each GRB. The average
number of a certain parameter is simply treated to be the mean value of the parameter without any weighting. Then, we compare the averaged episode parameters of the precursors to those of the main bursts.
The results are shown in Figure 9.
The averaged parameters of $\tau_{1}$, $\tau_{2}$, $\tau_{p}$, $\omega$, $\tau_{r}$, and $\tau_{d}$ can be redshift-corrected. We compare the
redshift-corrected parameters of the precursors and those of the main bursts. The results are shown in Figure 10. We perform the correlation
analysis to quantitatively
examine the relation between the precursor and the main burst. The correlation coefficient and the $p$-value are given in each panel of Figures 9 and 10. Although it is shown in a few panels that a loose correlation exists, we note that
the data have large dispersions.

We further select 11 GRBs, and each GRB contains only one precursor episode and only one main burst episode. The GRBs are GRB 050726, GRB 060604, GRB 060605, GRB 060607,
GRB 060923A, GRB 071010B, GRB 091208B, GRB 120211A, GRB 120224A, GRB 140311A, and GRB 140515A.
We can compare the episode parameters of the precursors to those of the main bursts. The results are shown in Figure 11.
We also perform the redshift-correction to the parameters of $\tau_{1}$, $\tau_{2}$, $\tau_{p}$, $\omega$, $\tau_{r}$, and $\tau_{d}$.
We can compare the redshift-corrected parameters of the precursors to those of the main bursts. The results are shown in Figure 12.
The correlation coefficient and the p-value are given in each panel of Figures 11 and 12. We do not find any reliable correlation.

We can investigate the correlation between the width and the peak time for all the episodes of all the GRBs in our sample.
We show the width versus the peak time for all the episodes in the left panel of Figure 13.
We also show the redshift-corrected width versus the redshift-corrected peak time for all the episodes in the right panel of Figure 13.
We perform the maximum likelihood method given by Amati et al. (2008) to fit the data. This method was also applied in our former works (Liu \& Mao 2019).
We use the linear function $\rm{log}~\omega=a\times \rm{log}~\tau_p+b$ to fit the dataset ($\tau_p$, $\omega$). Each data pair is ($\tau_{p,i}$, $\omega_i$), where $i=1,...,n$. The error of each data pair
is ($\sigma_{\tau_p,i}$, $\sigma_{\omega_i}$). We build the likelihood function as
\begin{equation}
\begin{aligned}
\rm{log}~p(a,b,\sigma_{\tau_p},\sigma_\omega|{\tau_{p,i},\omega_i,\sigma_{\tau_p,i},\sigma_{\omega,i}})
& = \frac{1}{2}\sum_{i=1}^{n}\{\rm~{log}[\frac{1}{2\pi(\sigma_\omega^2+a^2\sigma_{\tau_p}^2+\sigma_{\omega,i}^2+a^2\sigma^2_{\tau_p,i})}] \\
& -\frac{(\omega_i-a\tau_{p,i}-b)^2}{\sigma_\omega^2+a^2\sigma_{\tau_p}^2+\sigma_{\omega,i}^2+a^2\sigma^2_{\tau_p,i}}\}.
\end{aligned}
\end{equation} 
We then minimize it to obtain the fitting parameters of $a$, $b$, and $\sigma_\omega$. In particular, we simply take the external error of the fitting $\sigma_{ext}$ to be $\sigma_{ext}=\sigma_\omega$, and we set $\sigma_{\tau_p}=0$ in Equation (1). The fitting results with the parameters are shown in Figure 13. We obtain the redshift-uncorrected relation of $\rm{log}\omega \sim(0.58\pm 0.04)\rm{log}\tau_p$ with the external error of $\sigma=0.24\pm 0.02$ and the redshift-corrected relation of $\rm{log}\omega \sim(0.61\pm 0.04)\rm{log}\tau_p$ with the external error of $\sigma=0.25\pm 0.01$. 
The correlation exists although the data dispersion is large.
The redshift-uncorrected relation and the redshift-corrected relation have no major difference.
However, we cannot distinguish the difference between precursor and main burst in the two panels.
In other words, both precursor and main burst follow the same $\tau_p-\omega$ relation.
We further investigate the $t_p-\omega$ relation that is shown in the GRB X-ray flares.
The GRB early X-ray flare sample given by
Chincarini et al. (2010) provides the relation of $\rm{log}\omega\sim (0.9\pm 0.4)\rm{log}t_p$. The GRB late X-ray flare sample given by
Bernardini et al. (2011) provides the relation of $\rm{log}\omega\sim (1.1\pm 0.2)\rm{log}t_p$.  
The relation of the early X-ray flare and that of the late X-ray flare are similar. 
However, 
we note that the start time of the X-ray flare temporal profile provided by Chincarini et al. (2010) and Bernardini et al. (2011) was not subtracted in the data analysis. In their works, they used $t_p=\tau_p+t_s$ to investigate the $t_p-\omega$ relation. In order to compare the temporal profile of the X-ray flare and that of the prompt emission in a self-consistent way, we further examine the $\tau_p-\omega$ relation for both the X-ray flares and the prompt emissions. The results are presented in Section 4.

The photon count of an episode can be obtained by the integral of the photon-count rate in the episode. We can calculate the total photon count for both the precursors and the main burst in one GRB. Thus, for each GRB, we have a photon-count number of all the precursors and a photon-count number for the main burst. We statistically examine the correlation between the photon count of the precursors and the photon count of the main burst. The results are shown in Figure 14. The correlation exists as log $C_p$ = ($-$0.29$^{+0.08}_{-0.08}$) + (0.66$^{+0.10}_{-0.10}$)log $C_m$, where $C_m$ indicates the photon count of the main burst and $C_p$ indicates the photon count of the precursor, and the extrinsic scatter $\sigma$ = 0.41$^{+0.04}_{-0.03}$. We conclude that the energy release in the precursor and the energy release in the main burst can be linked by a certain mechanism in general.
We caution that the spectral property of the precursor and that of the main burst may have a difference. In general, spectral evolution may occur between precursor and main burst in a GRB lightcurve. When we compare the energy accumulation between the precursor and the main burst derived from the photon count, the effect of the spectral evolution should be considered. We will further study this issue as one part of the future work on the spectral analysis of the GRB precursor.
It is interesting to identify five special GRBs, which are GRB 050319, GRB 050401, GRB 080602, GRB 120211A, and GRB 120922A. In each GRB mentioned above, the photon count of the precursor is larger than that of the main burst.

One may expect that a precursor is the initial energy release of a GRB. It takes time to have the energy accumulation for the subsequent main burst. Therefore, we first examine if the ratio between the photon count of each precursor and the photon count of the main burst can be related to the observed quiescent time $\Delta t_{q,i}$ or the redshift-corrected quiescent time $\Delta t_{q,i}/(1+z)$ in one GRB. The results are shown in the upper-left and the upper-right panels of Figure 15. We then examine if the ratio between the photon count of all the precursors and the photon count of the main burst can be related to the total observed quiescent time
$\Delta t_q$ or the redshift-corrected quiescent time $\Delta t_q/(1+z)$ in one GRB. The results are shown in the middle-left and the middle-right panels of Figure 15. Finally, we examine if the ratio between the photon count of one pulse and the photon count of the subsequent pulse can be related to the observed quiescent time $\Delta t_{q,i}$ or the redshift-corrected quiescent time
$\Delta t_{q,i}/(1+z)$ in one GRB. In this case, both pulses may be precursors. The results are shown in the lower-left and the lower-right panels of Figure 15. We cannot find any relation between the energy release and the quiescent time.

We can further investigate if the energy accumulation of one pulse is related to the peak-time interval. We first examine if the ratio between the photon count of each precursor and the photon count of the main burst can be related to the observed peak-time interval $\Delta t_{p,i}$ or the redshift-corrected peak-time interval $\Delta t_{p,i}/(1+z)$ in one GRB. The results are shown in the upper-left and the upper-right panels of Figure 16. We then examine if the ratio between the photon count of all the precursors and the photon count of the main burst can be related to the total observed peak-time interval
$\Delta t_p$ or the redshift-corrected quiescent time $\Delta t_p/(1+z)$ in one GRB. The results are shown in the middle-left and the middle-right panels of Figure 16. Finally, we examine if the ratio between the photon count of one pulse and the photon count of the subsequent pulse can be related to the observed peak-time interval $\Delta t_{p,i}$ or the redshift-corrected peak-time interval
$\Delta t_{p,i}/(1+z)$ in one GRB. In this case, both pulses may be precursors. The results are shown in the lower-left and the lower-right panels of Figure 16. We cannot find any relation between the energy release and the peak-time interval.
Thus, from the results shown in Figures 15 and 16, we cannot confirm that the energy accumulation or the energy release is related to
the time. It is suggested that the energy accumulation or the energy release of each GRB pulse is random.

\section{DISCUSSION}
GRB precursor has different definitions presented in some literatures.
Here, we list three major issues.
First, we visually identify precursors. The precursors identified in this paper are very bright. The percentage of the GRBs with the precursors in our sample is 13.8\%. While many precursors identified by Coppin et al. (2020) are very faint. Even though the sample of Coppin et al. (2020) includes the GRB sources without redshift examination, the long GRBs with the precursors have the fraction of 10.5\% in the total sources. We also note that the precursor feature is identified in 10\% of the GRBs in the sample of Hu et al. (2014). This percentage is roughly consistent with our results.
Second, we do not have any duration limit to the quiescent time.
However, Koshut et al. (1995) required the quiescent time that should be at least as long as the duration of the main burst. This selection limitation provides that the GRBs with the precursors only take 2\% in their sample.
Third, we do not have any requirement that is related to the GRB trigger time to identify a precursor.
However, Lazzati et al. (2005) required a precursor that cannot be a GRB trigger. The sample selected by Lazzati et al. (2005) has the precursors that are in 20\% total GRBs. 
To have a comparison, in Table 1, we label the GRBs with the precursor signature identified by other researchers.

X-ray flash (XRF) is a kind of GRB that has the main energy release in the soft energy band (Sakamoto et al. 2005). 
Bi et al. (2018) classified Swift-observed GRBs into three subclasses: XRF, X-ray rich (XRR), and classical GRB (C-GRB). The ratio between the XRF number and the total GRB number is about 8\% in that sample. The classification was based on the ratio between the fluence of 20-50 keV and that of 50-100 keV (Sakamoto et al. 2008). In this paper, we use the sample given by Bi et al. (2018).
We find 18 C-GRBs and 33 XRRs that have precursors, and we label each C-GRB/XRR in Table 1. However, the GRBs with the precursor activity in our sample are not XRFs. It is indicated that the strong precursor signature does not refer to the XRF case. 
However, we note that the absence of the precursor signature in the XRFs could be due to a certain observational bias. For example, one of the selection criteria to identify the GRB precursor in this paper is that the precursor can be visually shown in the lightcurves in all the BAT energy channels.
But, compared with the emissions of the C-GRBs, the emissions of the XRFs are mainly detected in the low-energy channel. Thus, the precursors of some XRFs may not be seen in the high-energy channel. The selection criterion of seeing a precursor in all the BAT-channels becomes a major reason for the discrepancy of the precursor identification between XRFs and C-GRBs.
In this paper, we only take the precursor sample in the total energy band of $15-350$ keV. The detailed analysis in the sub-energy bands ($15-25$, $25-50$, $50-100$, and $100-350$ keV) can be helpful to further identify this property. A large sample is required to confirm this preliminary result in the future.

In order to explore the similarity of the temporal profile between GRB prompt emission and GRB X-ray flare, we reexamine the $\tau_p-\omega$ relation among precursor, main burst, early X-ray flare, and late X-ray flare. 
The values of $\tau_p$ are obtained from Chincarini et al. (2010) and Bernardini et al. (2011), and the relation of $\tau_p=t_p-t_s$ is used. Thus, the start time of the X-ray flare temporal profile is subtracted in the data analysis.
The redshift-uncorrected results are shown in Figure 17(a) and the redshift-corrected results are shown in Figure 17(b). The relations can be presented by log$\omega$ = (0.44$^{+0.03}_{-0.03}$) + (0.81$^{+0.02}_{-0.03}$)log$\tau_{p}$,
and log$\omega$ = (0.34$^{+0.02}_{-0.02}$) + (0.83$^{+0.02}_{-0.03}$)log$\tau_{p}$, respectively. It is clearly shown that precursor, main burst, early X-ray flare, and late X-ray flare follow the same $\tau_p-\omega$ relation. We further examine the
$\tau_p-\omega/\tau_p$ relation. The redshift-uncorrected plot and the redshift-corrected plot are shown in Figure 17(c) and Figure 17(d), respectively. We can see that the relations of precursor, main burst, and early X-ray flare are clustered, while the relation of the late X-ray flare has a certain dispersion. We note that a larger late X-ray flare sample is required to further identify the dispersion property.

Overall, we do not find any significant difference between the precursor and the main burst when we perform the temporal analysis to the episode profiles.
It is suggested that the precursor and the main burst have a common physical origin. This conclusion is consistent to those given by Koshut et al. (1995) and Charisi et al. (2015).
Burlon et al. (2008) and Hu et al. (2014) performed the spectral analysis to the precursors and the main bursts. The PL model was adopted to fit the GRB spectra.
It was found, from the statistics, that the photon index of the precursor and that of the main burst are similar.
Hence, the precursor and the main burst are originated from the same physical mechanism.
It is naturally considered that the GRB emission begins at the precursor stage (Ramirez-Ruiz et al. 2002). In the GRB fireball model, the precursor is the remnant of the thermal emission trapped in the fireball (Li 2007). Thus, the precursor can be dominated by the thermal radiation. This is not consistent to the GRB spectral analysis given by Bulton (2008) and Hu et al. (2004).
Lazzati (2005) fitted the episode spectra by both PL model and BB model. It was found that most precursors are favorable to the nonthermal emission.
However, the precursors are softer than the main bursts by the the softness ratio comparison.
Thus, the precursor and the main burst may have the spectral evolution.

The GRB emission may have two stages. The collapsar may first turn into a kind of GRB progenitor with the precursor activity; then the progenitor may further collapse and release the prompt emission. The two-stage emission suggested by Lipunova et al. (2009)
can be adopted to explain the relatively long quiescent time of about $10^2$ s. This is consistent to the data analysis in this paper. However, the model predicted that the quiescent time can be longer than $10^3$ s, and we do not find such GRBs in our sample. In addition, the model given by Lipunova et al. (2009) suggested that the fraction of the long GRBs with the precursors is about 10\%. This fraction number is consistent to the fraction number in this paper.
A jet can be produced during a core collapse of a GRB progenitor. Sometimes, the jet has interactions with the surrounding cocoon (Lazzati \& Begelman 2005). The interactions may produce the precursor activity. The jet breakout and the evolution of the jet opening angle may take effects on the long quiescent time. The radiation from the cocoon is thermal. Thus, the jet-cocoon model cannot be directly used to explain the nonthermal precursor. However, the jet initially out of the cocoon may have the nonthermal emission, although the thermal component from the cocoon still exists. 
This can explain the nonthermal emission of the precursors measured by Lazatti (2005).
Wang \& M{\'e}sz{\'a}ros (2007) proposed a collapsar fallback model. A weak jet due to the collapse can be formed and produce the precursors, and the main burst is generated later by a strong jet powered by the fallback accretion. We can use this model to explain the precursor activity and the main-burst emission. One advantage of the model is to produce the quiescent time that is longer than 10 s. In our work, we find that the quiescent time has the mean value of 14.6 s. The quiescent time in our sample can be explained by this model. However, in our sample, we identify 10 GRBs, and each GRB has two precursors. We also find one GRB with three precursors.
We note that the models mentioned above cannot be directly applied to explain the physical origin of the multiple precursors in one GRB lightcurve. Lipunova et al (2009) suggested two possibilities. First, due to the fallback procedure, the matter might be destroyed as fragmentation pieces. Each piece produces one precursor event. Thus, we may observe several precursors. Second, the preliminary GRB jet during the shock breakout may have instabilities. The dynamical instabilities can produce multiple precursors. In the two cases, the quiescent times among the multiple precursors and the main-burst pulse
are long. The theoretical details are required to have a full explanation on the multiple precursors.

If a GRB jet is magnetically dominated, the GRB central engine could be a magnetar (Bernaidni et al. 2013). The energy of the pulse is released by the accretion. If the accretion is paused by the centrifugal force, the GRB enters a quiescent phase.
During the quiescent time, the energy is accumulated. When the accretion begins again, the energy of the pulse can be released again. Thus, by the magnetar model, multiple precursors and long quiescent times in one GRB can be produced.
Because the energy is accumulated during the quiescent time, the subsequent pulse is stronger if the quiescent time is longer.
However, we cannot find the relation between the energy release and the quiescent time in our sample,
and the relation was also not obtained by Lan et al. (2018), Charisi et al. (2015), and Coppin et al. (2020), respectively. In the propeller model, it is assumed that both the central engine and the accretion flow have 
similar physical parameters to explain all the GRBs. However, the physical conditions of different GRBs may have large diversities. Thus, the unified correlation between the energy accumulation and the quiescent time in a GRB sample
may not be found.

We select long GRBs in our sample. Short GRBs are excluded in this paper. However, we think that the short GRBs with the precursor activity can also provide some hints on the physical origin of the GRB precursor.
Troja et al. (2010) identified the precursor activity of the short GRBs in the Swift-BAT catalog.
There is no major difference between the precursor and the main burst.
Zhong et al. (2019) performed the spectral analysis to the short GRB precursors that were identified in the Fermi and the Swift samples.
It was found that the precursors are only slightly softer than the main bursts.
Li et al. (2020) obtained a short GRB precursor sample in the BATSE catalog. The angular spreading timescale and the cooling timescale were considered in the jet-cocoon scenario. Furthermore, Li et al. (2021) investigated the precursors, main bursts, and extended emissions of the short GRBs in the Swift catalog as a whole system. Here, we have a simple comparison between the short GRB precursor and the long GRB precursor. The methods to do the temporal and spectral analysis to the short and long GRB precursors are similar. The temporal profile of the long GRB precursor and that of the short GRB precursor can be well described by the FRED feature. The $\tau_p-\omega$ relation is common to both short GRB precursors and long GRB precursors. Therefore, we simply conclude that the precursor is one common signature that is shown in both short and long GRBs. It indicates that the physical origin of the short GRB precursor and that of the 
long GRB precursor are same, although short GRB and long GRB have different progenitors.
In the future, it will be interesting to further compare the precursor properties of long GRBs to those of short GRBs from a large sample.

\section{CONCLUSIONS}

We summarize the main findings obtained in our GRB-precursor sample below.
(1) It is examined by the K-S test that the temporal profile of the main burst and that of the precursor have no major difference.
(2) In most cases, the quiescent time in the rest frame
 between the precursor and the main burst has a range of 10$-$100 s. The mean value is 14.6 s with the deviation of 23.9 s. A few GRBs with the precursor activity have the quiescent time longer than 100 s.
(3) There is a correlation between the photon count of the precursor and the photon count of the main burst. It is presented by $\rm{log} C_p\sim 0.66 \rm{log} C_m$ with the extrinsic scatter of 0.41.
(4) There is no correlation between the energy release of the precursor estimated by the total number of photon counts and the quiescent time.

We find that the distribution of the precurosr FRED parameters and that of the main-burst FRED parameters are similar. Both the precursor and the main burst follow the same $\tau_p-\omega$ relation. 
It is suggested that the precursor and the
main burst may have the same physical origin. However, we do not find any correlation in the FRED-fitting parameters between the precursor and the main burst.
If the physical conditions during the dissipation process are different from one case to the other, the lack of the systematical correlations among the episodes' physical quantities between precursors and main event can be reasonable. In addition, the lack of the correlations also indicates that some stochastic processes are strongly involved in the GRB central engine activity. The stochastic processes may make the former episode lose {\it memory} of the previous one.
 
We derived the quiescent time between the precursor and the main burst from the GRB lightcurve analysis. 
The collapsar fallback model and the jet-cocoon model can roughly explain the quiescent time between the precursor and the main burst. We are inclined to choose the collapsar fallback model or the jet-cocoon model to explain the GRB-precursor phenomena in this paper. If GRB jet is magnetized, the magnetar model is also a choice to explain the precursor features.

\acknowledgments
We are grateful to the referee for the careful review and very helpful suggestions.
J.M. is supported by the National Natural Science Foundation of China 11673062 and the Oversea Talent Program of Yunnan Province.

\clearpage


\clearpage
\begin{deluxetable}{cccccccccccccccc}
\setlength{\tabcolsep}{0.9mm}
\rotate
\tabletypesize{\scriptsize}
\tablecaption{The fitting parameters of the GRB Episodes. ``p" indicates the precursor and ``m" indicates the main pulse in the clolumn of ``p/m". The coloumn of ``Mark" indicates that the GRB precursor is also identified in other references. ``B08" indicates Burlon et al. (2008), ``H14" indicates Hu et al. (2014), ``L18" indicates Lan et al.(2018), and ``C20" indicates Coppin et al. (2020). \label{chartable}}
\tablewidth{0pt}
\tabletypesize{\tiny}

\tablehead{
\colhead{GRB}                  &  \colhead{Type}        &  \colhead{$z$}         &  \colhead{$T_{90}$}        &  \colhead{p/m}
&  \colhead{$A$}               &  \colhead{$t_{s}$}     &  \colhead{$\tau_{1}$}  &  \colhead{$\tau_{2}$}      &  \colhead{$\tau_{p}$}
&  \colhead{$\omega$}          &  \colhead{$\kappa$}    &  \colhead{$\tau_{r}$}  &  \colhead{$\tau_{d}$}      &  \colhead{Mark}
&  \colhead{}
\\
\colhead{}                     &  \colhead{}            &  \colhead{}            &  \colhead{(s)}             &  \colhead{}
&  \colhead{($count~s^{-1}~det^{-1}$)}  &  \colhead{(s)}         &  \colhead{(s)}         &  \colhead{(s)}             &  \colhead{(s)}
&  \colhead{(s)}               &  \colhead{}            &  \colhead{(s)}         &  \colhead{(s)}             &  \colhead{}
&  \colhead{}
}
\startdata
050315  & XRR   & 1.949                           & 95.40  & p  & 0.04 $\pm$ 0.01 & -60.43  $\pm$ 2.04  & 23.36  $\pm$ 4.00  & 3.22  $\pm$ 2.02 & 8.68  $\pm$ 2.81  & 11.06 $\pm$ 7.11  & 0.29 $\pm$ 0.04 & 3.92  $\pm$ 1.67 & 7.14  $\pm$ 3.68  &B08  \\
        &       &                                 &        &    & 0.03 $\pm$ 0.01 & -42.96  $\pm$ 2.17  & 20.24  $\pm$ 2.97  & 3.05  $\pm$ 2.02 & 7.86  $\pm$ 2.67  & 10.26 $\pm$ 6.98  & 0.30 $\pm$ 0.05 & 3.60  $\pm$ 1.62 & 6.66  $\pm$ 3.64  &H14  \\
        &       &                                 &        & m  & 0.14 $\pm$ 0.00 & -10.16  $\pm$ 0.55  & 10.37  $\pm$ 1.54  & 17.41 $\pm$ 0.68 & 13.44 $\pm$ 1.03  & 35.20 $\pm$ 1.72  & 0.49 $\pm$ 0.01 & 8.89  $\pm$ 0.54 & 26.31 $\pm$ 1.03  \\
        &       &                                 &        &    & 0.09 $\pm$ 0.01 & 19.05   $\pm$ 0.39  & 29.32  $\pm$ 2.00  & 1.06  $\pm$ 0.13 & 5.57  $\pm$ 0.40  & 4.97  $\pm$ 0.66  & 0.21 $\pm$ 0.01 & 1.96  $\pm$ 0.18 & 3.01  $\pm$ 0.31  \\
\hline
050318  & XRR   & 1.44                            & 16.12  & p  & 0.23 $\pm$ 0.02 & -1.80   $\pm$ 0.34  & 12.88  $\pm$ 5.95  & 0.47  $\pm$ 0.12 & 2.45  $\pm$ 0.65  & 2.18  $\pm$ 0.63  & 0.21 $\pm$ 0.03 & 0.86  $\pm$ 0.20 & 1.32  $\pm$ 0.30  &B08  \\
        &       &                                 &        & m  & 0.28 $\pm$ 0.01 & 19.52   $\pm$ 0.16  & 49.99  $\pm$ 0.61  & 0.87  $\pm$ 0.05 & 6.58  $\pm$ 0.20  & 4.85  $\pm$ 0.29  & 0.18 $\pm$ 0.00 & 1.99  $\pm$ 0.08 & 2.86  $\pm$ 0.13  \\
        &       &                                 &        &    & 0.26 $\pm$ 0.02 & 28.51   $\pm$ 0.04  & 0.02   $\pm$ 0.02  & 3.97  $\pm$ 0.40 & 0.29  $\pm$ 0.15  & 4.52  $\pm$ 0.53  & 0.88 $\pm$ 0.05 & 0.27  $\pm$ 0.13 & 4.25  $\pm$ 0.44  \\
\hline
050319  & XRR   & 3.24                            & 151.58 & p1 & 0.14 $\pm$ 0.02 & -134.24 $\pm$ 0.11  & 0.34   $\pm$ 0.31  & 0.89  $\pm$ 0.17 & 0.55  $\pm$ 0.25  & 1.66  $\pm$ 0.42  & 0.54 $\pm$ 0.09 & 0.39  $\pm$ 0.14 & 1.28  $\pm$ 0.25  &H14  \\
\enddata
\end{deluxetable}

\begin{deluxetable}{ccccc}
\tabletypesize{\footnotesize}
\tablecaption{The K-S test for the Distributions of the Episode Parameters. ``D" is defined to be the maximum value of the absolute difference between two cumulative distributions. \label{tab:sample}}
\tablewidth{0pt}
\tablehead{ \multirow{2}{*}{Parameters} & \multicolumn{2}{c}{D} & \multicolumn{2}{c}{p-value} \\\cmidrule(lr){2-3}\cmidrule(lr){4-5}
& without redshift correction & redshift correction  & without redshift correction & redshift correction }
\startdata
 $A$             &0.33    &$-$        &$<10^{-4}$     &$-$        \\
 $t_s$       &0.62    &0.33       &$<10^{-4}$     &$<10^{-4}$ \\
 ${\tau_{1}}$    &0.12    &0.15       &0.39           &0.15       \\
 $\tau_{2}$      &0.13    &0.15       &0.27           &0.12       \\
 $\tau_{p}$      &0.11    &0.11       &0.47           &0.44       \\
 $\omega$        &0.11    &0.13       &0.51           &0.28       \\
 $\kappa$        &0.17    &$-$        &0.07           &$-$        \\
 $\tau_{r}$      &0.11    &0.09       &0.51           &0.71       \\
 $\tau_{d}$      &0.12    &0.13       &0.35           &0.26       \\
\enddata
\end{deluxetable}
\clearpage

\begin{deluxetable}{cccccccccccc}
\tablecaption{The Quiescent Time and the Peak Time}
\tablewidth{0pt}
\tabletypesize{\footnotesize}

\tablehead{
\colhead{GRB}                  &  \colhead{Type}        &  \colhead{$z$}         &  \colhead{$T_{90}$}        &  \colhead{$\Delta t_{q,1}$} &  \colhead{$\Delta t_{q,2}$}
&  \colhead{$\Delta t_{q,3}$} &  \colhead{$\Delta t_{p,1}$}               &  \colhead{$\Delta t_{p,2}$}     &  \colhead{$\Delta t_{p,3}$}
\\
\colhead{}         &  \colhead{}     &  \colhead{}      &  \colhead{(s)}     &  \colhead{(s)}   &  \colhead{(s)}    &  \colhead{(s)}
&  \colhead{(s)}  &  \colhead{(s)}         &  \colhead{(s)}
}
\startdata
050315  & XRR   & 1.949  & 95.40  & 22.83  & $-$    & $-$   & 55.03  & $-$    & $-$   \\
050318  & XRR   & 1.44   & 16.12  & 22.12  & $-$    & $-$   & 28.15  & $-$    & $-$   \\
050319  & XRR   & 3.24   & 151.58 & 10.48  & 95.65  & $-$   & 31.37  & 103.10 & $-$   \\
050401  & C-GRB & 2.9    & 32.09  & 15.46  & $-$    & $-$   & 24.83  & $-$    & $-$   \\
050502B & XRR   & 5.2    & 17.72  & 4.93   & $-$    & $-$   & 9.49   & $-$    & $-$   \\
050726  & C-GRB & 3.5    & 45.50  & 168.71 & $-$    & $-$   & 178.86 & $-$    & $-$   \\
050820A & C-GRB & 2.612  & 240.77 & 204.96 & $-$    & $-$   & 231.06 & $-$    & $-$   \\
050822  & XRR   & 1.434  & 104.29 & 24.12  & $-$    & $-$   & 45.58  & $-$    & $-$   \\
\enddata
\end{deluxetable}
\clearpage

\begin{figure}[!htp]
\centering
\includegraphics[height=8cm,width=16cm]{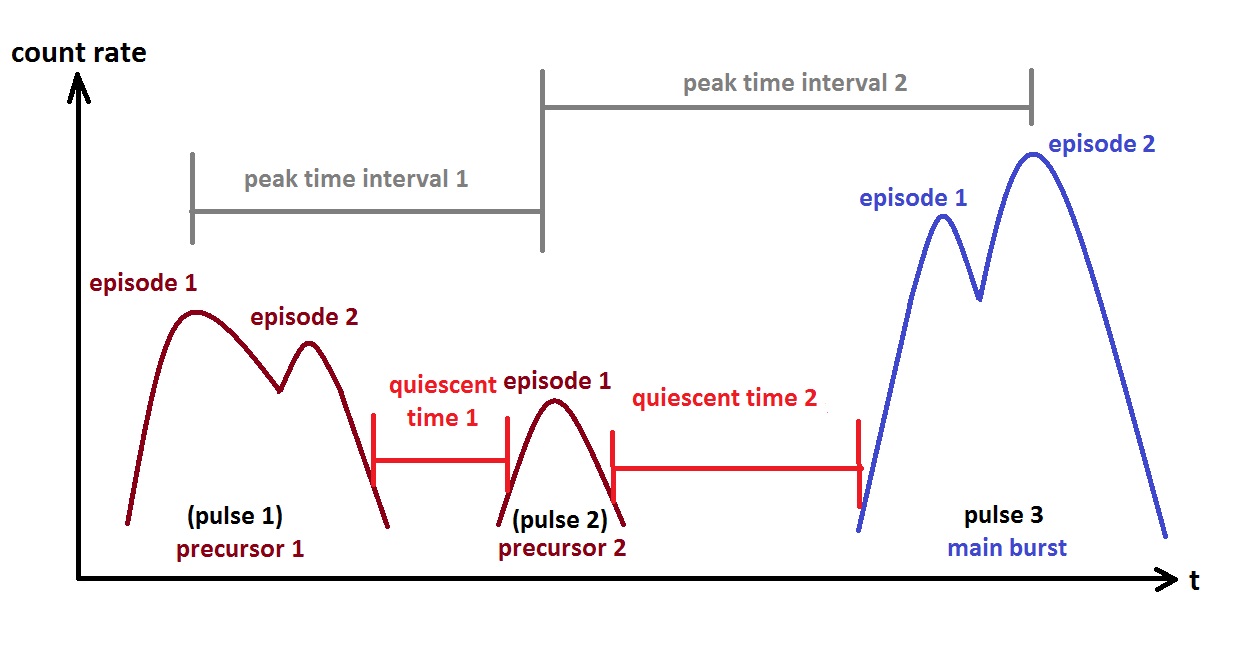}
\caption{We illustrate a GRB lightcurve. Three pulses are shown. The first pulse and the second pulse are the precursors, and the third pulse is the main burst.
The first precursor and the second precursor contain two episodes and one episode, respectively. The main burst contains two episodes. Each quiescent time (indicated by $\Delta t_{q,i}$ in the paper) and each peak-time interval (indicated by $\Delta t_{p,i}$ in the paper) are also labeled.}
\end{figure}
\clearpage

\begin{figure}[!htp]
\centering
\subfigure{
\includegraphics[height=6.5cm,width=7cm]{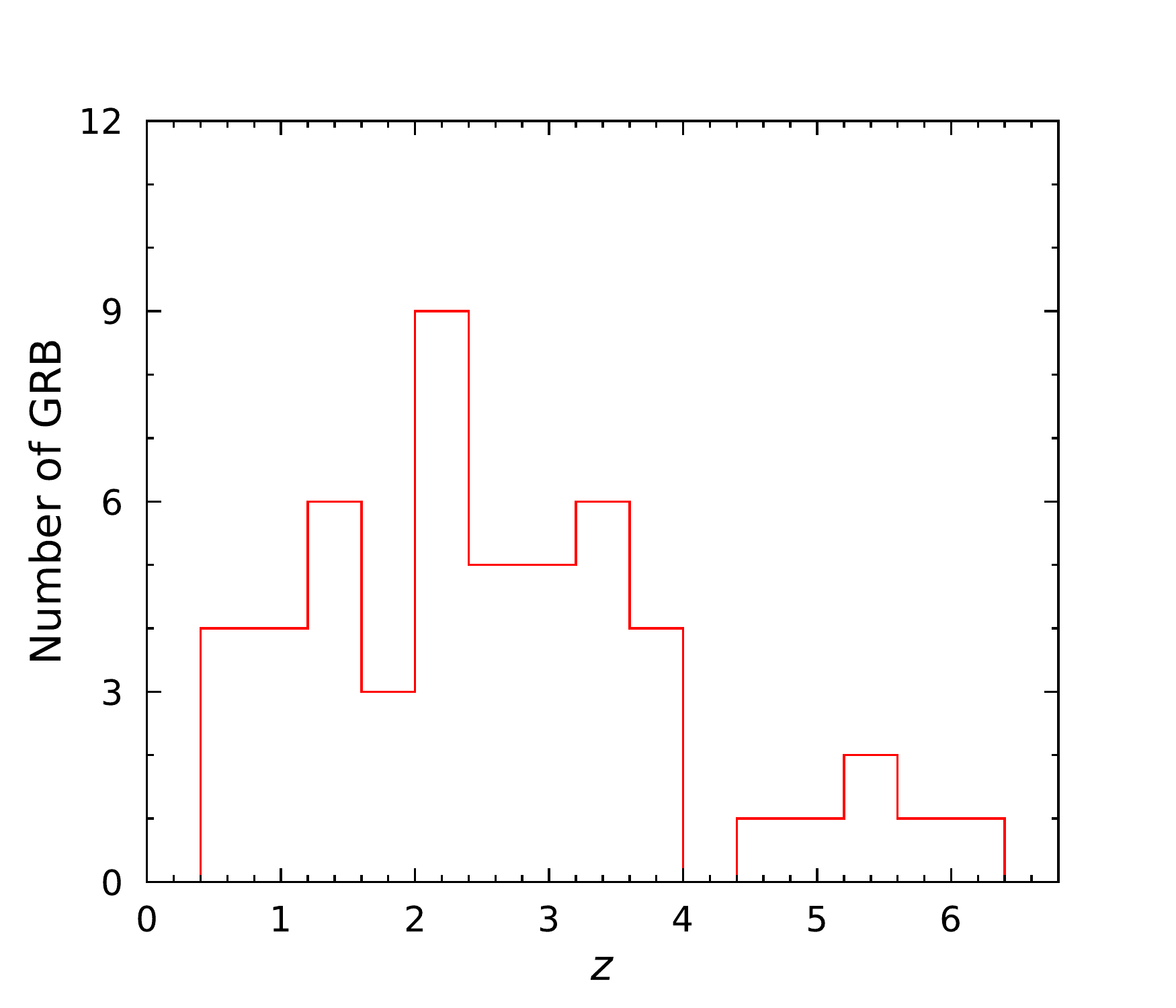}}
\subfigure{
\includegraphics[height=6.5cm,width=7cm]{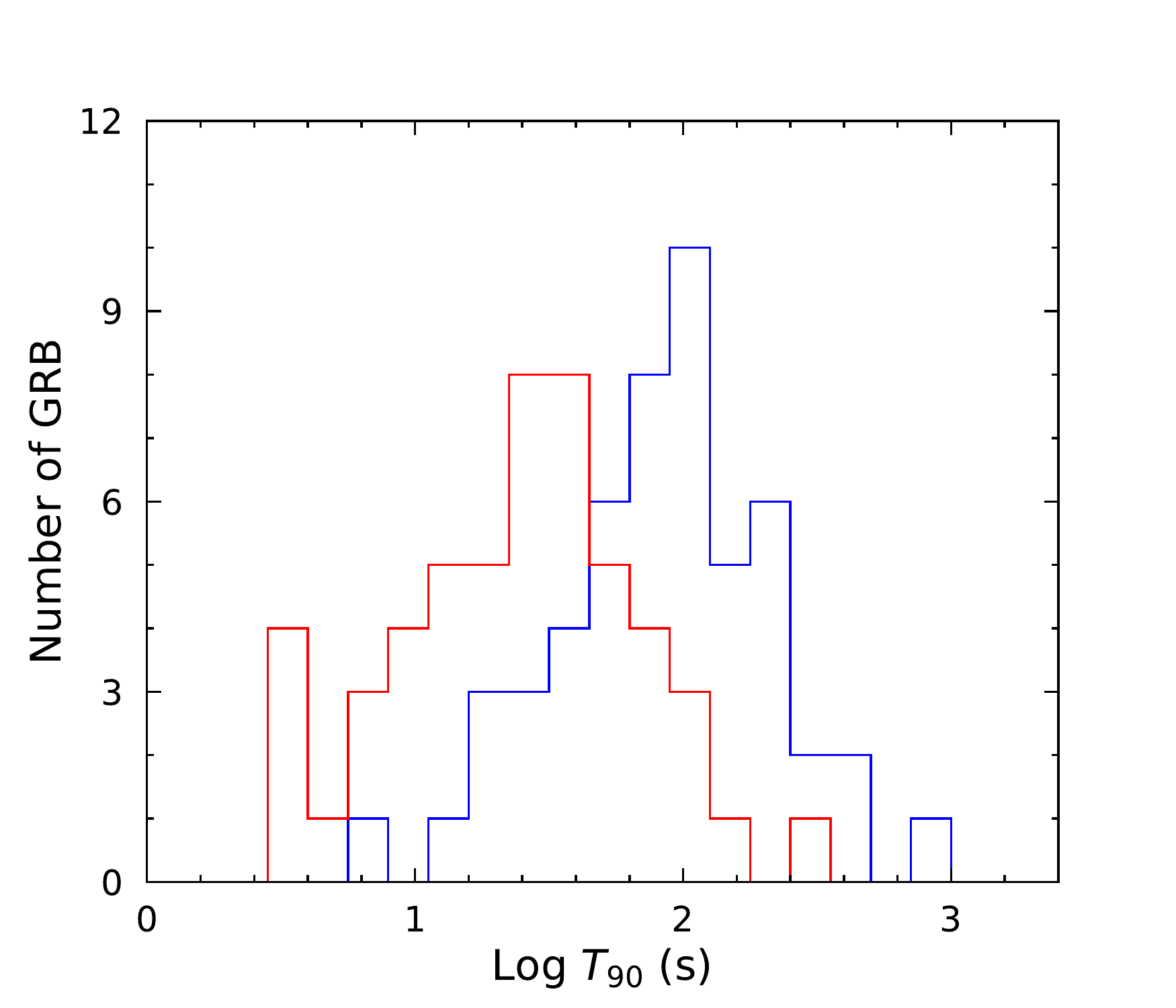}}
\caption{Left panel: the distribution of the GRB redshift in our sample. Right panel: the distribution of the GRB duration in our sample. The solid line with the blue color represents the distribution of $T_{90}$, and
the solid line with the red color represents the distribution of $T_{90}/(1+z)$.}
\end{figure}

\begin{figure}[!htp]
\centering
\subfigure[]{
\includegraphics[height=4.5cm,width=5cm]{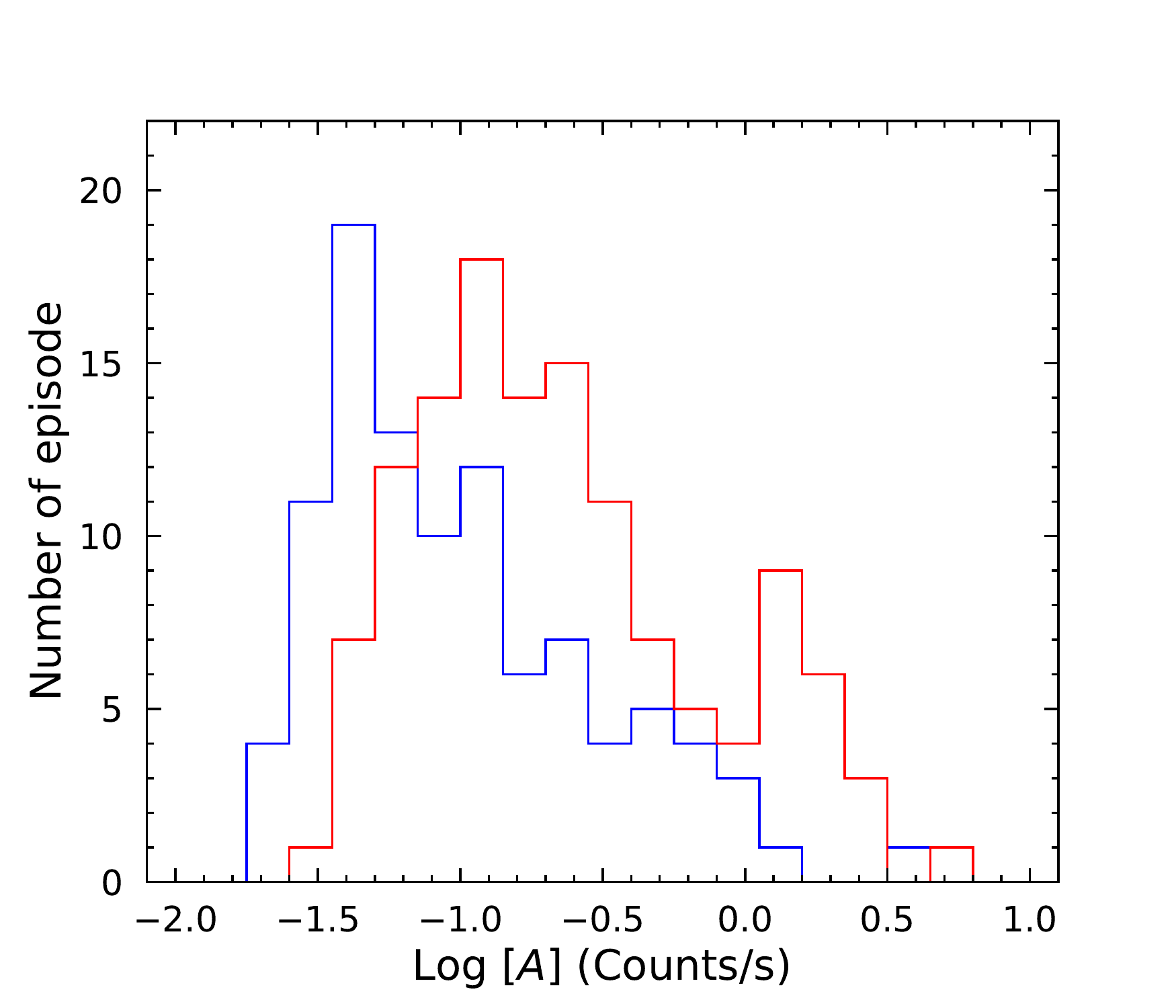}}
\subfigure[]{
\includegraphics[height=4.5cm,width=5cm]{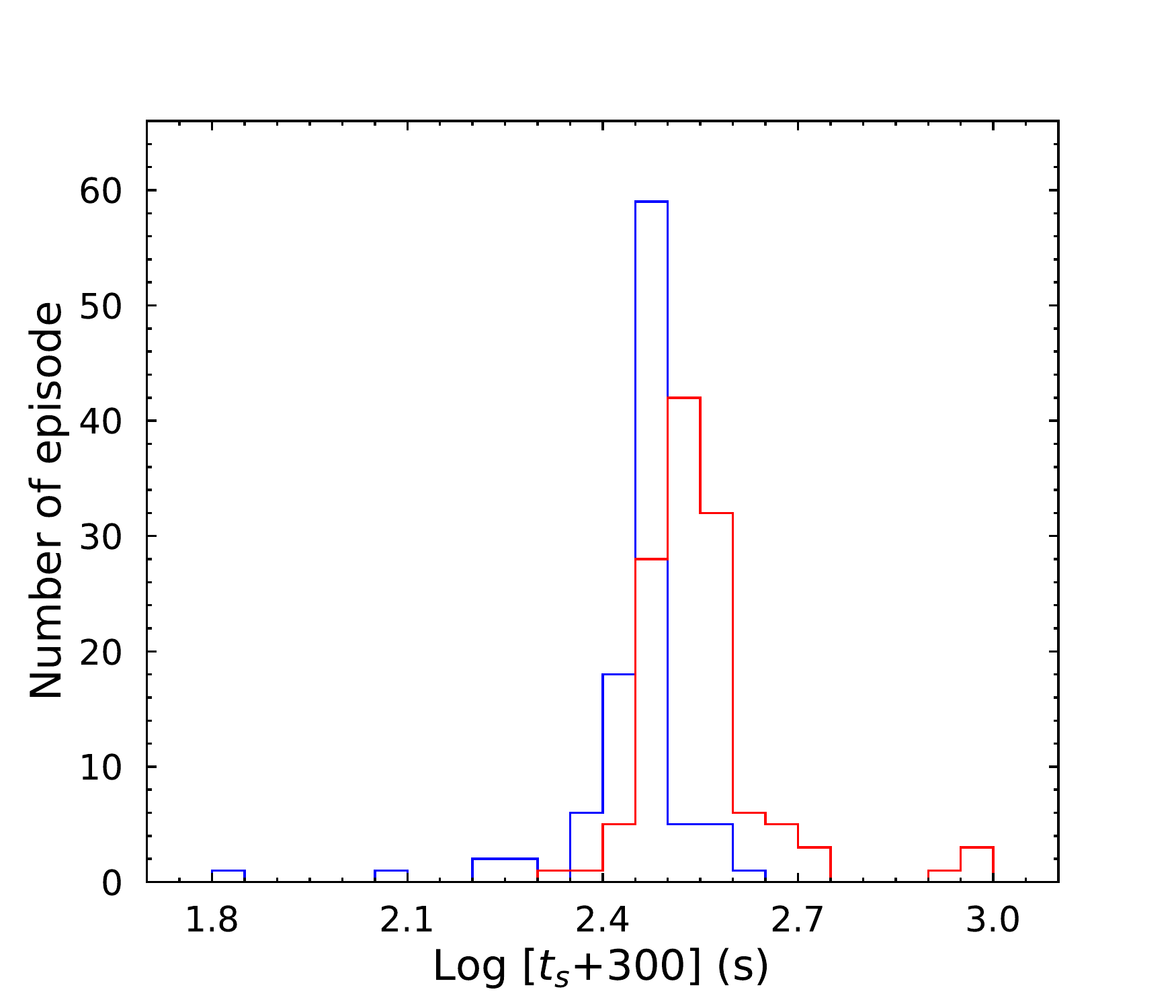}}
\subfigure[]{
\includegraphics[height=4.5cm,width=5cm]{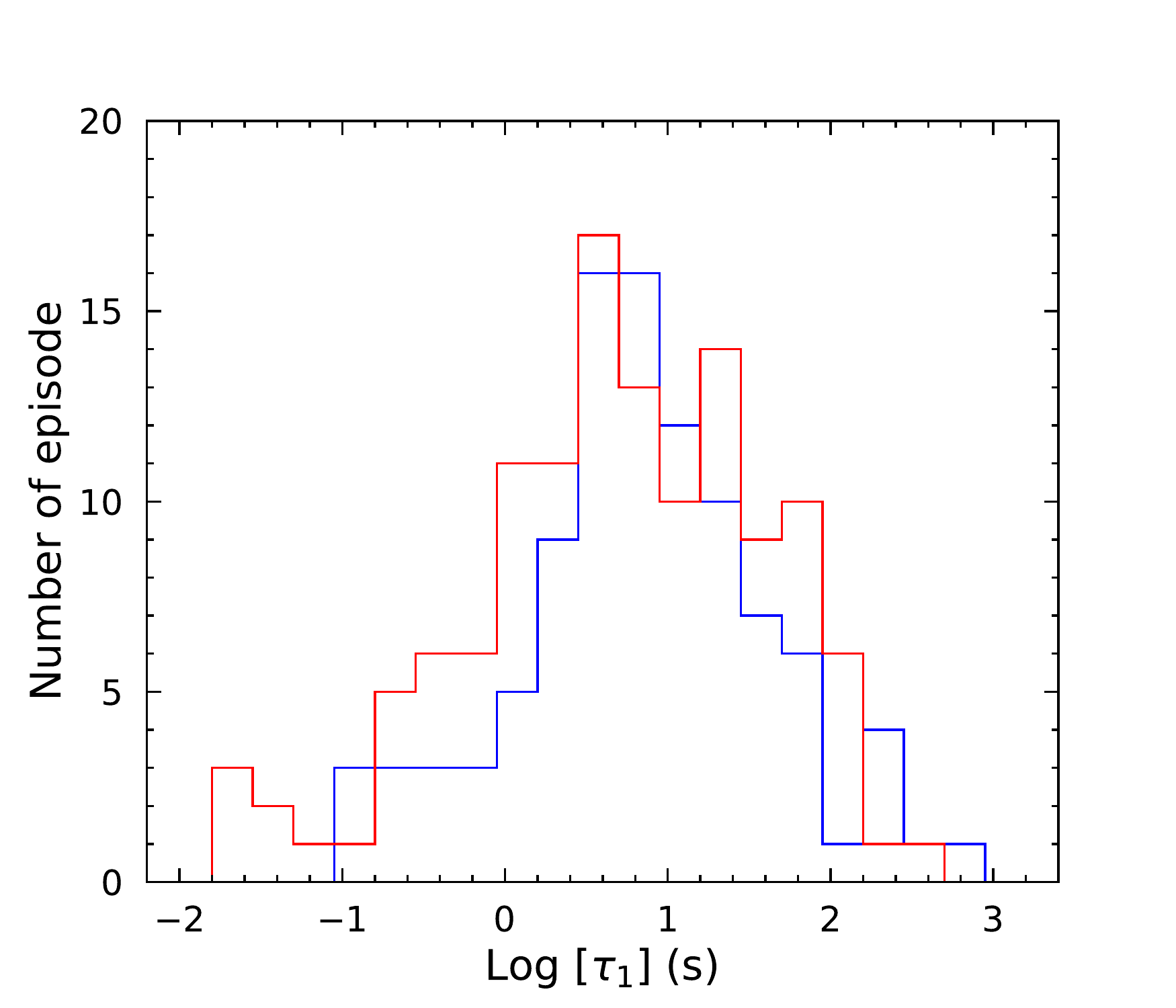}}
\subfigure[]{
\includegraphics[height=4.5cm,width=5cm]{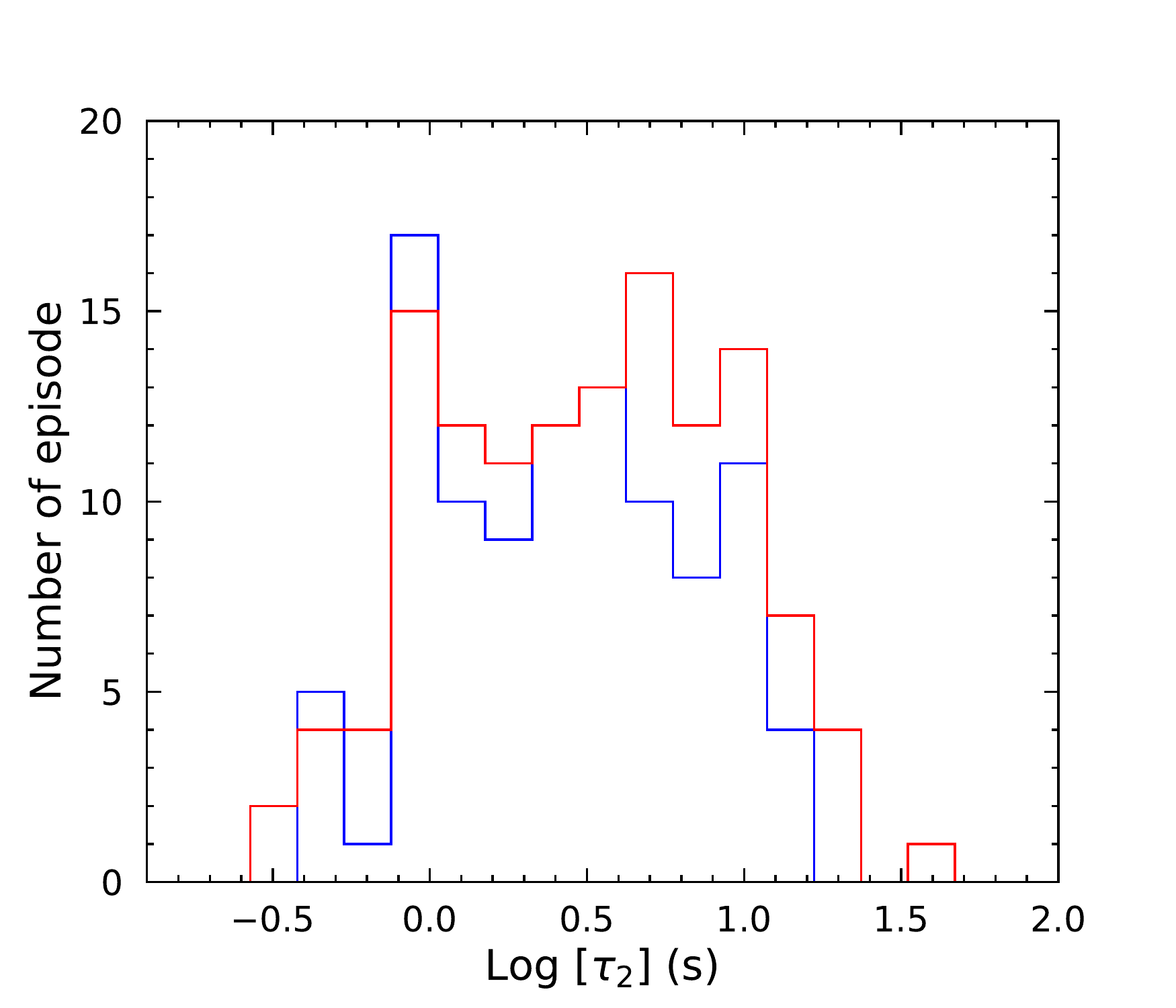}}
\subfigure[]{
\includegraphics[height=4.5cm,width=5cm]{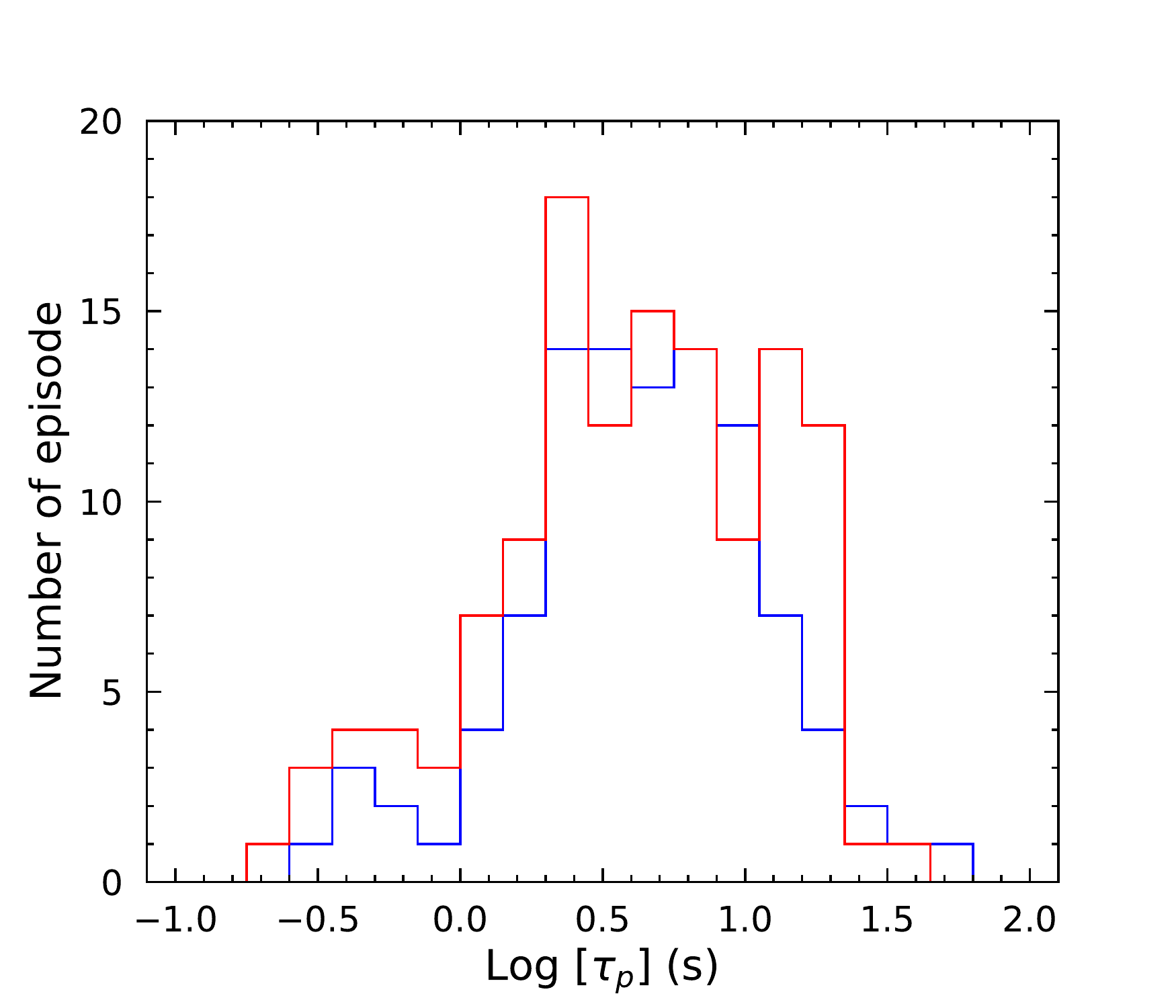}}
\subfigure[]{
\includegraphics[height=4.5cm,width=5cm]{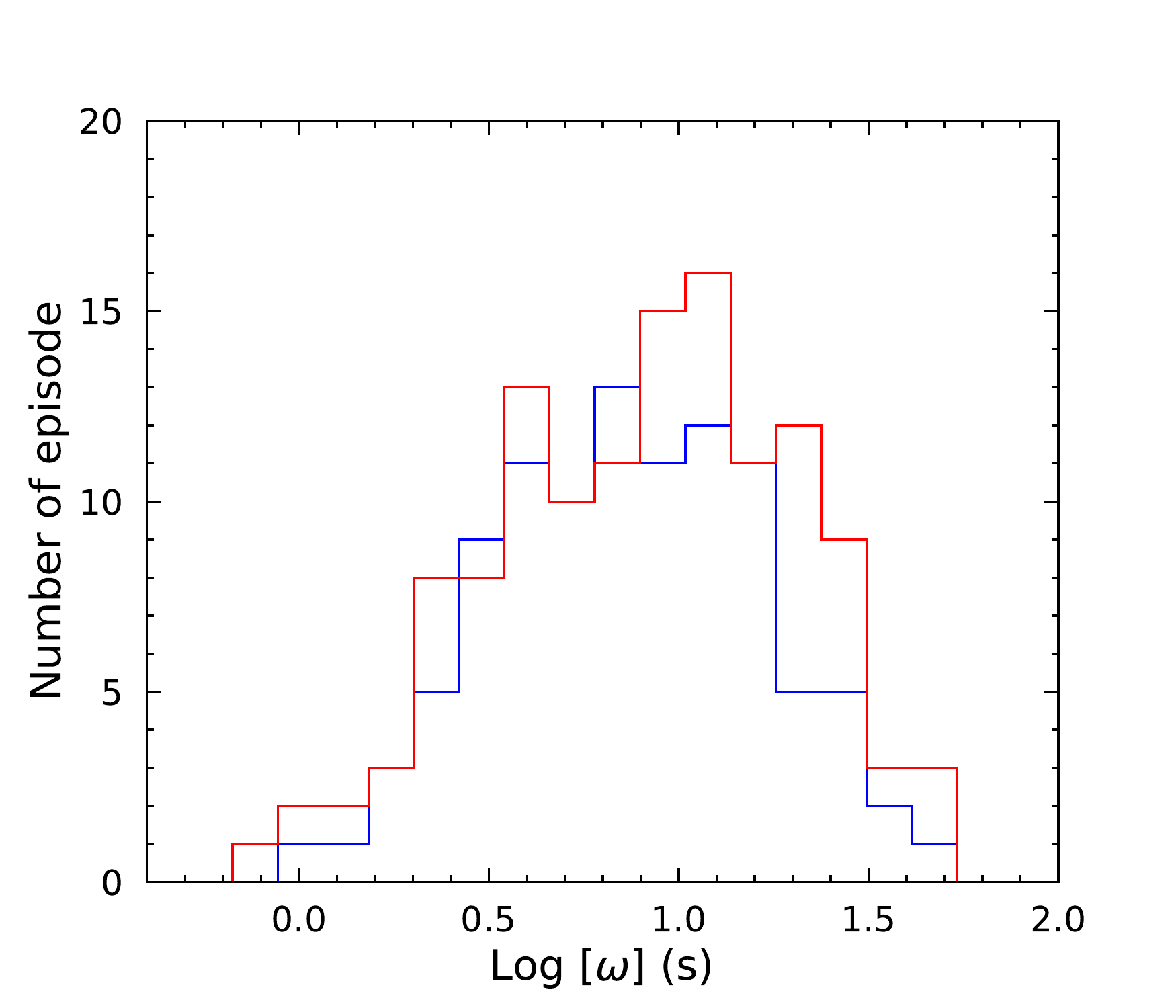}}
\subfigure[]{
\includegraphics[height=4.5cm,width=5cm]{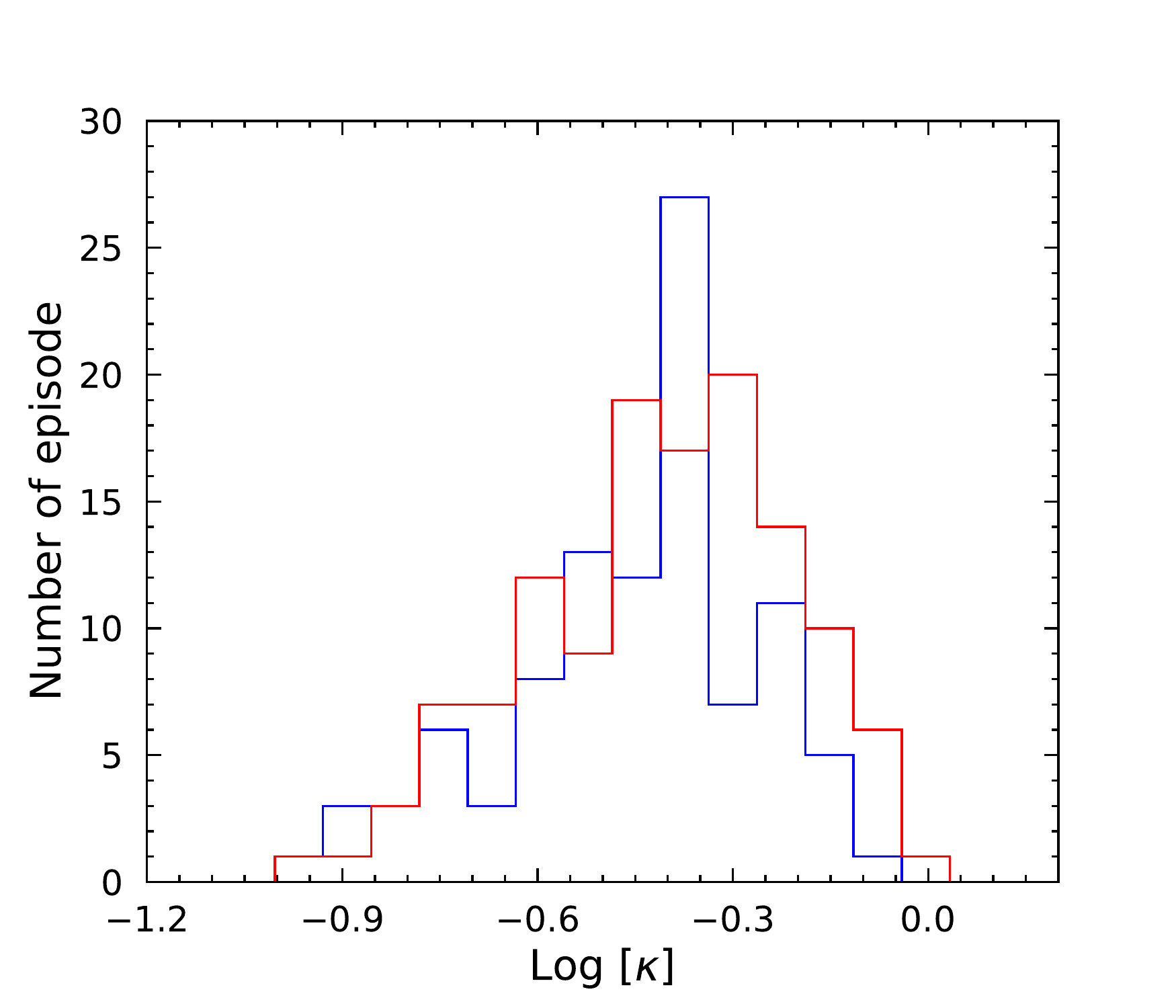}}
\subfigure[]{
\includegraphics[height=4.5cm,width=5cm]{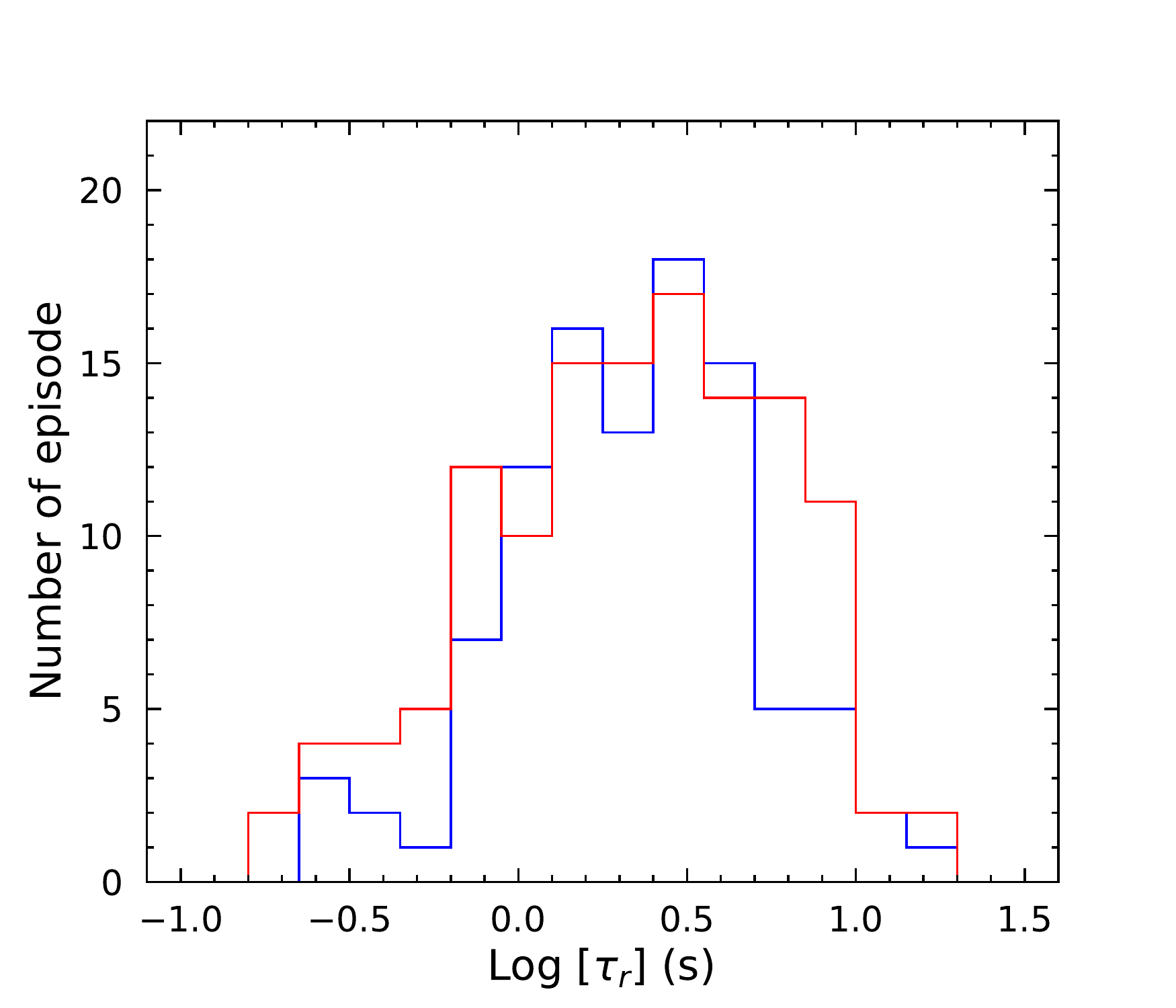}}
\subfigure[]{
\includegraphics[height=4.5cm,width=5cm]{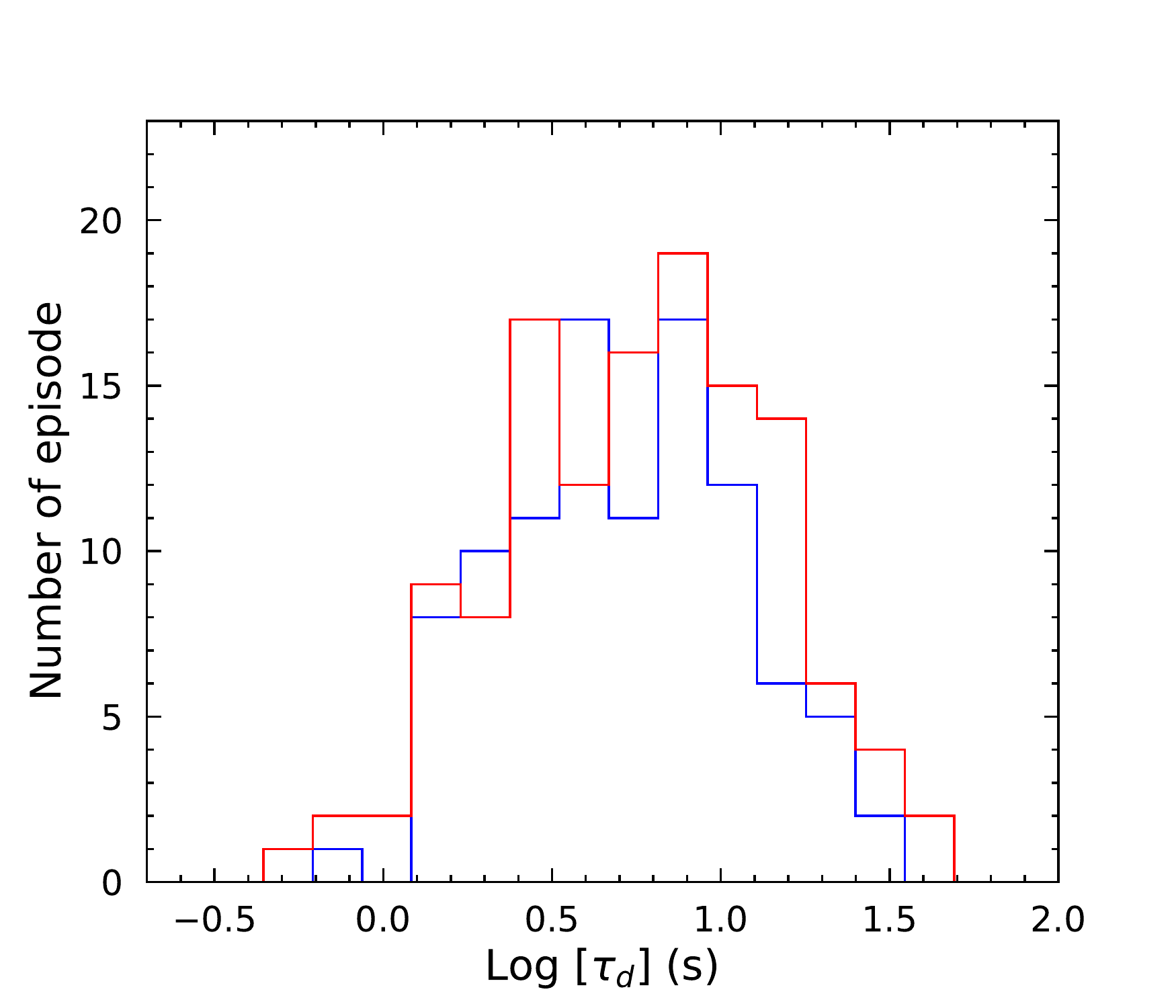}}
\caption{The distributions of the observed episode parameters.
The blue solid line represents the precursors, while the red solid line represents the main bursts.}
\end{figure}
\clearpage

\begin{figure}[!htp]
\centering
\subfigure[]{
\includegraphics[height=4.5cm,width=5cm]{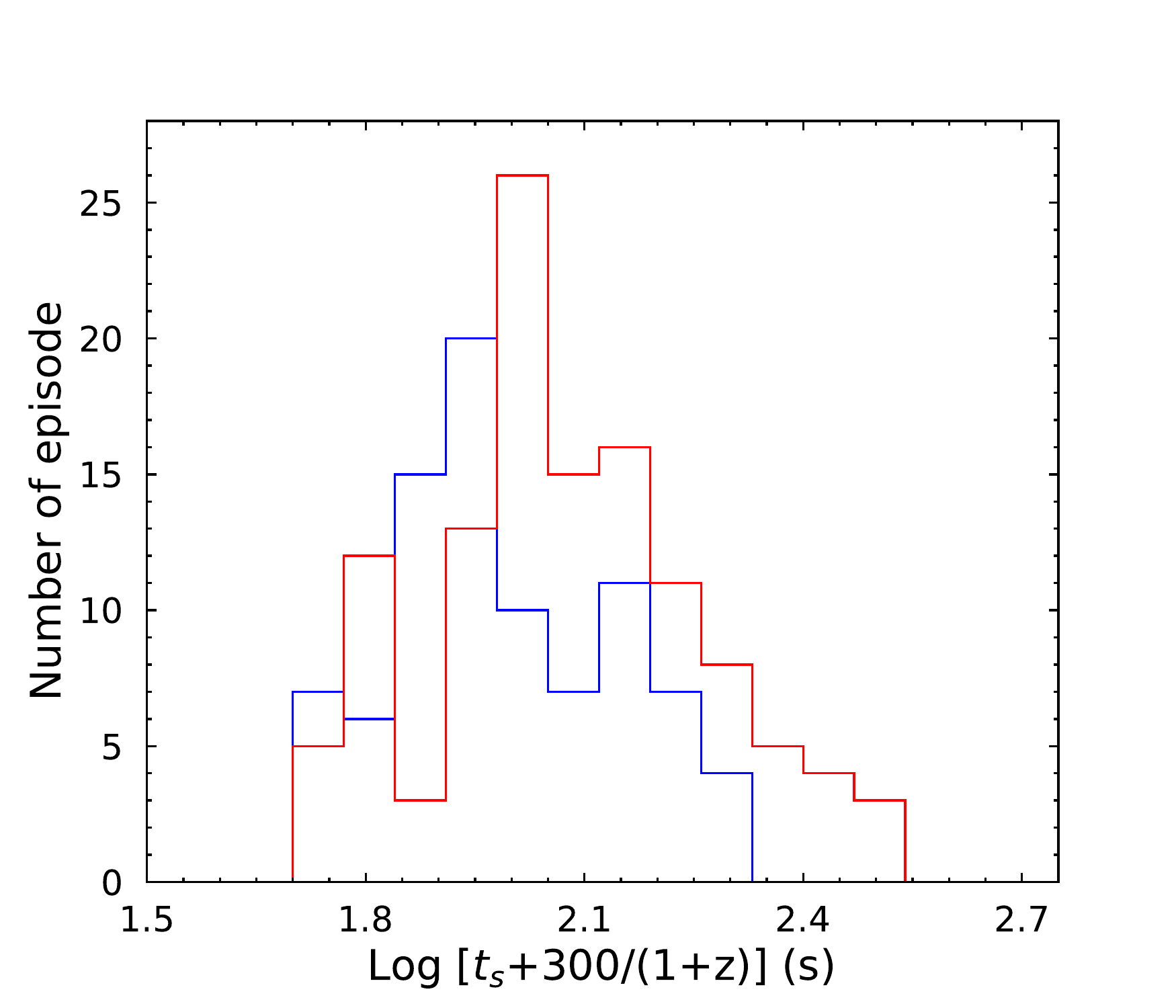}}
\subfigure[]{
\includegraphics[height=4.5cm,width=5cm]{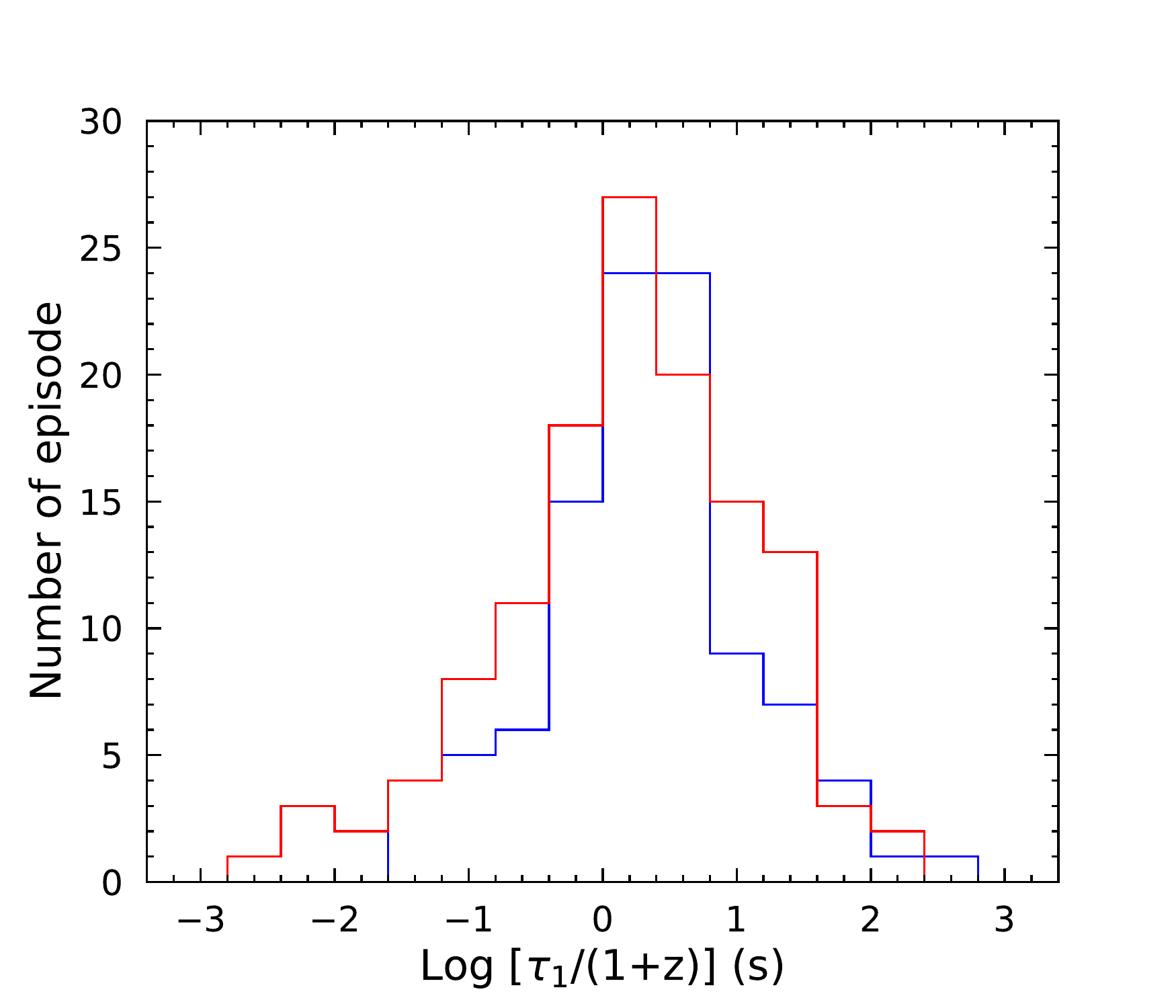}}
\subfigure[]{
\includegraphics[height=4.5cm,width=5cm]{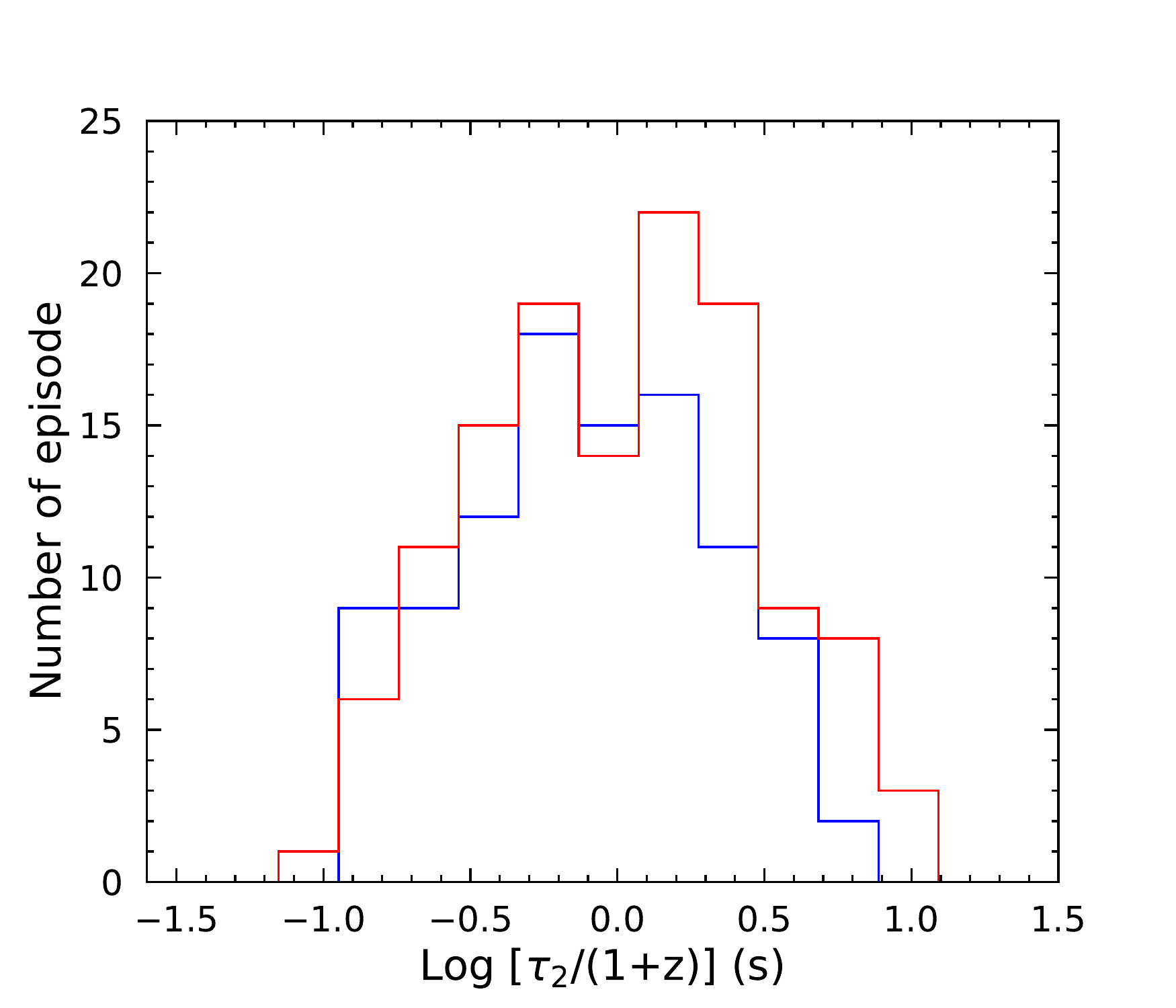}}
\subfigure[]{
\includegraphics[height=4.5cm,width=5cm]{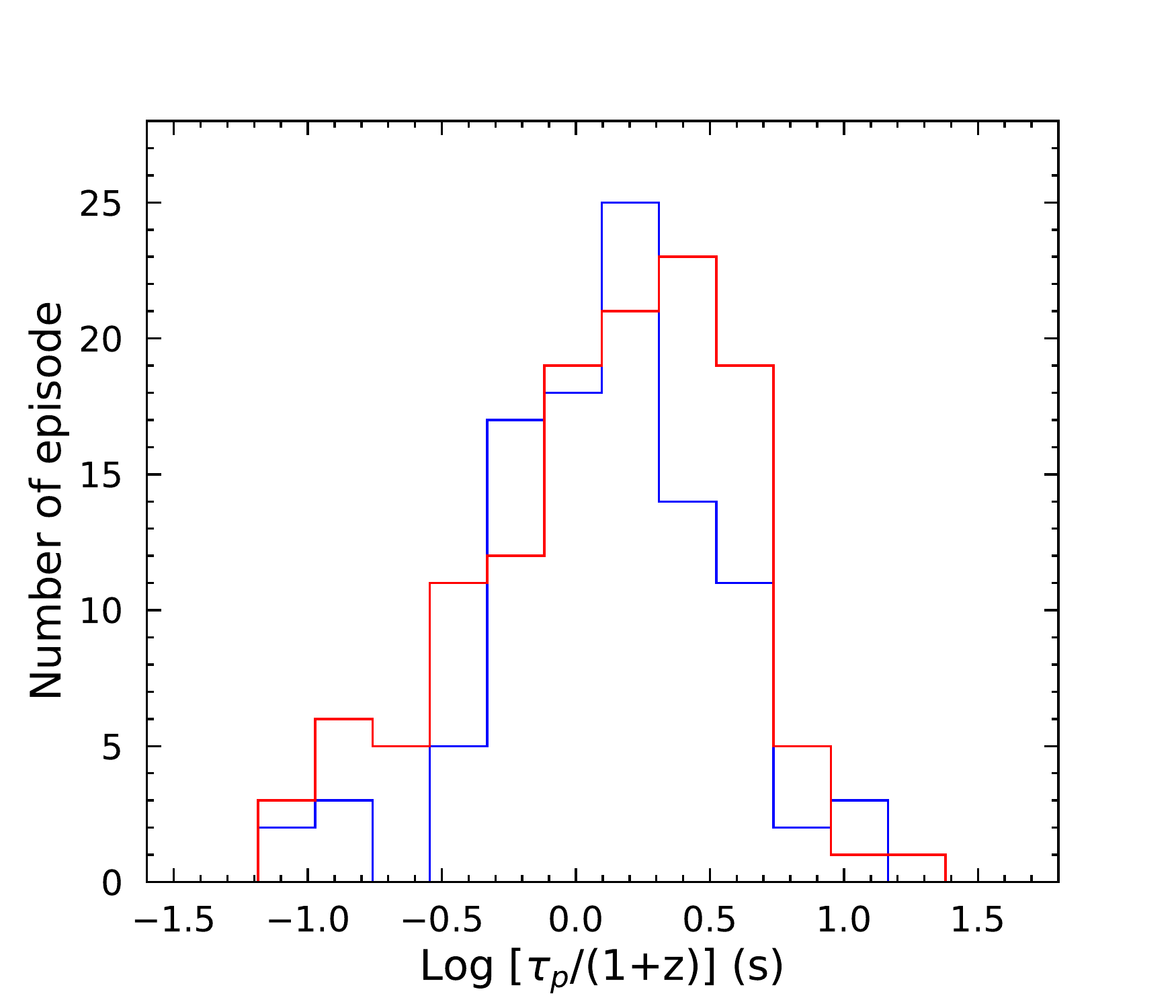}}
\subfigure[]{
\includegraphics[height=4.5cm,width=5cm]{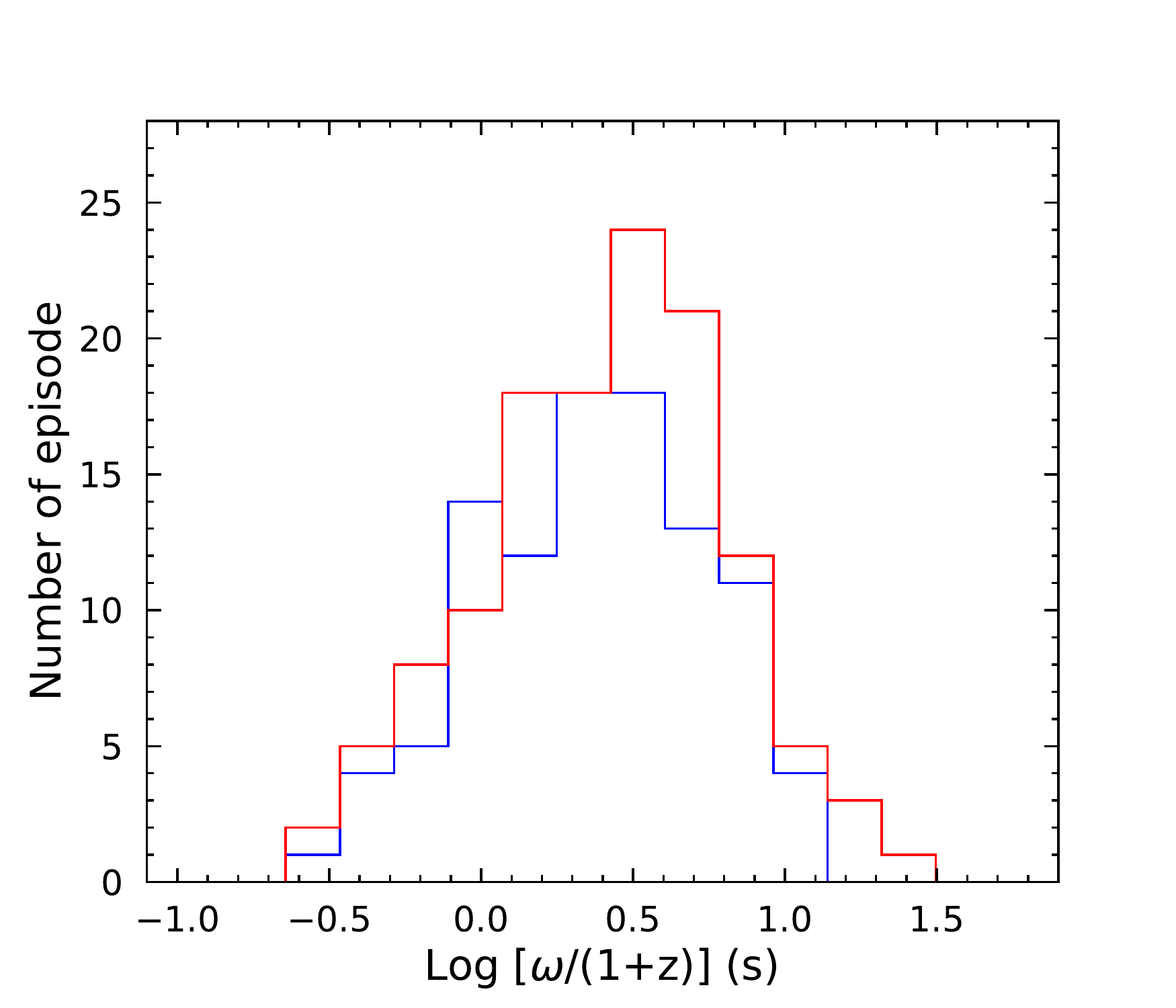}}
\subfigure[]{
\includegraphics[height=4.5cm,width=5cm]{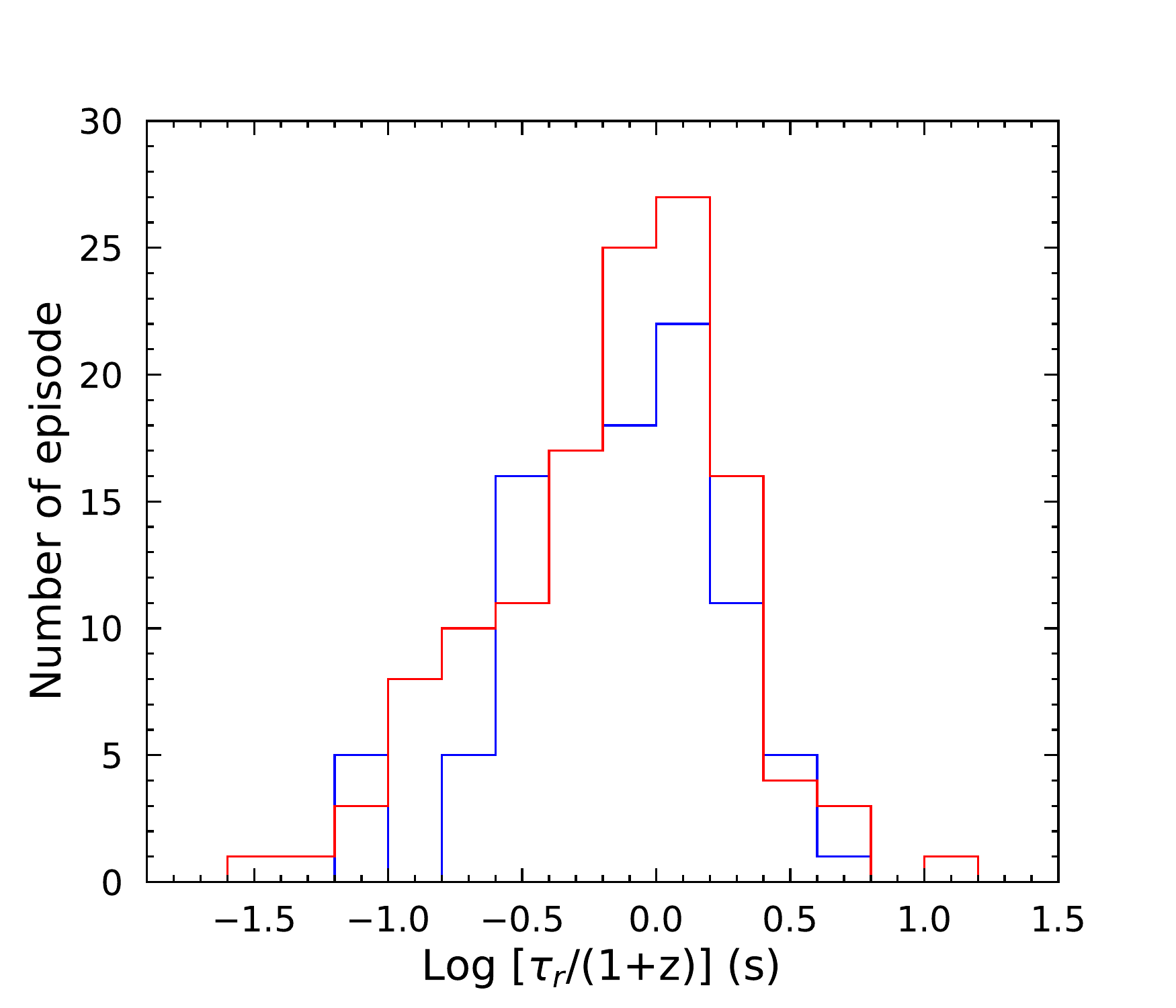}}
\subfigure[]{
\includegraphics[height=4.5cm,width=5cm]{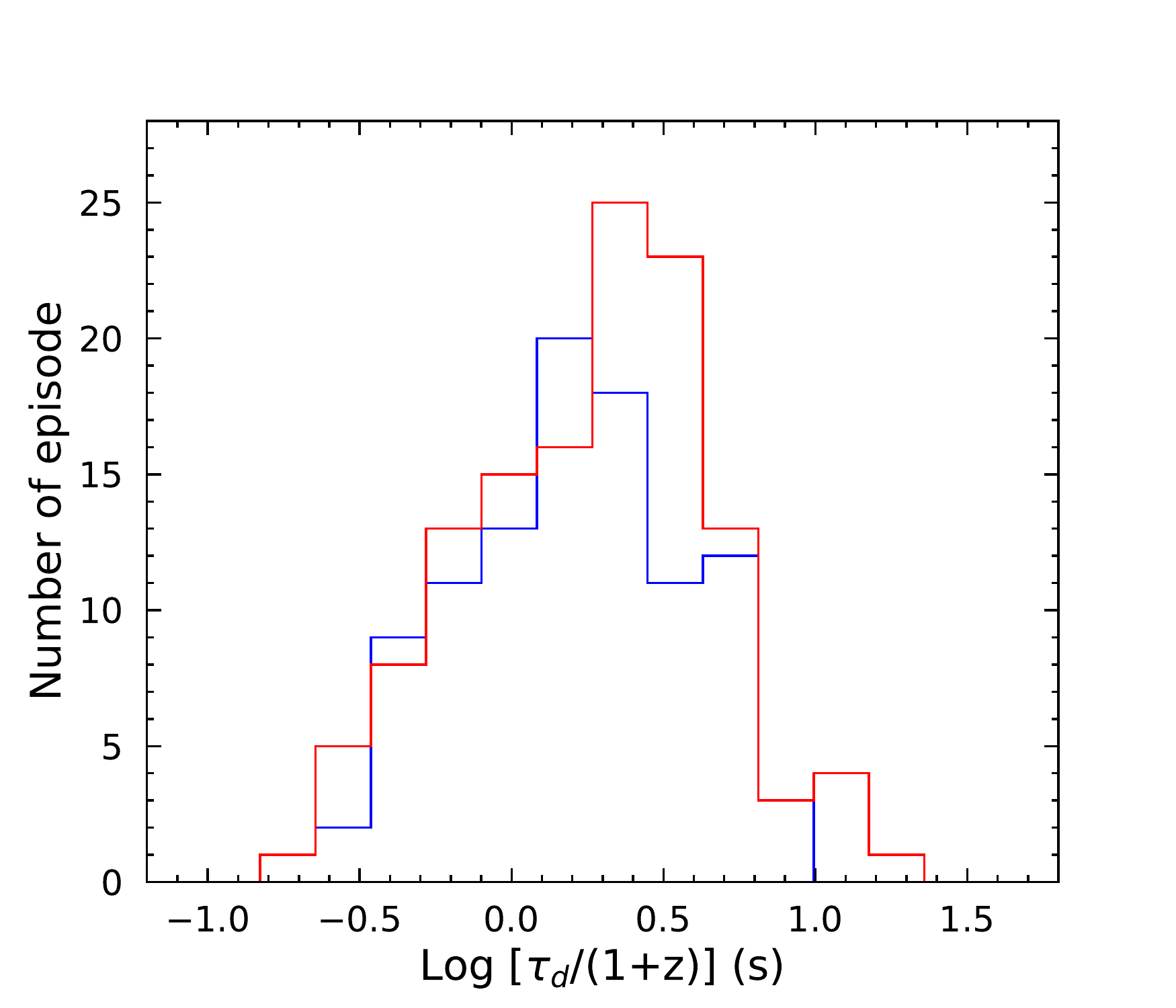}}
\caption{The distributions of the redshift-corrected episode parameters.
The blue solid line represents the precursors, while the red solid line represents the main bursts.}
\end{figure}
\clearpage

\begin{figure}[!htp]
\centering
\subfigure{
\includegraphics[height=6.5cm,width=7cm]{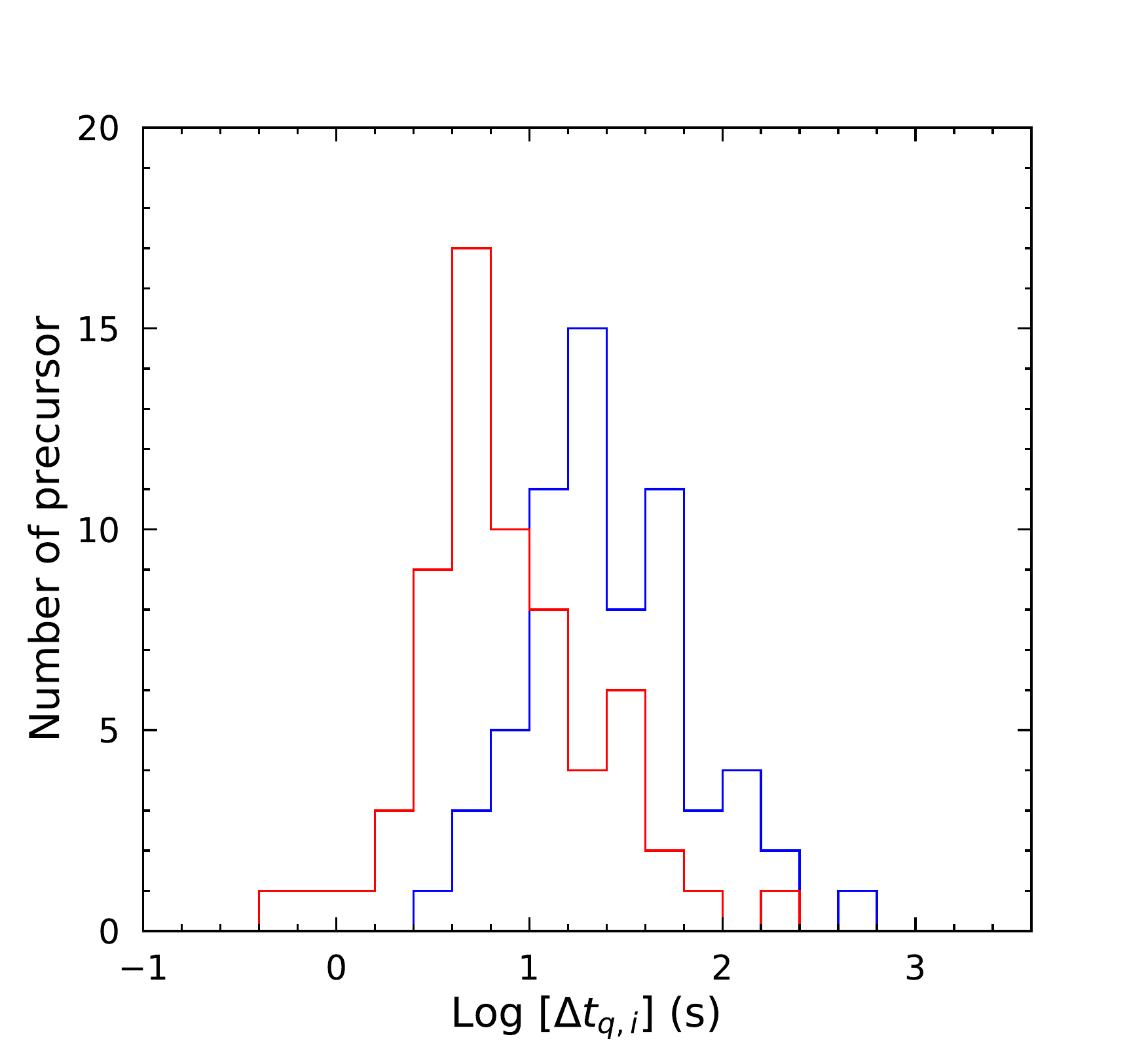}}
\subfigure{
\includegraphics[height=6.5cm,width=7cm]{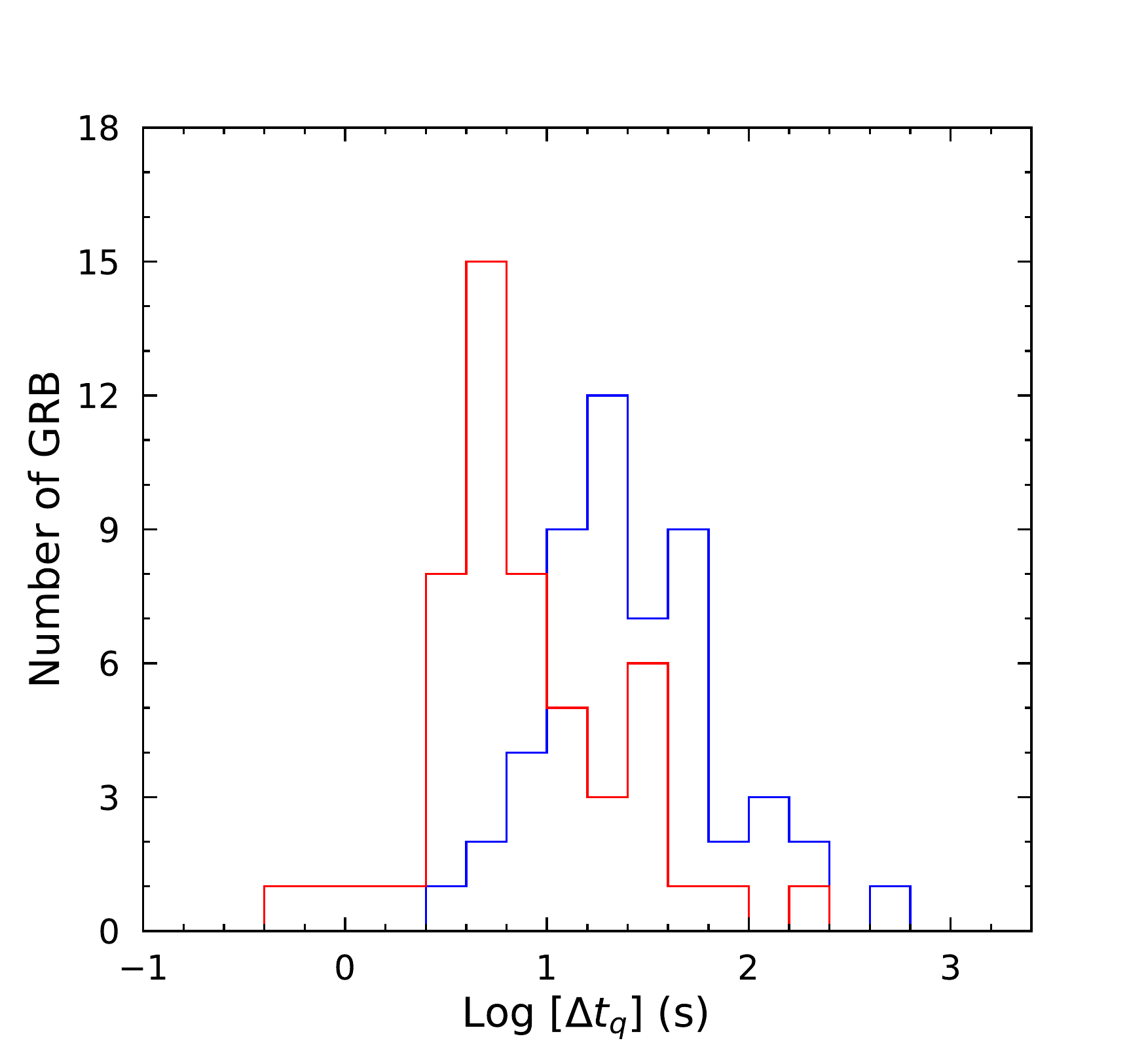}}
\caption{Left panel:
The distribution of $\Delta t_{q,i}$.
The blue solid line represents for the observed quiescent time,
and the red solid line represents for the quiescent time after the redshift correction. 
Right panel: The distribution of $\Delta t_q$.
The blue solid line represents for the observed quiescent time,
and the red solid line represents for the quiescent time after the redshift correction. 
}
\end{figure}
\clearpage

\begin{figure}[!htp]
\centering
\subfigure{
\includegraphics[height=6.5cm,width=7cm]{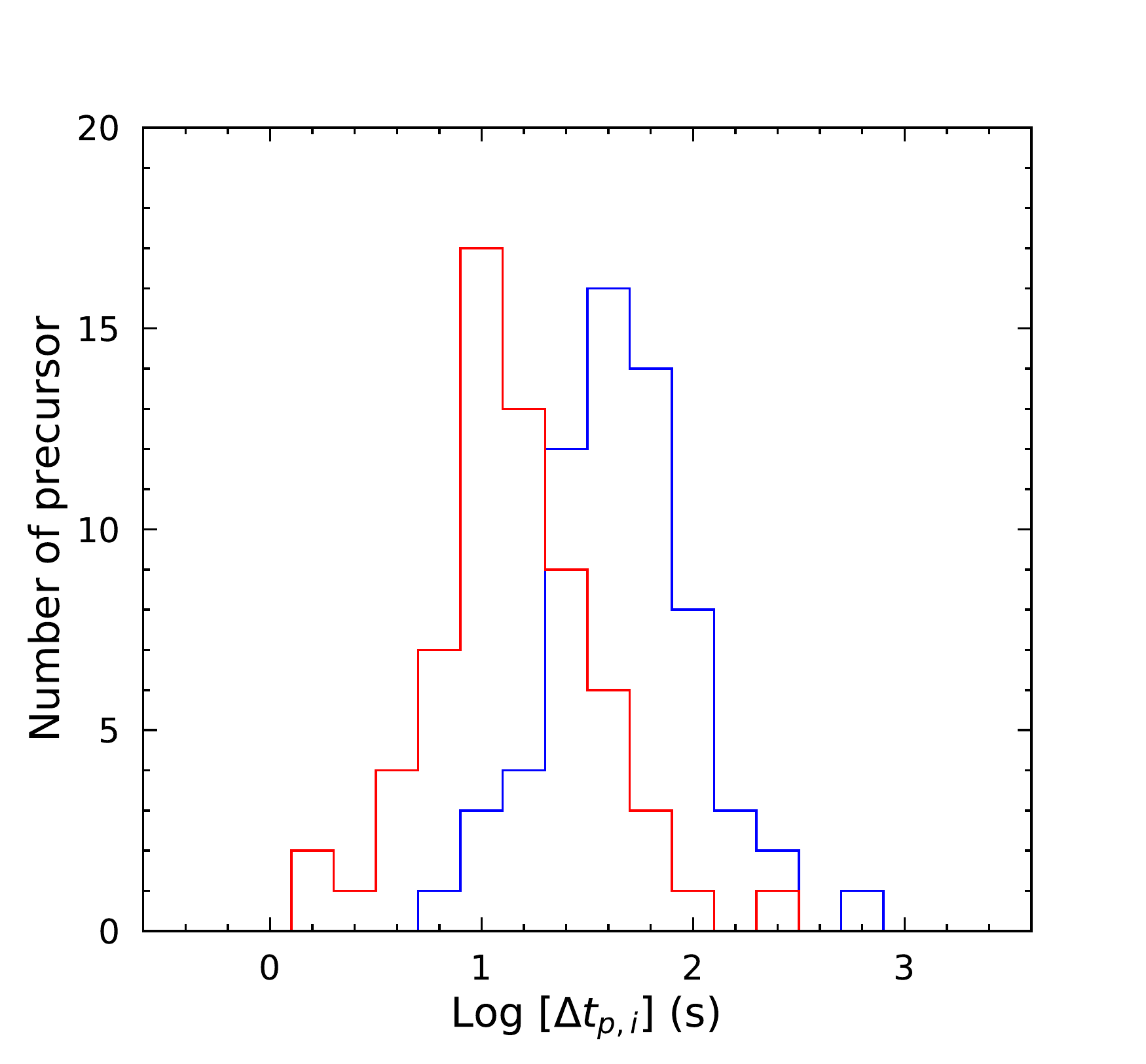}}
\subfigure{
\includegraphics[height=6.5cm,width=7cm]{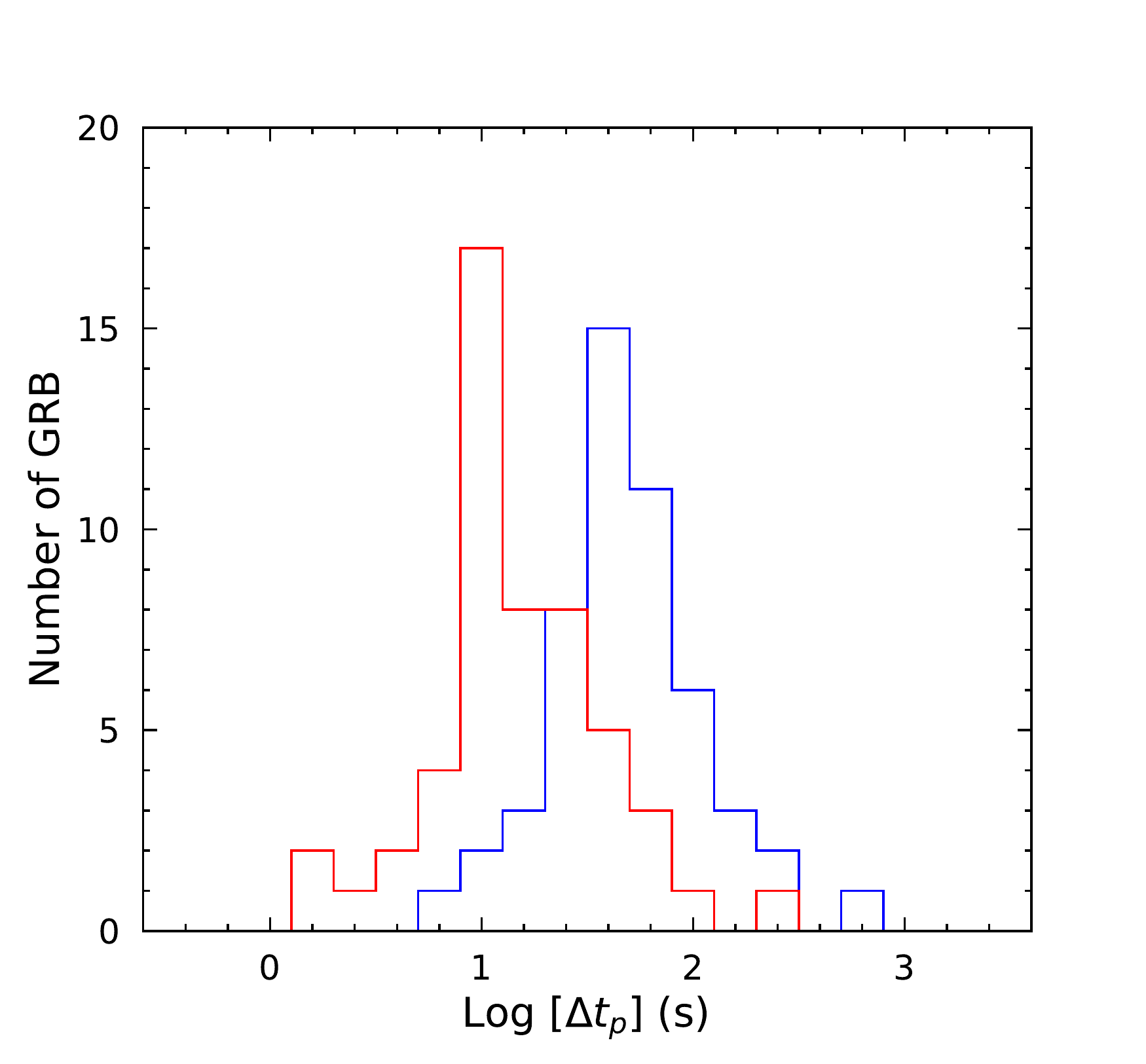}}
\caption{Left panel:
the distribution of $\Delta t_{p,i}$.
The blue solid line represents for the observed peak time,
and the red solid line represents for the peak time after the redshift correction. 
Right panel: the distribution of $\Delta t_p$.
The blue solid line represents for the observed peak time,
and the red solid line represents for the peak time after the redshift correction. 
}
\end{figure}
\clearpage

\begin{figure}[!htp]
\centering
\subfigure[$r=0.69,P<10^{-4}$]{
\includegraphics[height=4.5cm,width=5cm]{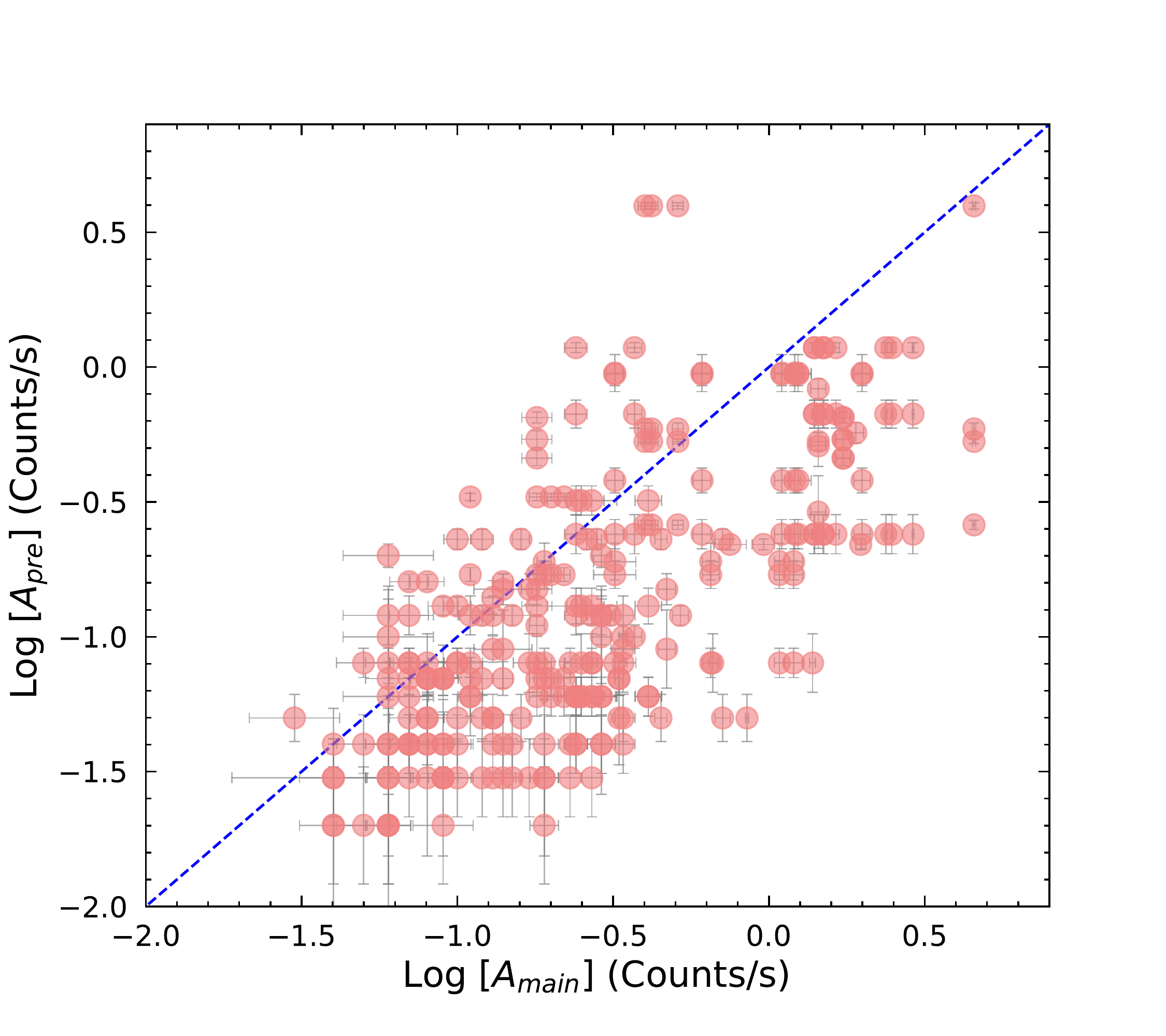}}
\subfigure[$r=0.14,P=0.02$]{
\includegraphics[height=4.5cm,width=5cm]{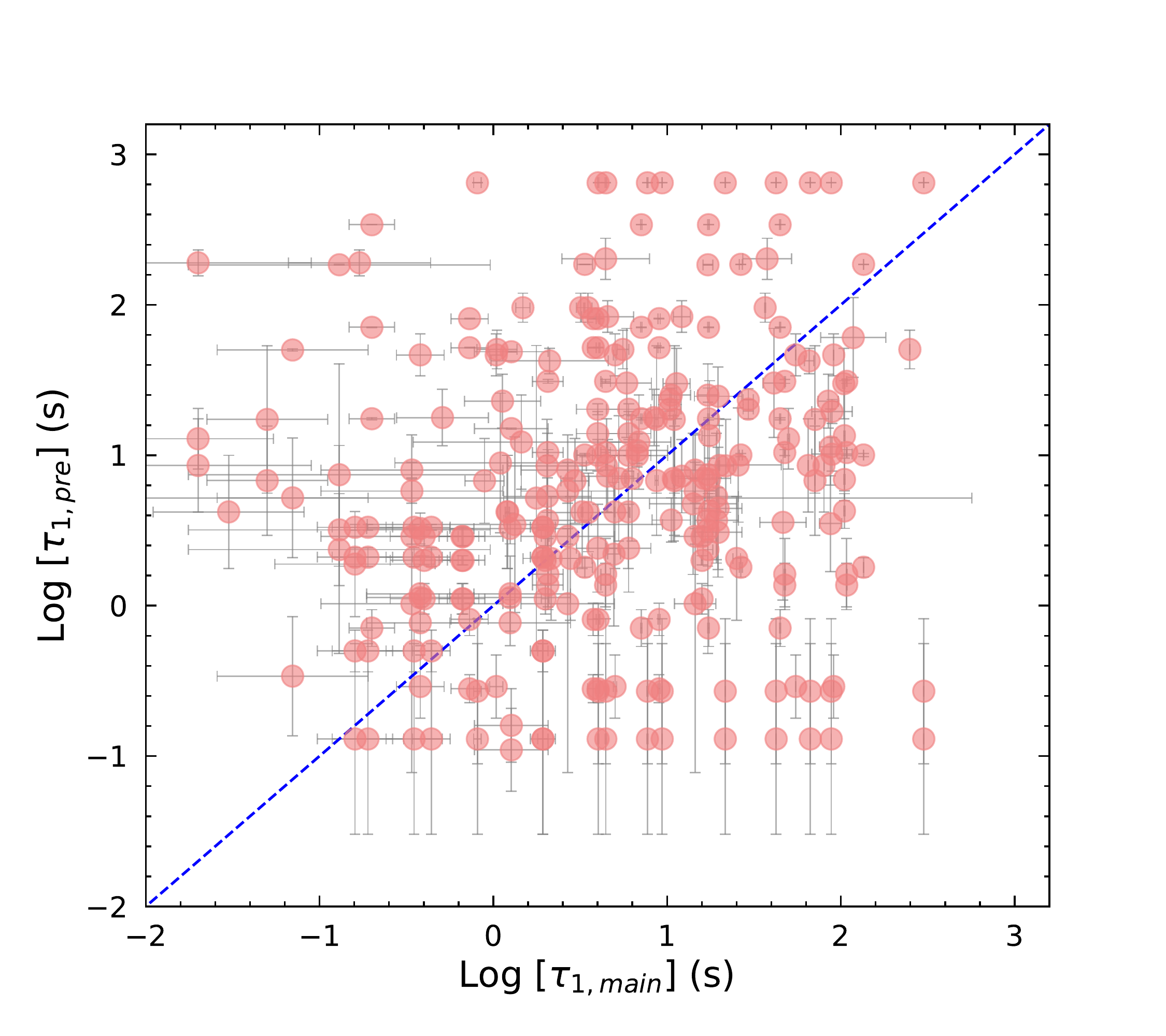}}
\subfigure[$r=0.34,P<10^{-4}$]{
\includegraphics[height=4.5cm,width=5cm]{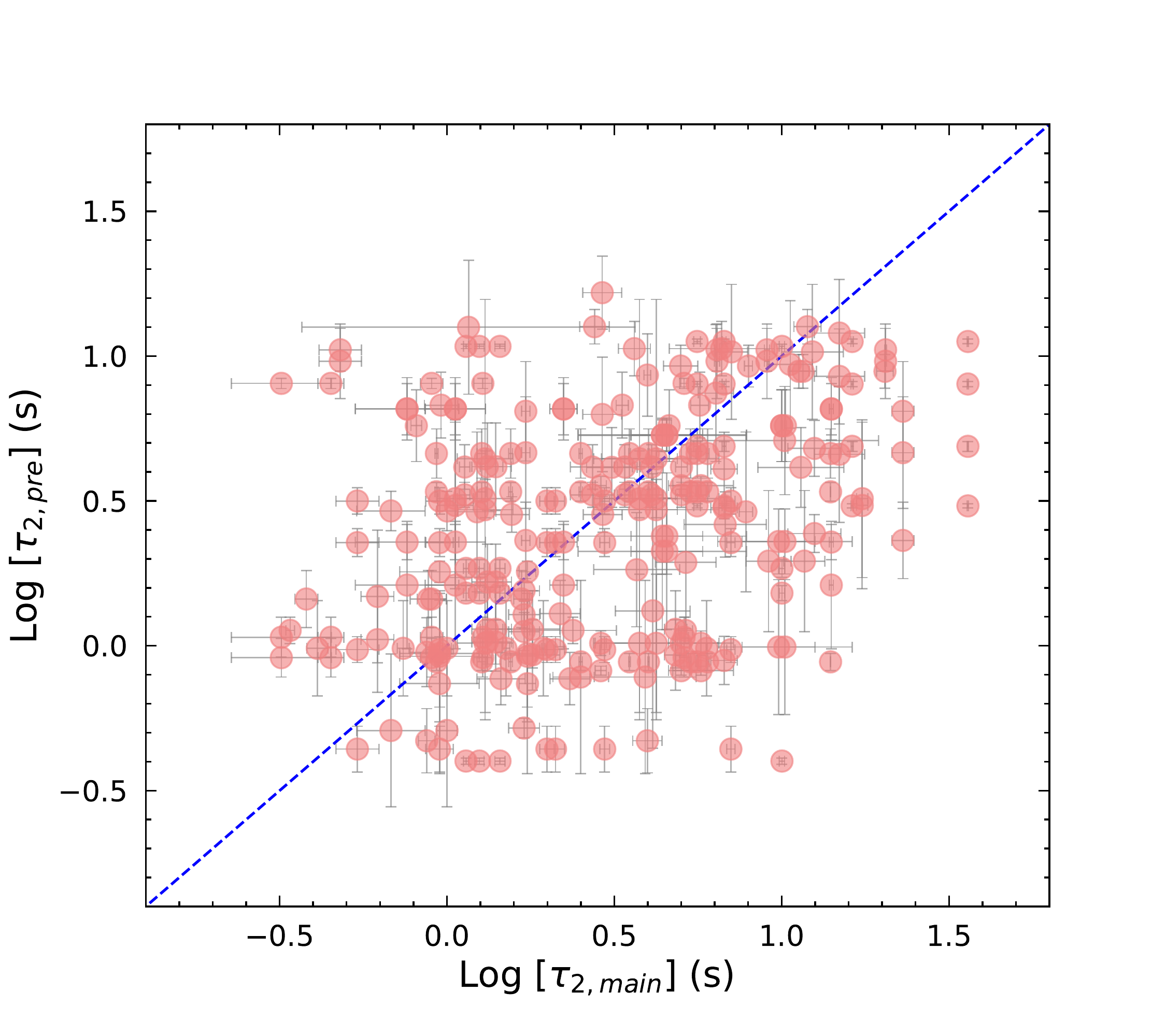}}
\subfigure[$r=0.31,P<10^{-4}$]{
\includegraphics[height=4.5cm,width=5cm]{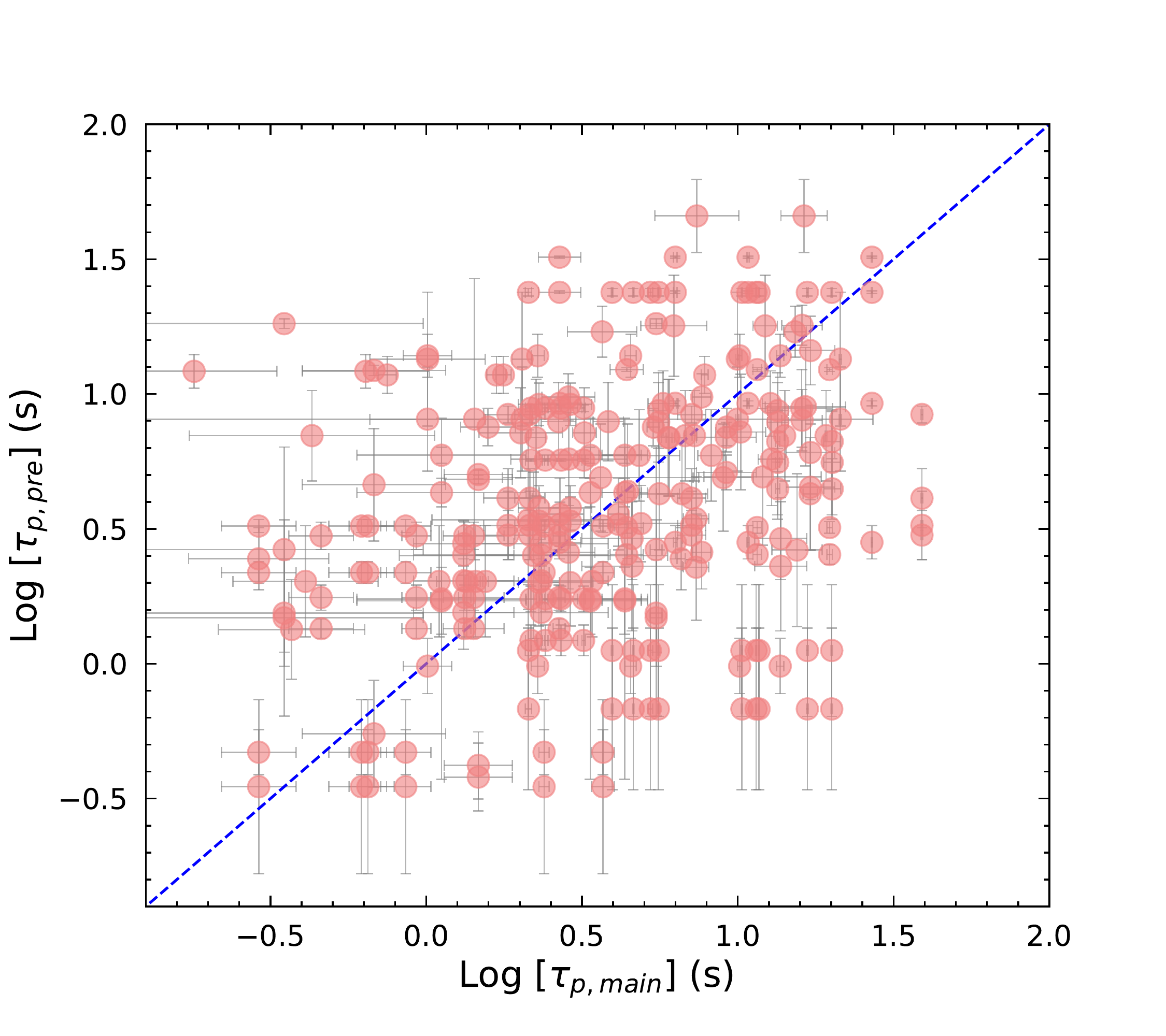}}
\subfigure[$r=0.46,P<10^{-4}$]{
\includegraphics[height=4.5cm,width=5cm]{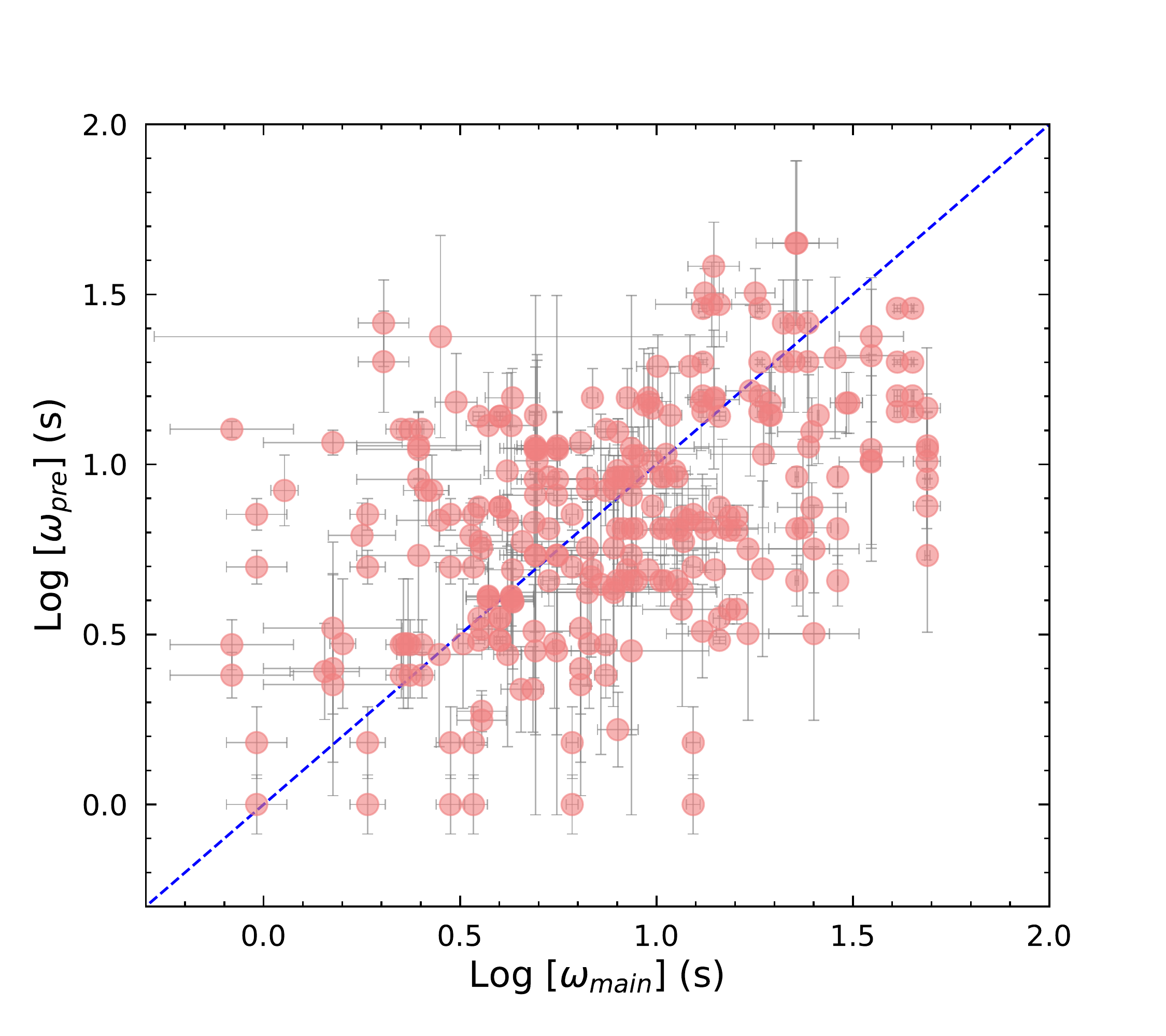}}
\subfigure[$r=0.09,P=0.12$]{
\includegraphics[height=4.5cm,width=5cm]{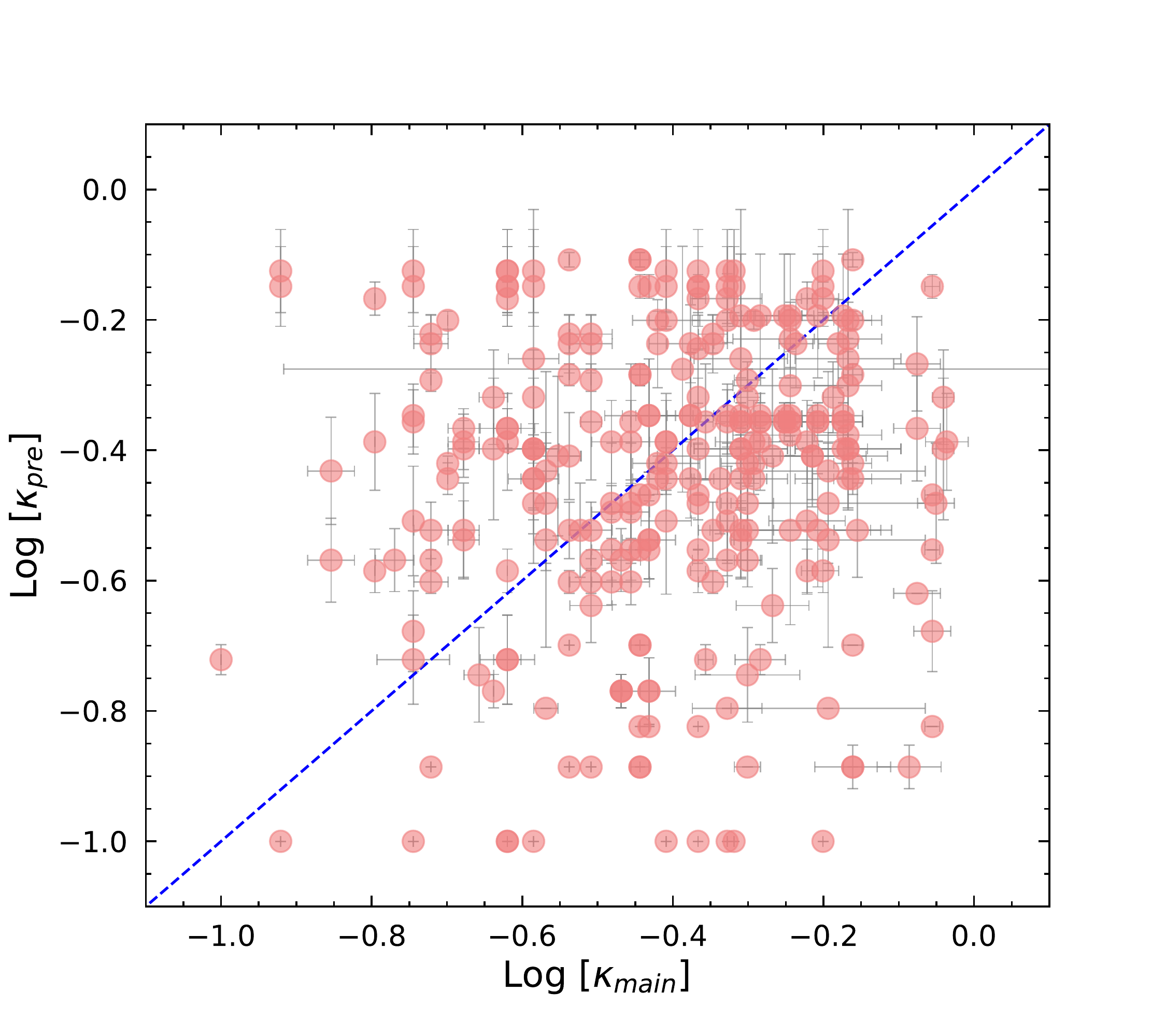}}
\subfigure[$r=0.38,P<10^{-4}$]{
\includegraphics[height=4.5cm,width=5cm]{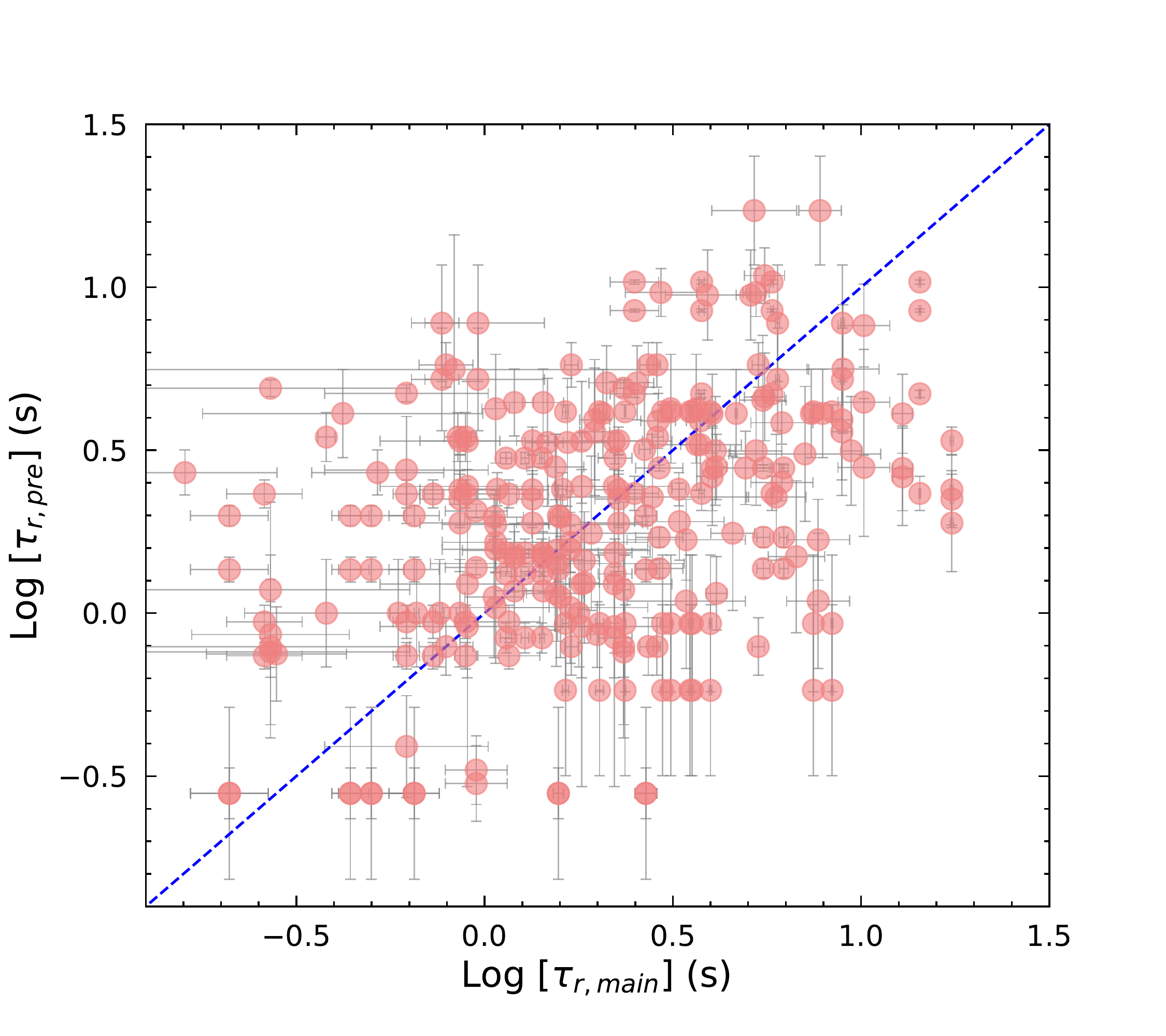}}
\subfigure[$r=0.44,P<10^{-4}$]{
\includegraphics[height=4.5cm,width=5cm]{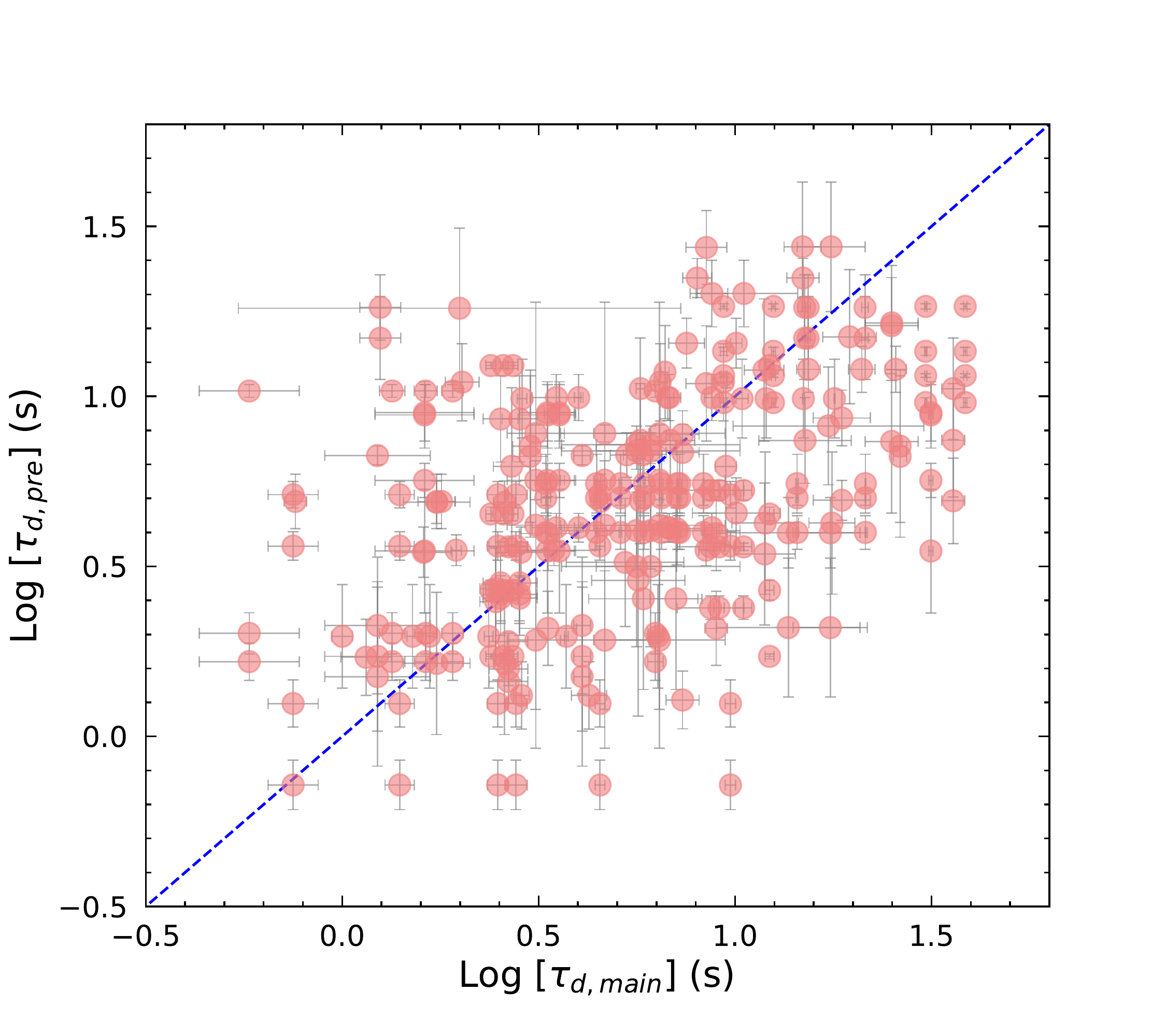}}
\caption{Comparisons of the observed episode parameters of the precursors to those of the main bursts.
The blue dashed lines represent for the case that the number in the X-axis is equal to the number in the Y-axis.
The subscript note ``main" indicates the main burst, and the subscript note ``pre" indicates the precursor. $r$ and $P$ represent for the Pearson correlation coefficient and the probability of chance correlation, respectively.}
\end{figure}
\clearpage

\begin{figure}[!htp]
\centering
\subfigure[$r=0.17, P = 0.01$]{
\includegraphics[height=4.5cm,width=5cm]{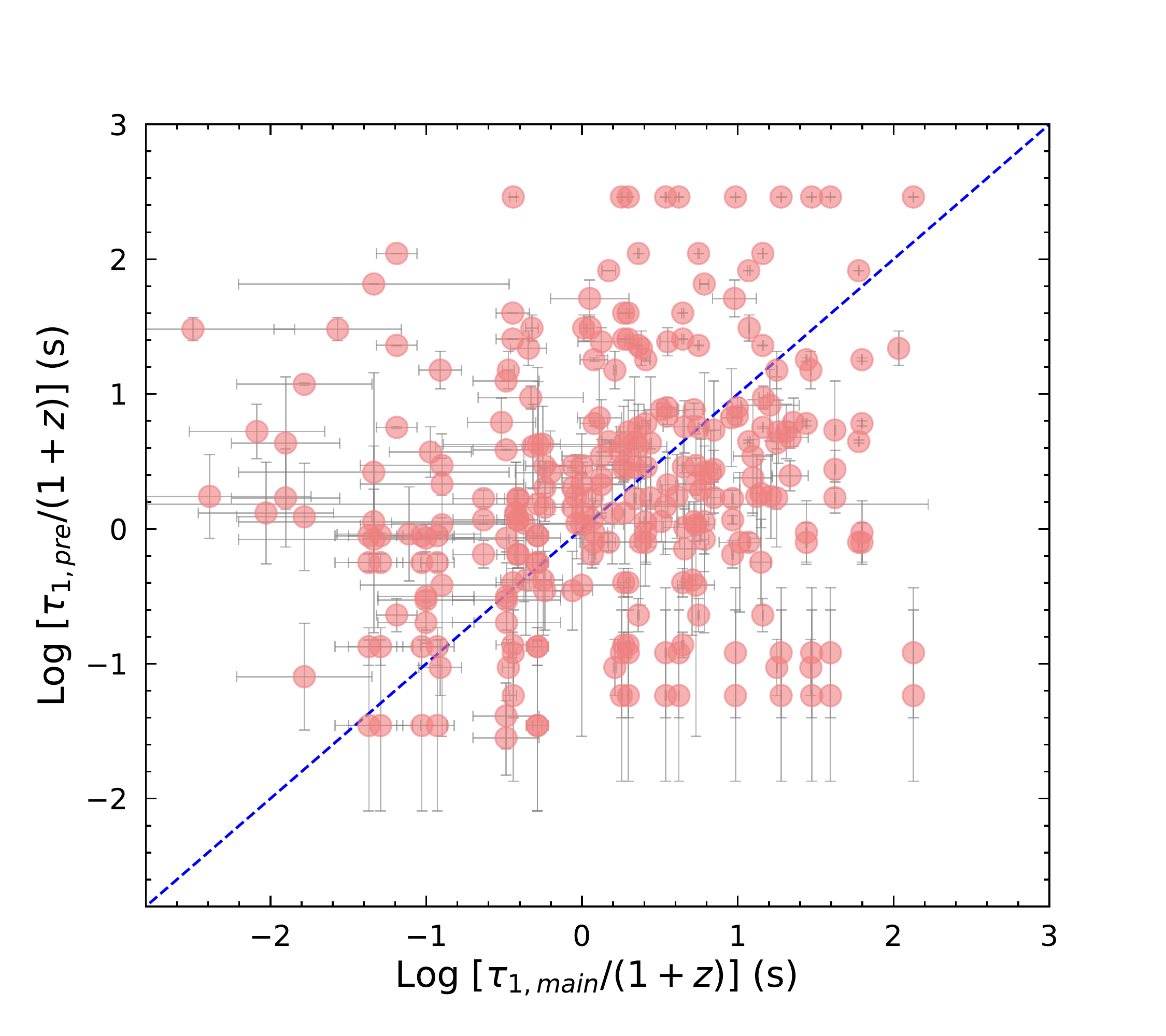}}
\subfigure[$r=0.39, P<10^{-4}$]{
\includegraphics[height=4.5cm,width=5cm]{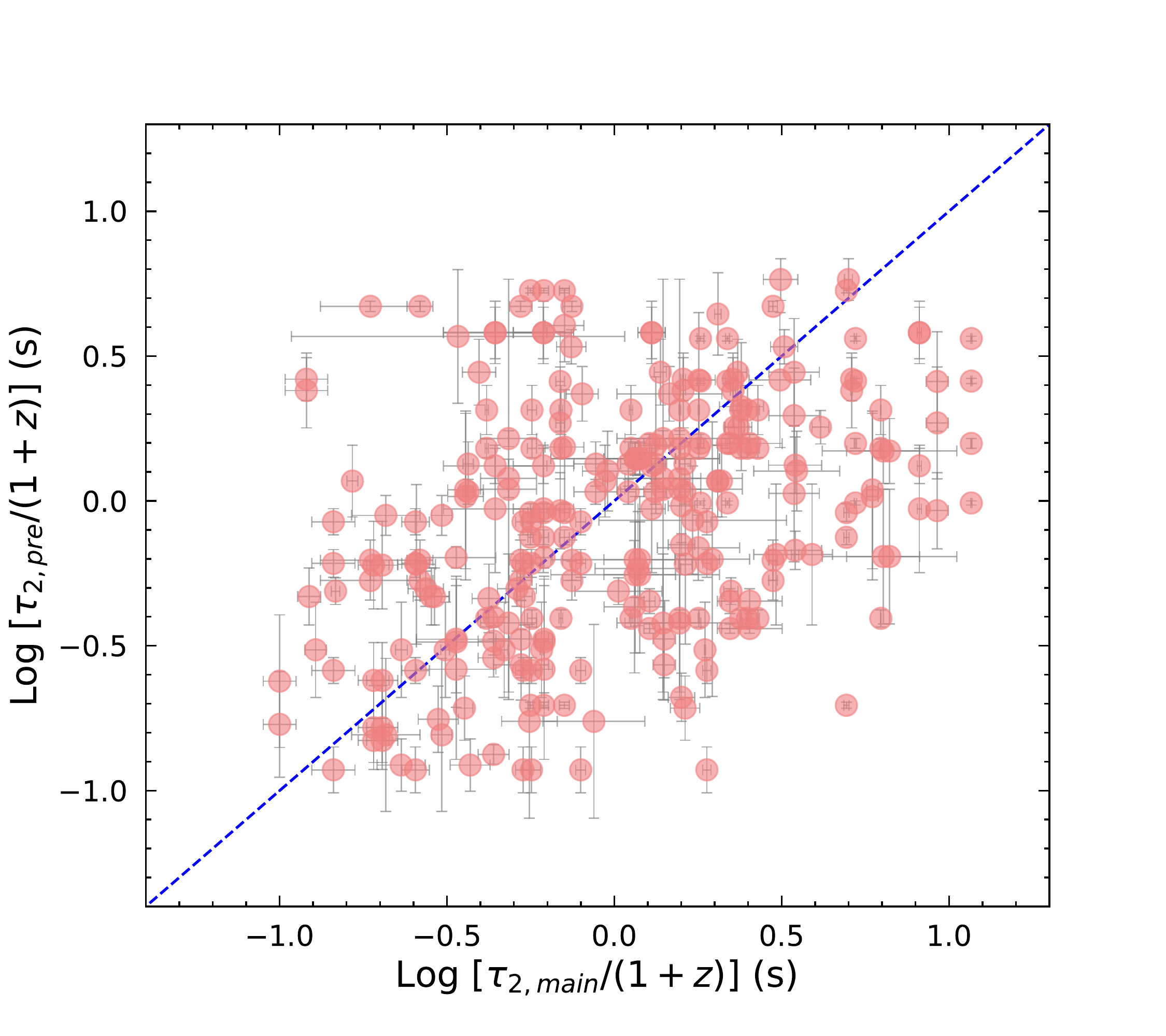}}
\subfigure[$r=0.36, P<10^{-4}$]{
\includegraphics[height=4.5cm,width=5cm]{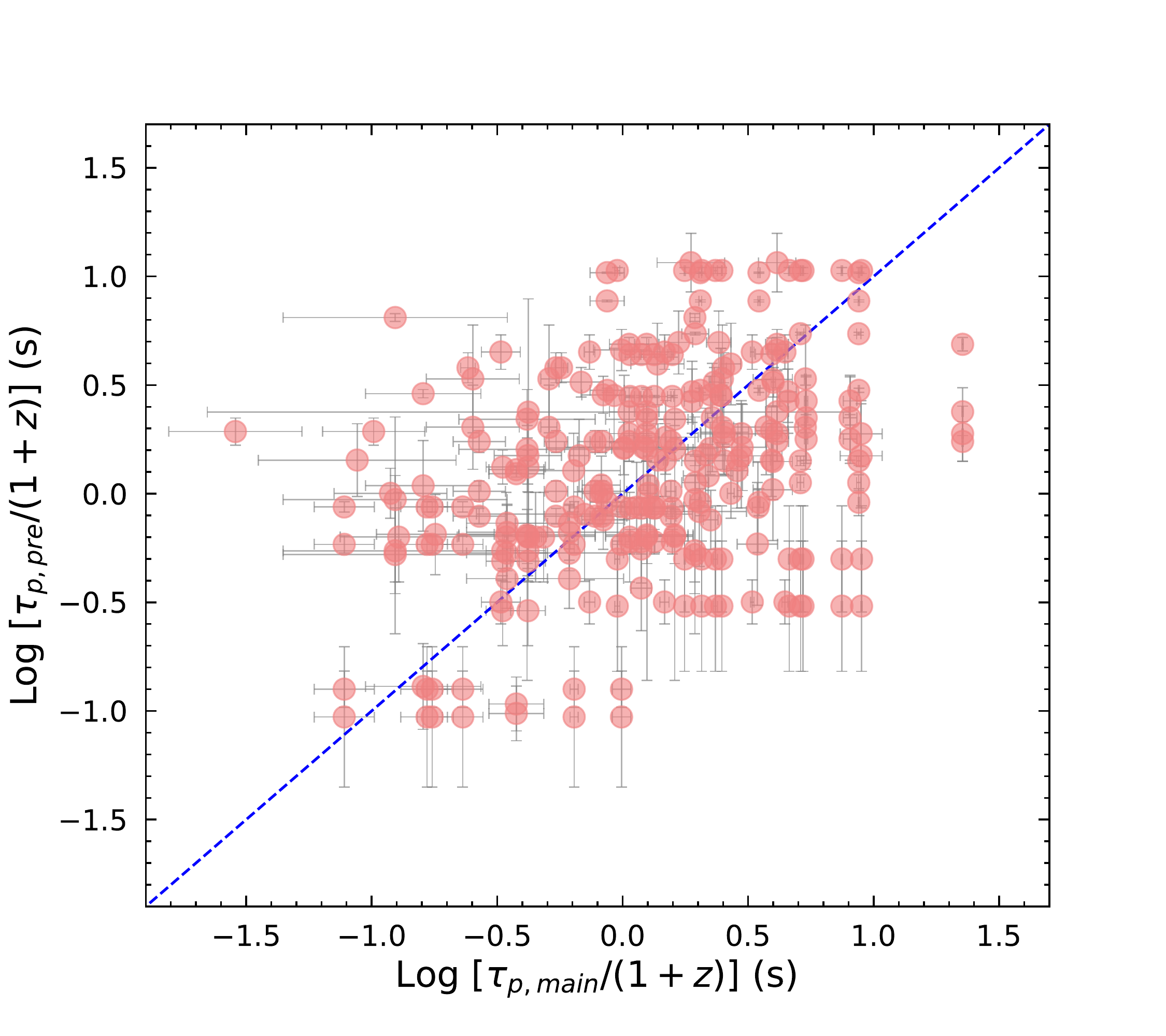}}
\subfigure[$r=0.52,P<10^{-4}$]{
\includegraphics[height=4.5cm,width=5cm]{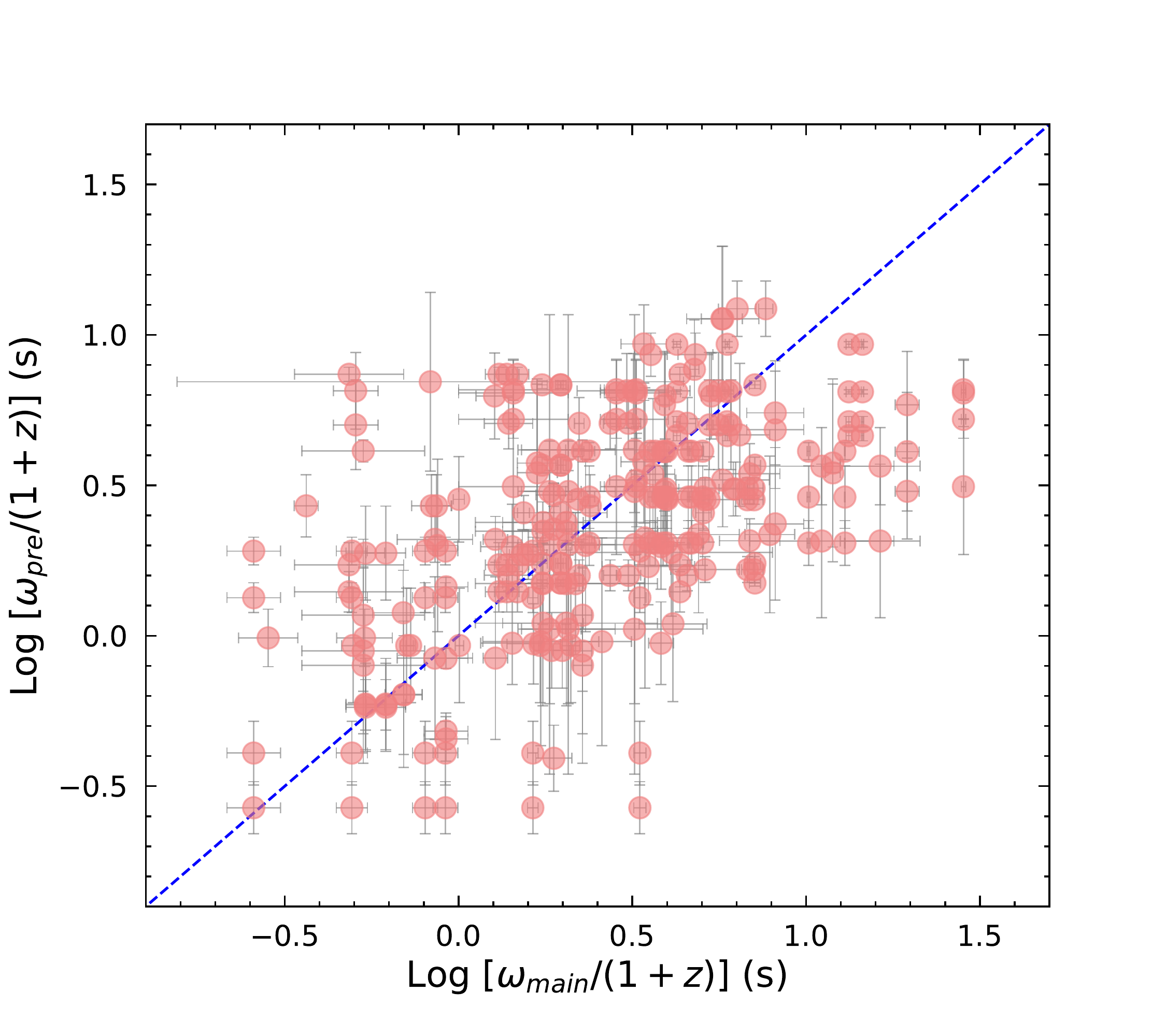}}
\subfigure[$r=0.45,P<10^{-4}$]{
\includegraphics[height=4.5cm,width=5cm]{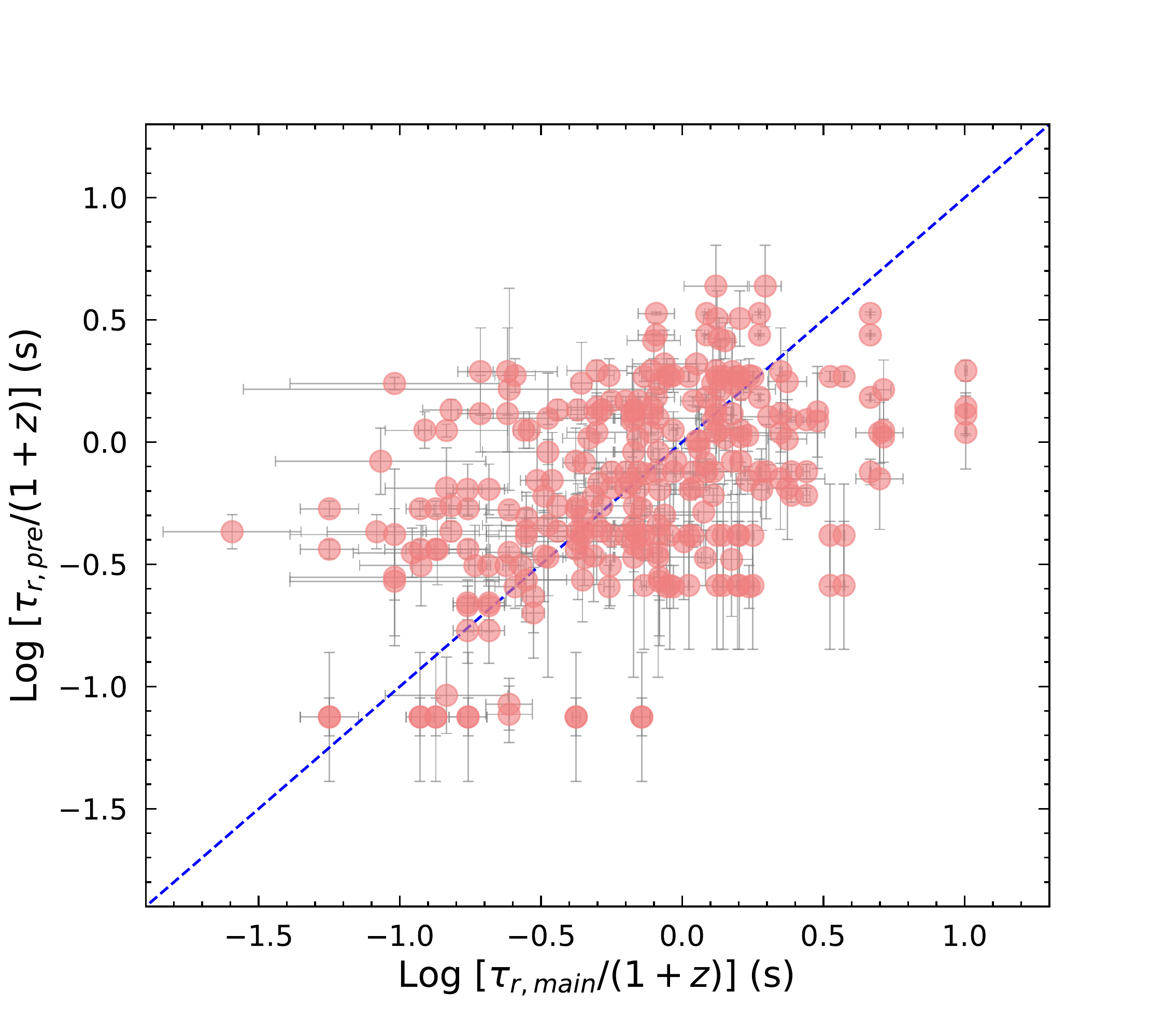}}
\subfigure[$r=0.50, P<10^{-4}$]{
\includegraphics[height=4.5cm,width=5cm]{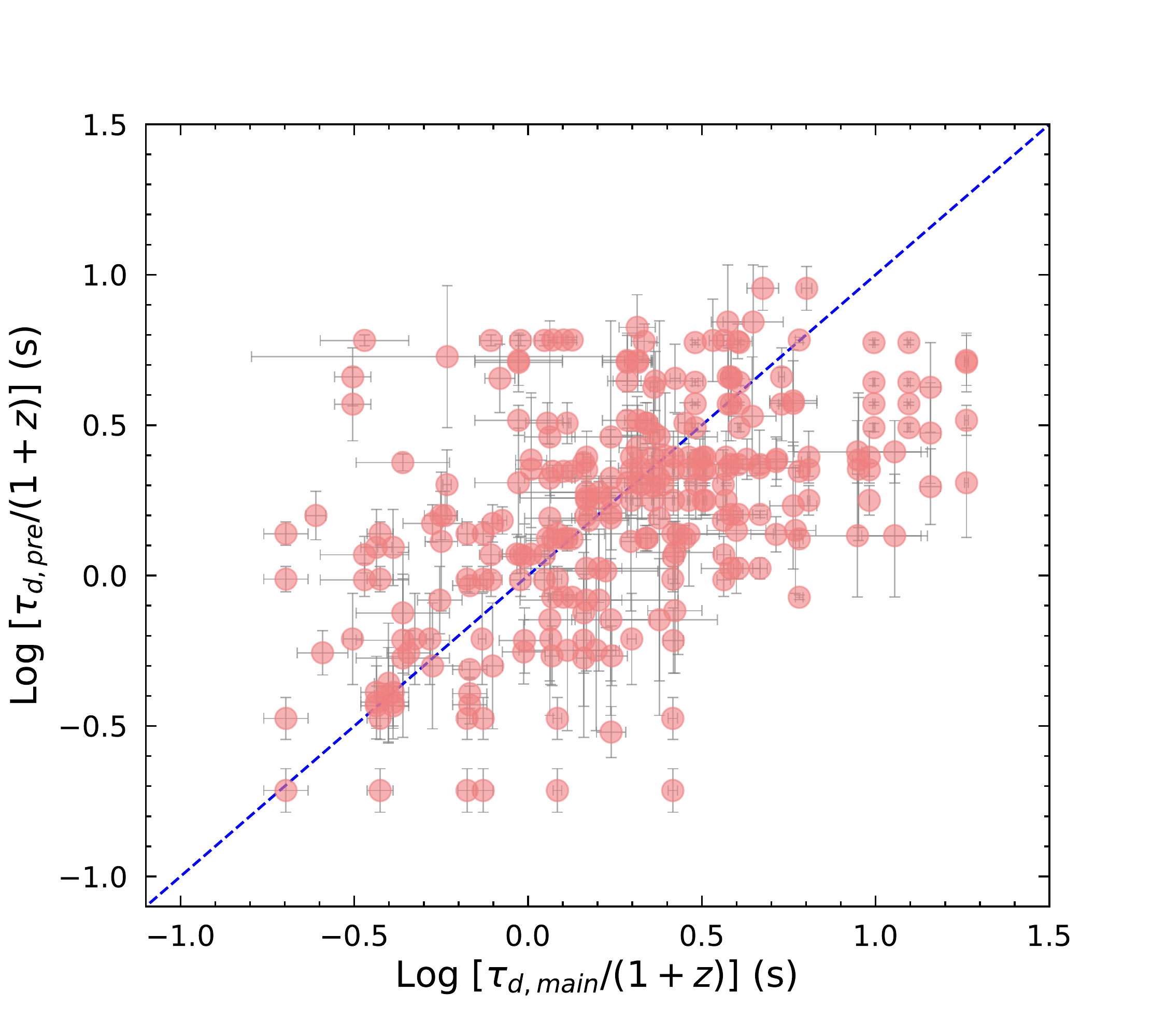}}
\caption{Comparisons of the redshift-corrected episode parameters of the precursors to those of the main bursts.
The blue dashed lines represent the case that the number in the X-axis is equal to the number in the Y-axis.
The subscript note ``main" indicates the main burst, and the subscript note ``pre" indicates the precursor. $r$ and $P$ represent the Pearson correlation coefficient and the probability of chance correlation, respectively.}
\end{figure}
\clearpage

\begin{figure}[!htp]
\centering
\subfigure[$r = 0.81, P < 10^{-4}$]{
\includegraphics[height=4.5cm,width=5cm]{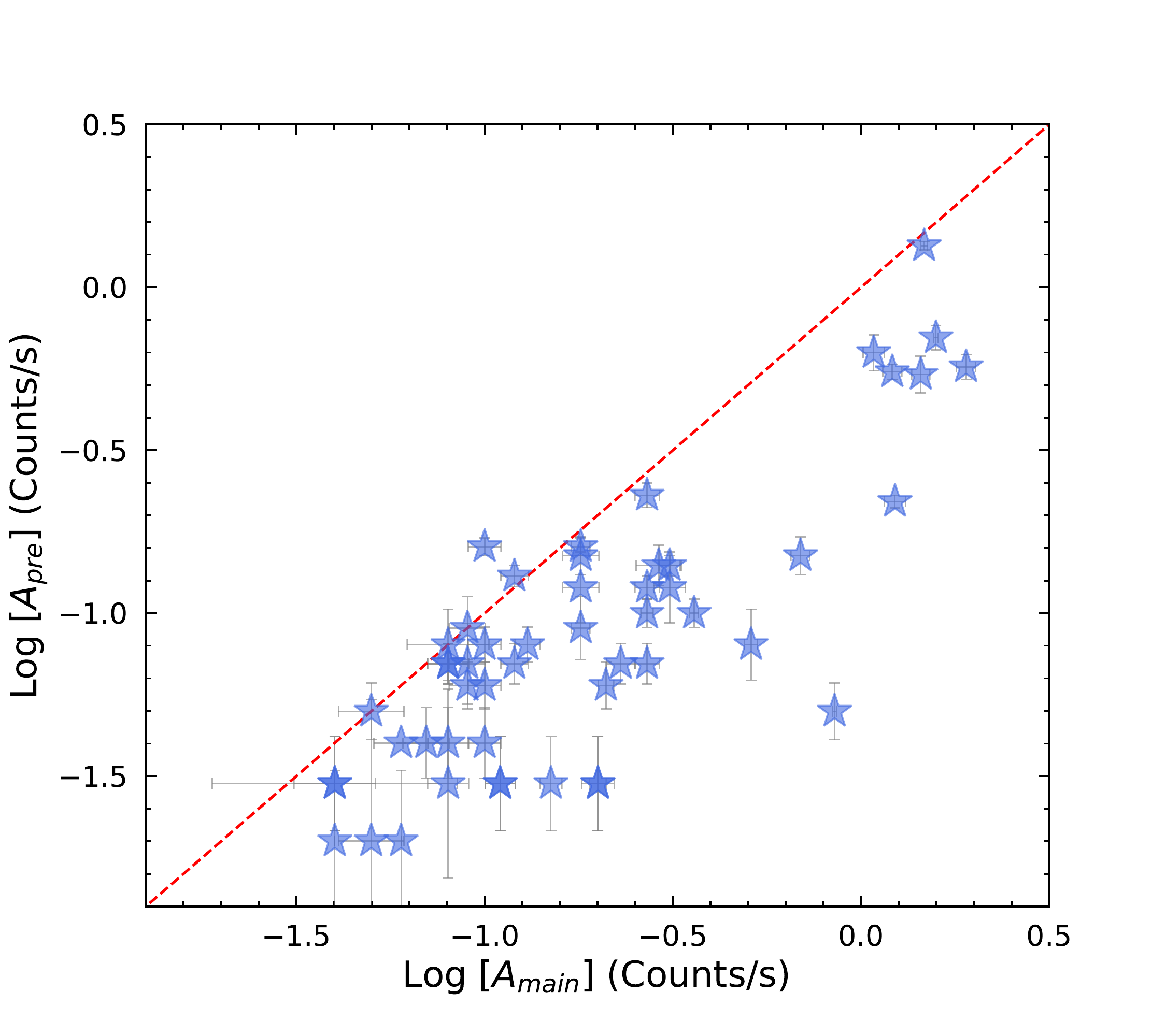}}
\subfigure[$r = 0.30, P = 0.03$]{
\includegraphics[height=4.5cm,width=5cm]{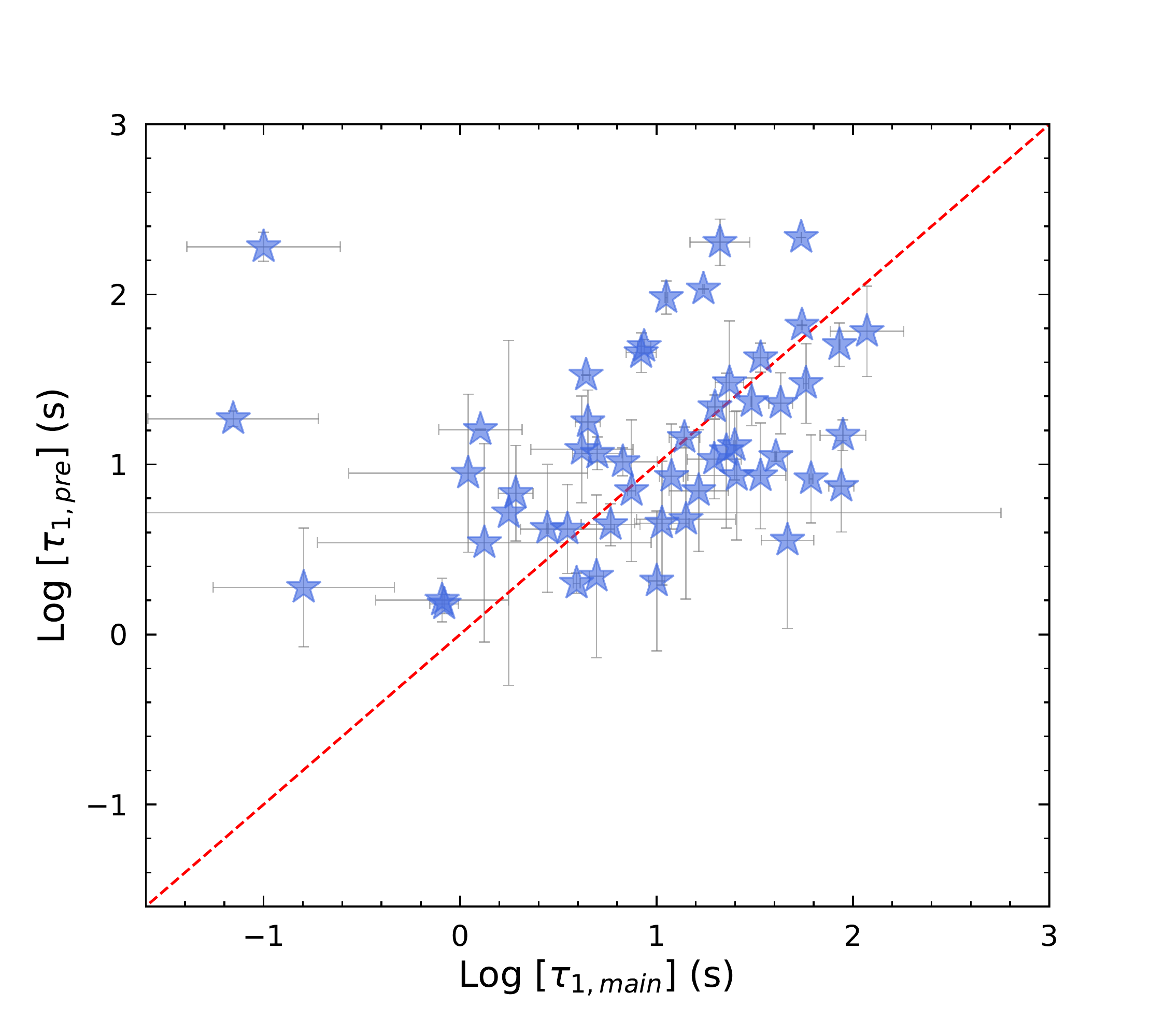}}
\subfigure[$r = 0.53, P < 10^{-4}$]{
\includegraphics[height=4.5cm,width=5cm]{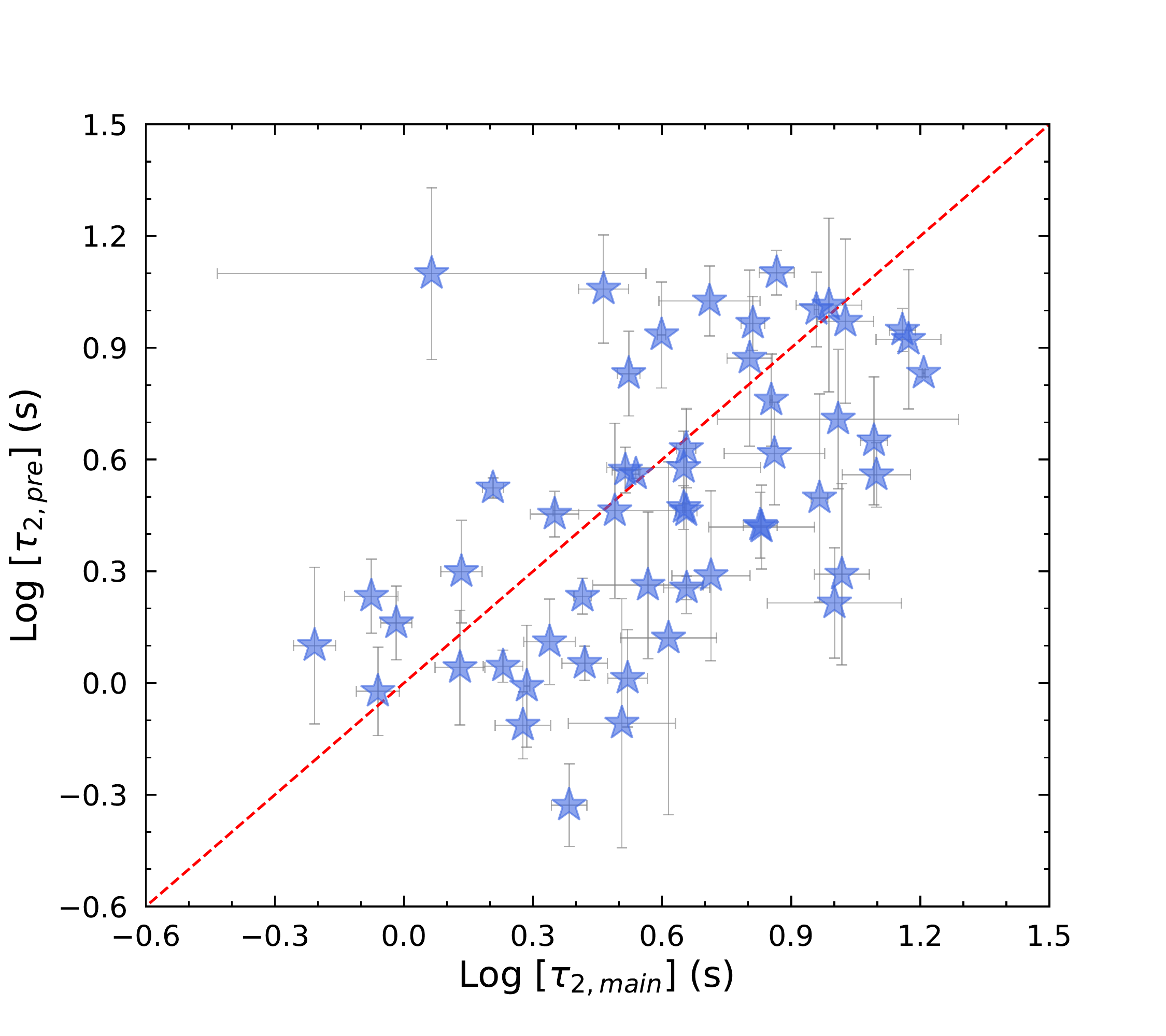}}
\subfigure[$r = 0.45, P = 7.74 \times 10^{-4}$]{
\includegraphics[height=4.5cm,width=5cm]{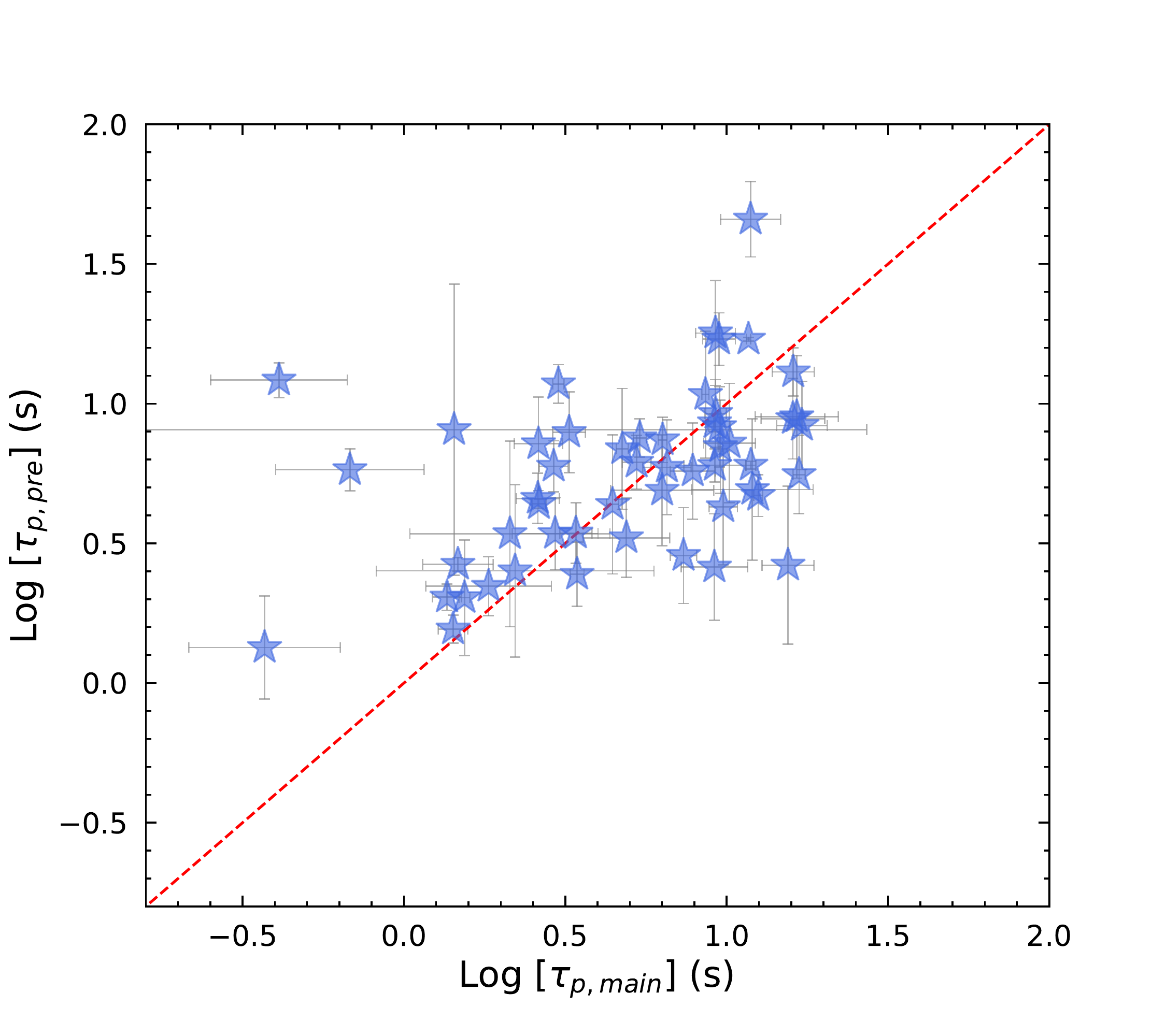}}
\subfigure[$r = 0.57, P < 10^{-4}$]{
\includegraphics[height=4.5cm,width=5cm]{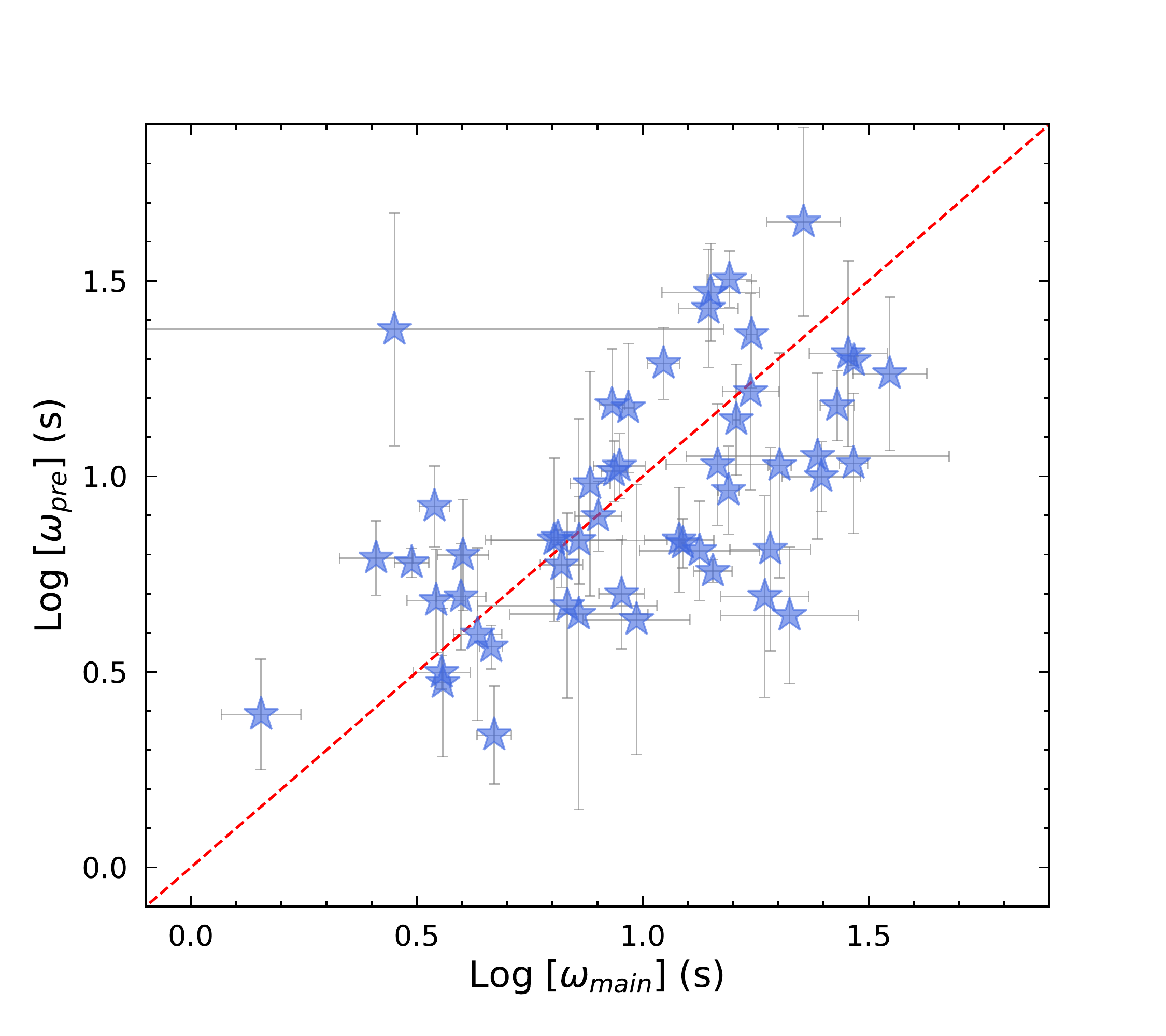}}
\subfigure[$r = 0.16, P = 0.25$]{
\includegraphics[height=4.5cm,width=5cm]{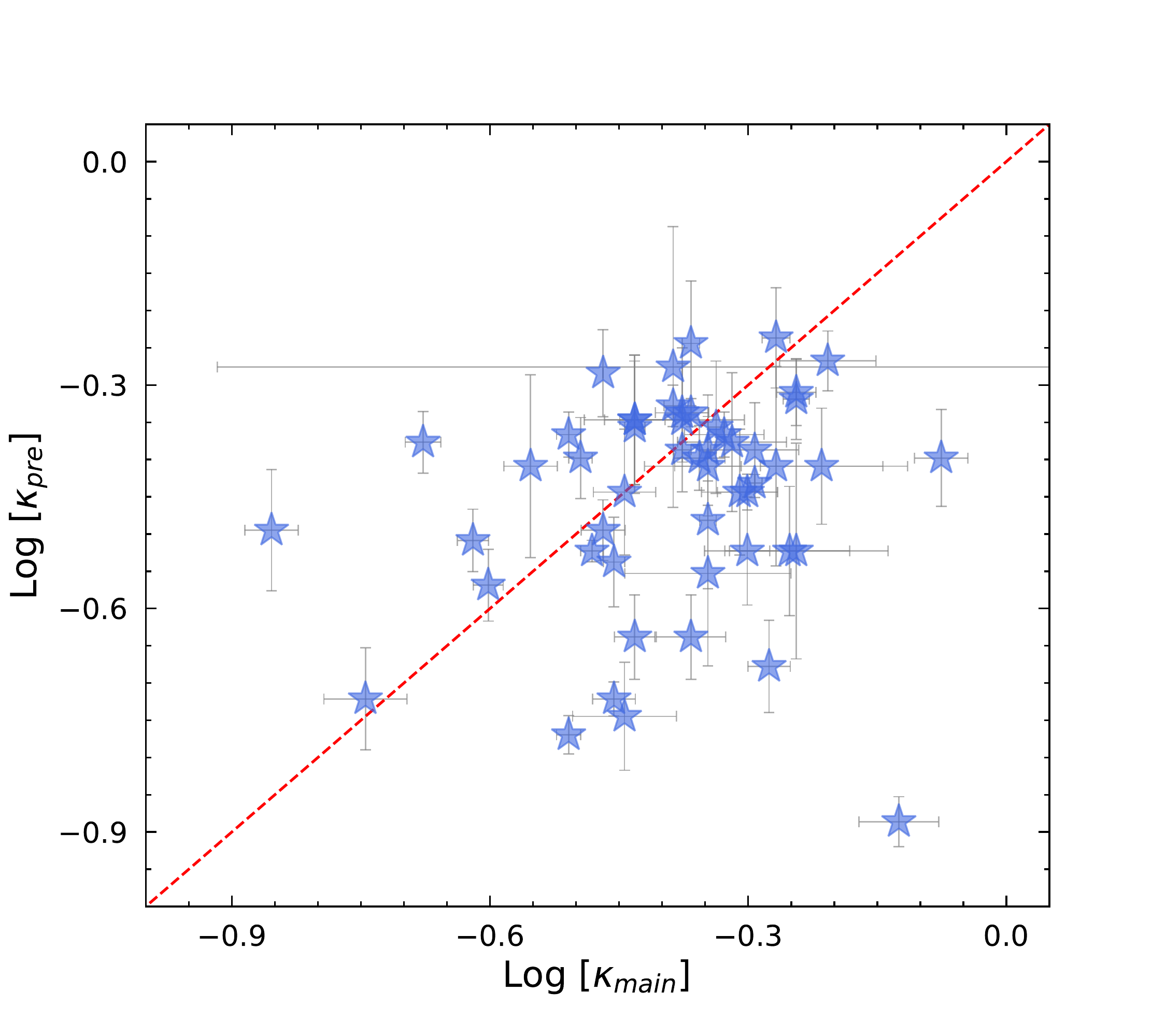}}
\subfigure[$r = 0.54, P < 10^{-4}$]{
\includegraphics[height=4.5cm,width=5cm]{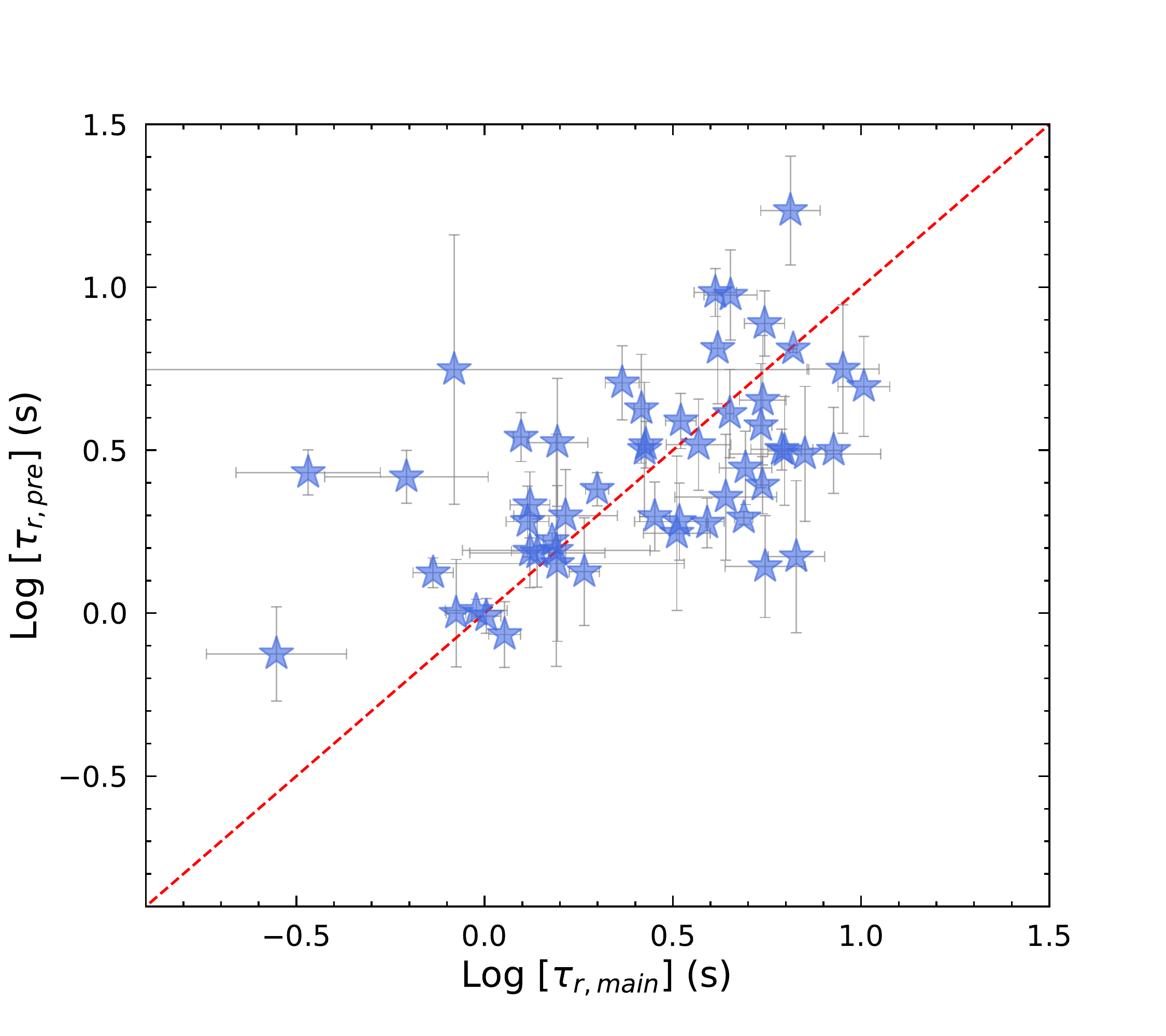}}
\subfigure[$r = 0.57, P < 10^{-4}$]{
\includegraphics[height=4.5cm,width=5cm]{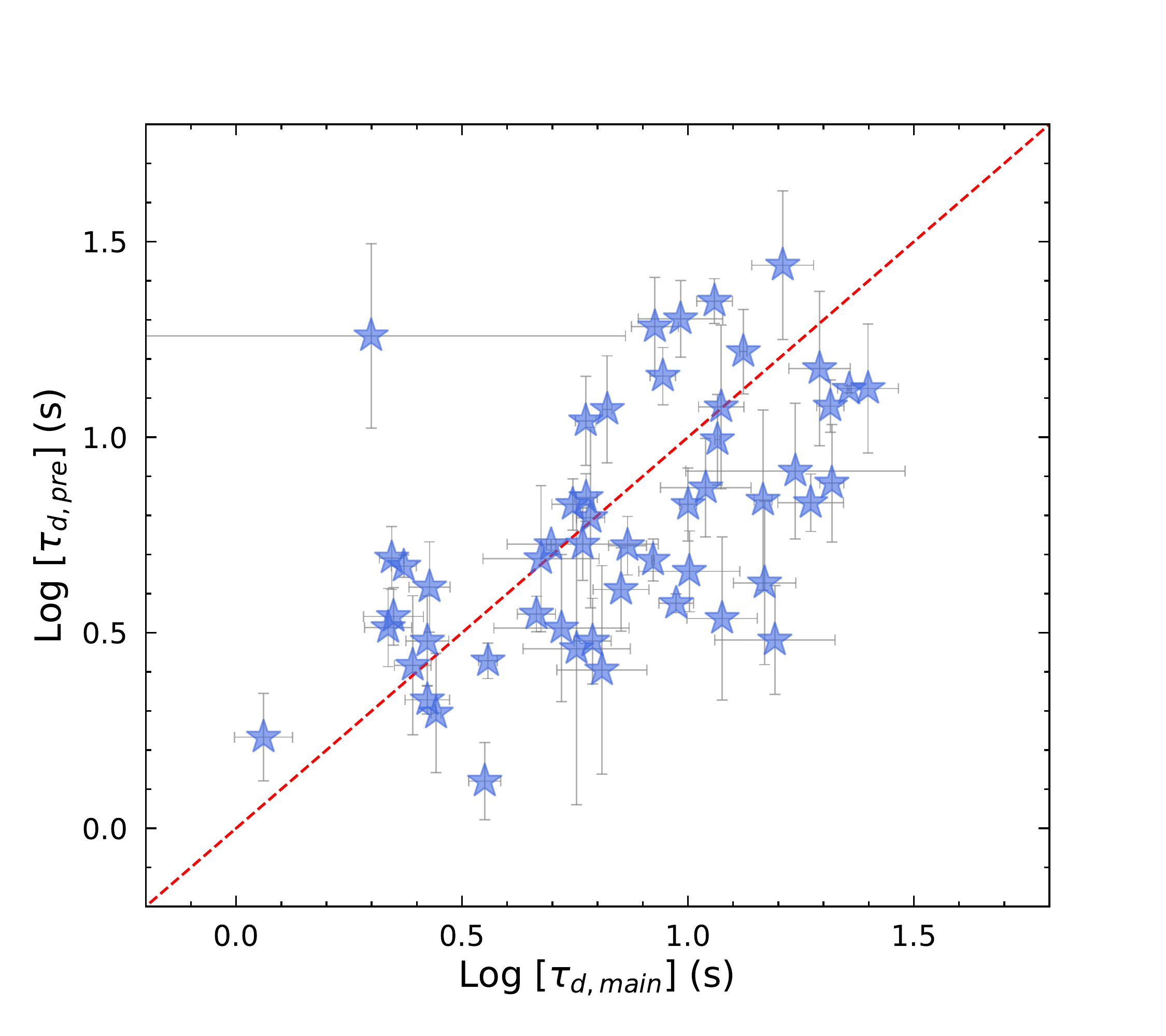}}
\caption{Comparisons of the averaged episode parameters of the precursors to those of the main bursts. The numbers have no redshift-correction.
The red dashed lines represent the case that the number in the X-axis is equal to the number in the Y-axis.
The subscript note ``main" indicates the main burst, and the subscript note ``pre" indicates the precursor.
$r$ and $P$ represent the Pearson correlation coefficient and the probability of chance correlation, respectively.}
\end{figure}
\clearpage

\begin{figure}[!htp]
\centering
\subfigure[$r = 0.30, P = 0.03$]{
\includegraphics[height=4.5cm,width=5cm]{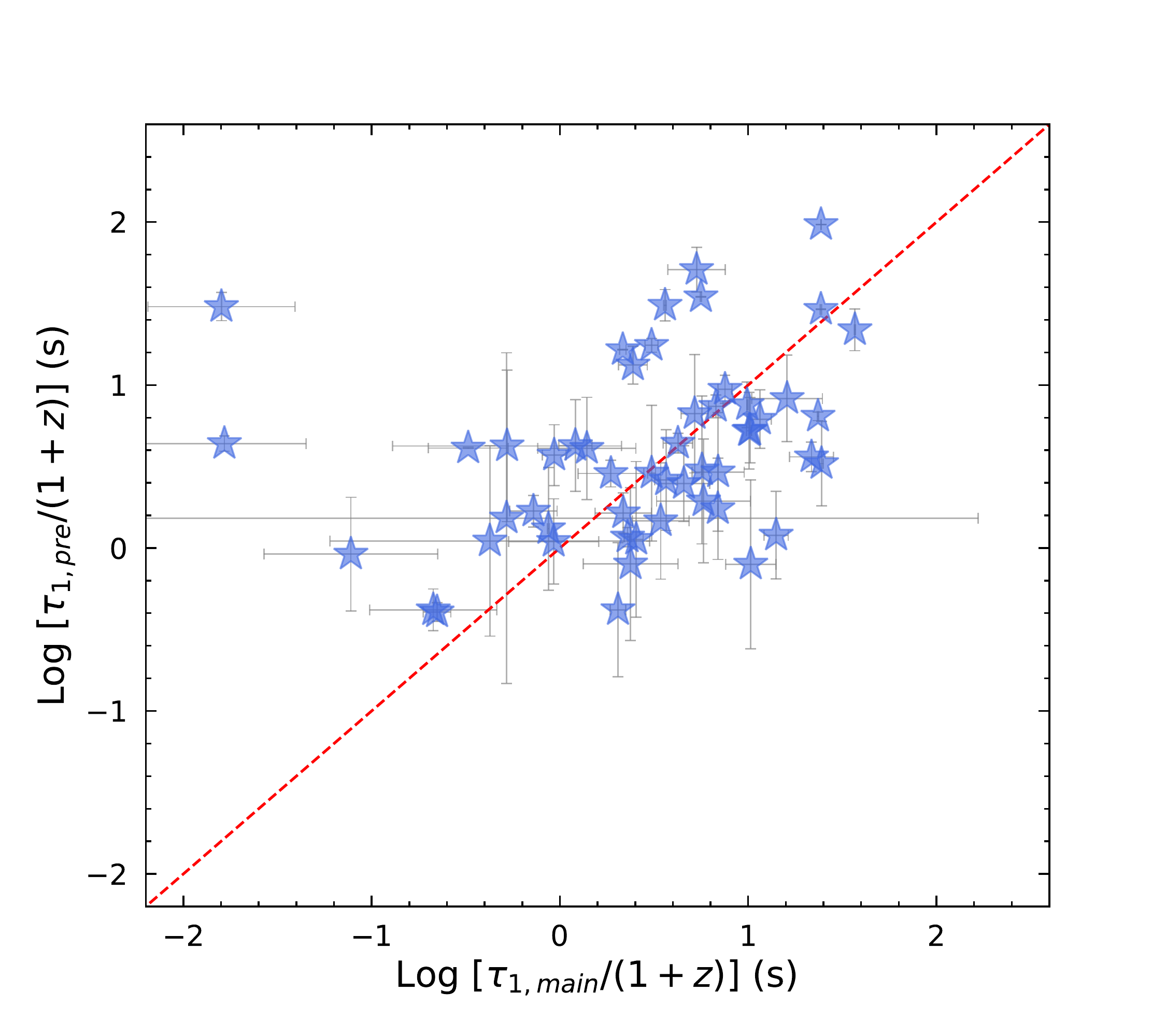}}
\subfigure[$r = 0.61, P < 10^{-4}$]{
\includegraphics[height=4.5cm,width=5cm]{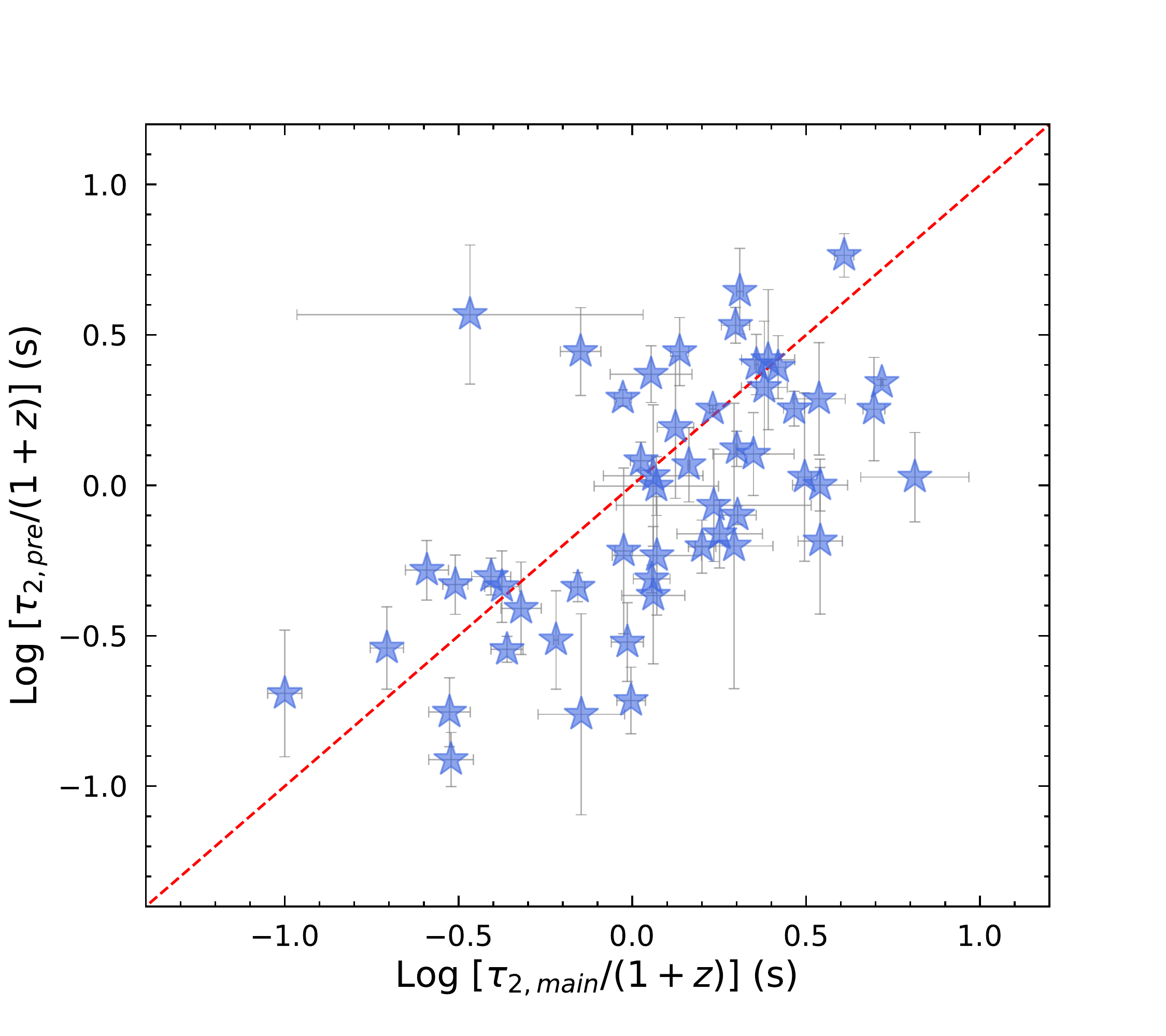}}
\subfigure[$r = 0.47, P = 3.99 \times 10^{-4}$]{
\includegraphics[height=4.5cm,width=5cm]{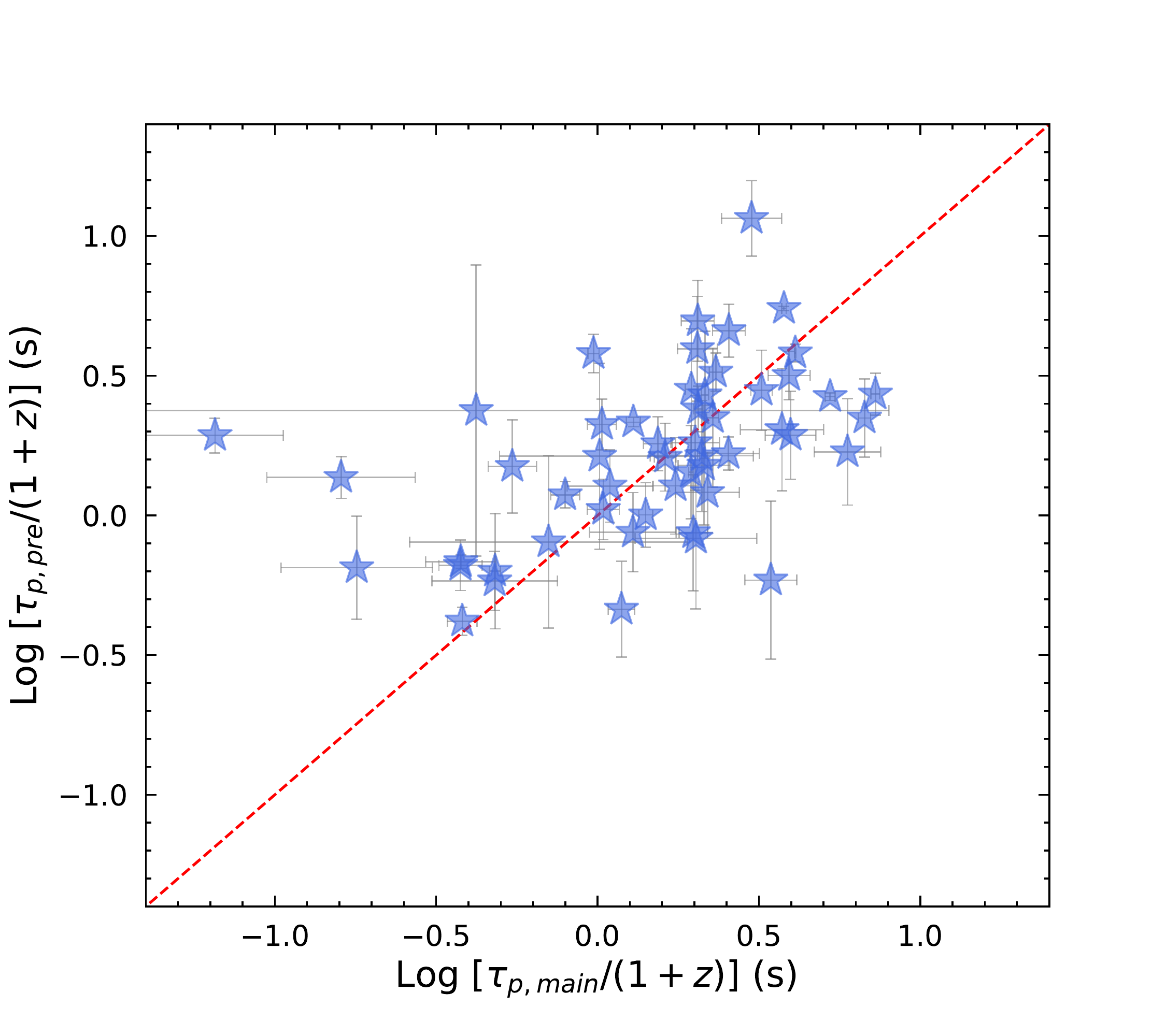}}
\subfigure[$r = 0.63, P < 10^{-4}$]{
\includegraphics[height=4.5cm,width=5cm]{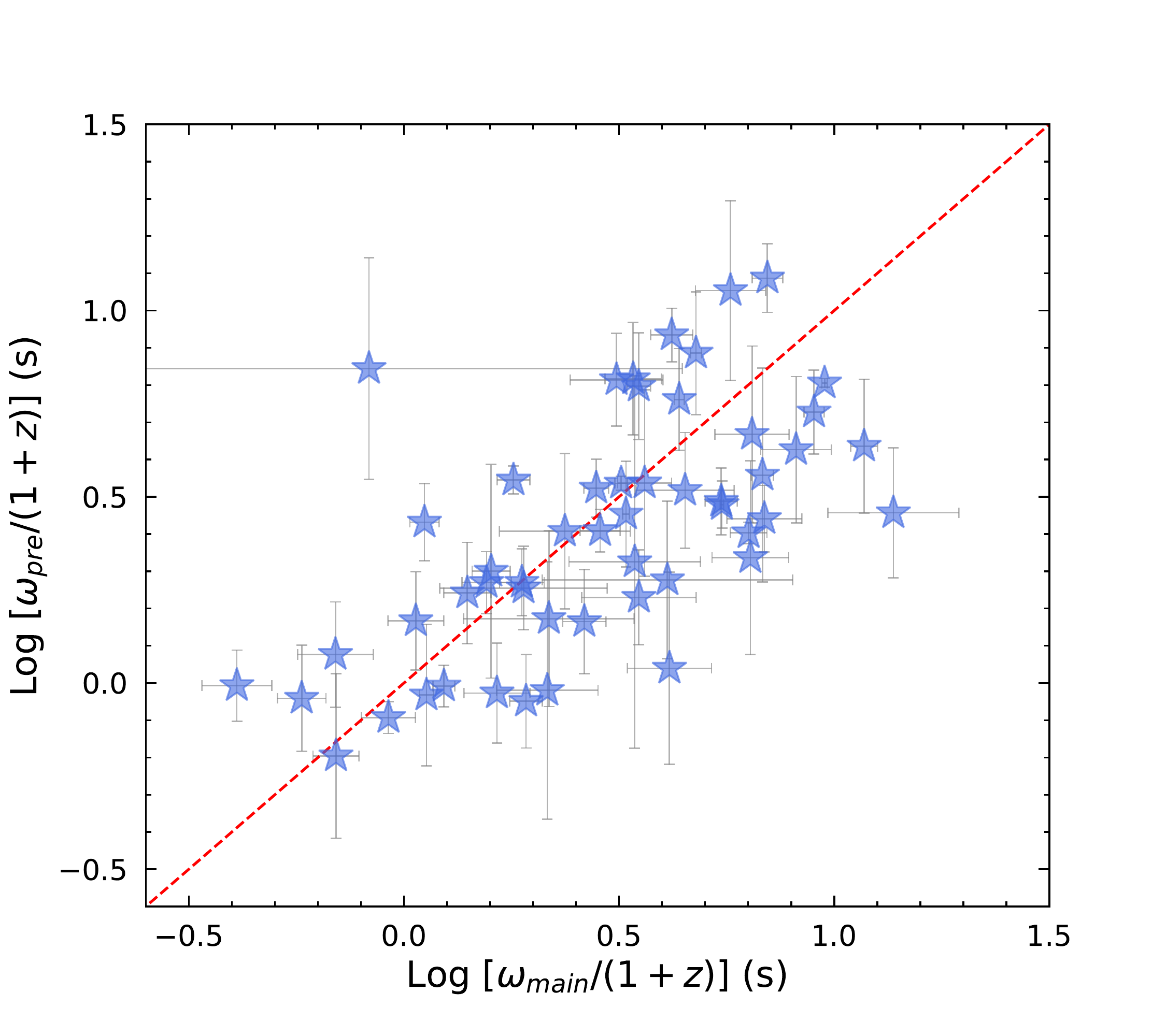}}
\subfigure[$r = 0.58, P < 10^{-4}$]{
\includegraphics[height=4.5cm,width=5cm]{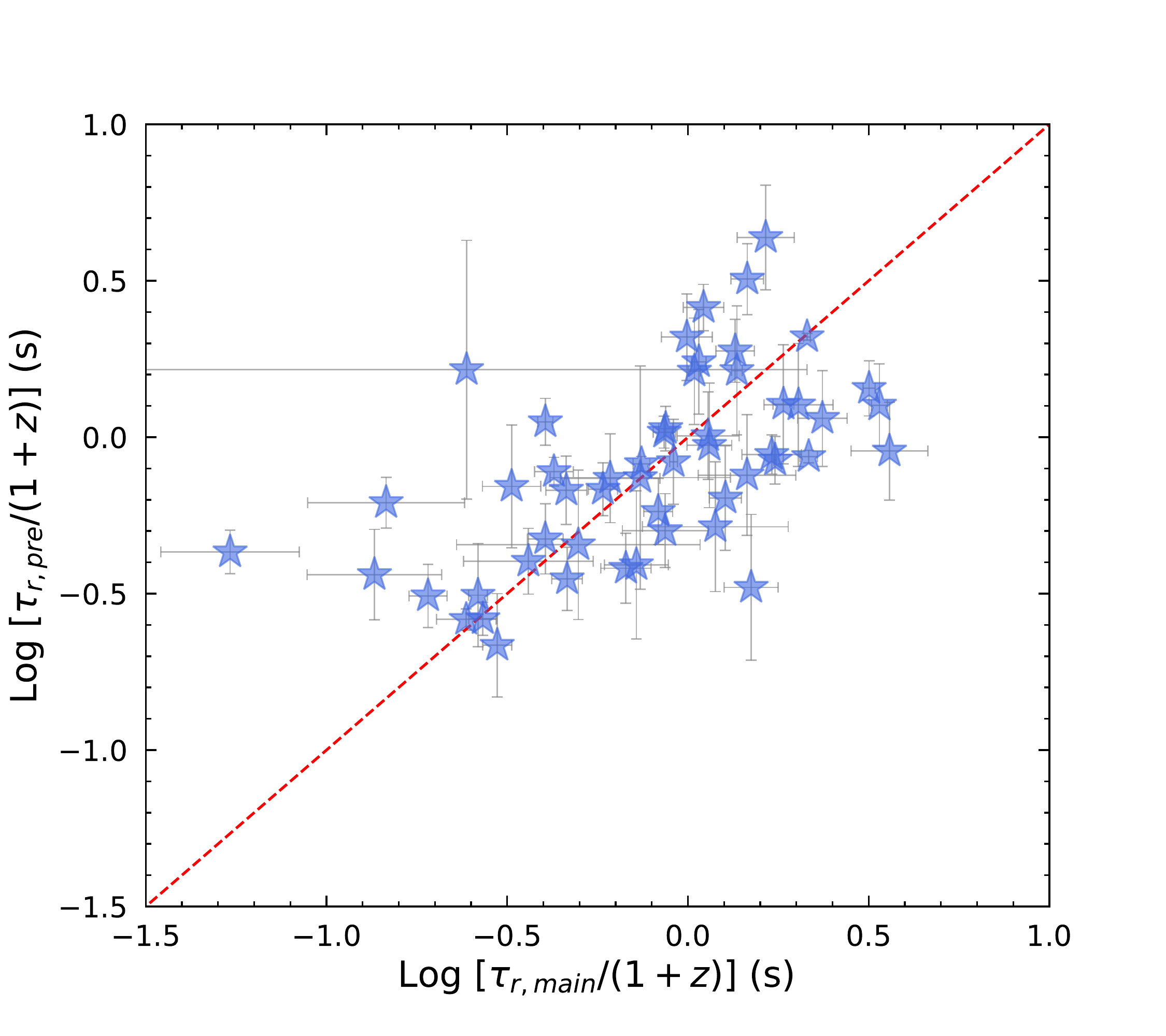}}
\subfigure[$r = 0.63, P < 10^{-4}$]{
\includegraphics[height=4.5cm,width=5cm]{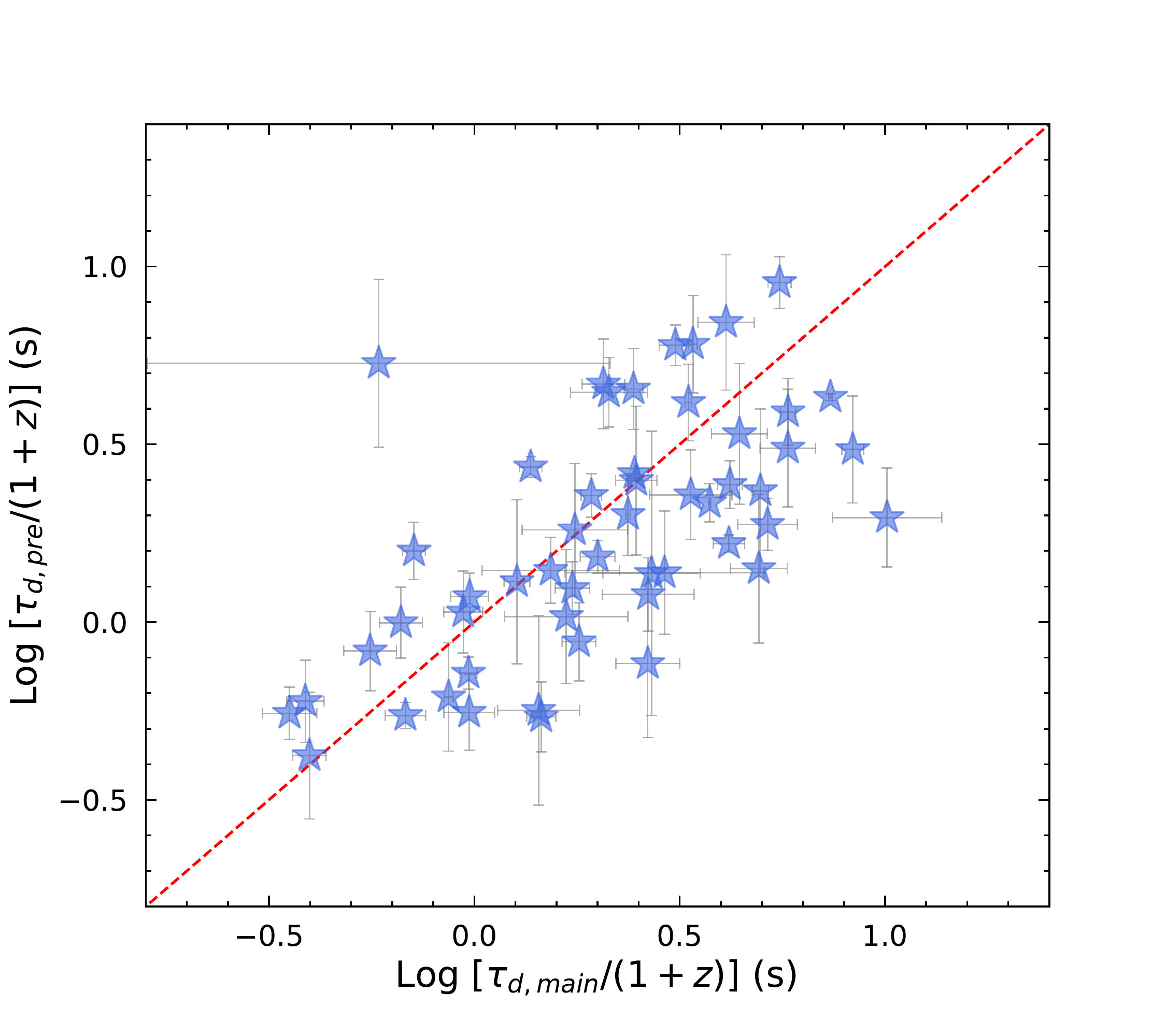}}
\caption{
Comparisons of the averaged episode parameters of the precursors to those of the main bursts. The numbers have redshift correction.
The red dashed lines represent for the case that the number in the X-axis is equal to the number in the Y-axis.
The subscript note ``main" indicates the main burst, and the subscript note ``pre" indicates the precursor.
$r$ and $P$ represent the Pearson correlation coefficient and the probability of chance correlation, respectively.}
\end{figure}
\clearpage

\begin{figure}[!htp]
\centering
\subfigure[$r = 0.76, P = 0.01$]{
\includegraphics[height=4.5cm,width=5cm]{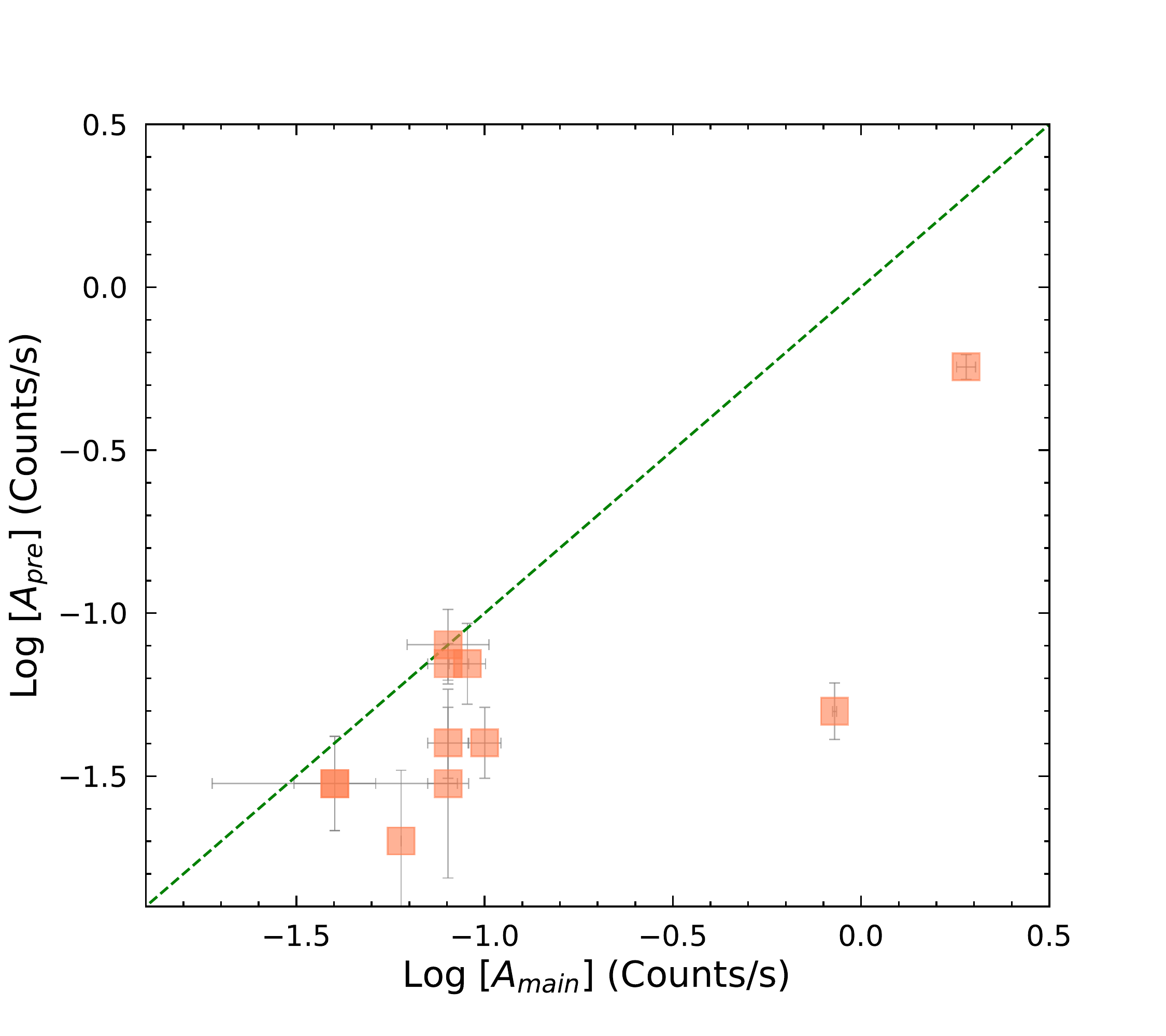}}
\subfigure[$r = 0.62, P = 0.04$]{
\includegraphics[height=4.5cm,width=5cm]{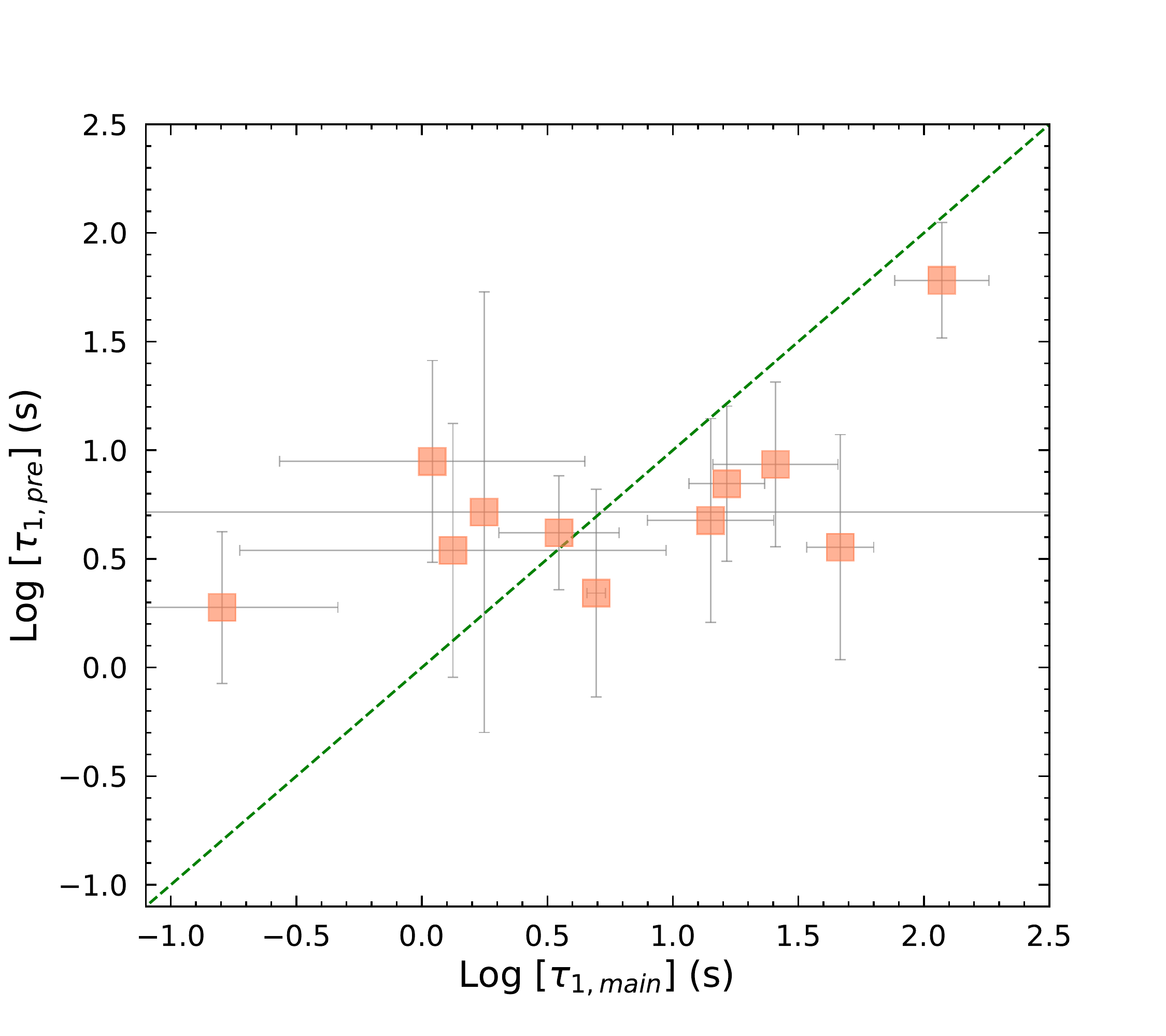}}
\subfigure[$r = 0.32, P = 0.35$]{
\includegraphics[height=4.5cm,width=5cm]{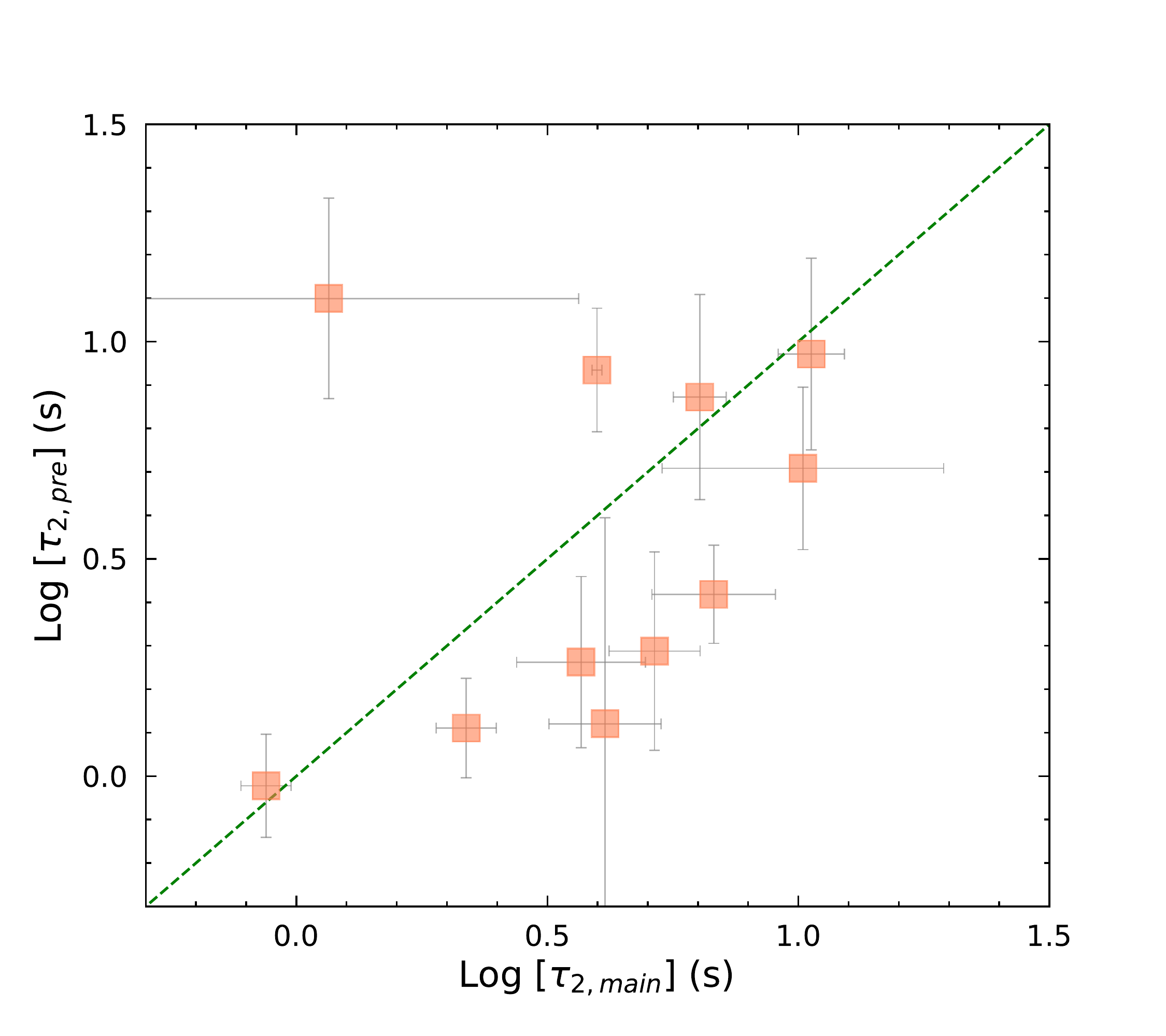}}
\subfigure[$r = 0.60, P = 0.05$]{
\includegraphics[height=4.5cm,width=5cm]{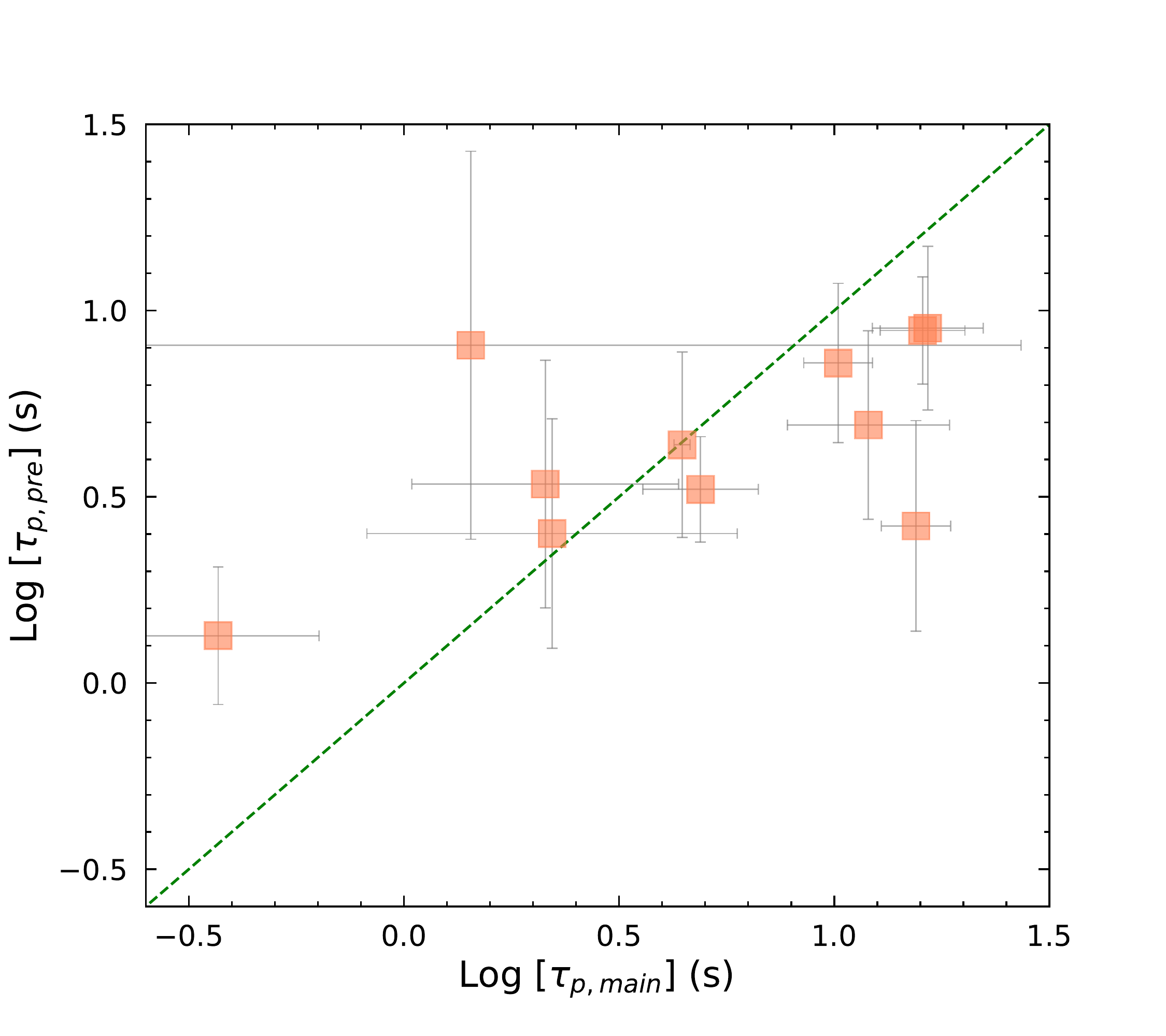}}
\subfigure[$r = 0.39, P = 0.24$]{
\includegraphics[height=4.5cm,width=5cm]{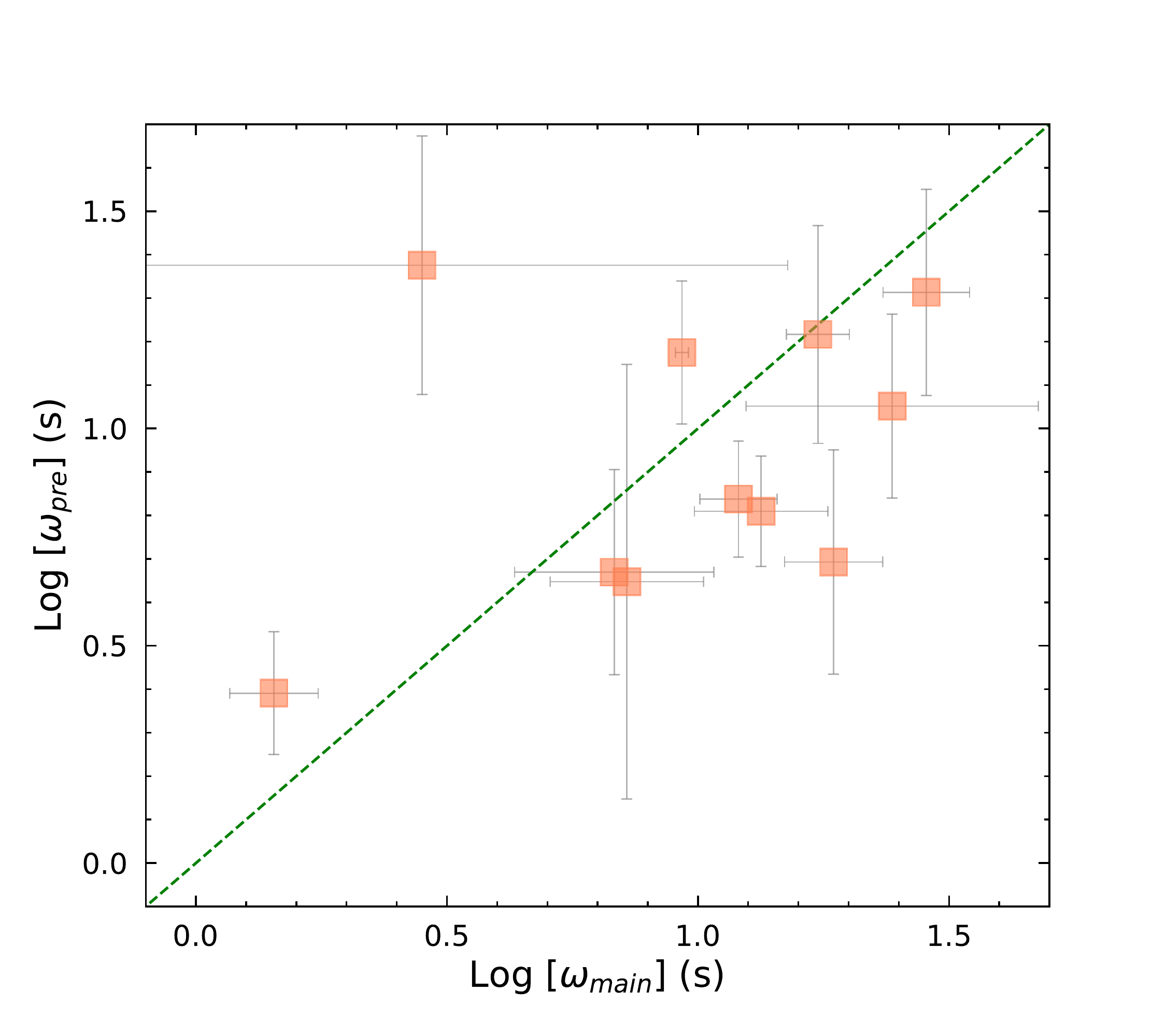}}
\subfigure[$r = 0.49, P = 0.13$]{
\includegraphics[height=4.5cm,width=5cm]{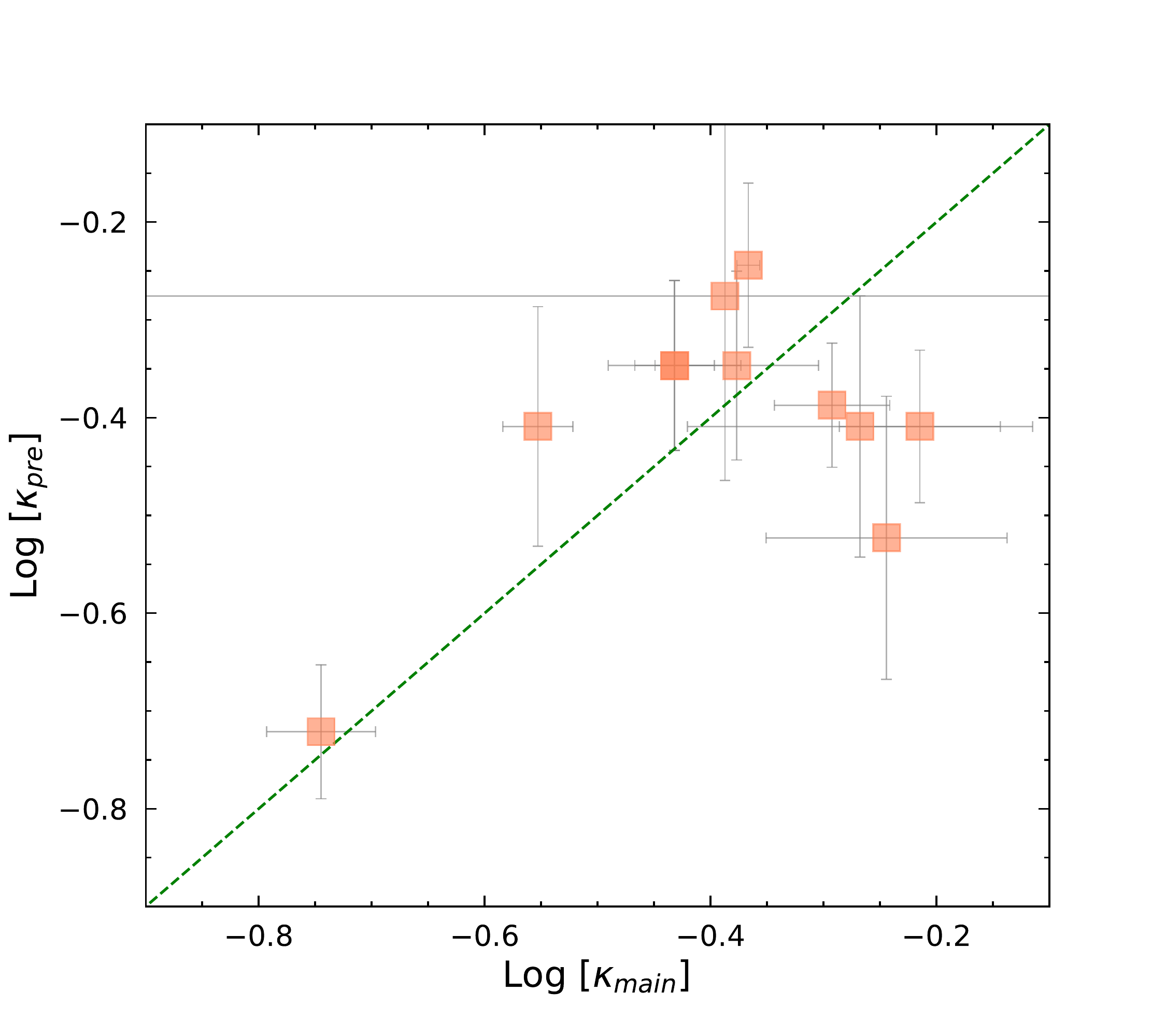}}
\subfigure[$r = 0.51, P = 0.11$]{
\includegraphics[height=4.5cm,width=5cm]{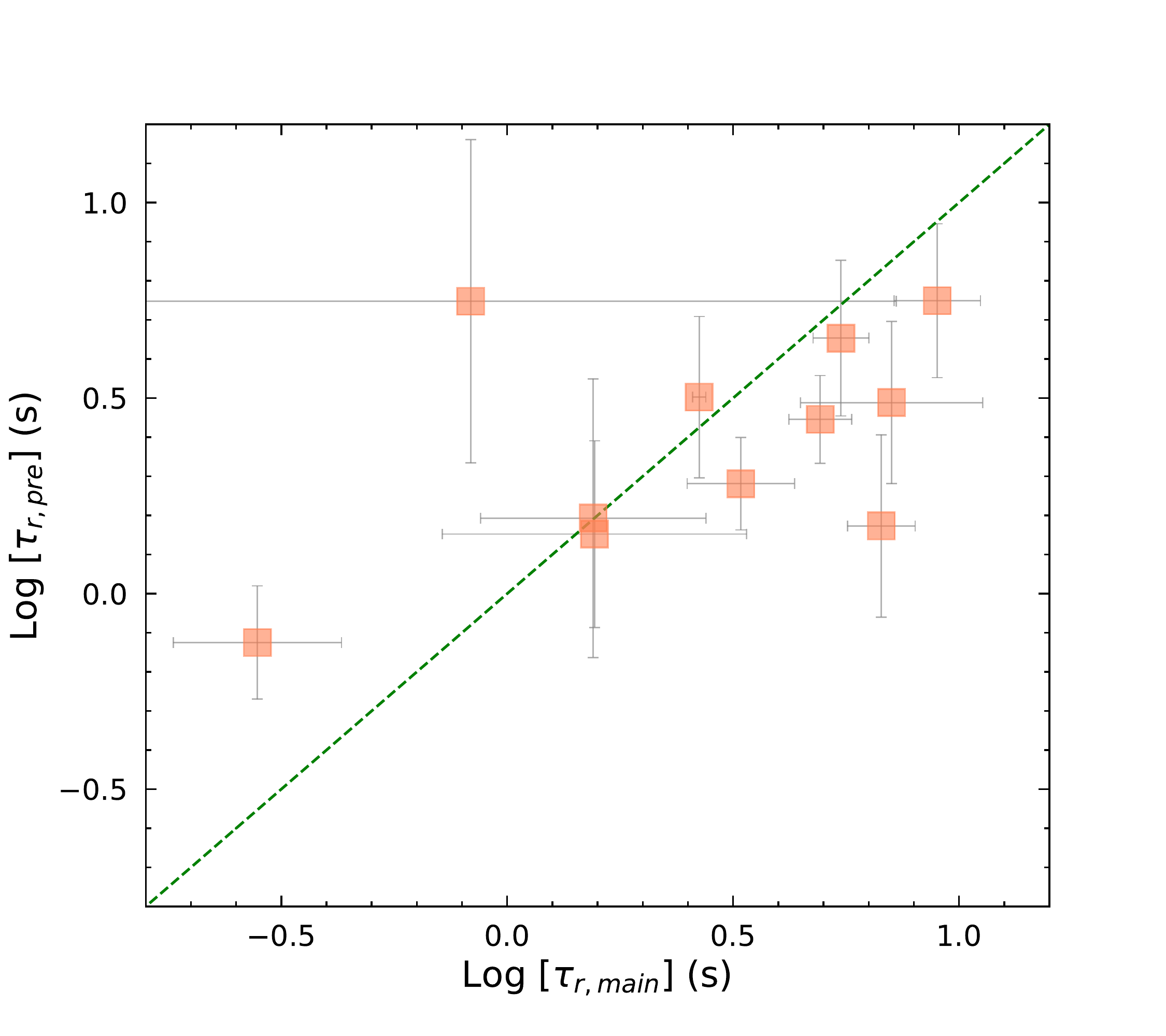}}
\subfigure[$r = 0.35, P = 0.30$]{
\includegraphics[height=4.5cm,width=5cm]{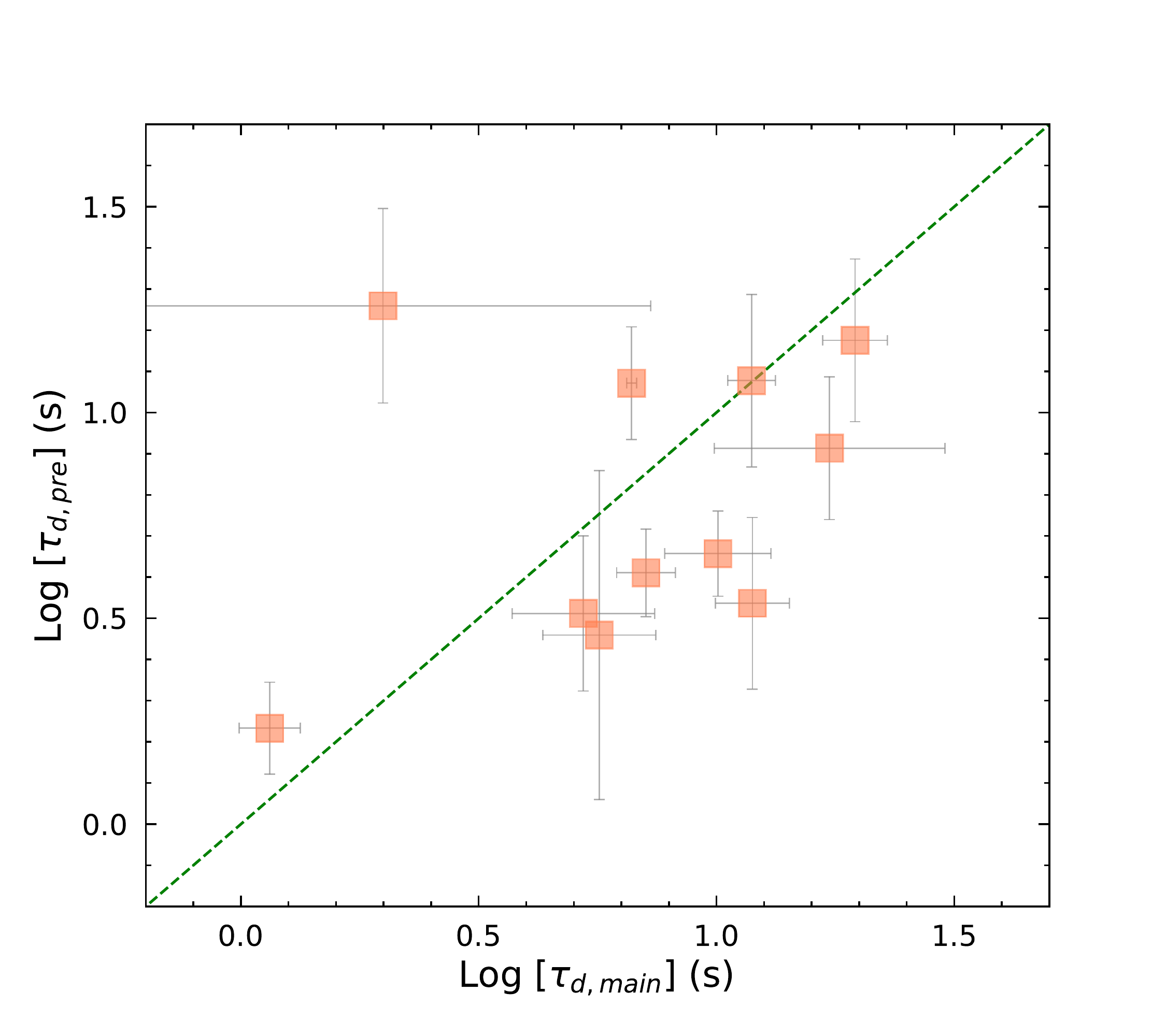}}
\caption{Comparisons of the episode parameters of the precursors to those of the main bursts for 14 special GRBs. Each GRB contains only one precursor episode and only one main-burst episode. The numbers have no redshift correction.
The green dashed lines represent for the case that the number in the X-axis is equal to the number in the Y-axis.
The subscript note ``main" indicates the main burst, and the subscript note ``pre" indicates the precursor.
$r$ and $P$ represent the Pearson correlation coefficient and the probability of chance correlation, respectively.}
\end{figure}
\clearpage

\begin{figure}[!htp]
\centering
\subfigure[$r = 0.31, P = 0.35$]{
\includegraphics[height=4.5cm,width=5cm]{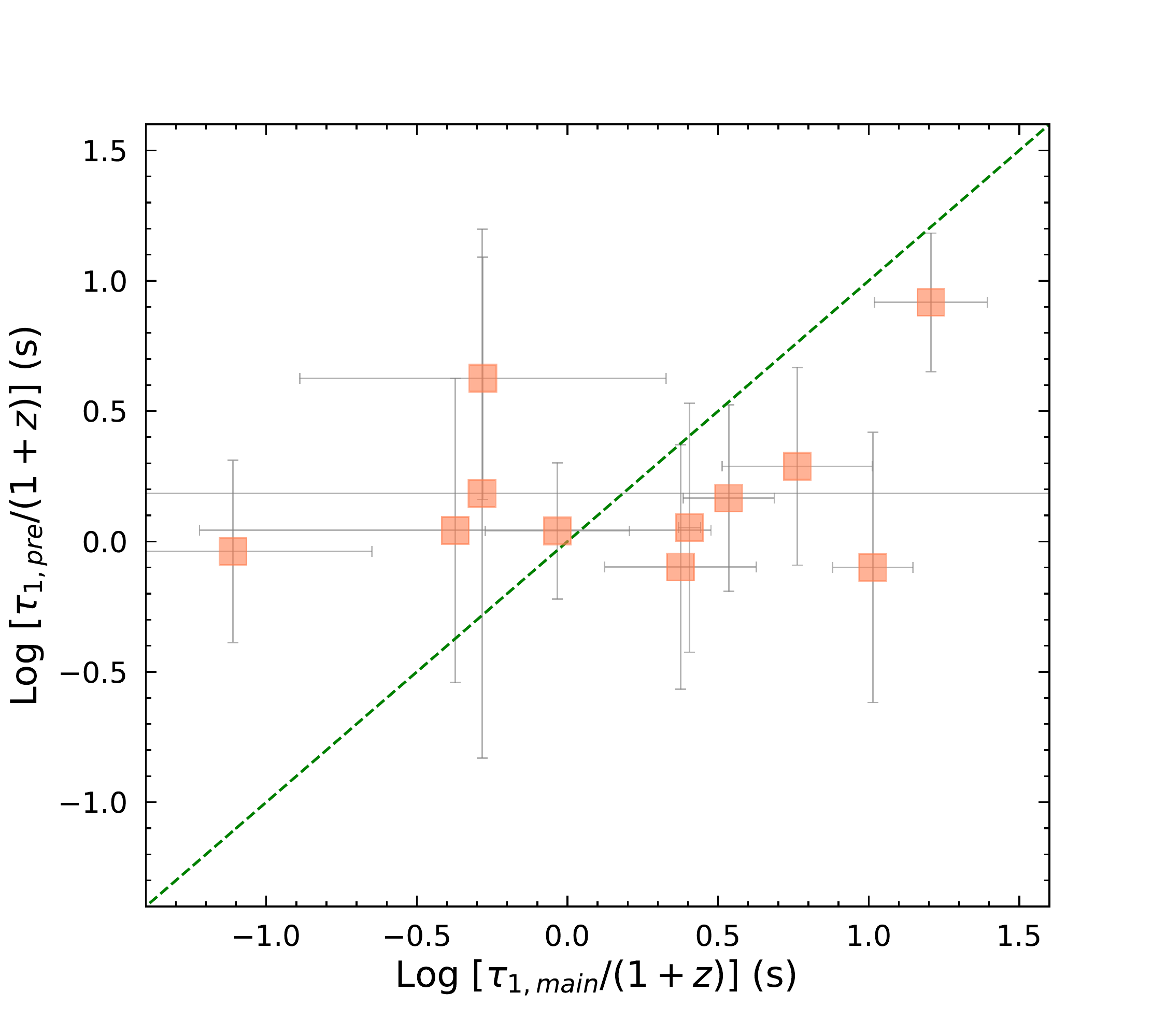}}
\subfigure[$r = 0.33, P = 0.32$]{
\includegraphics[height=4.5cm,width=5cm]{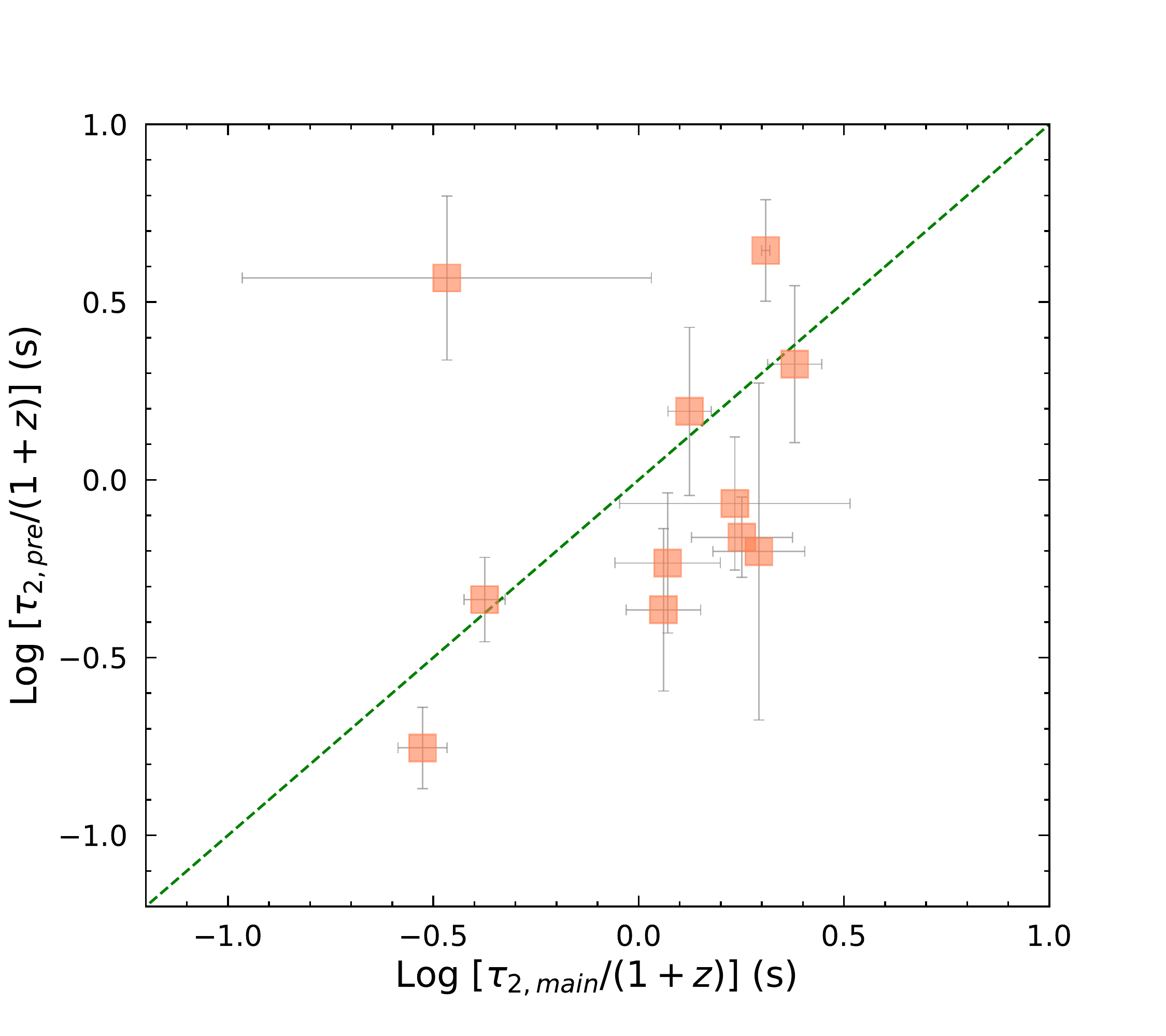}}
\subfigure[$r = 0.16, P = 0.63$]{
\includegraphics[height=4.5cm,width=5cm]{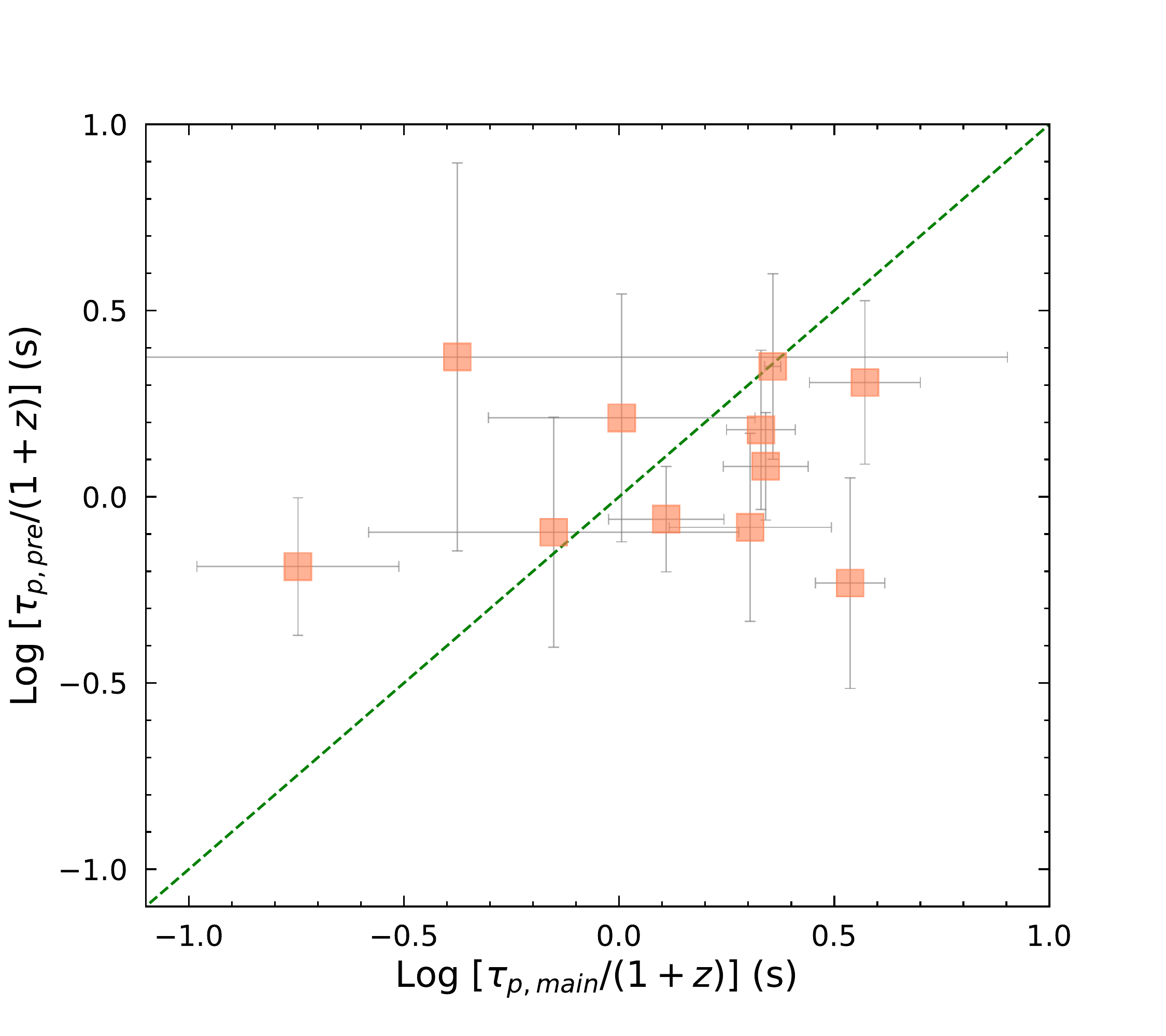}}
\subfigure[$r = 0.20, P = 0.56$]{
\includegraphics[height=4.5cm,width=5cm]{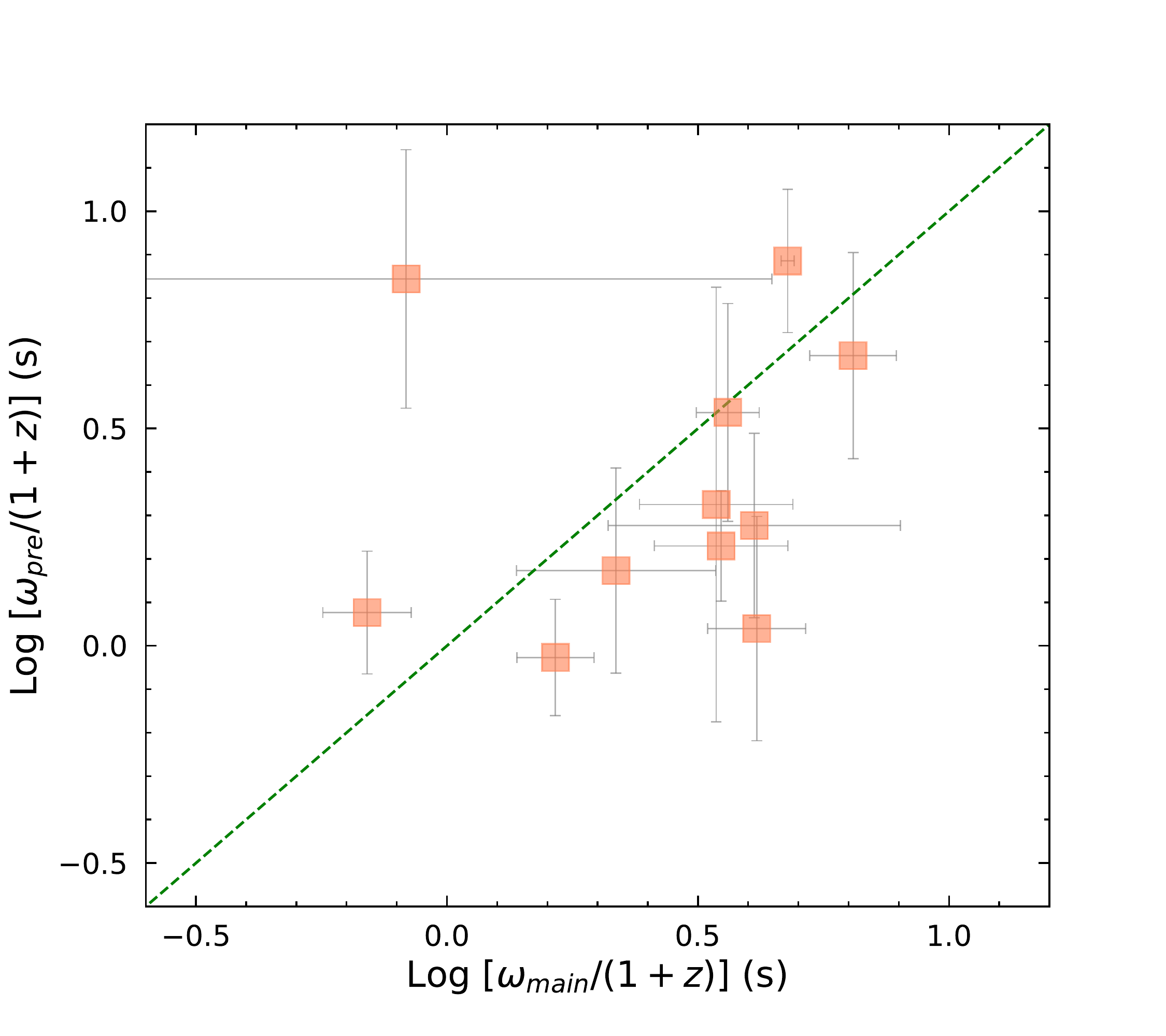}}
\subfigure[$r = 0.19, P = 0.58$]{
\includegraphics[height=4.5cm,width=5cm]{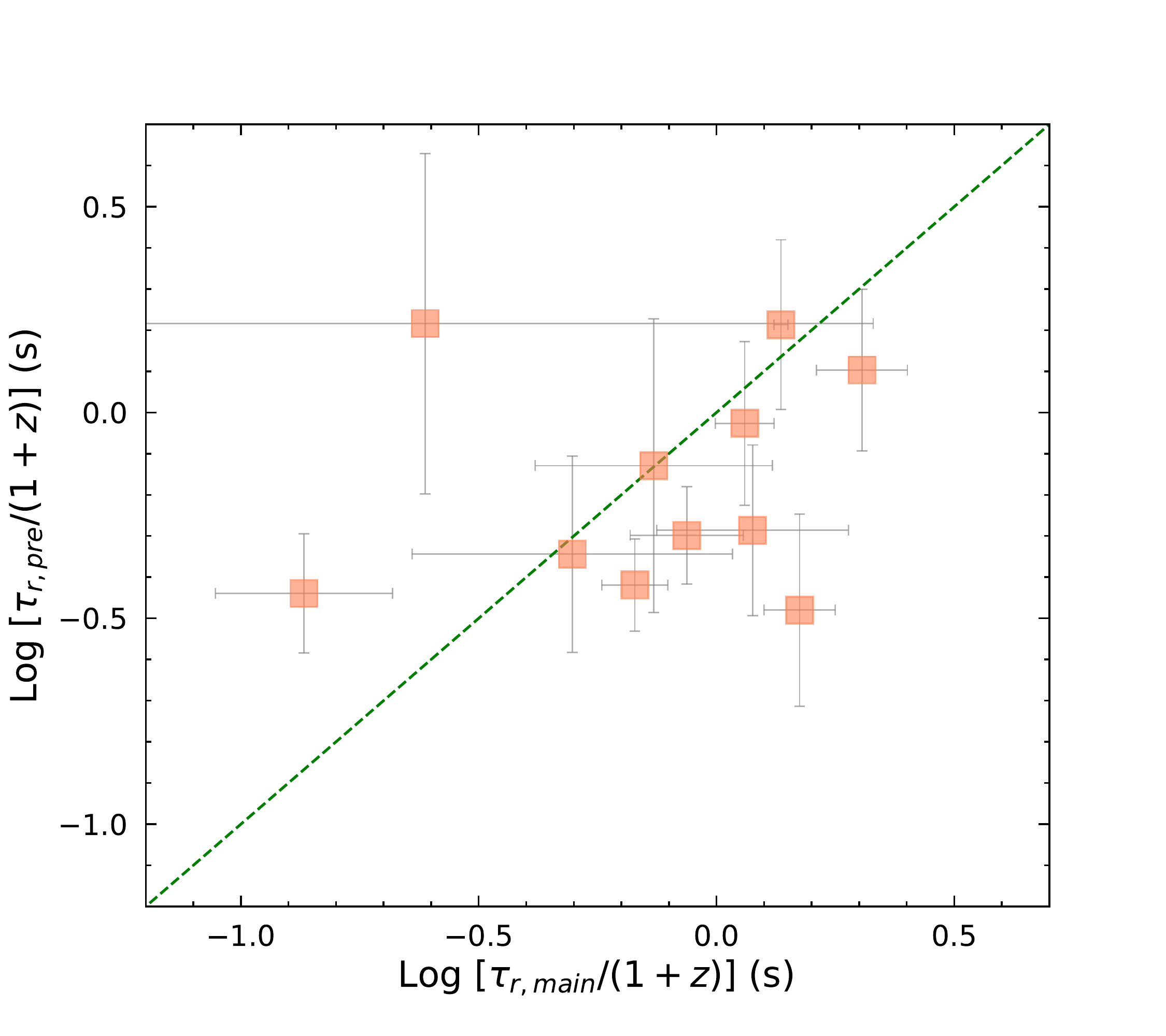}}
\subfigure[$r = 0.22, P = 0.52$]{
\includegraphics[height=4.5cm,width=5cm]{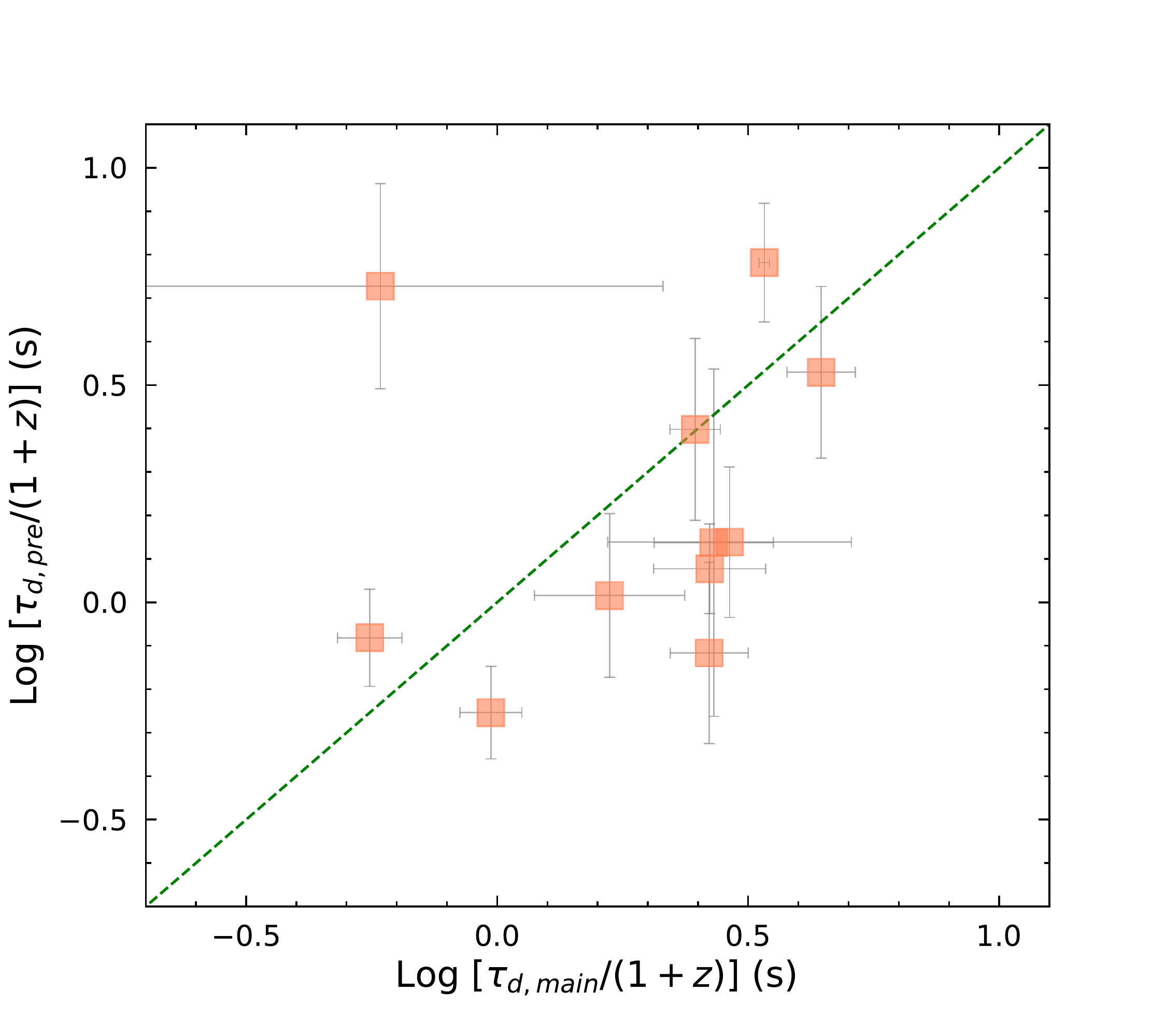}}
\caption{Comparisons of the episode parameters of the precursors to those of the main bursts for 14 special GRBs. Each GRB contains only one precursor episode and only one main-burst episode. The numbers have redshift correction.
The green dashed lines represent for the case that the number in the X-axis is equal to the number in the Y-axis.
The subscript note ``main" indicates the main burst, and the subscript note ``pre" indicates the precursor.
$r$ and $P$ represent for the Pearson correlation coefficient and the probability of chance correlation, respectively.}
\end{figure}
\clearpage

\begin{figure}[!htp]
\centering
\subfigure{
\includegraphics[height=6.5cm,width=7cm]{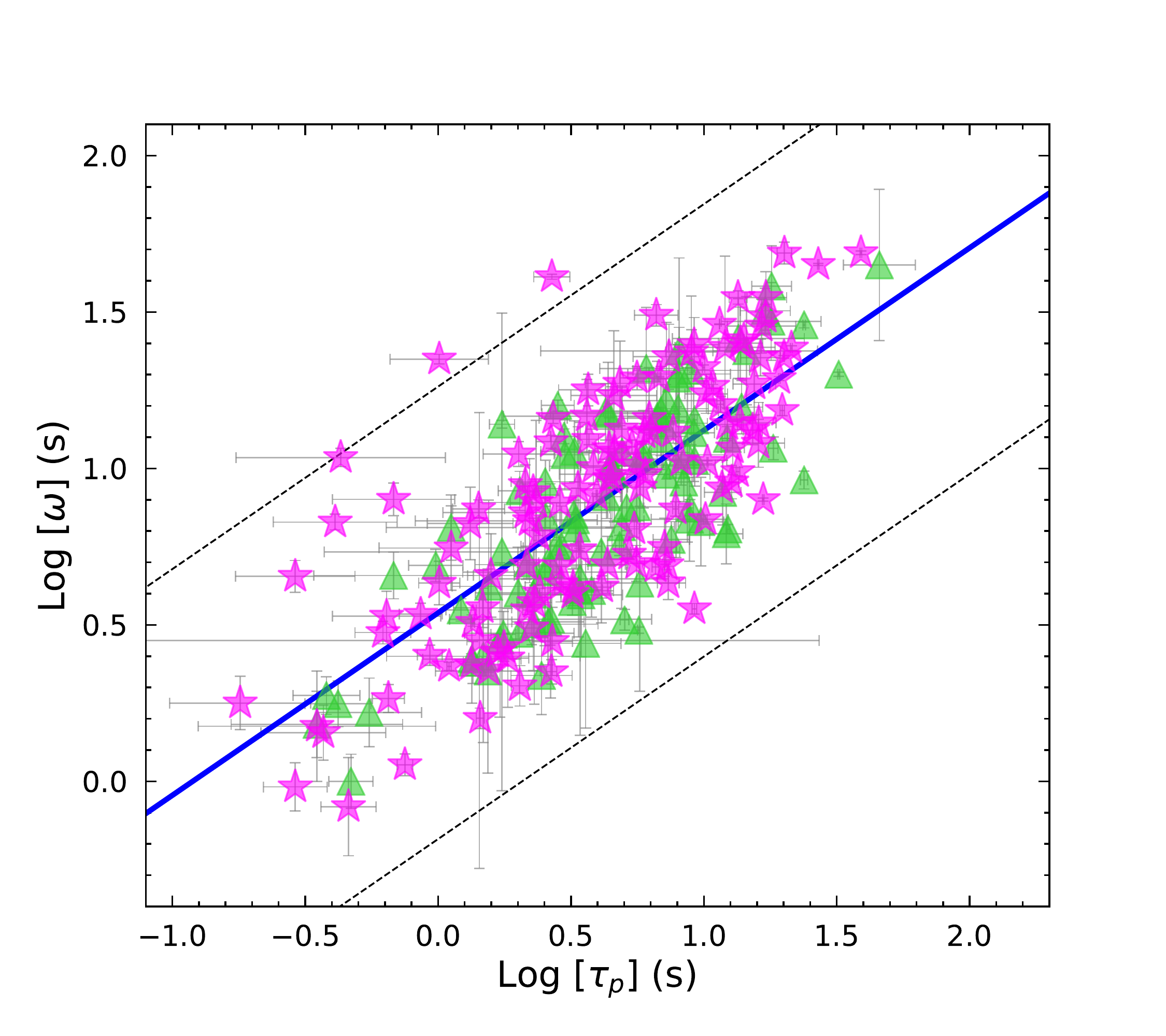}}
\subfigure{
\includegraphics[height=6.5cm,width=7cm]{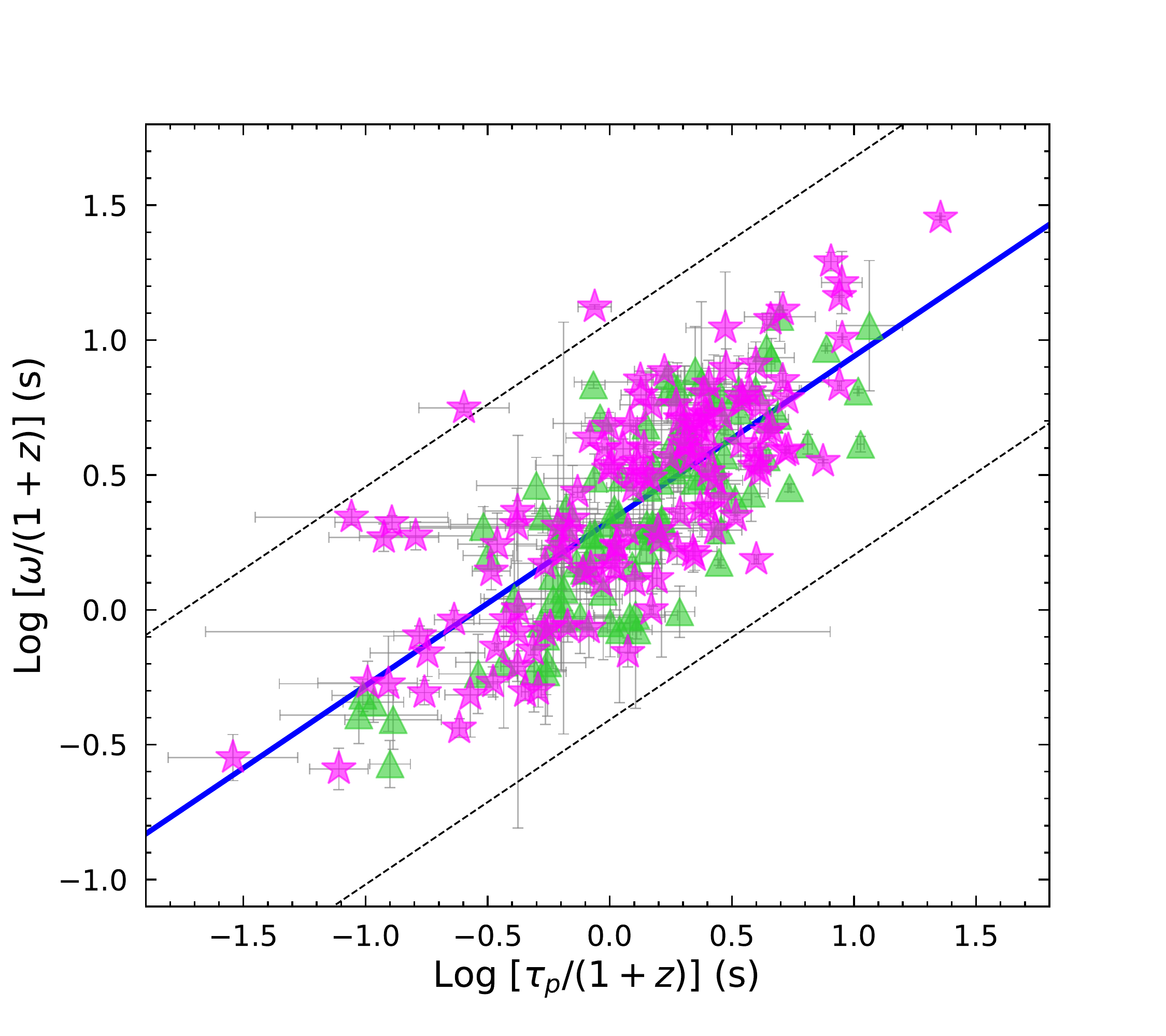}}
\caption{Left panel: the correlation between the width and the peak time without redshift correction.
The blue solid line represents for the best fitting
as log $\omega$ = (0.54$^{+0.03}_{-0.03}$) + (0.58$^{+0.04}_{-0.04}$)log $\tau_{p}$.
The black dashed lines enclose the data within 3$\sigma$, and $\sigma$ = 0.24$^{+0.02}_{-0.02}$ is
the extrinsic scatter.
Green triangle and pink star in the panel correspond to precursor and main burst, respectively.
Right panel: the correlation between the pulse width and the peak time after redshift correction.
The blue solid line represents for the best fitting as 
log $\omega$ = (0.33$^{+0.02}_{-0.02}$) + (0.61$^{+0.04}_{-0.04}$)log $\tau_{p}$.
The black dashed lines enclose the data within 3$\sigma$, and $\sigma$ = 0.25$^{+0.01}_{-0.01}$ is the extrinsic scatter.
Green triangle and pink star in the panel correspond to precursor and main burst, respectively.}
\end{figure}
\clearpage

\begin{figure}[!htp]
\centering
\includegraphics[height=8.5cm,width=10cm]{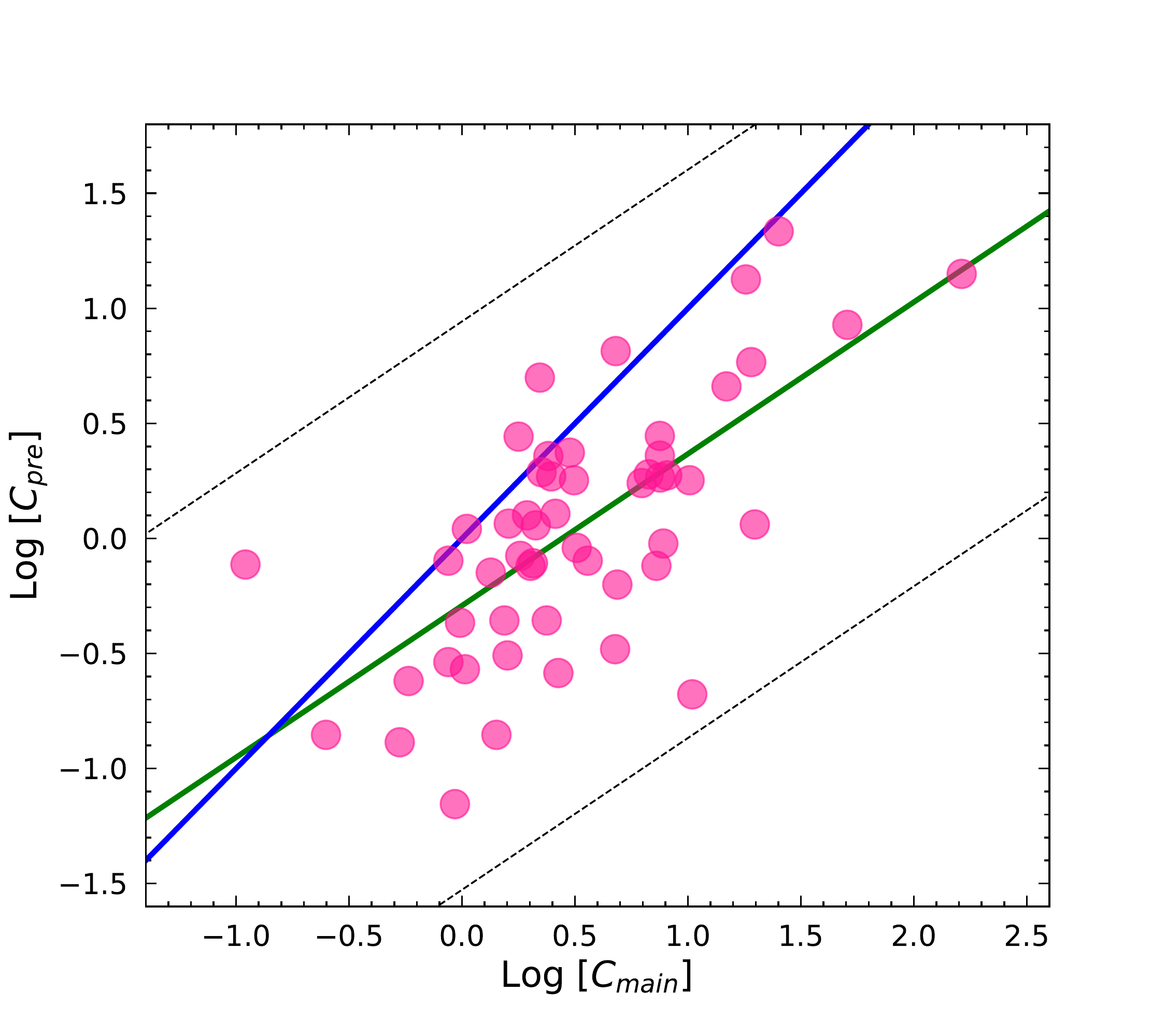}
\caption{The correlation between the total photon count of the precursor to the photon count of the main burst.
The blue solid line represents for the case that the number in the X-axis is equal to the number in the Y-axis.
The green solid line represents for the best fitting as 
log $C_p$ = ($-$0.29$^{+0.08}_{-0.08}$) + (0.66$^{+0.10}_{-0.10}$)log $C_m$.
The black dashed lines enclose the data within 3$\sigma$, and $\sigma$ = 0.41$^{+0.04}_{-0.03}$
is the extrinsic scatter. The subscript note ``main" indicates the main burst, and the subscript note ``pre" indicates the precursor.}
\end{figure}
\clearpage

\begin{figure}[!htp]
\centering
\subfigure{
\includegraphics[height=6.cm,width=7cm]{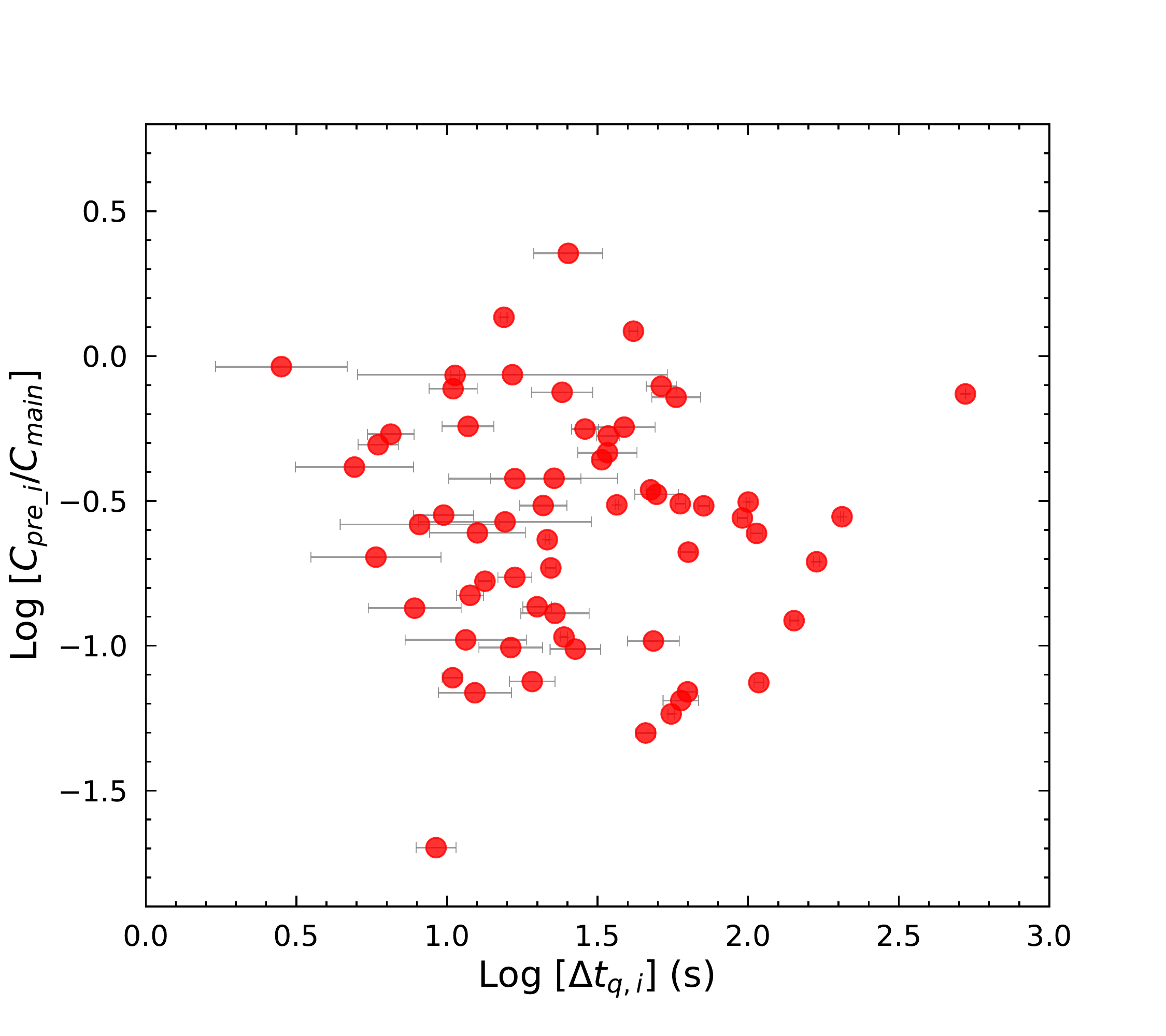}}
\subfigure{
\includegraphics[height=6.cm,width=7cm]{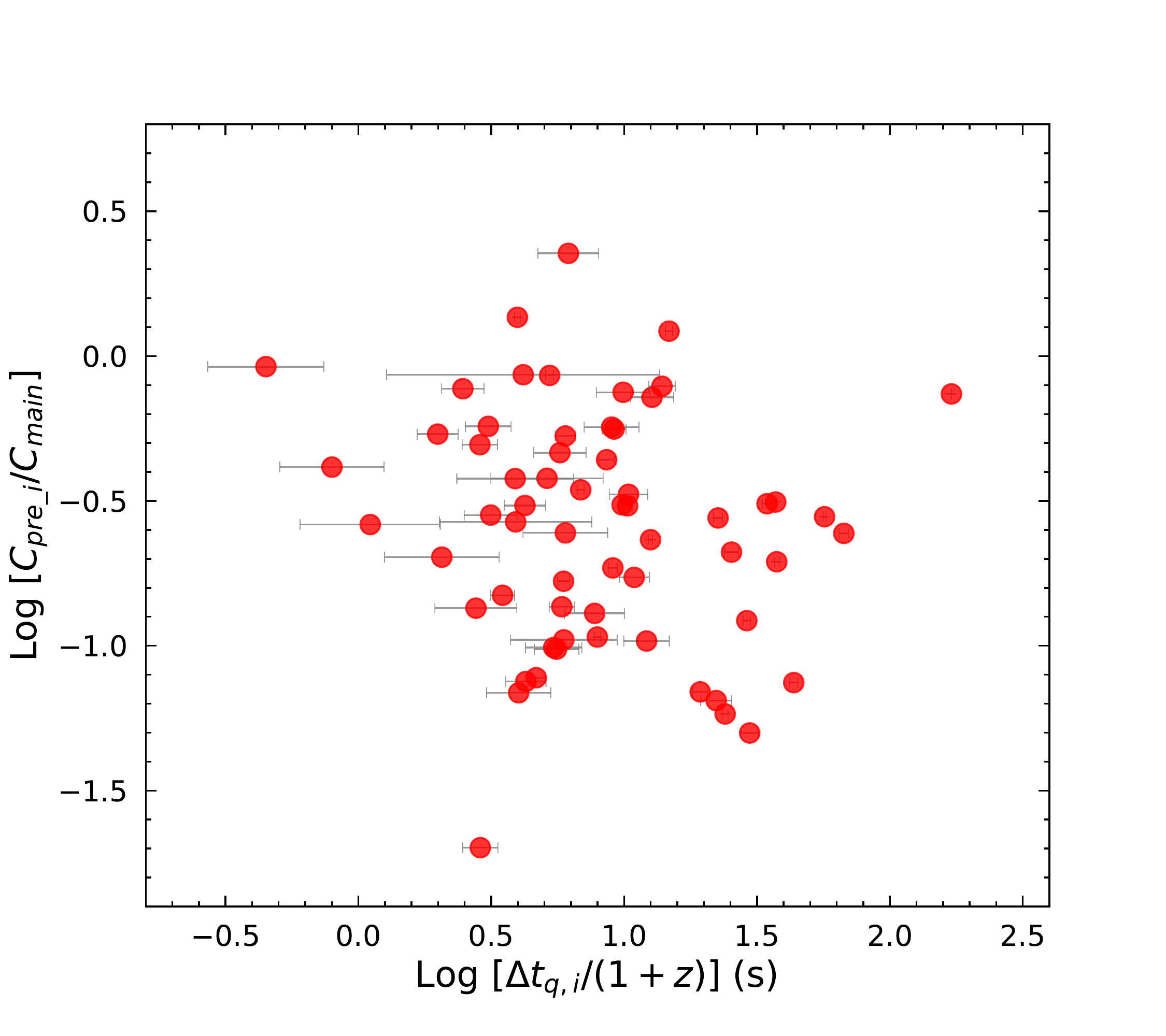}}
\subfigure{
\includegraphics[height=6.cm,width=7cm]{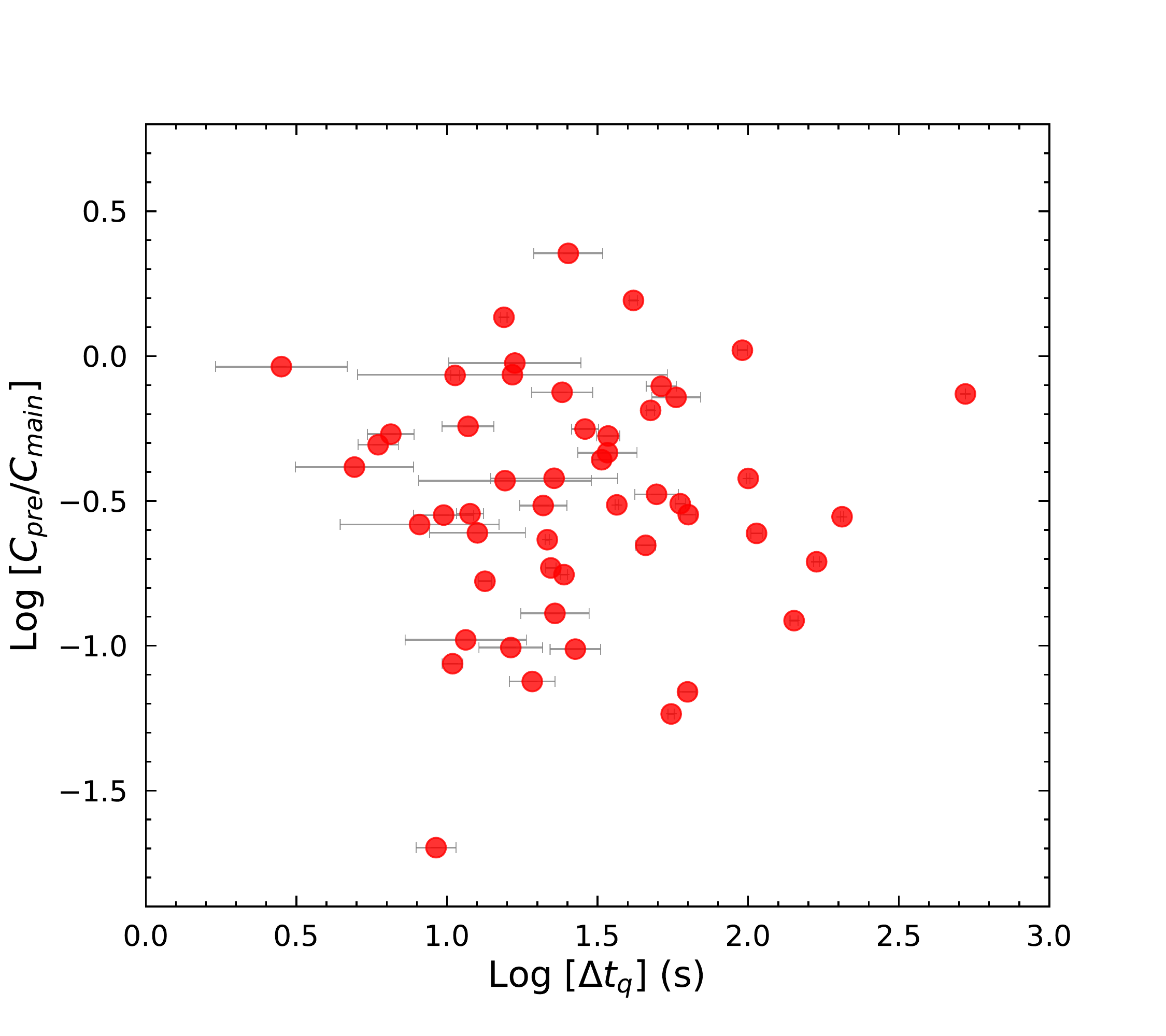}}
\subfigure{
\includegraphics[height=6.cm,width=7cm]{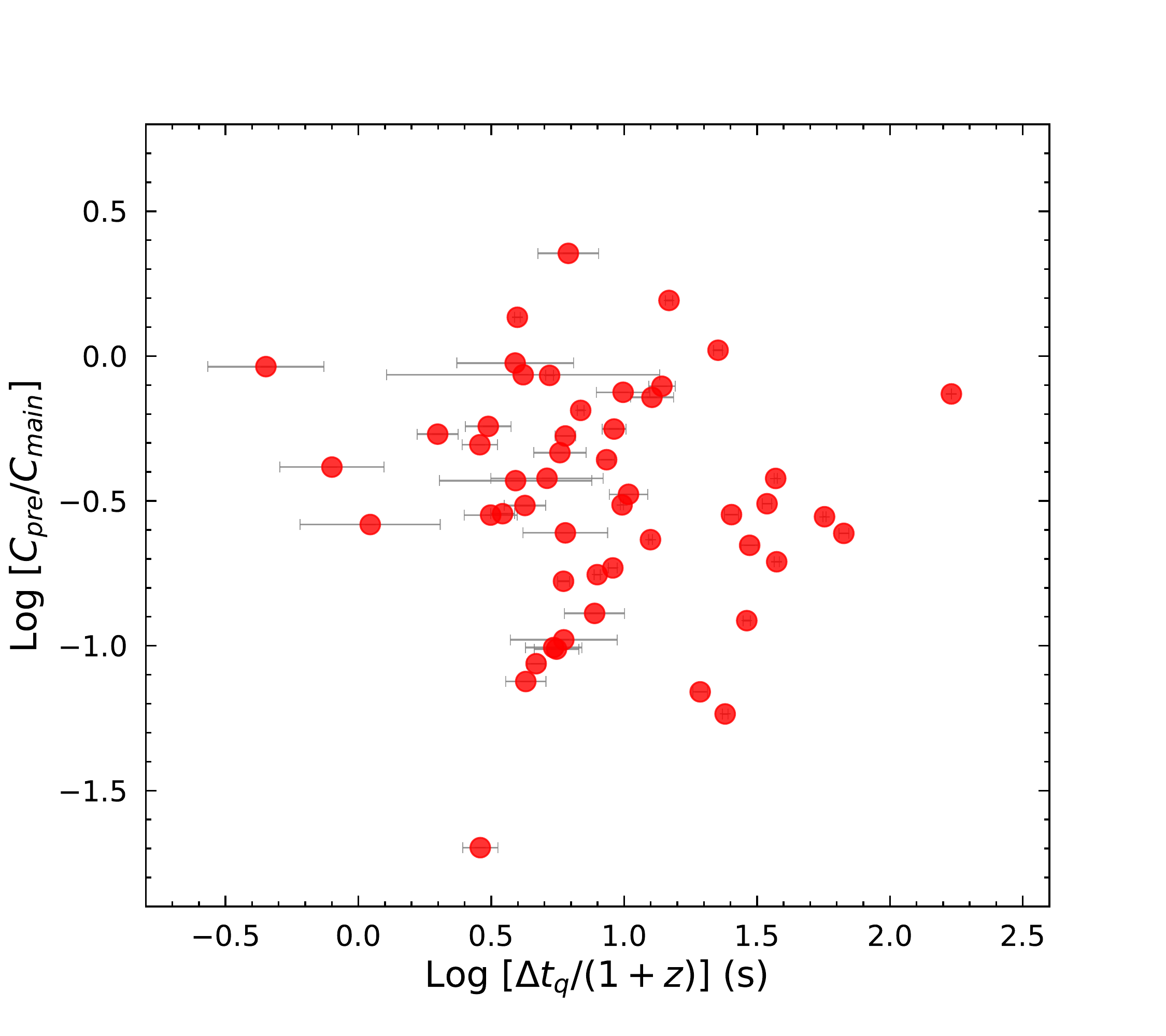}}
\subfigure{
\includegraphics[height=6.cm,width=7cm]{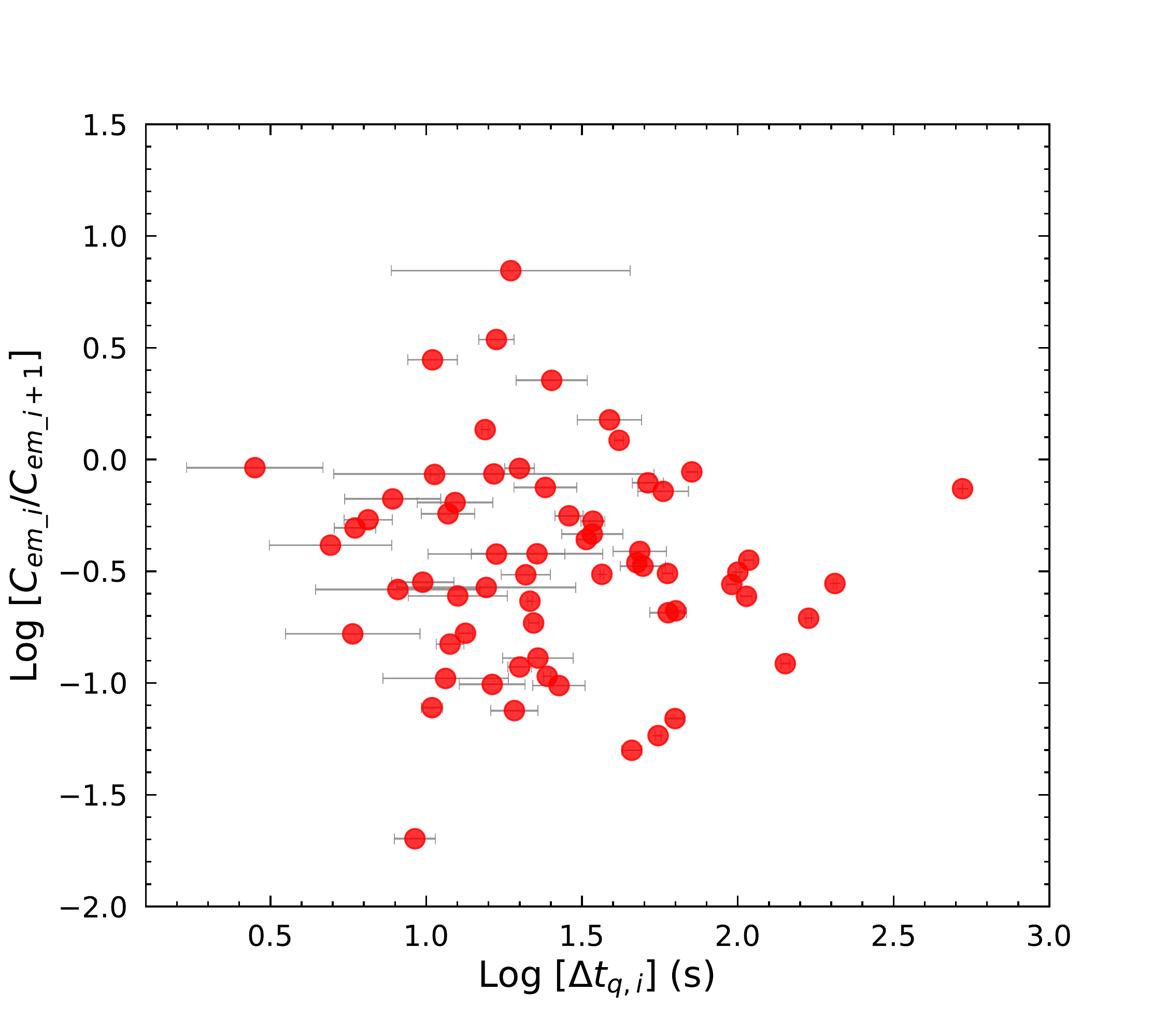}}
\subfigure{
\includegraphics[height=6.cm,width=7cm]{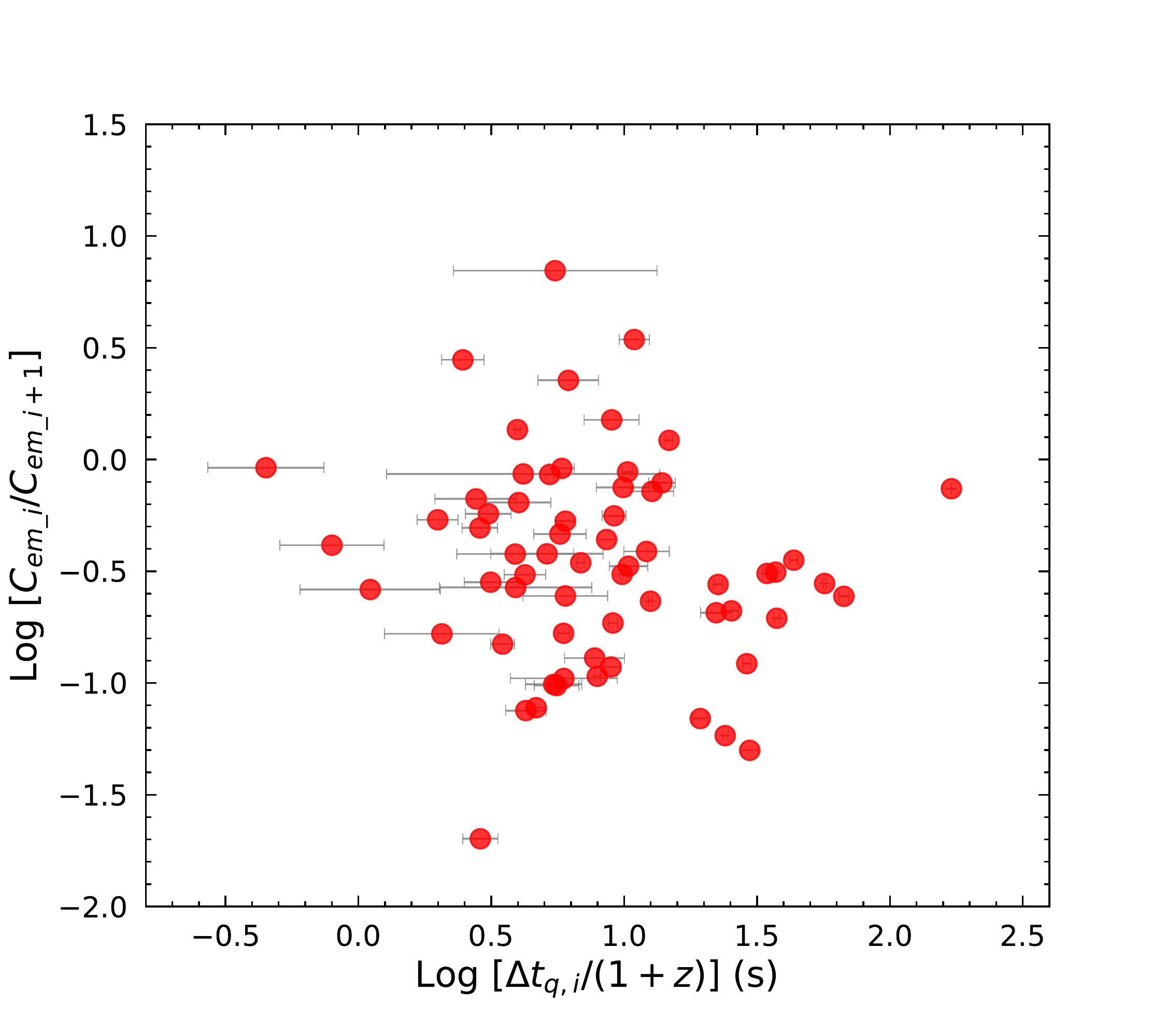}}
\caption{\footnotesize{Upper-left panel: the ratio between the photon count of each precursor and that of the main burst vs. the quiescent time
$\Delta_{q,i}$. Upper-right panel: the ratio between the photon count of each precursor and that of the main burst vs. the quiescent time $\Delta_{q,i}/(1+z)$. Middle-left panel: the ratio between the photon count of all the precursors and that of the main burst vs. the quiescent time $\Delta_{q}$. Middle-right panel: the ratio between the photon count of all the precursors and that of the main burst vs. the quiescent time $\Delta_{q}/(1+z)$.
Lower-left panel: the ratio between the photon count of one pulse and that of the subsequent pulse vs. the quiescent time
$\Delta_{q,i}$.
Lower-right panel: the ratio between the photon count of one pulse and that of the subsequent pulse vs. the quiescent time
$\Delta_{q,i}/(1+z)$.}}
\end{figure}
\clearpage

\begin{figure}[!htp]
\centering
\subfigure{
\includegraphics[height=6.cm,width=7cm]{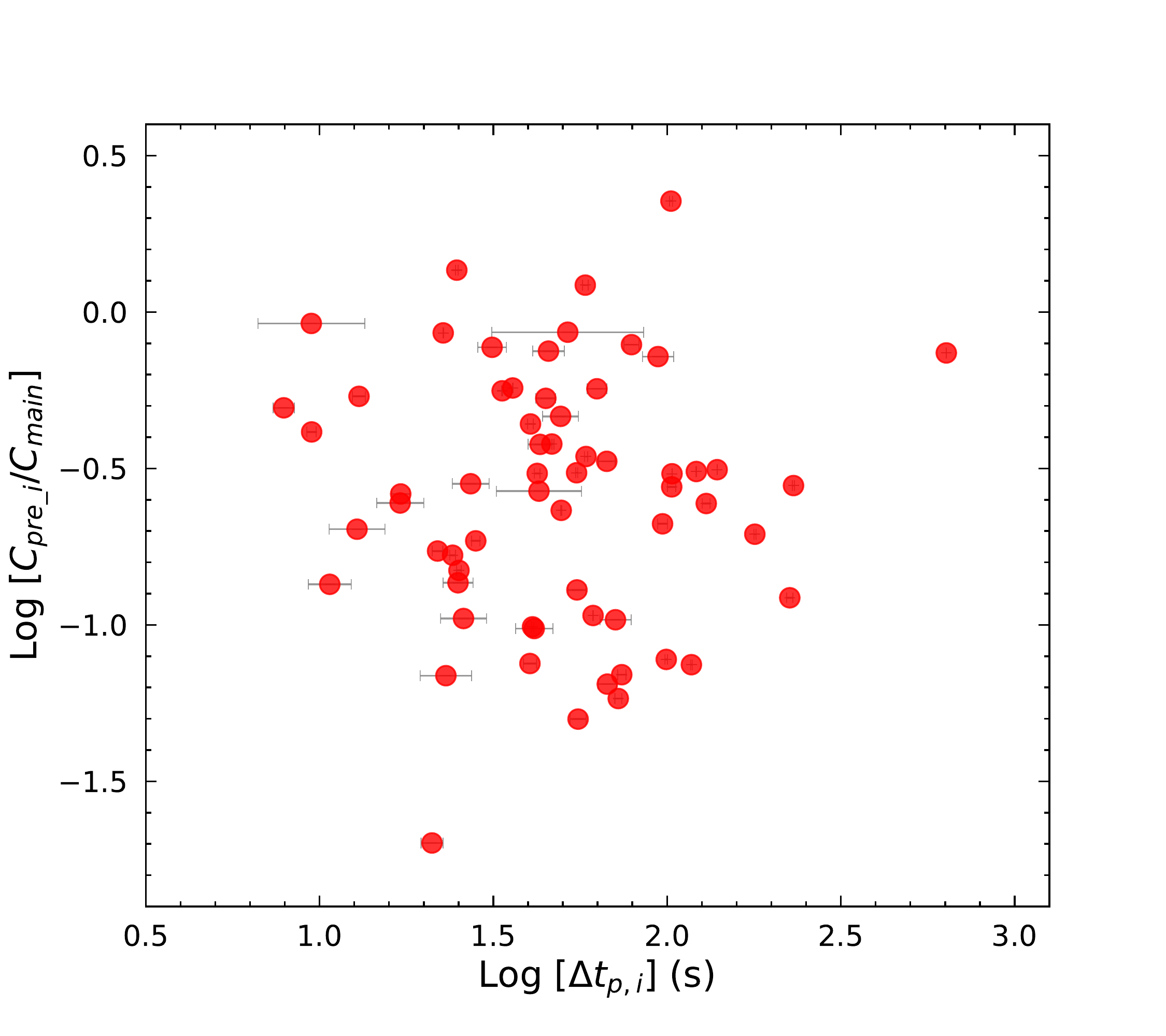}}
\subfigure{
\includegraphics[height=6.cm,width=7cm]{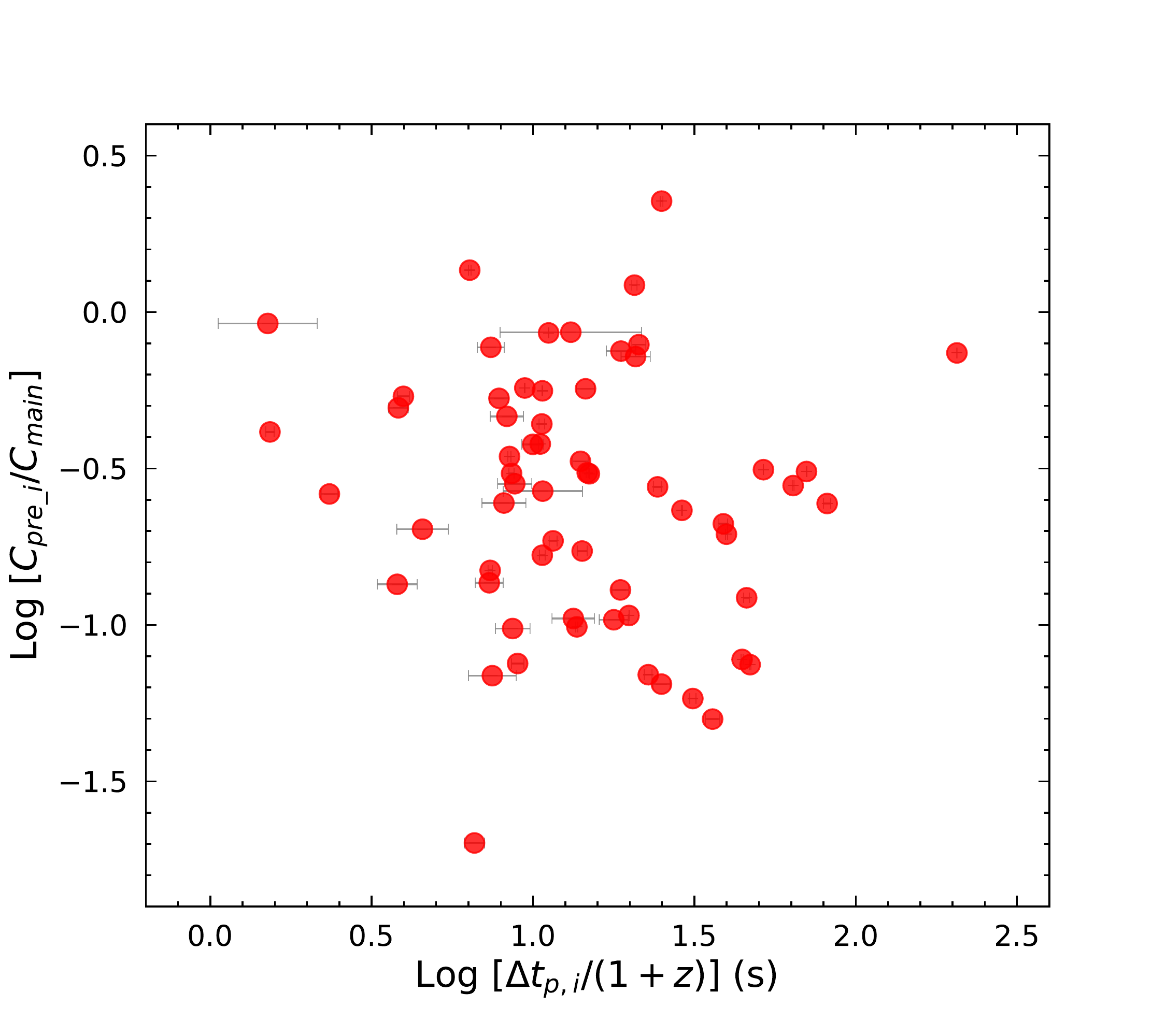}}
\subfigure{
\includegraphics[height=6.cm,width=7cm]{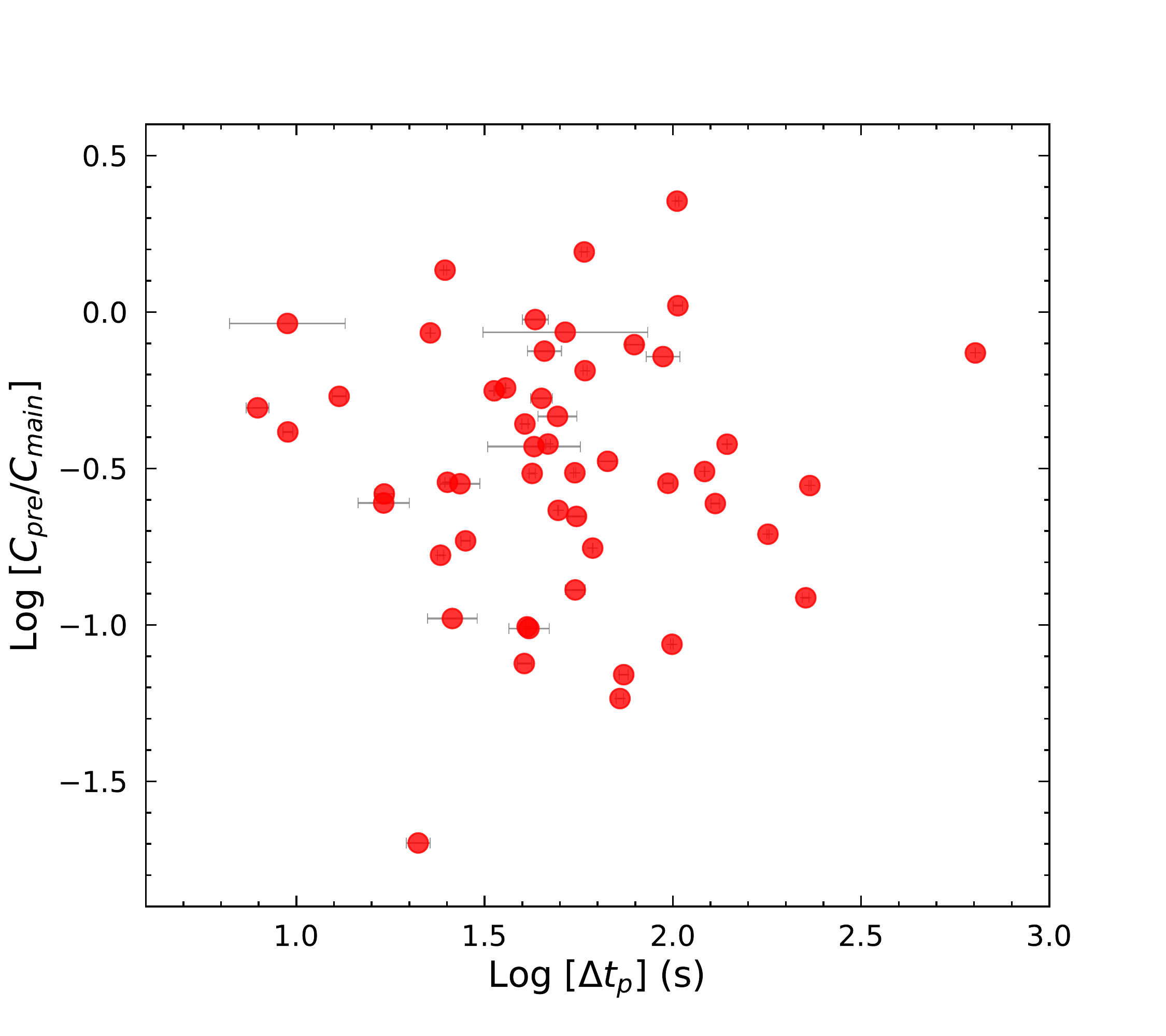}}
\subfigure{
\includegraphics[height=6.cm,width=7cm]{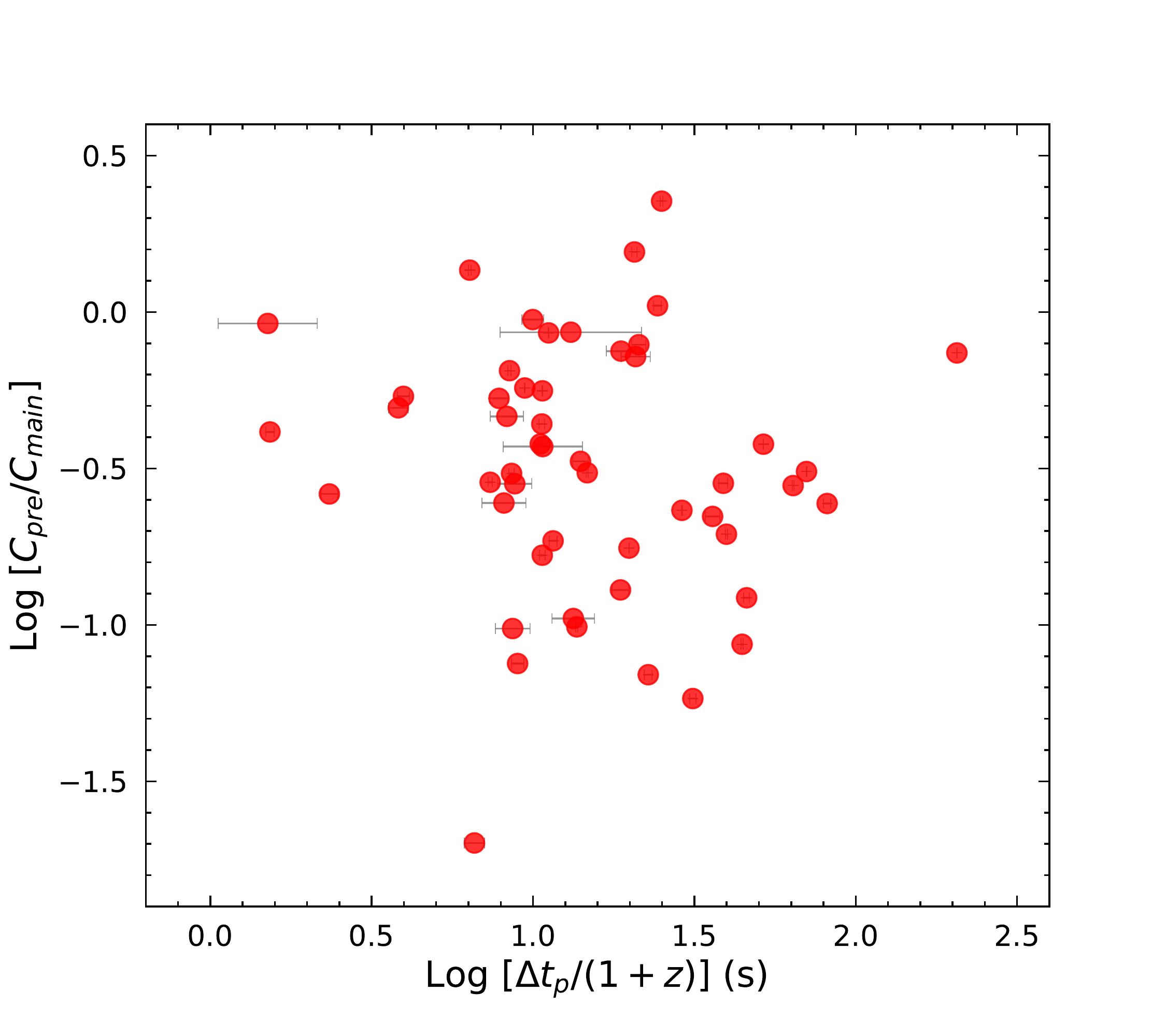}}
\subfigure{
\includegraphics[height=6.cm,width=7cm]{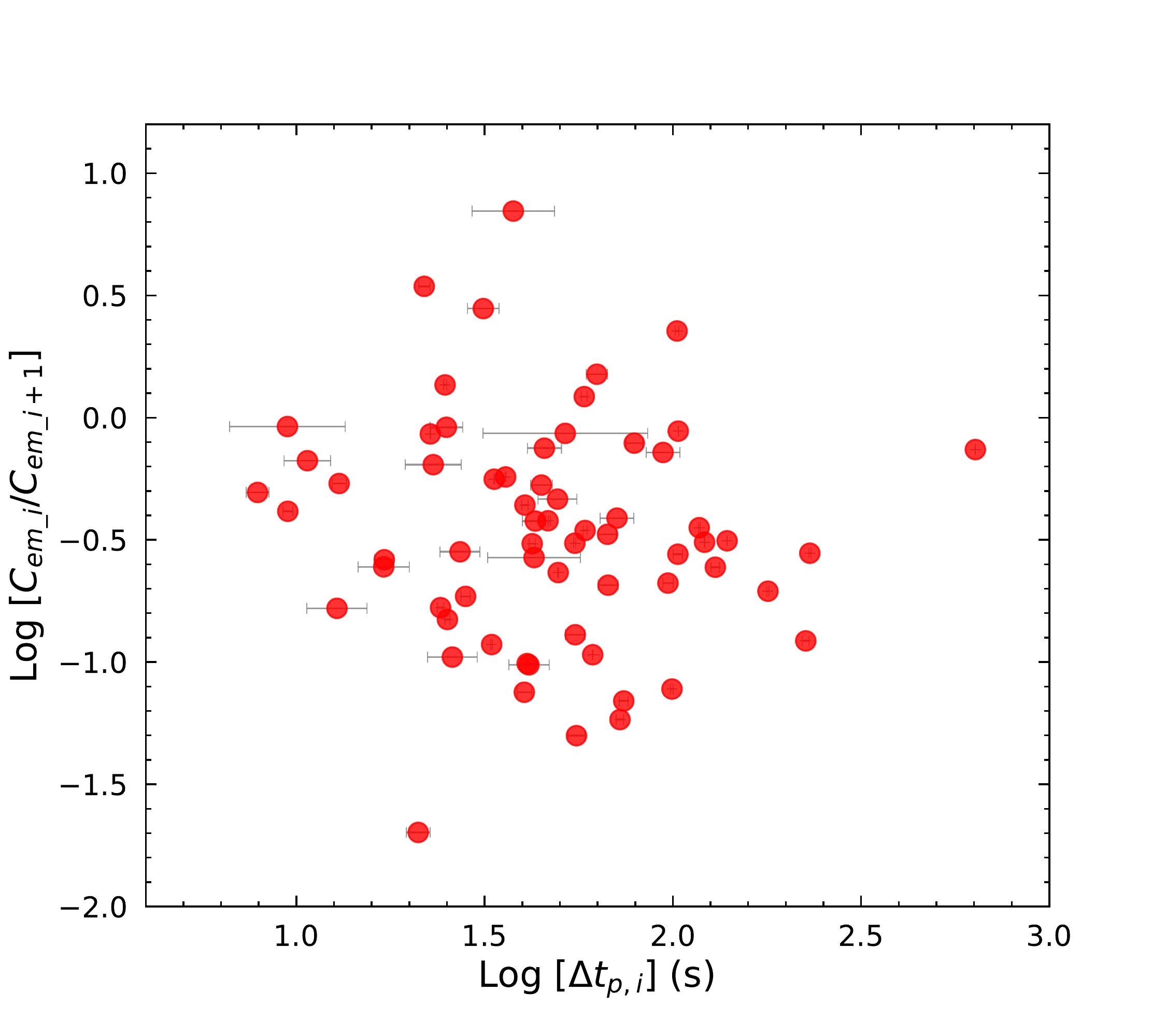}}
\subfigure{
\includegraphics[height=6.cm,width=7cm]{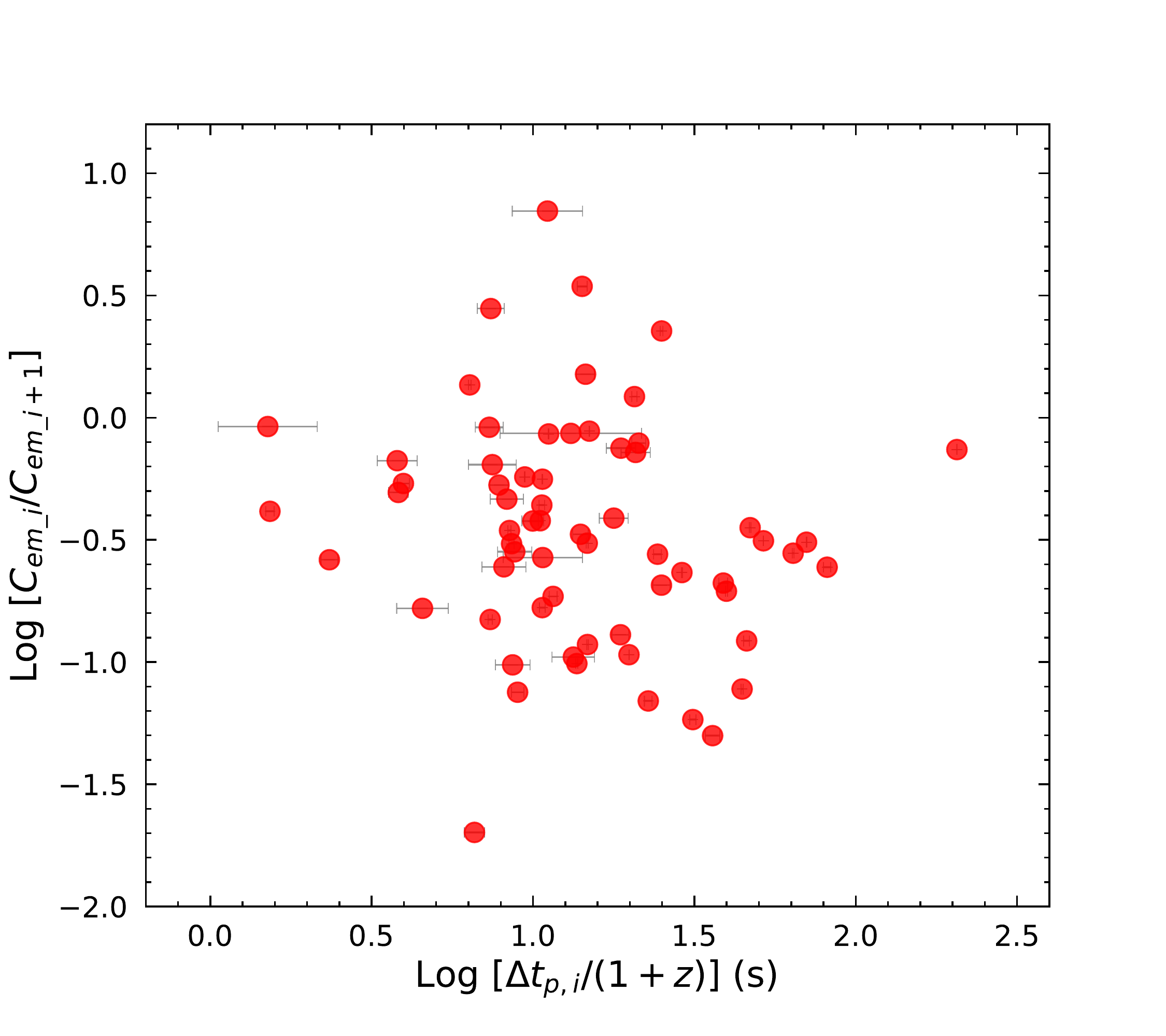}}
\caption{\footnotesize{Upper-left panel: the ratio between the photon count of each precursor and that of the main burst vs. the peak-time interval $\Delta_{p,i}$. Upper-right panel: the ratio between the photon count of each precursor and that of the main burst vs. the peak-time interval $\Delta_{p,i}/(1+z)$. Middle-left panel: the ratio between the photon count of all the precursors and that of the main burst vs. the peak-time interval $\Delta_{p}$. Middle-right panel: the ratio between the photon count of all the precursors and that of the main burst vs. the peak-time interval $\Delta_{p}/(1+z)$.
Lower-left panel: the ratio between the photon count of one pulse and that of the subsequent pulse vs. the peak-time interval
$\Delta_{p,i}$.
Lower-right panel: the ratio between the photon count of one pulse and that of the subsequent pulse vs. the peak-time interval
$\Delta_{p,i}/(1+z)$.}}
\end{figure}
\clearpage

\begin{figure}[!htp]
\centering
\subfigure{
\includegraphics[height=7.cm,width=8.cm]{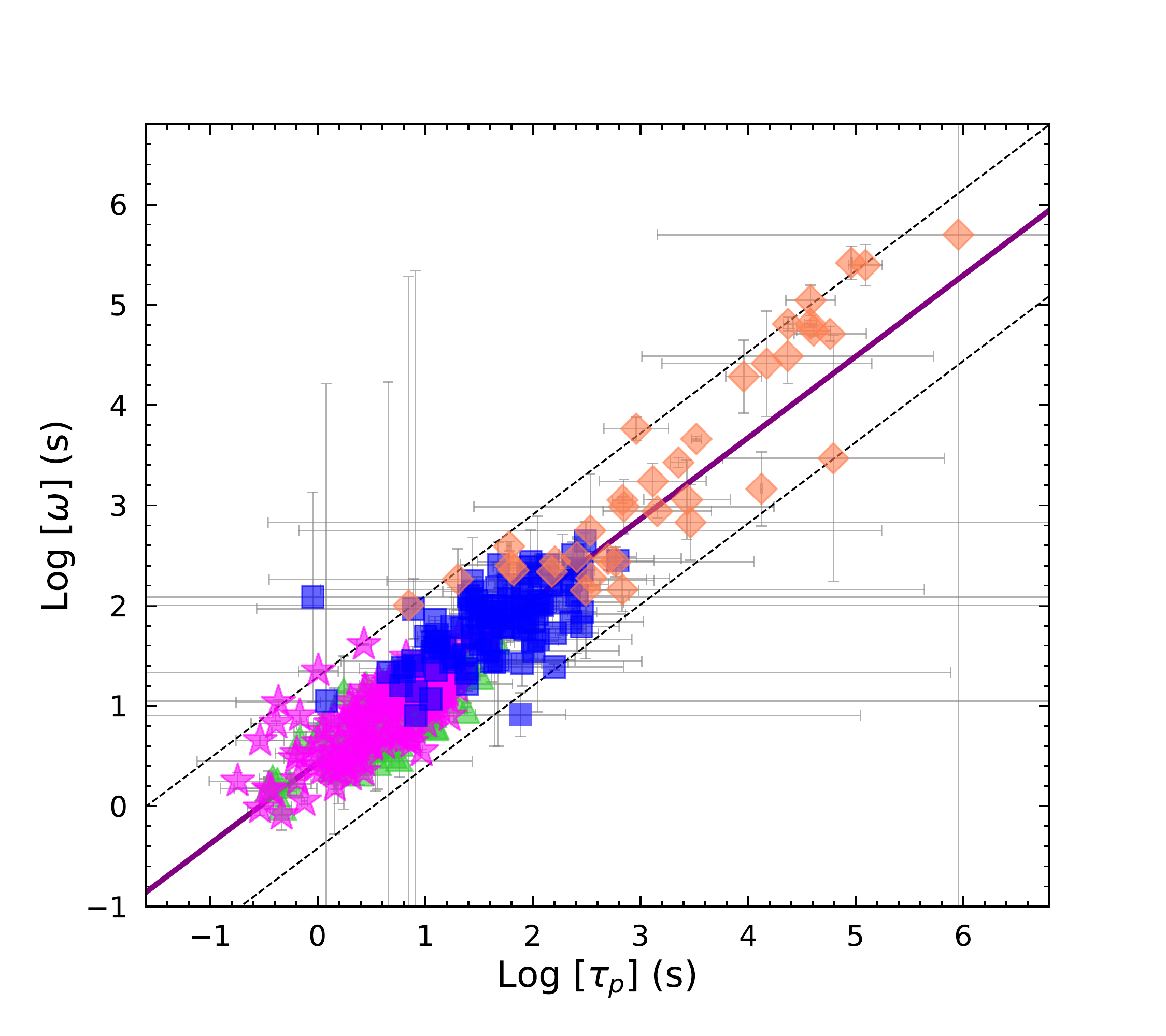}}
\subfigure{
\includegraphics[height=7.cm,width=8.cm]{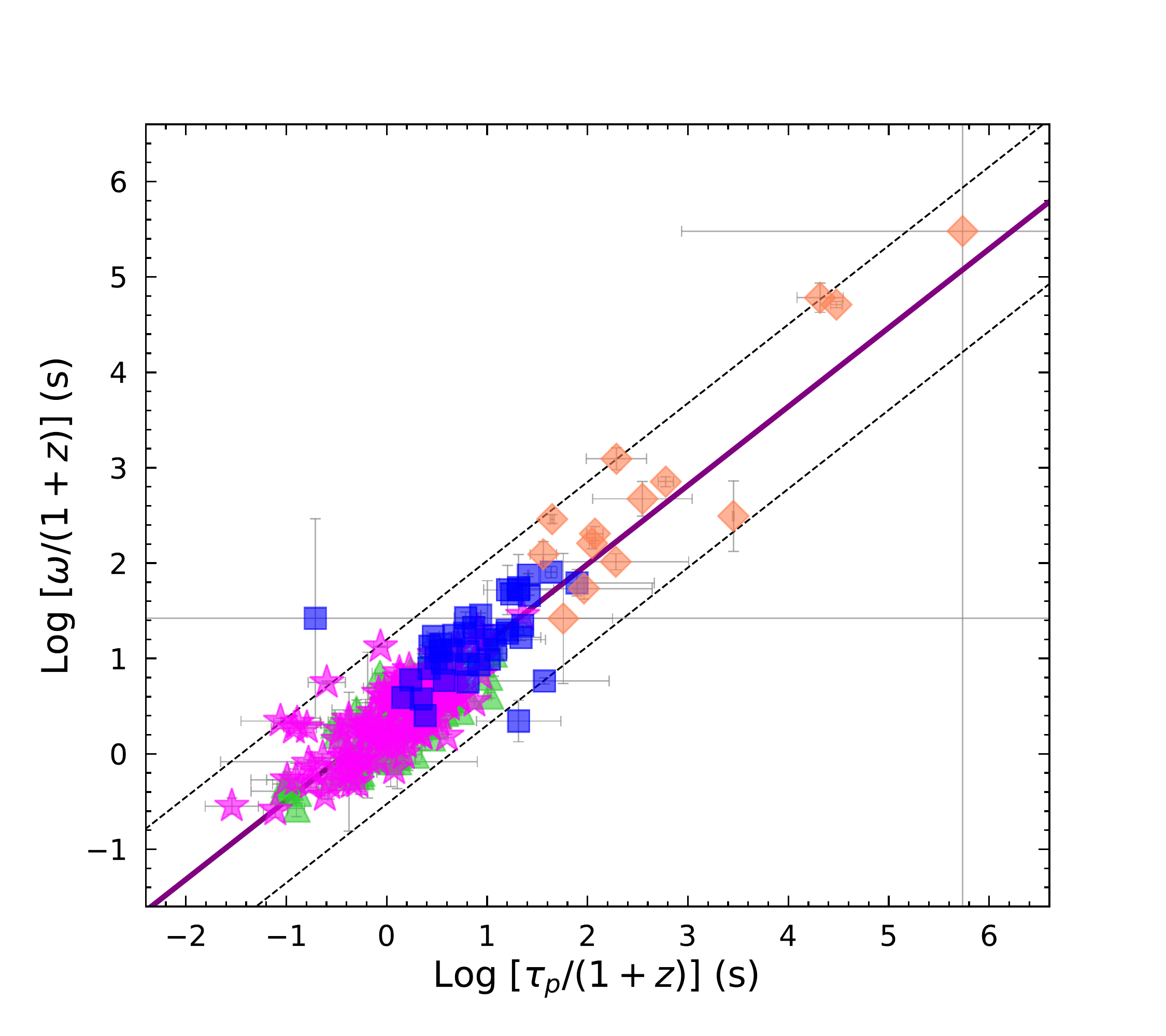}}
\subfigure{
\includegraphics[height=7.cm,width=8.cm]{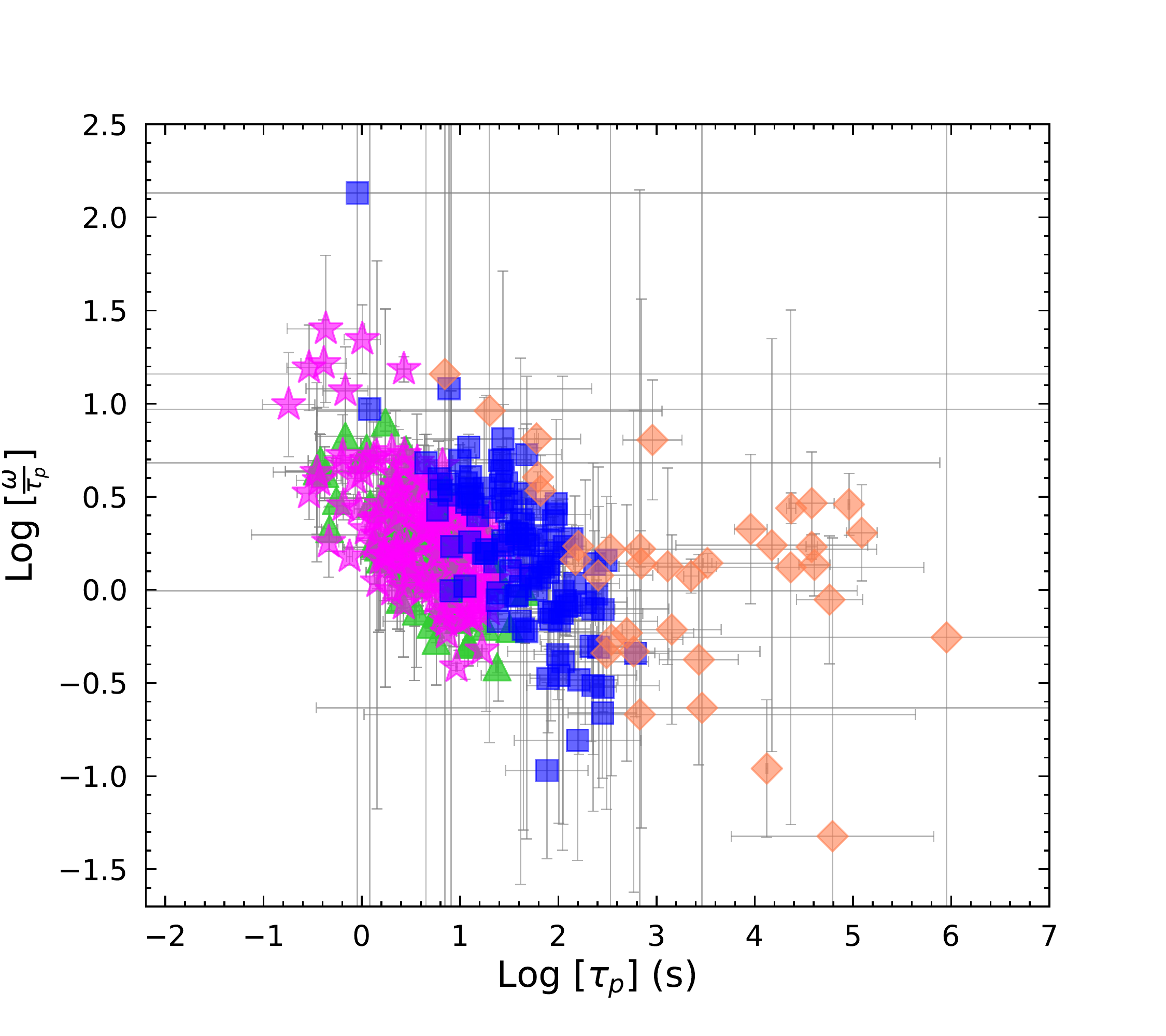}}
\subfigure{
\includegraphics[height=7.cm,width=8.cm]{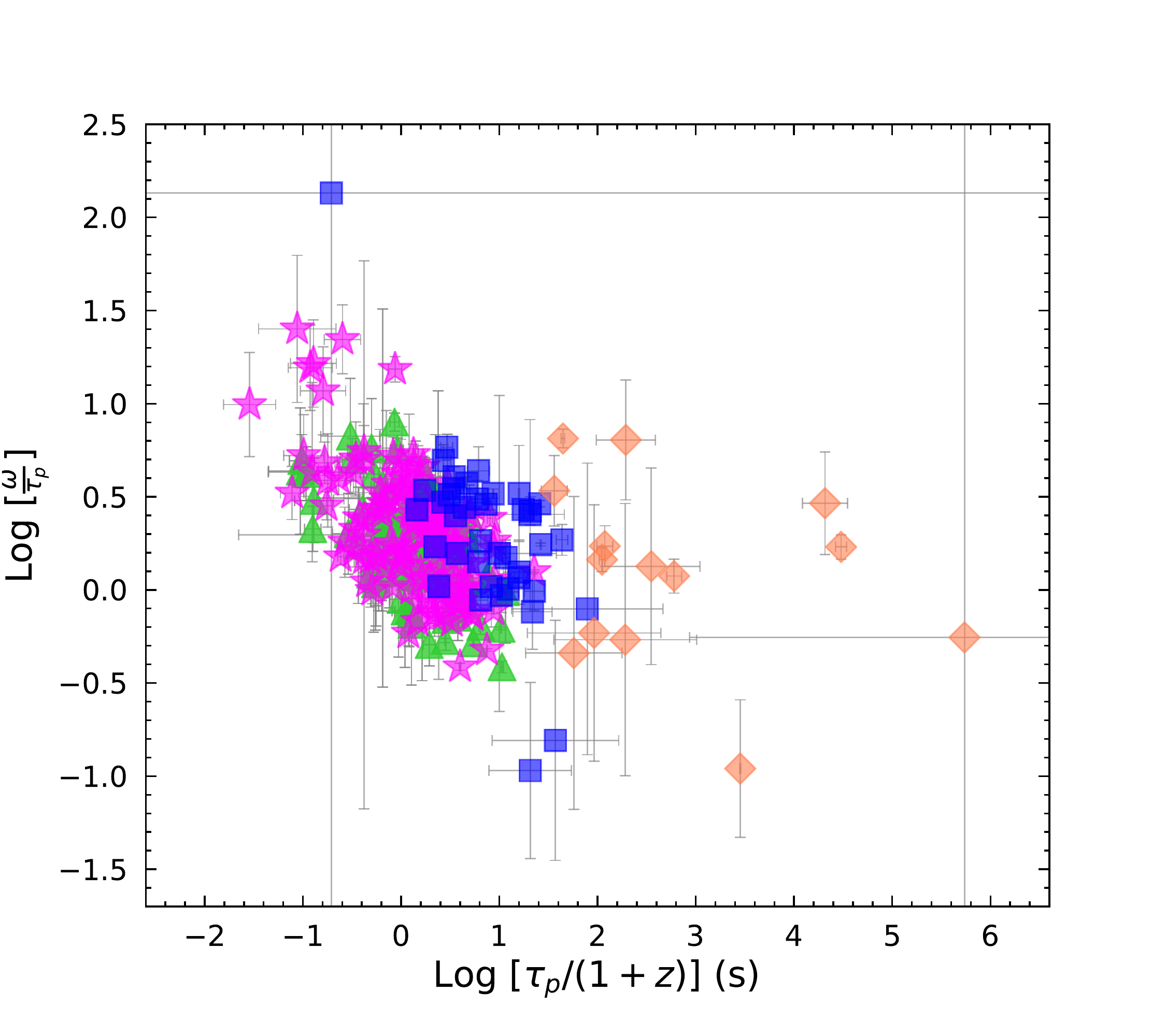}}
\caption{\footnotesize{Upper-left panel: the correlation between the width and the peak time without redshift correction. The purple solid line represents the best fitting as log $\omega$ = (0.44$^{+0.03}_{-0.03}$) + (0.81$^{+0.02}_{-0.03}$)log $\tau_{p}$.
The black dashed lines enclose the data within 3$\sigma$, and $\sigma$ = 0.28$^{+0.01}_{-0.01}$. Upper-right panel: the correlation between the width and the peak time with redshift correction. The purple solid line represents the best fitting as log $\omega$ = (0.34$^{+0.02}_{-0.02}$) + (0.83$^{+0.02}_{-0.03}$)log $\tau_{p}$. The black dashed lines enclose the data within 3$\sigma$, and $\sigma$ = 0.29$^{+0.01}_{-0.01}$.
Lower-left panel: the ratio of the width and the peak time as a function of the peak time without redshift correction.
Lower-right panel: the ratio of the width and the peak time as a function of the peak time with redshift correction. In the above four panels, green triangle and pink star correspond to precursor and main burst, respectively, while blue square and dark yellow diamond correspond to the early GRB flare identified in Chincarini et al. (2010) and the late X-ray flare identified in Bernardini et al. (2011), respectively.}}
\end{figure}
\clearpage

\appendix

\section{Fitting Profiles}

We use Norris function $I(t)$ = $A$$\lambda$$e^{-{\tau_{1}}/{(t-t_{s})}-{(t-t_{s})}/{\tau_{2}}}$ to fit the dataset ($t$, $I$). Each data pair is ($t_{i}$, $I_{i}$),
where $i = 1 , \ldots, n$. The error of each data pair is given by ($\sigma_{t,i}$, $\sigma_{I,i}$). Here, we take $\sigma_{t,i} = 0$. We then build the likelihood function as
\begin{equation}
{\rm log\,p}(A,t_{s},\tau_{1},\tau_{2},f|t_{i},I_{i},\sigma_{I,i}) = \frac{1}{2}\sum\limits^{n}_{i=1}[{\rm log}(\frac{1}{2\pi s^{2}_{i}})-\frac{(I_{i}-A\lambda e^{-{\tau_{1}}/{(t_{i}-t_{s})}-{(t_{i}-t_{s})}/{\tau_{2}}})^{2}}{s^{2}_{i}}]
\end{equation}
where, $s^{2}_{i} = \sigma^{2}_{I,i}+f^{2}(A\lambda e^{-{\tau_{1}}/{(t_{i}-t_{s})}-{(t_{i}-t_{s})}/{\tau_{2}}})^{2}$. Here, we simply take the external error of the fitting $\sigma_{ext}$ to be $\sigma_{ext} = f$.
We visually examine the fitting results. In general, the fitting is good with small residuals. However, we still find a few cases that the fitting to a certain spike structure or an overlapped structure has large residuals. In this paper, our aim to use the Norris function is for the fitting of the gross temporal profile. We note that the Norris function is too simple to be used for the fitting of a complex episode structure in GRB lightcurves.

We illustrate the fitting profiles for the precursors and the main burst in each selected GRB.
The GRB lightcurve has the time resolution of 64 ms in the 15-350 keV energy band.
The results are shown in Figures 18-69. The solid lines are the overall fitting curves.
The dashed lines are the fitting curves for the episodes.

\begin{figure}[!htp]
\centering
\subfigure{
\includegraphics[scale = 0.4]{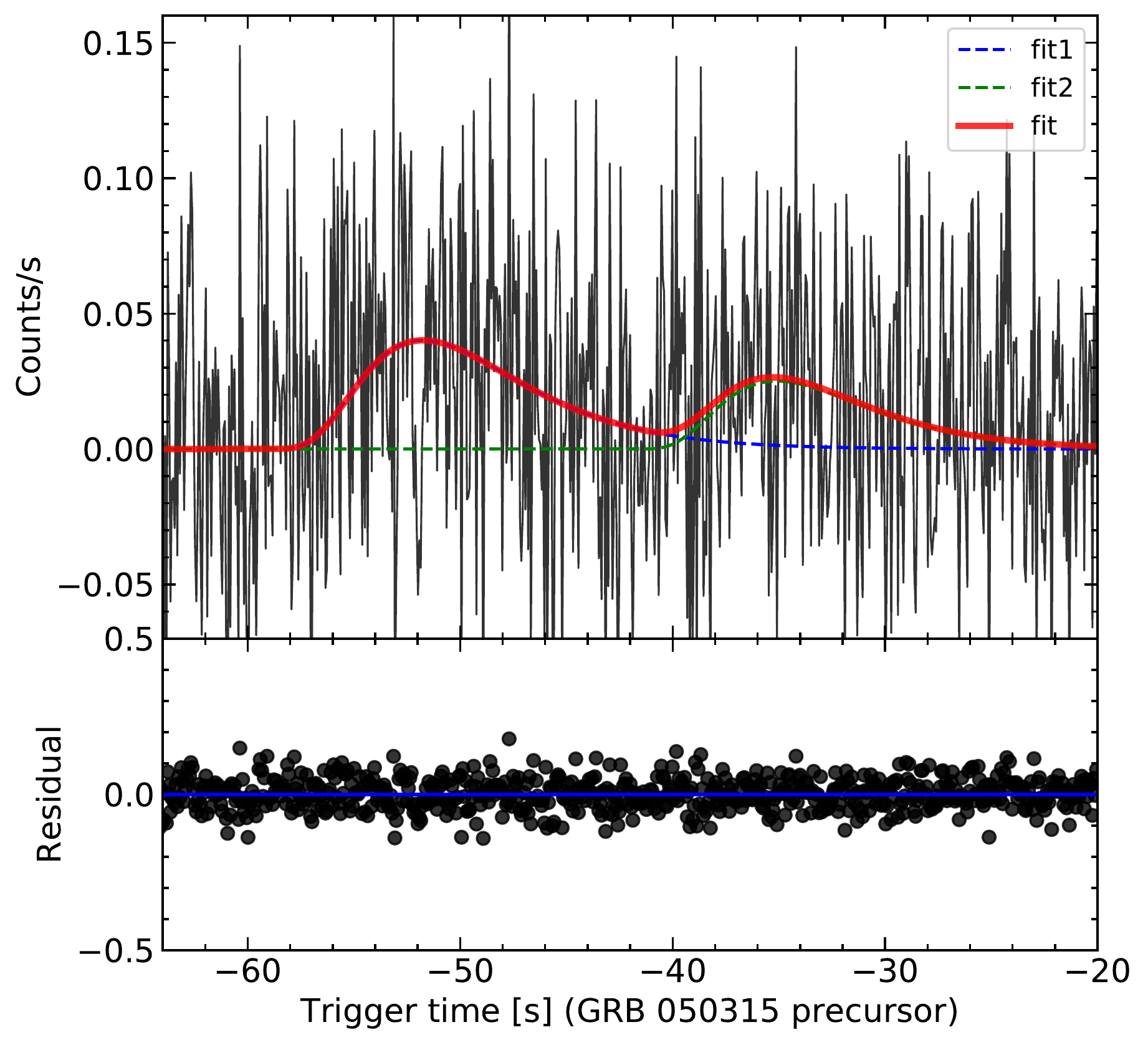}}
\subfigure{
\includegraphics[scale = 0.4]{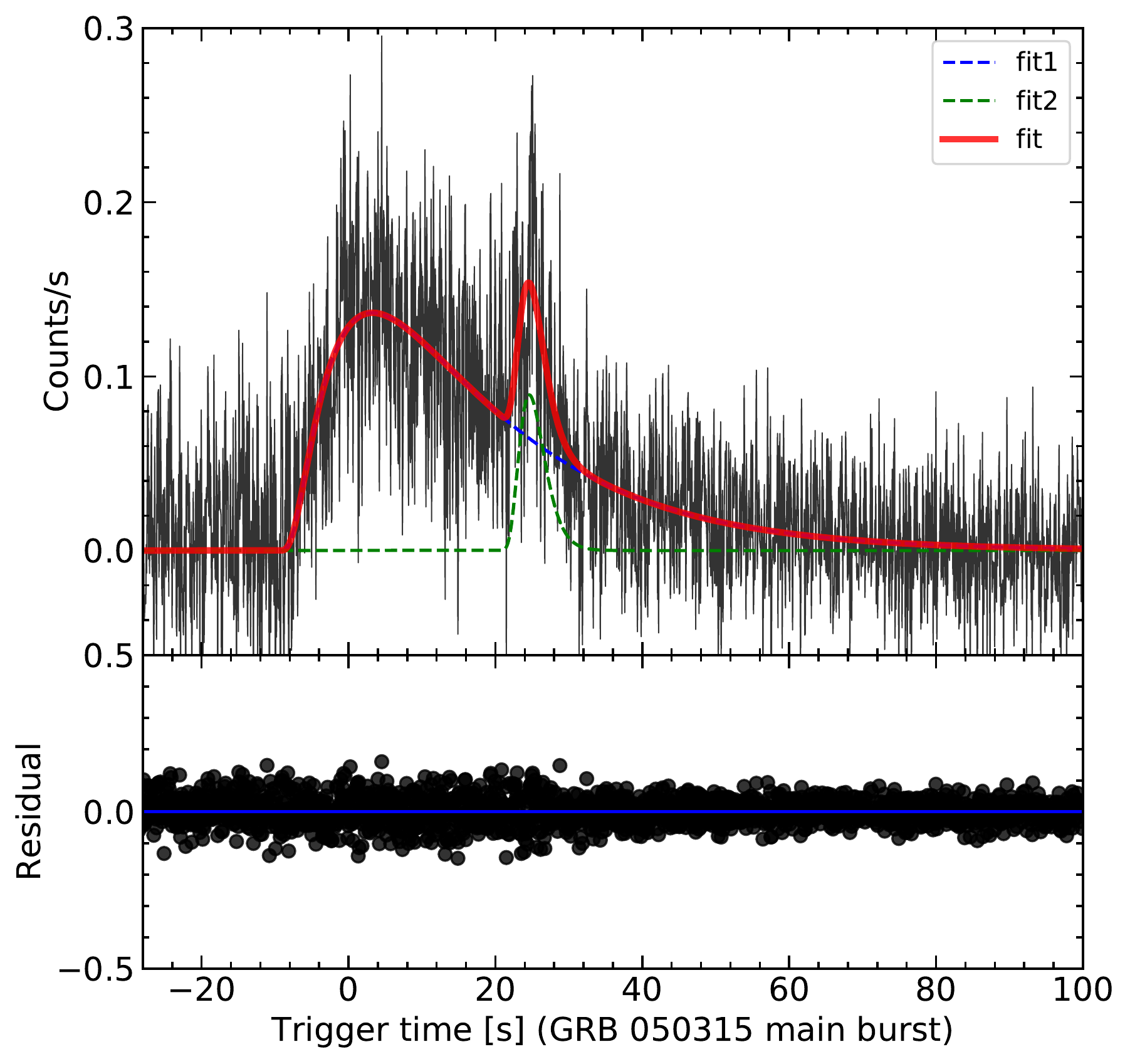}}
\caption{
The lightcurve of GRB 050315
}
\end{figure}

\begin{figure}[!htp]
\centering
\subfigure{
\includegraphics[scale = 0.4]{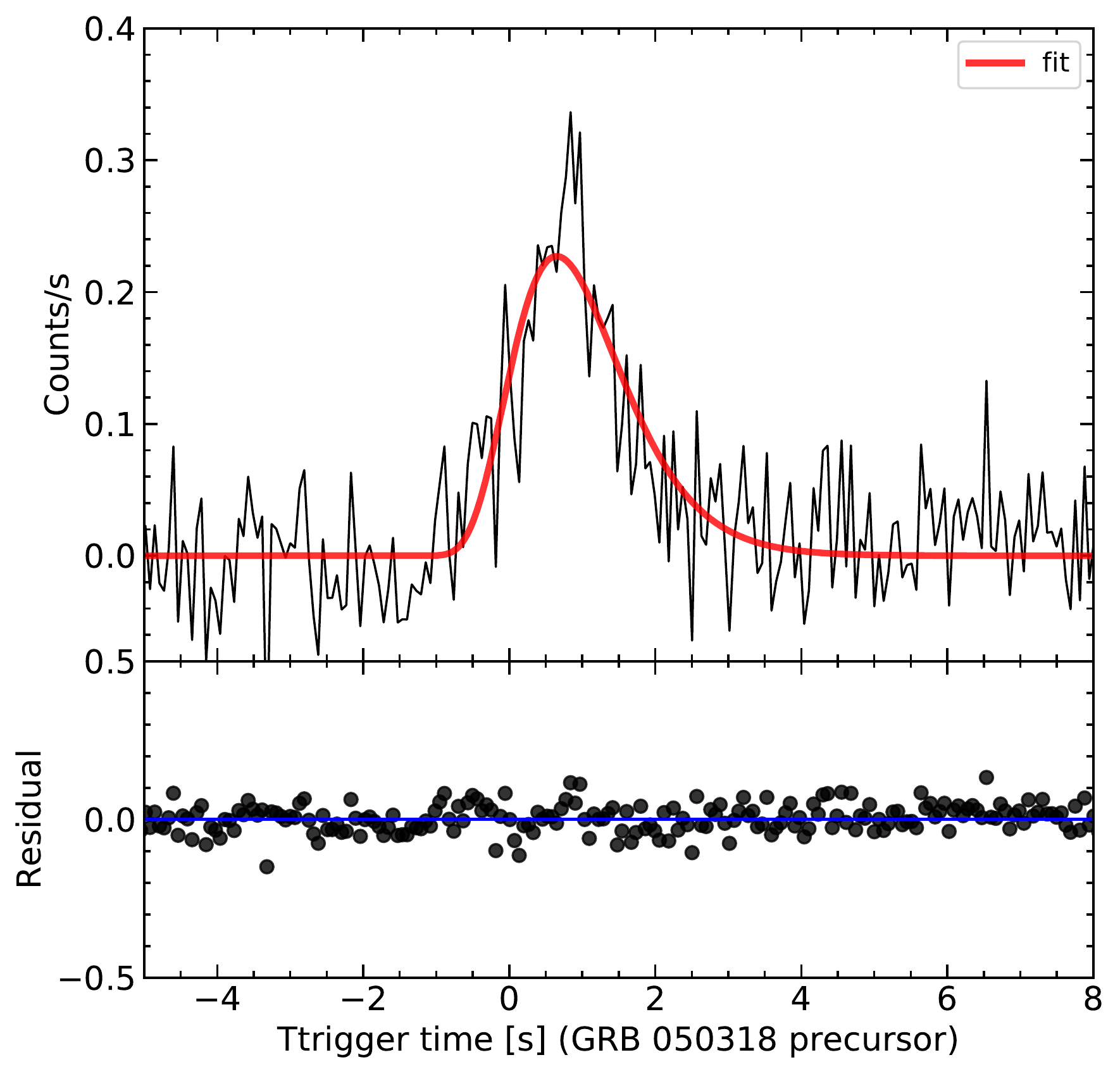}}
\subfigure{
\includegraphics[scale = 0.4]{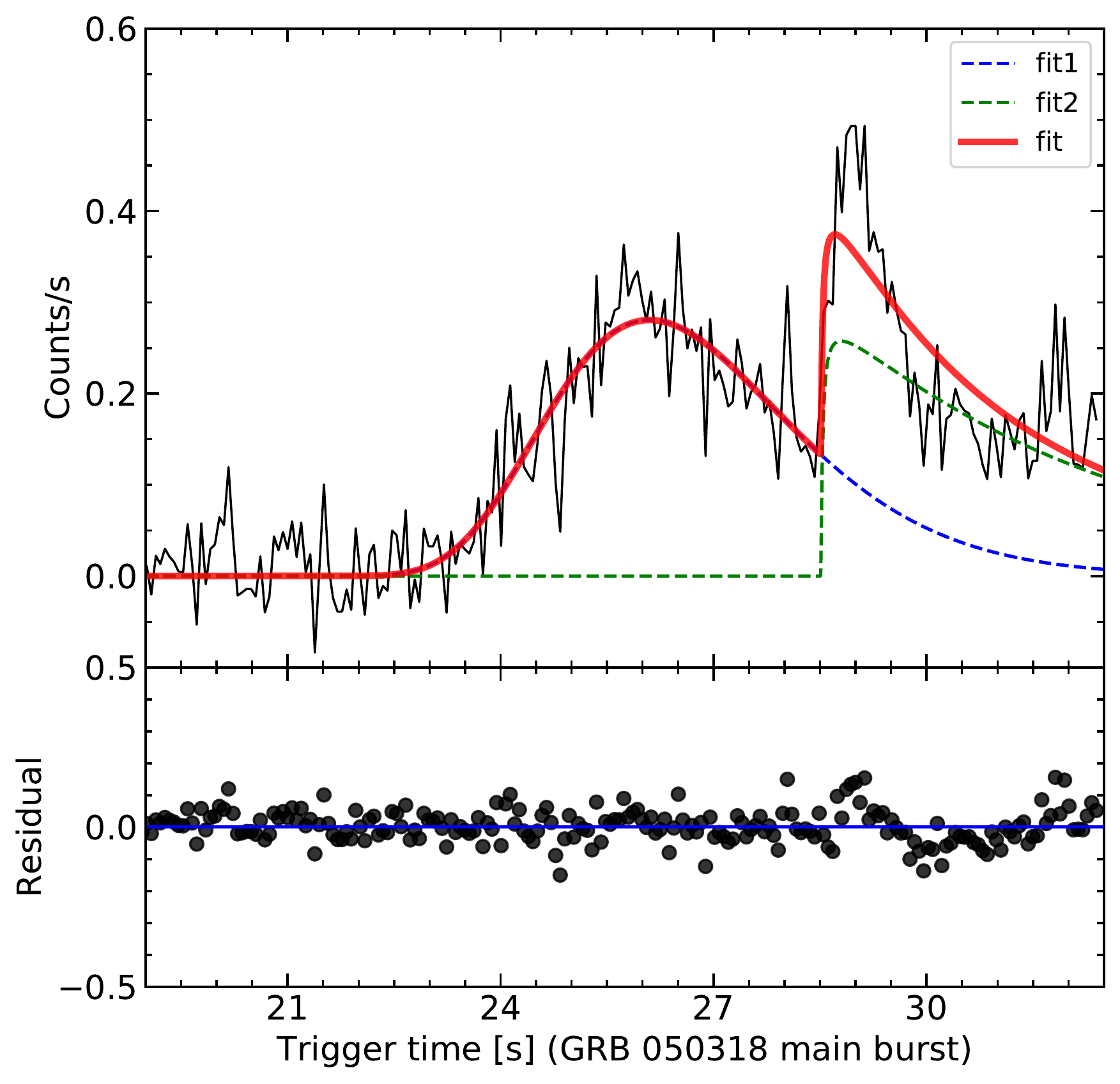}}
\caption{The lightcurve of GRB 050318
}
\end{figure}

\begin{figure}[!htp]
\centering
\subfigure{
\includegraphics[scale = 0.4]{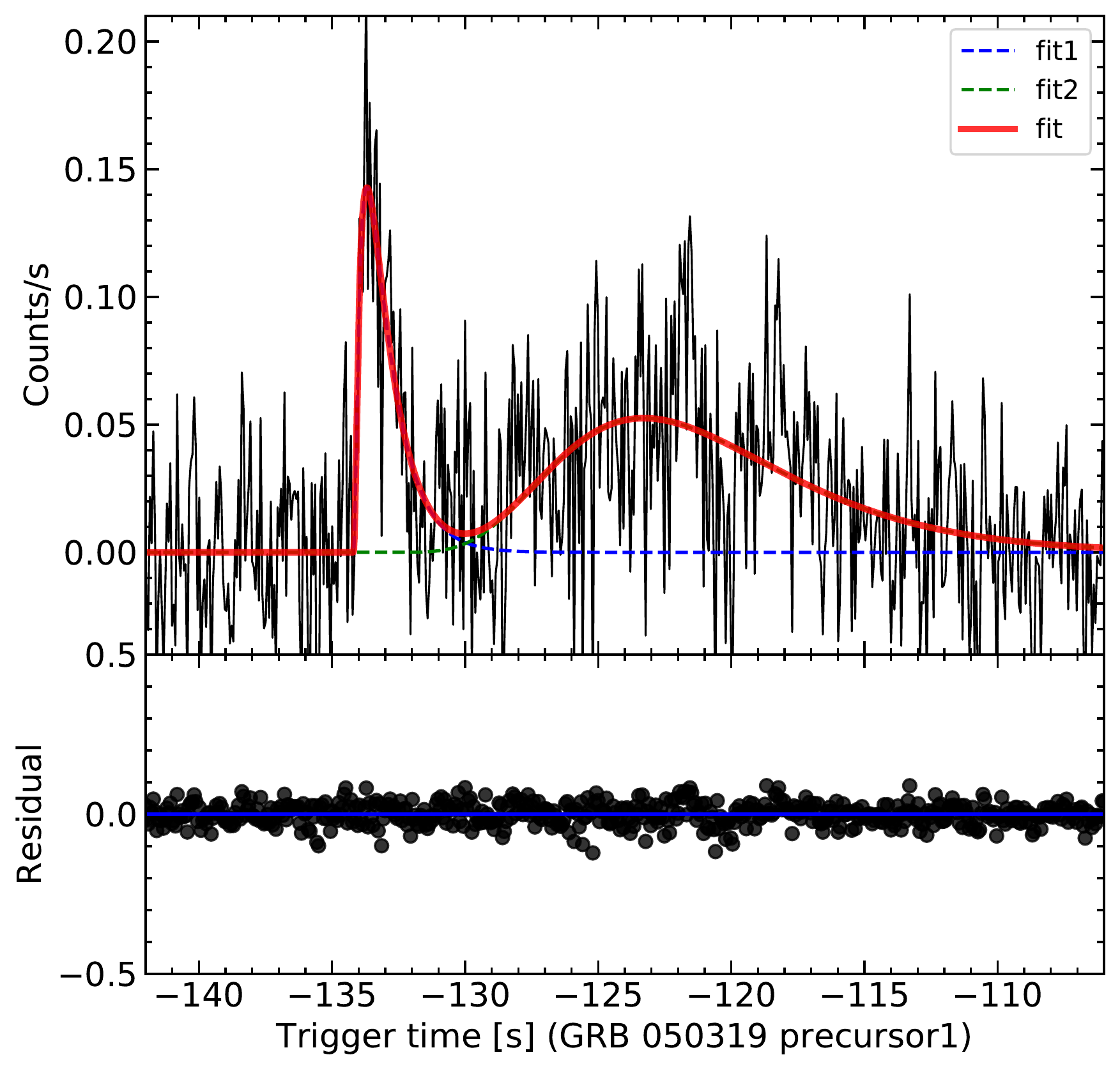}}
\subfigure{
\includegraphics[scale = 0.4]{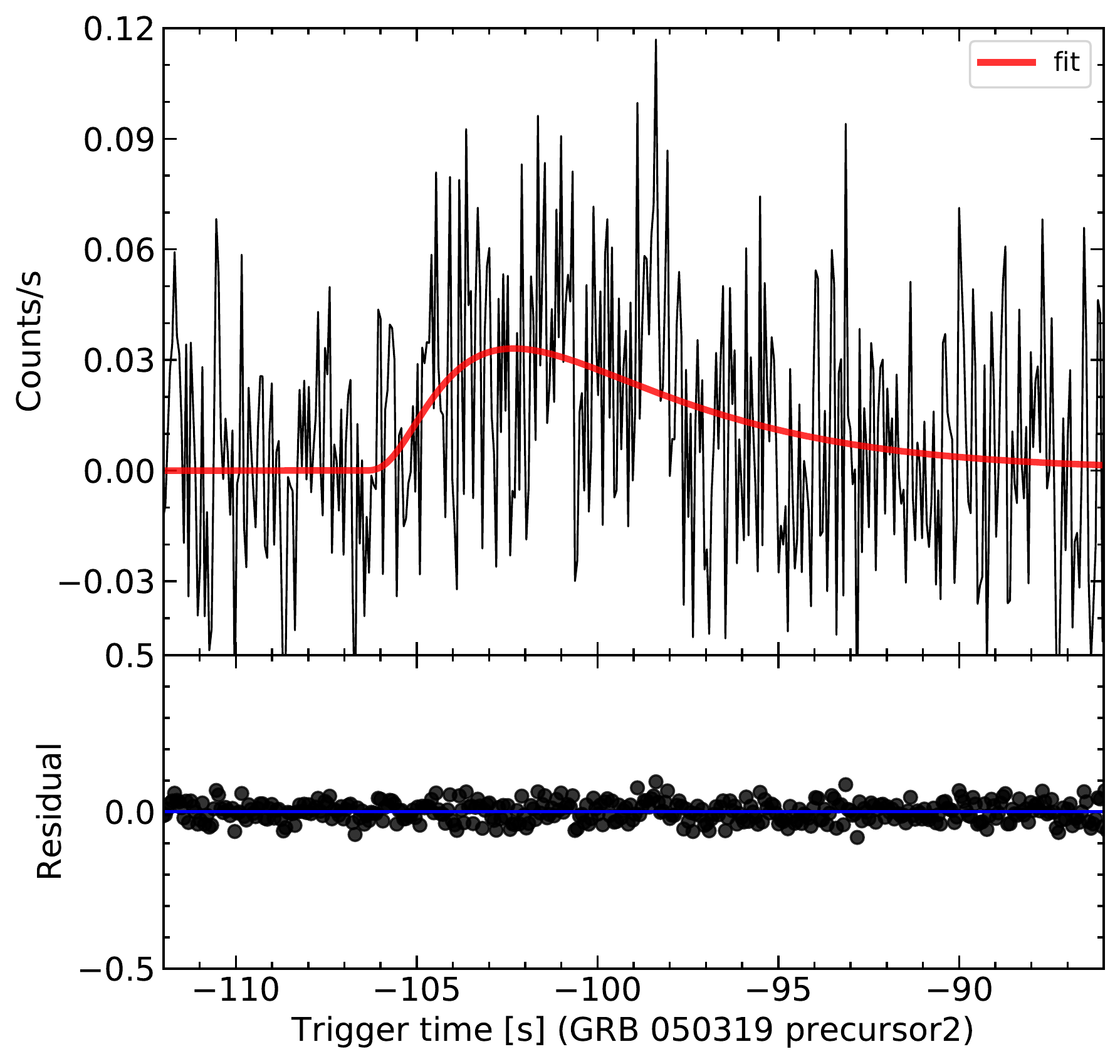}}
\subfigure{
\includegraphics[scale = 0.4]{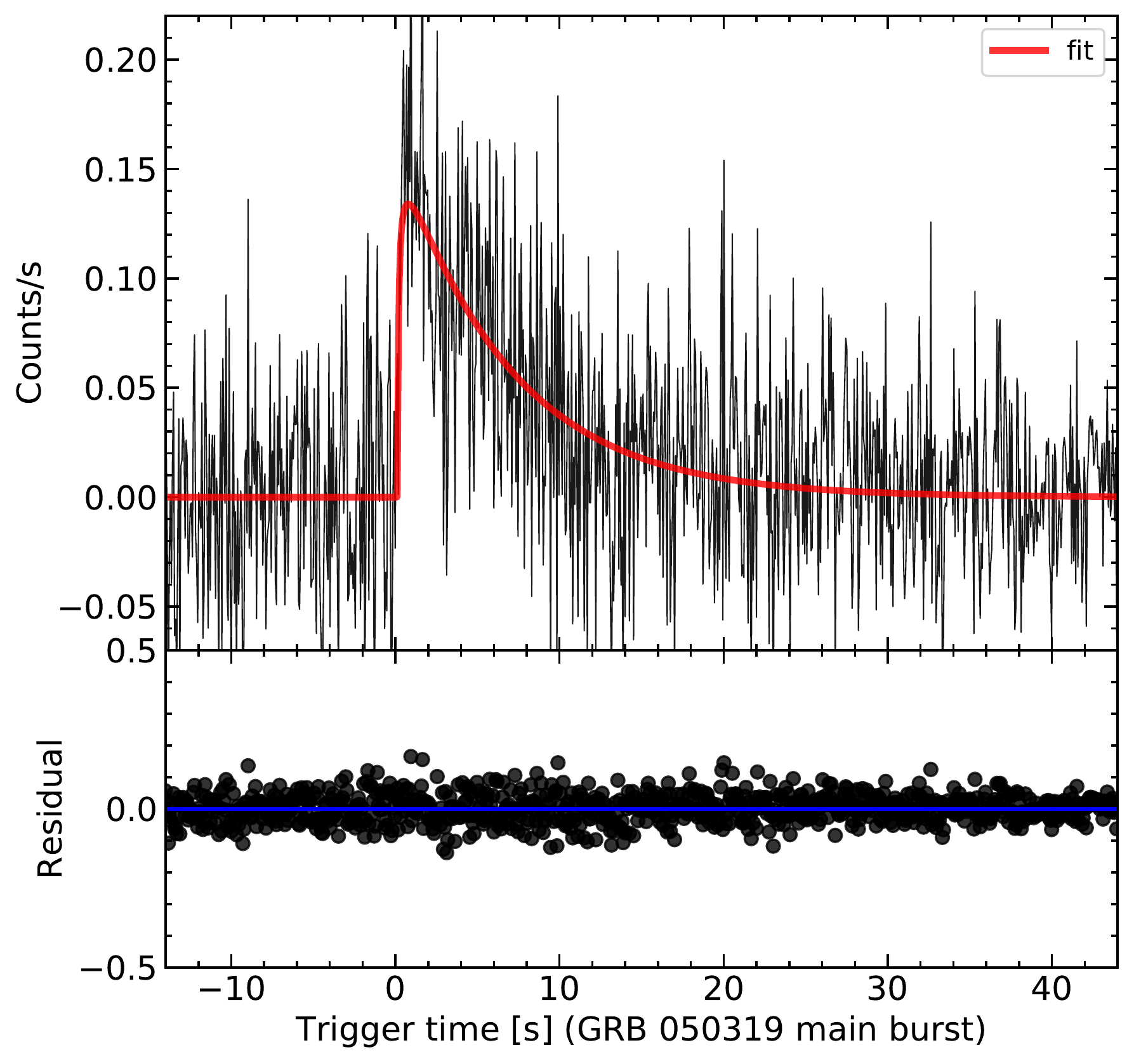}}
\caption{The lightcurve of GRB 050319
}
\end{figure}

\begin{figure}[!htp]
\centering
\subfigure{
\includegraphics[scale = 0.4]{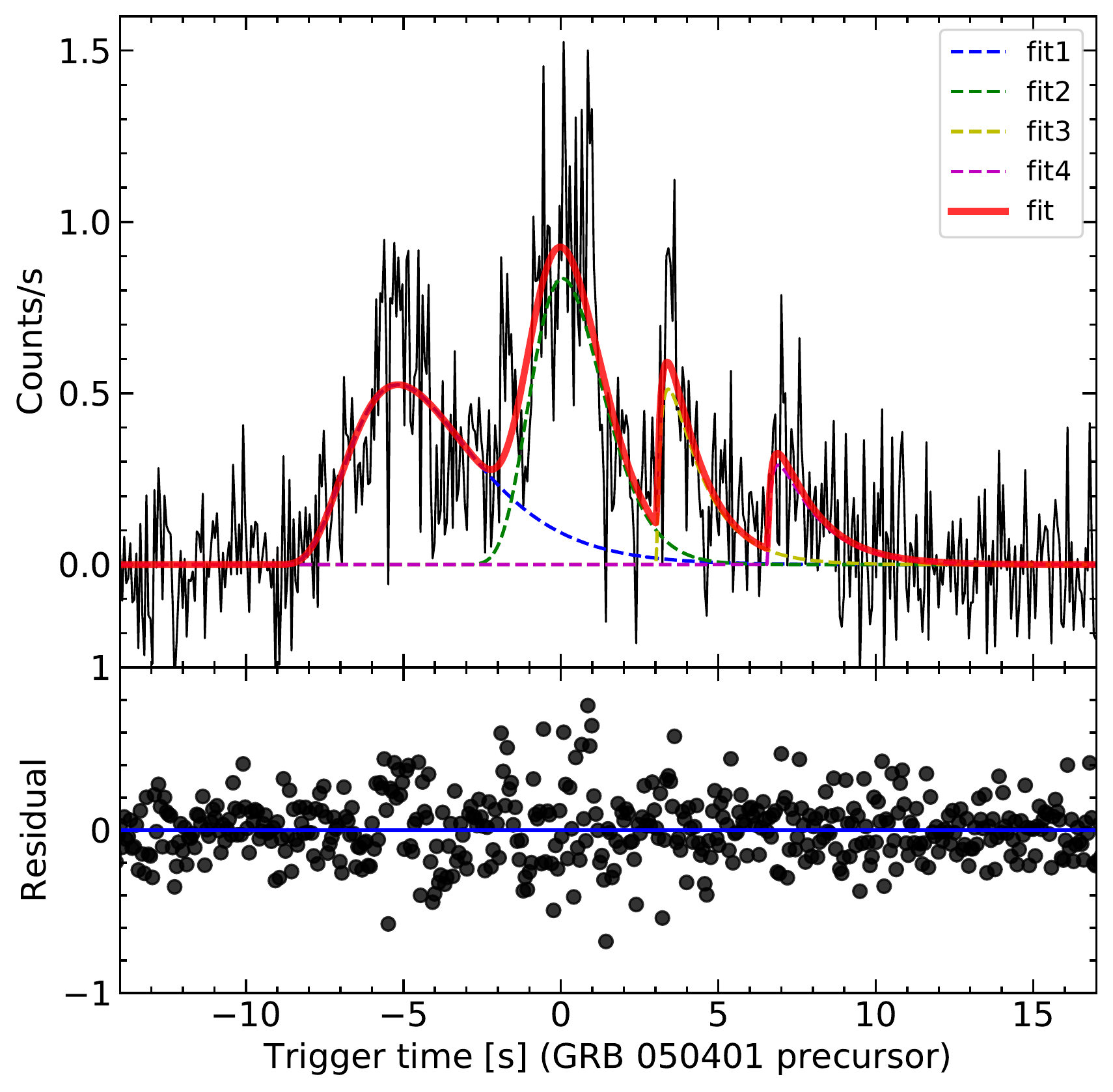}}
\subfigure{
\includegraphics[scale = 0.4]{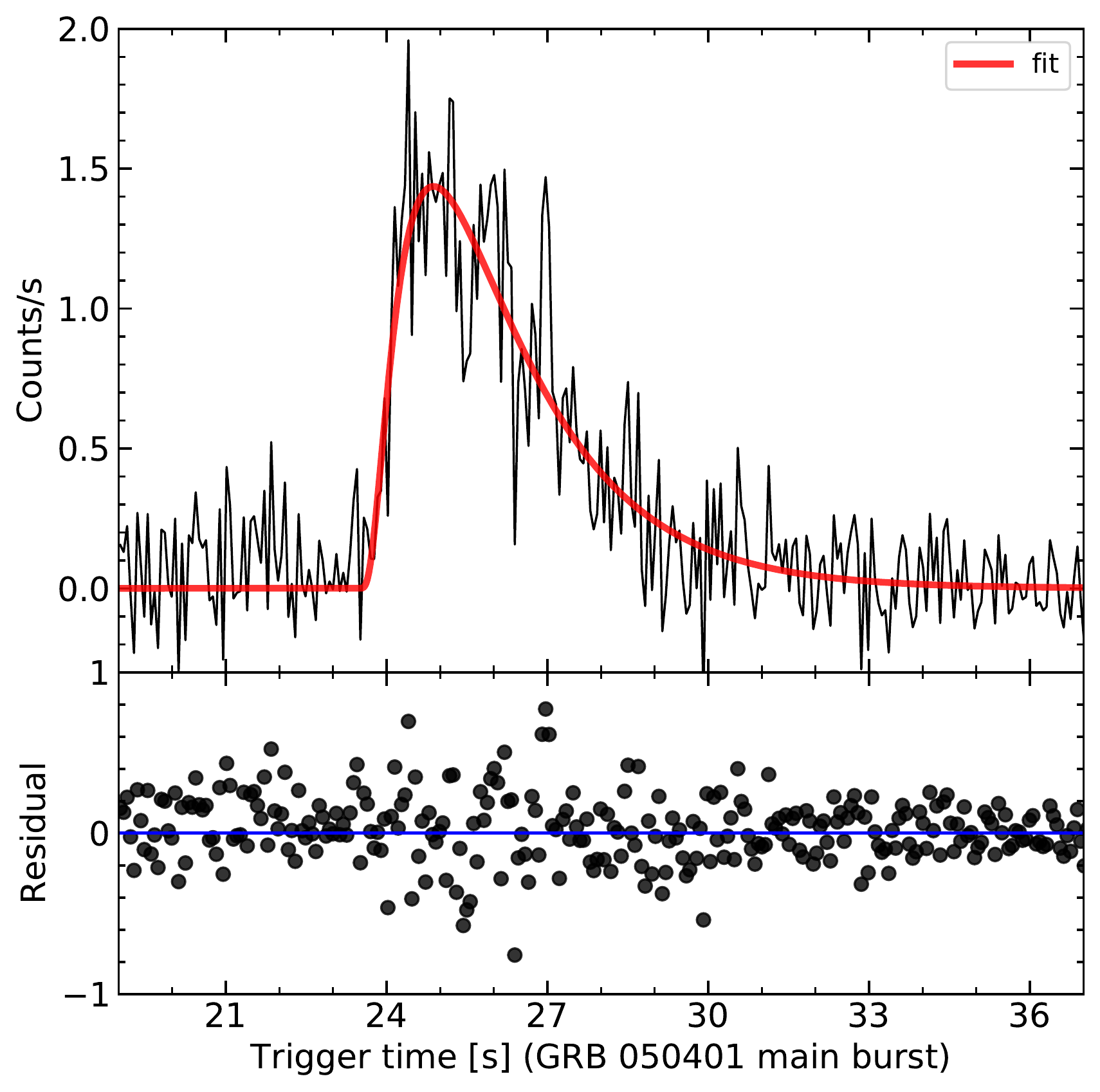}}
\caption{The lightcurve of GRB 050401
}
\end{figure}

\begin{figure}[!htp]
\centering
\subfigure{
\includegraphics[scale = 0.4]{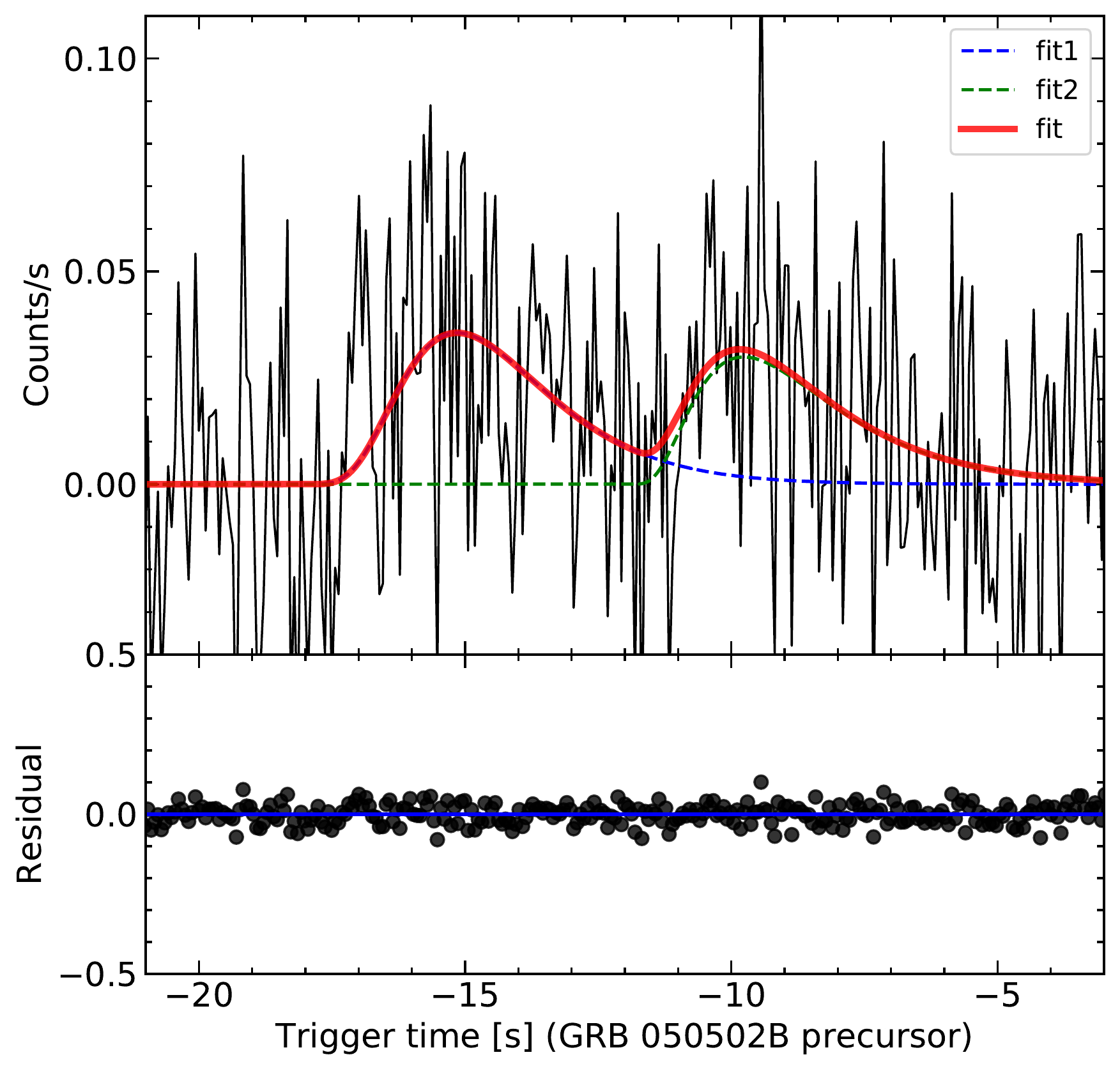}}
\subfigure{
\includegraphics[scale = 0.4]{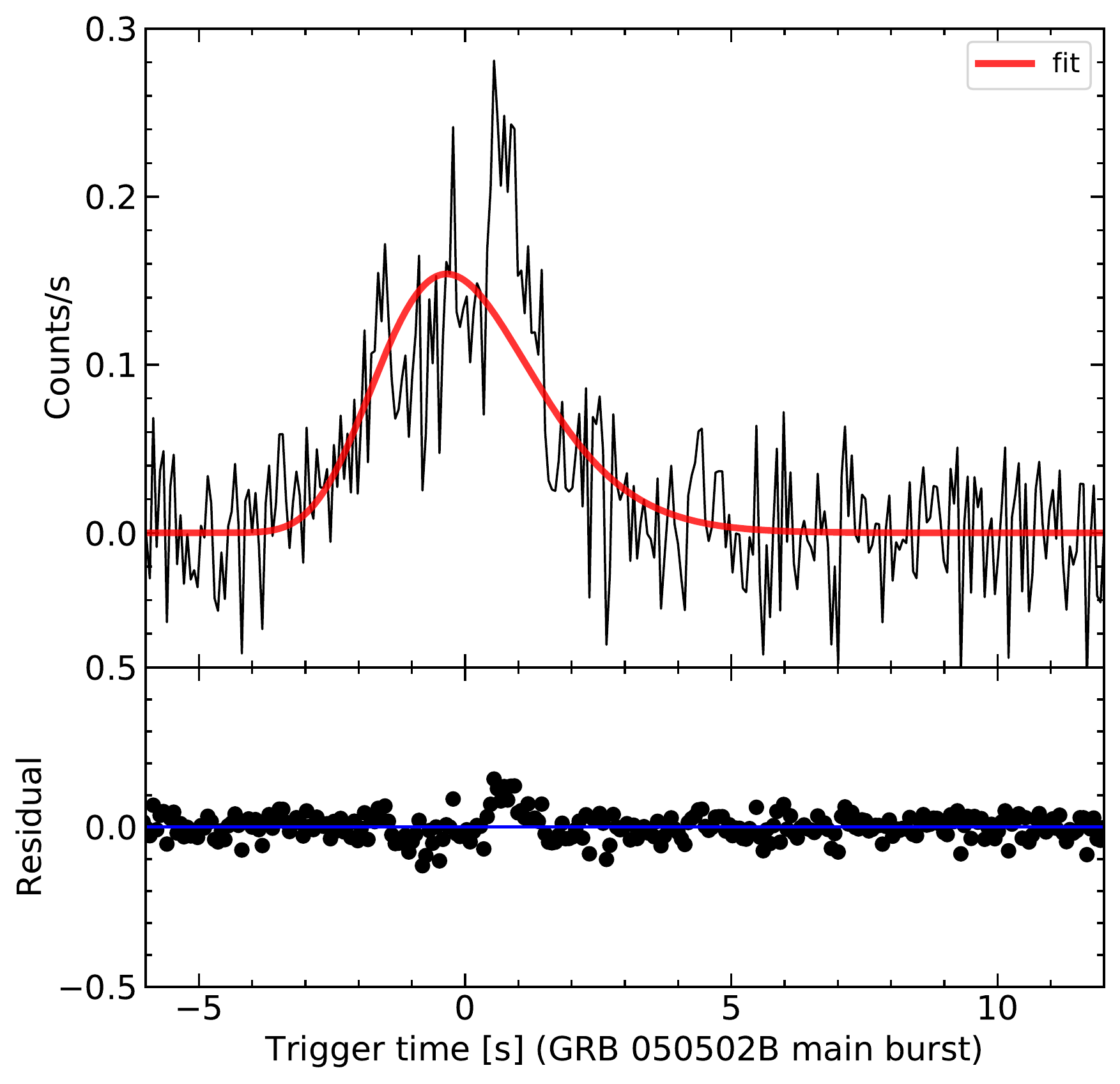}}
\caption{The lightcurve of GRB 050502B
}
\end{figure}

\begin{figure}[!htp]
\centering
\subfigure{
\includegraphics[scale = 0.4]{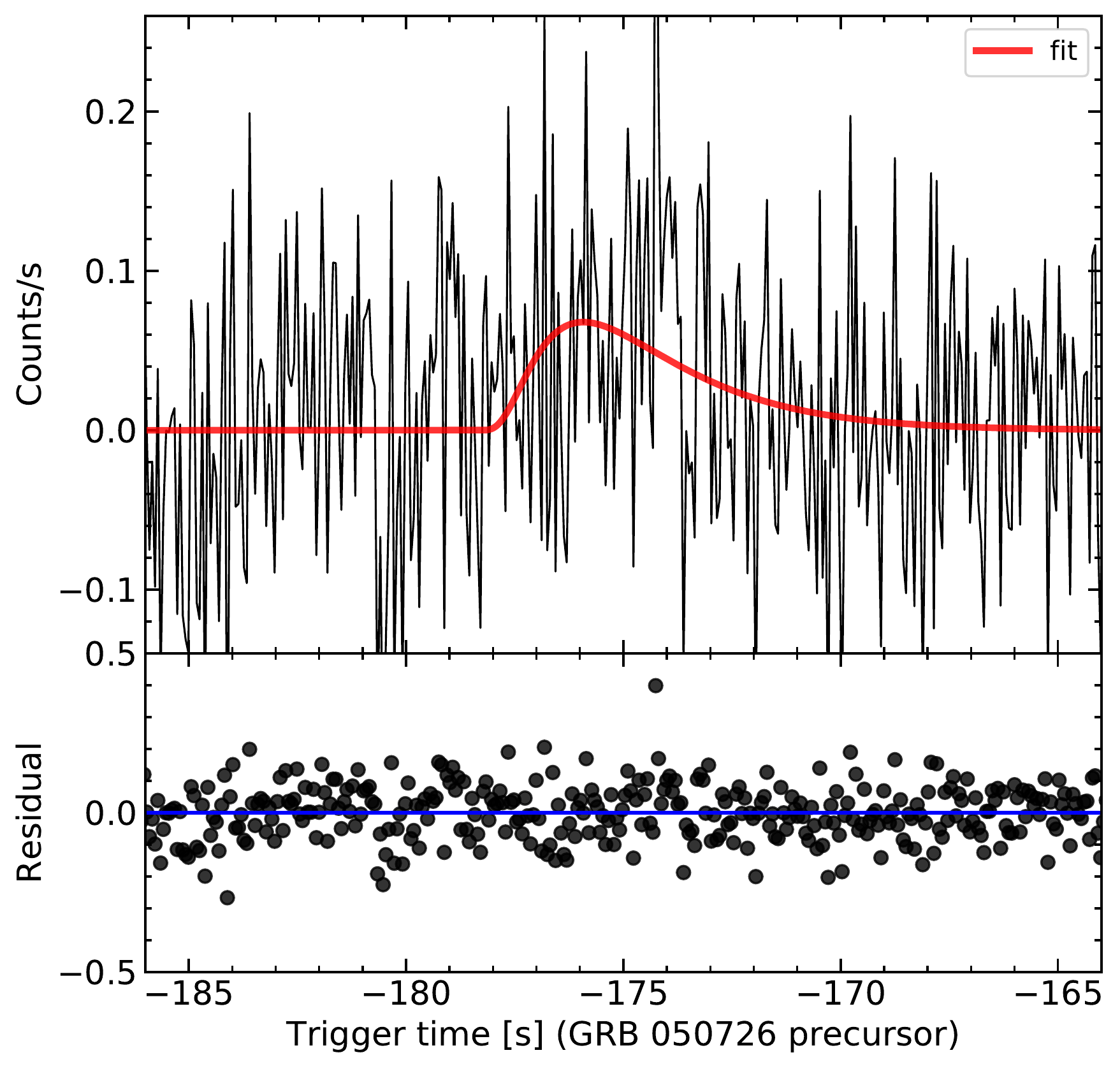}}
\subfigure{
\includegraphics[scale = 0.4]{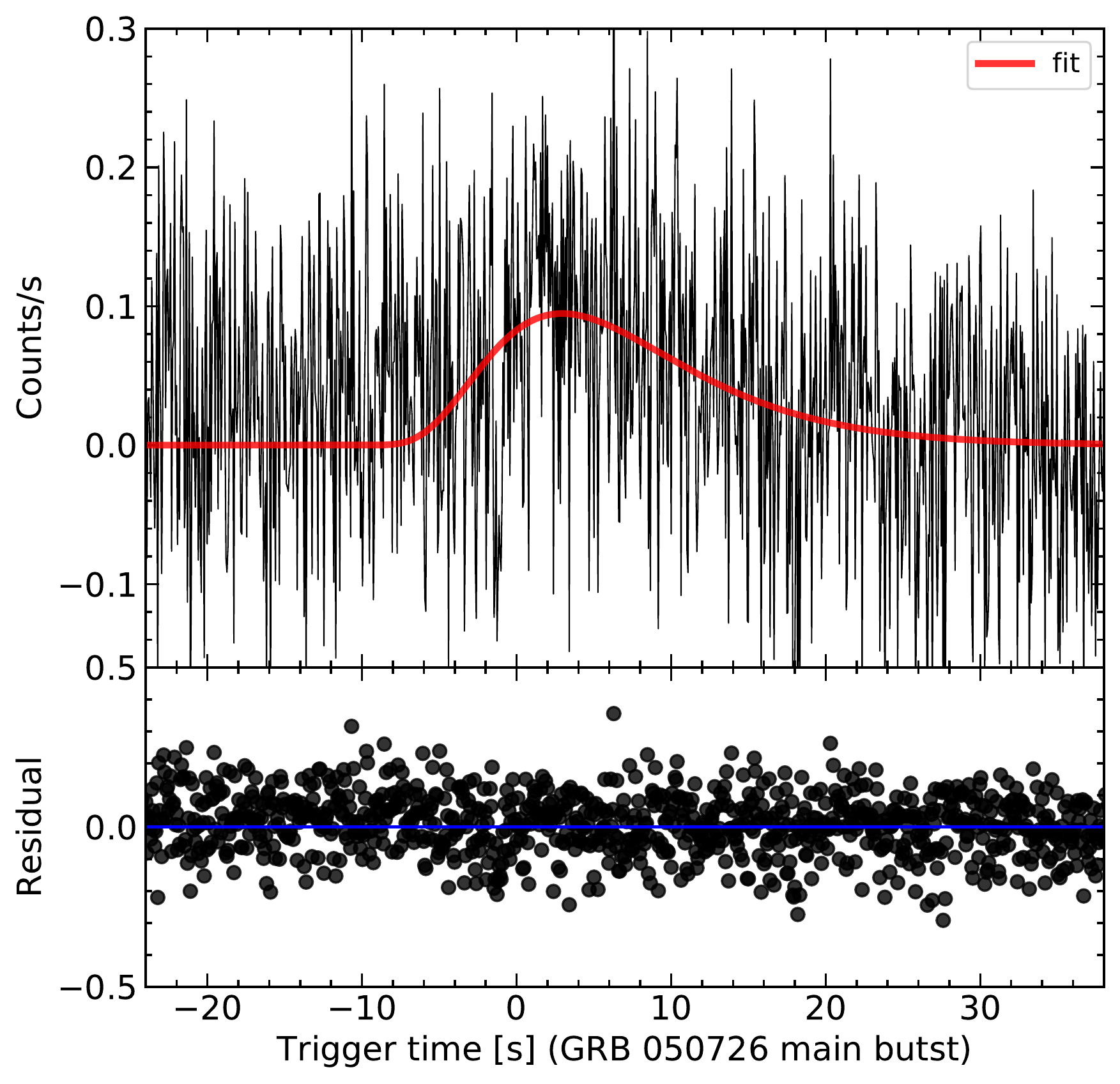}}
\caption{The lightcurve of GRB 050726
}
\end{figure}

\begin{figure}[!htp]
\centering
\subfigure{
\includegraphics[scale = 0.4]{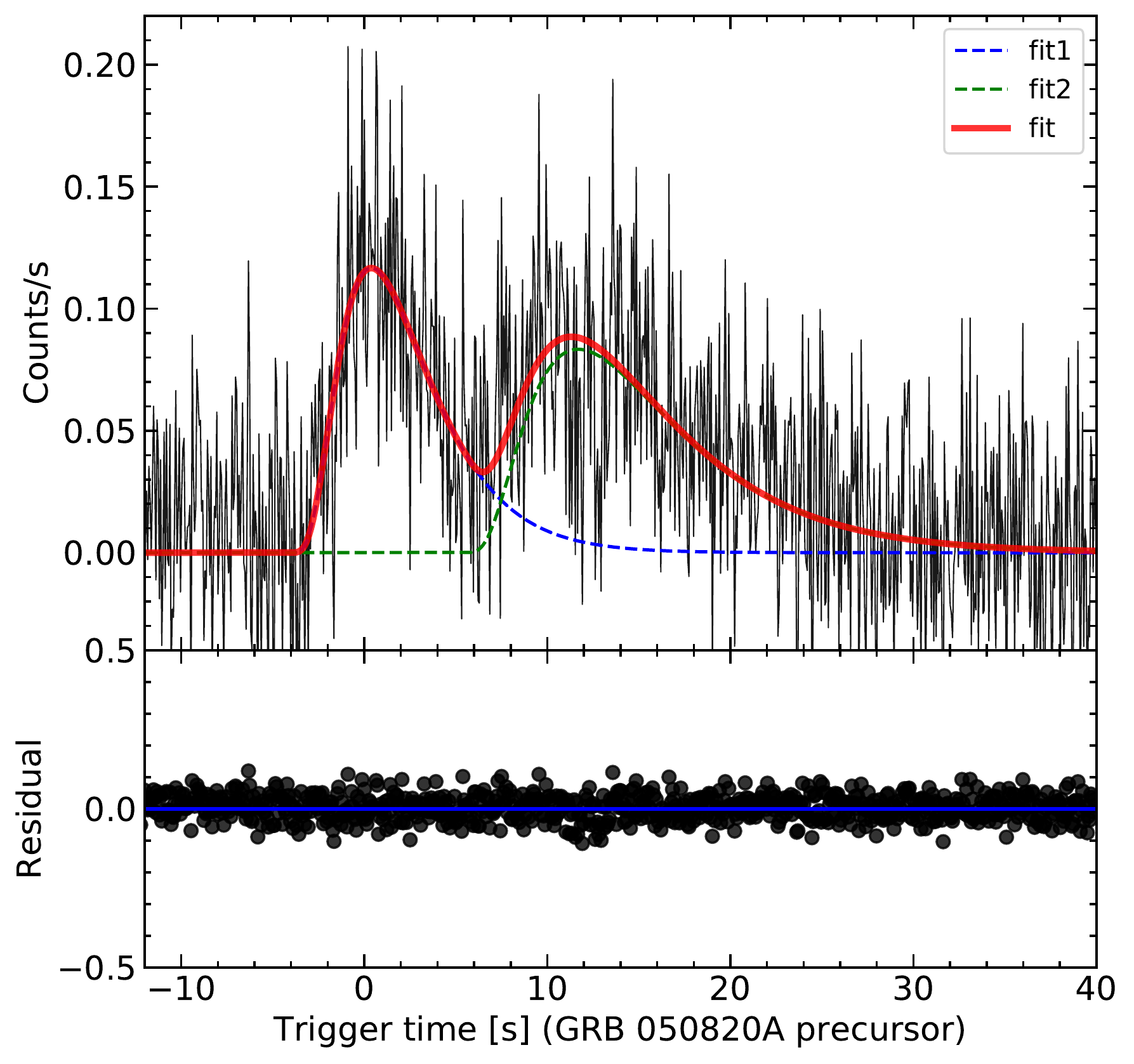}}
\subfigure{
\includegraphics[scale = 0.4]{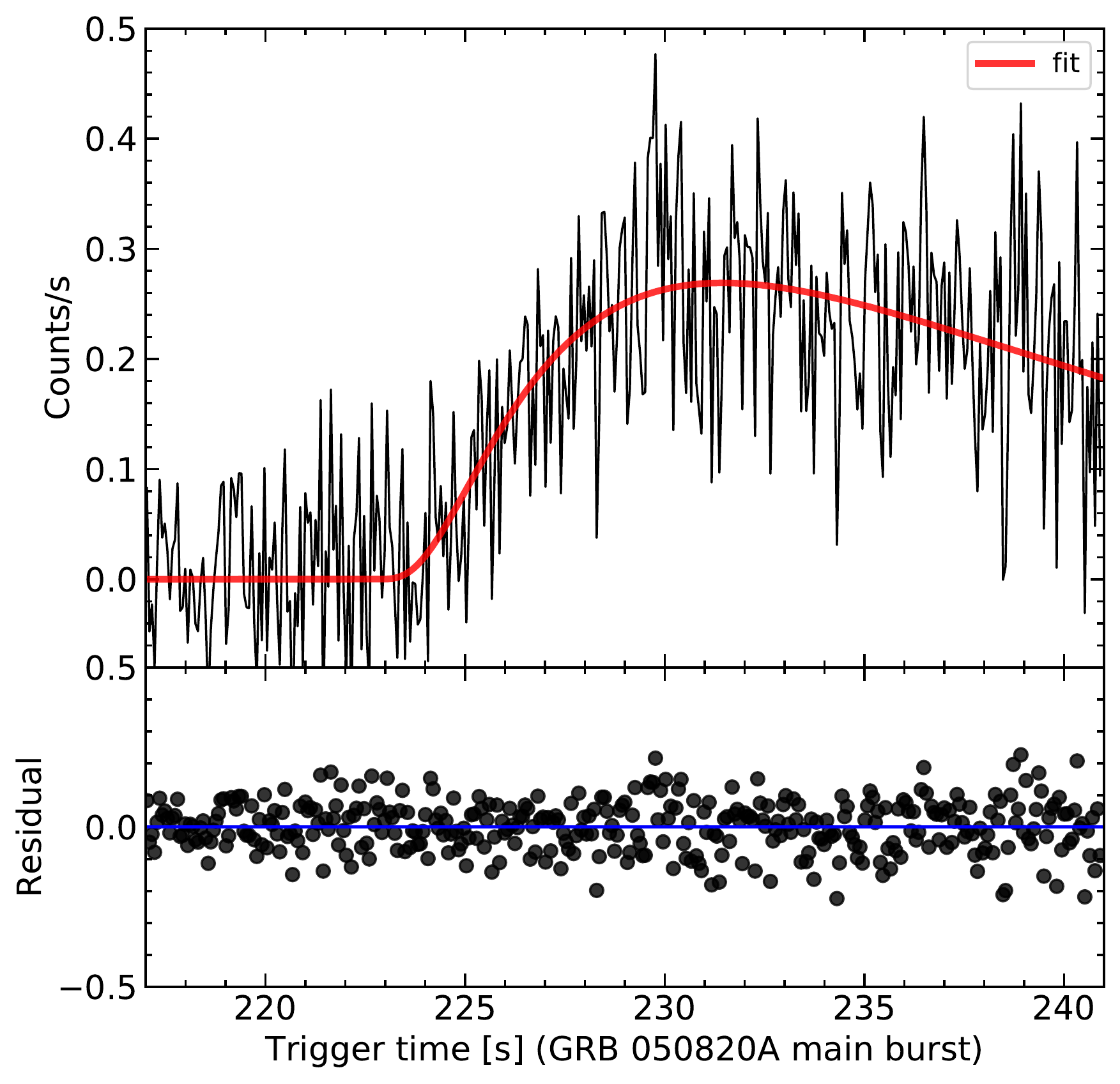}}
\caption{The lightcurve of GRB 050820A
}
\end{figure}

\begin{figure}[!htp]
\centering
\subfigure{
\includegraphics[scale = 0.4]{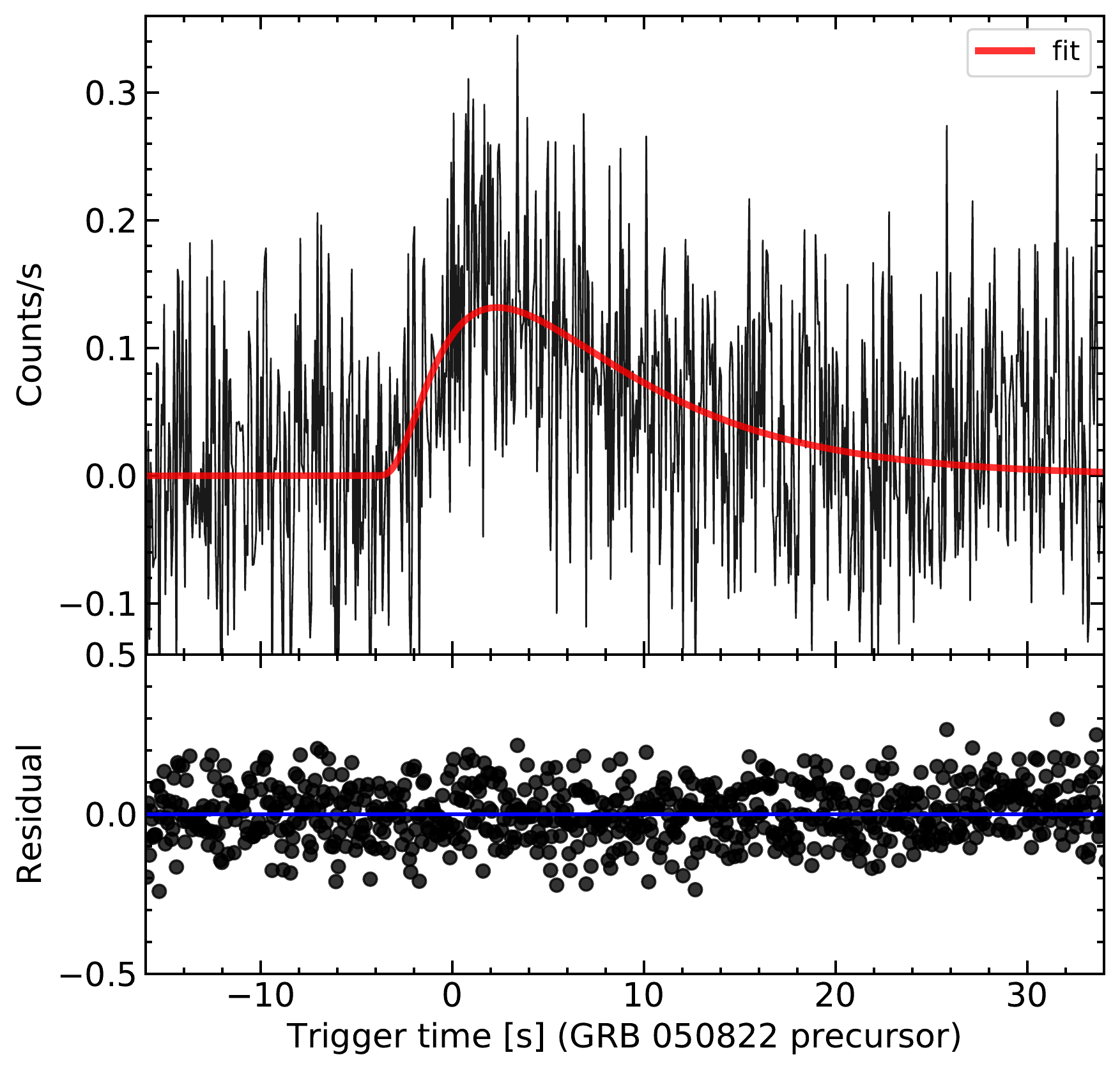}}
\subfigure{
\includegraphics[scale = 0.4]{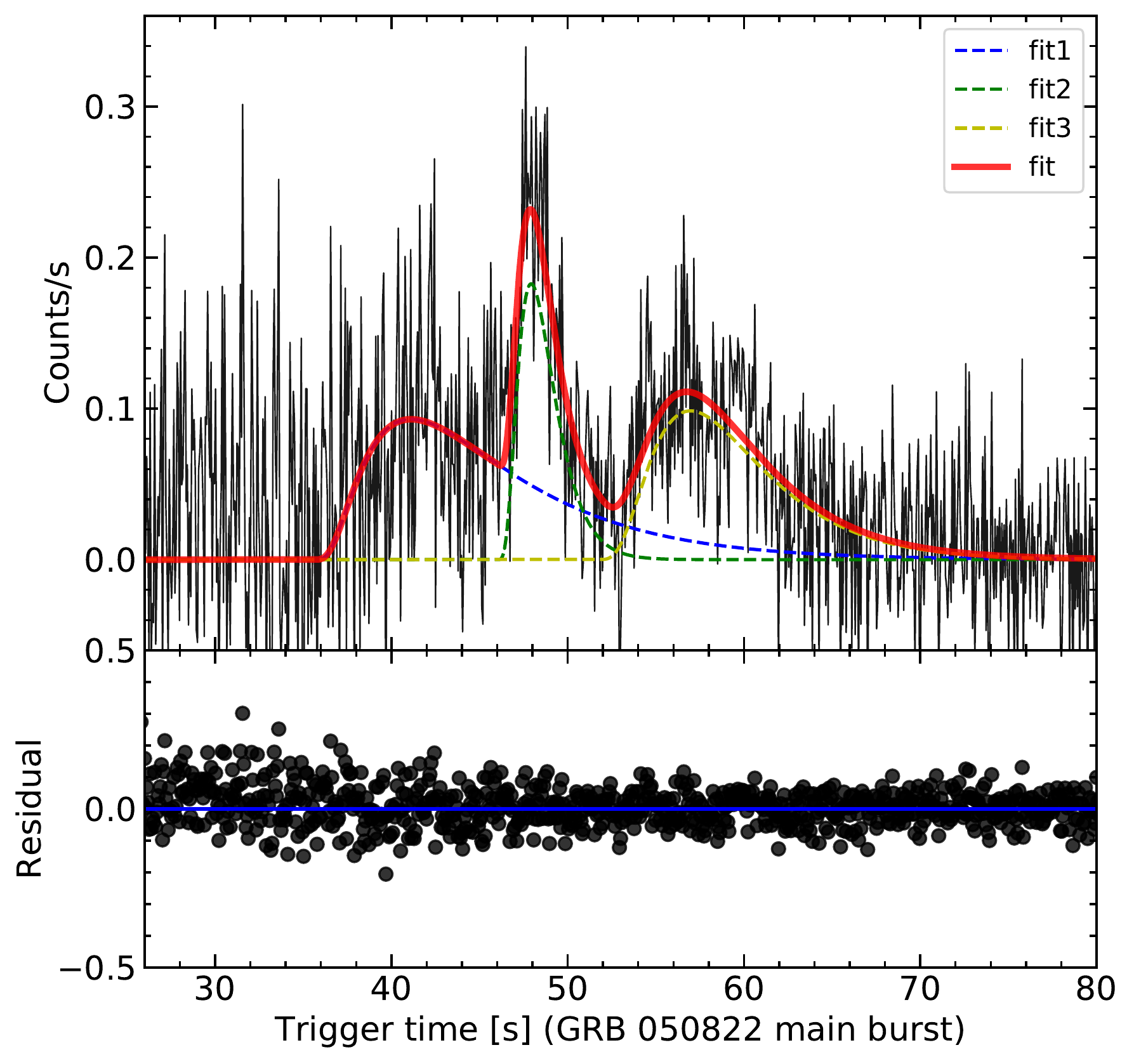}}
\caption{The lightcurve of GRB 050822
}
\end{figure}

\begin{figure}[!htp]
\centering
\subfigure{
\includegraphics[scale = 0.4]{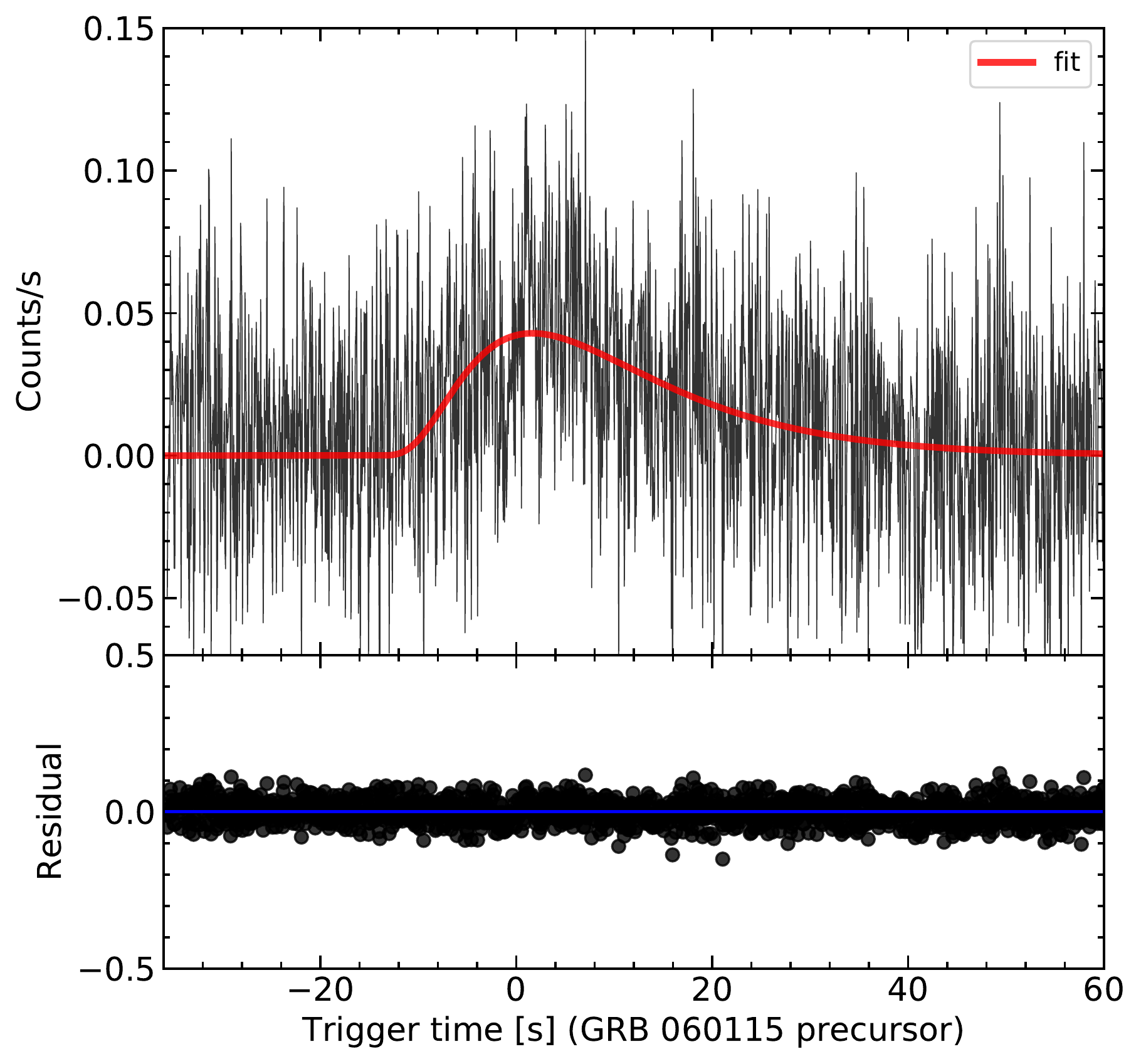}}
\subfigure{
\includegraphics[scale = 0.4]{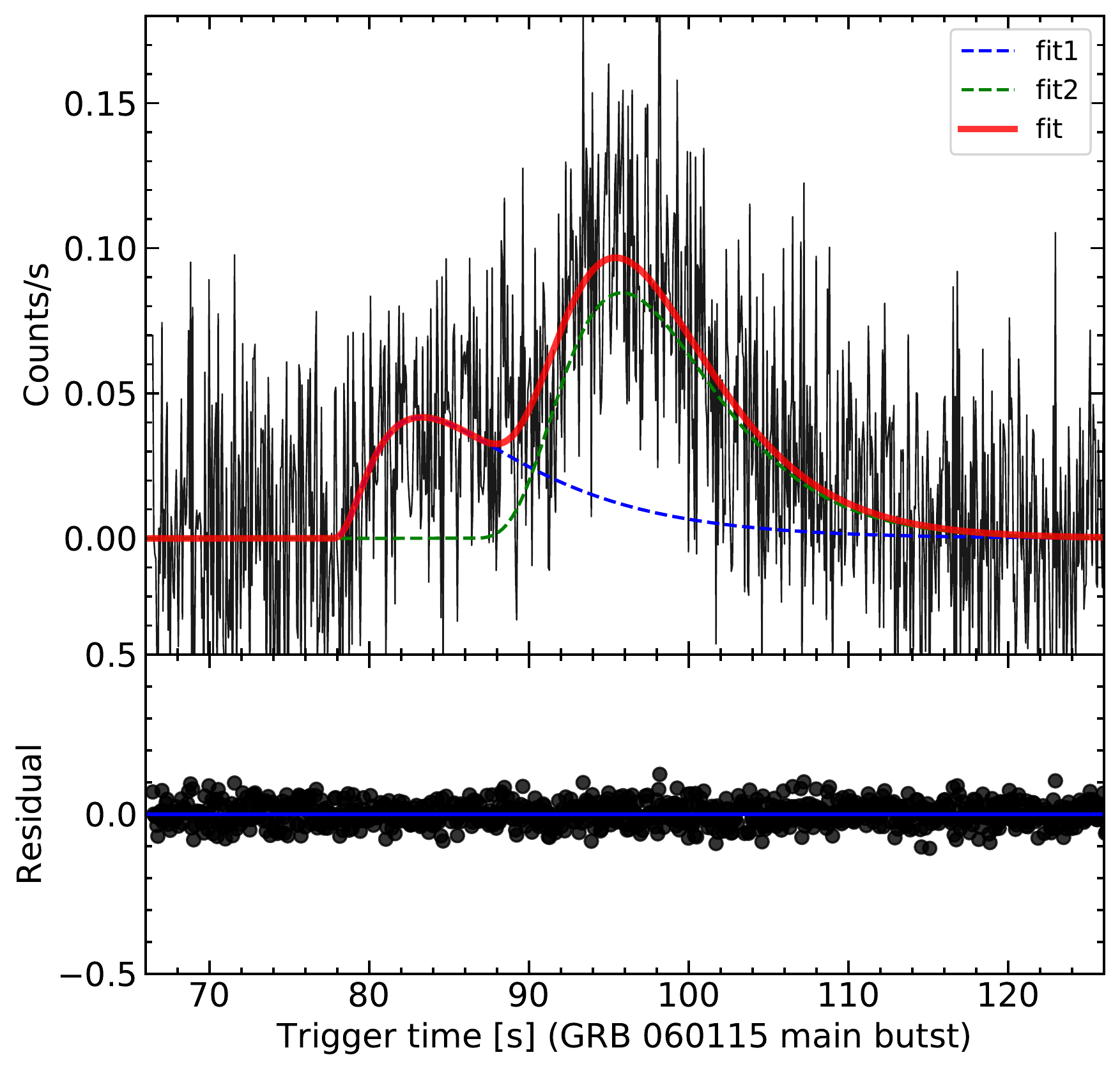}}
\caption{The lightcurve of GRB 060115
}
\end{figure}

\begin{figure}[!htp]
\centering
\subfigure{
\includegraphics[scale = 0.4]{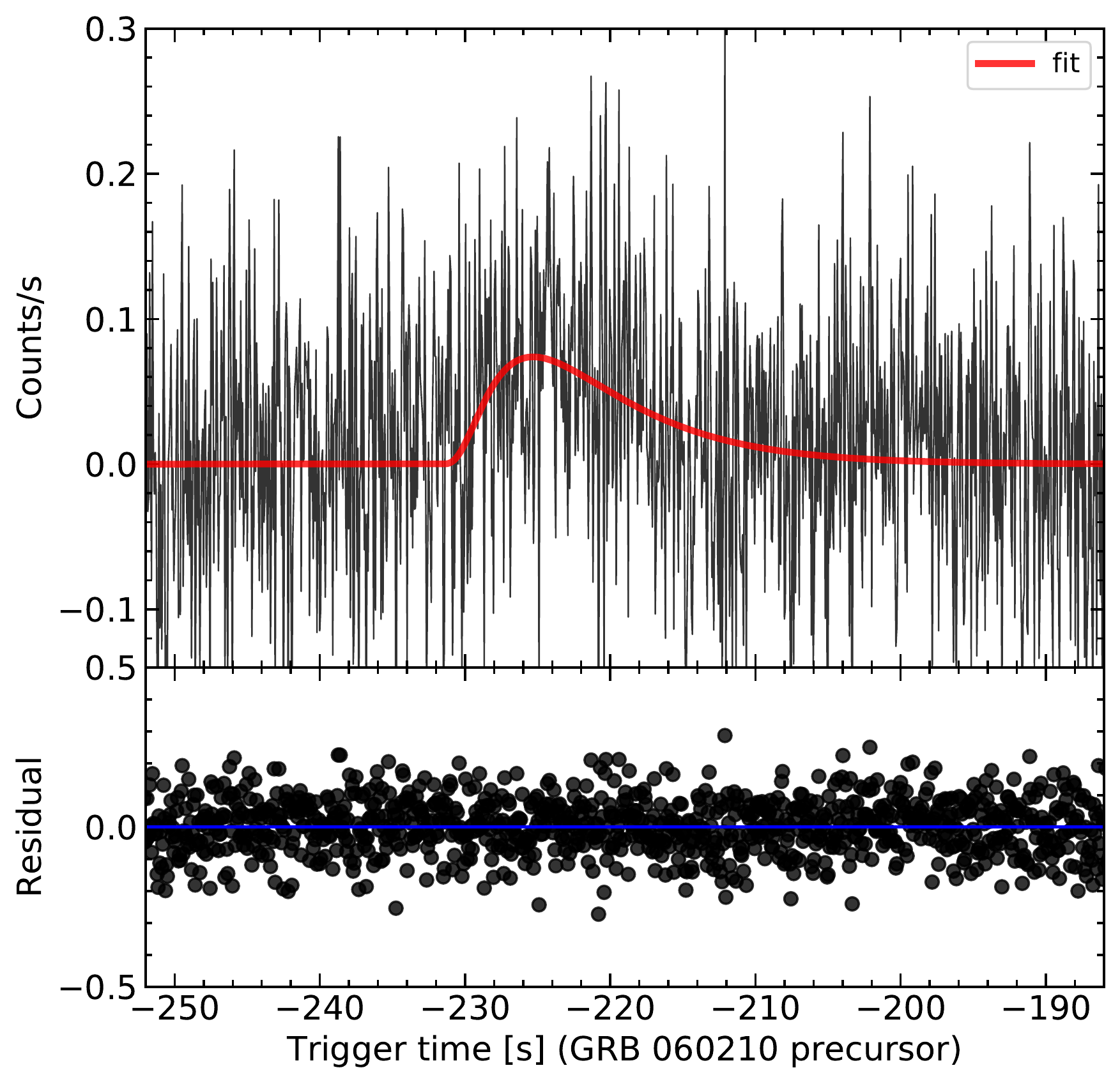}}
\subfigure{
\includegraphics[scale = 0.4]{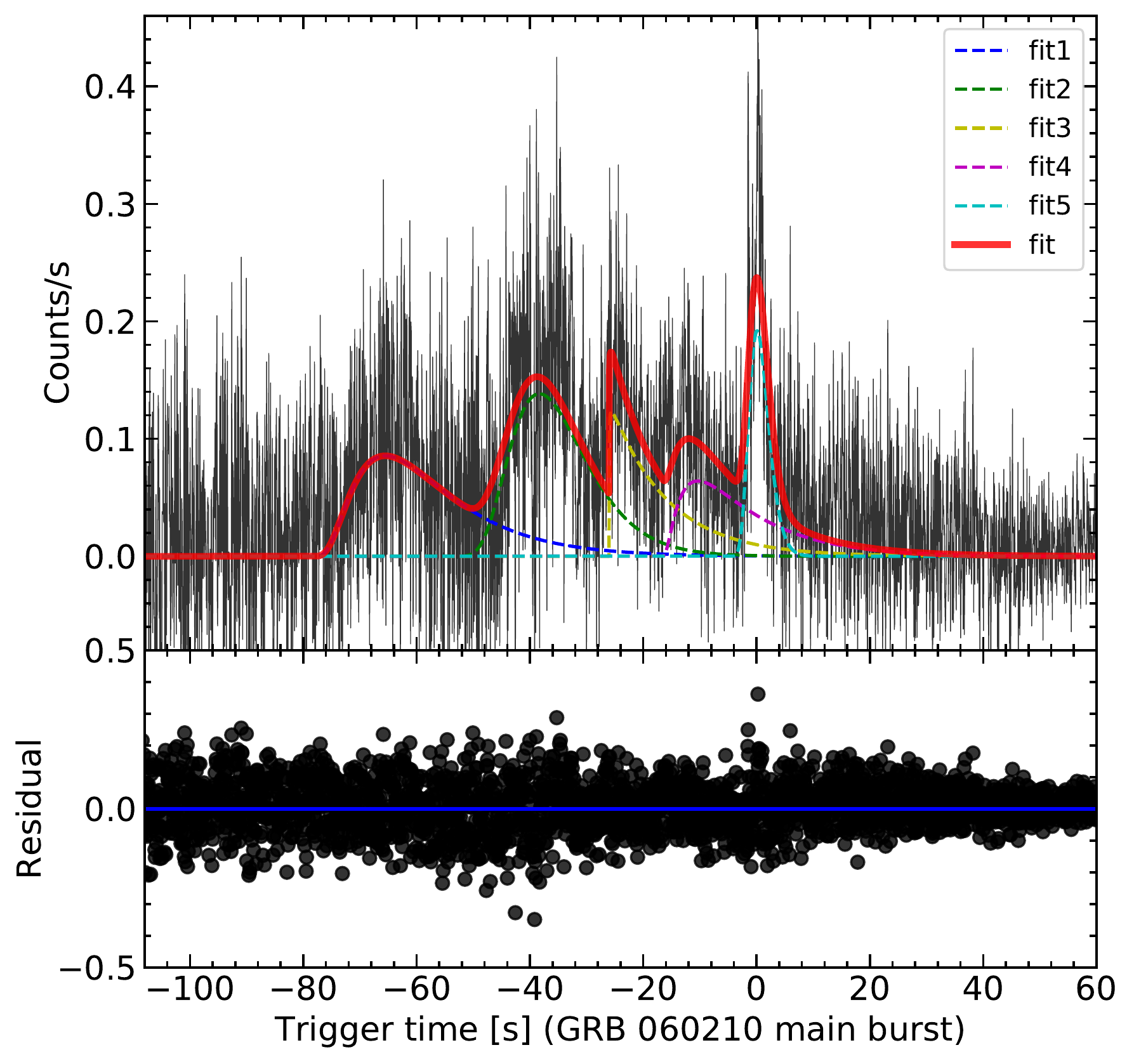}}
\caption{The lightcurve of GRB 060210
}
\end{figure}

\begin{figure}[!htp]
\centering
\subfigure{
\includegraphics[scale = 0.4]{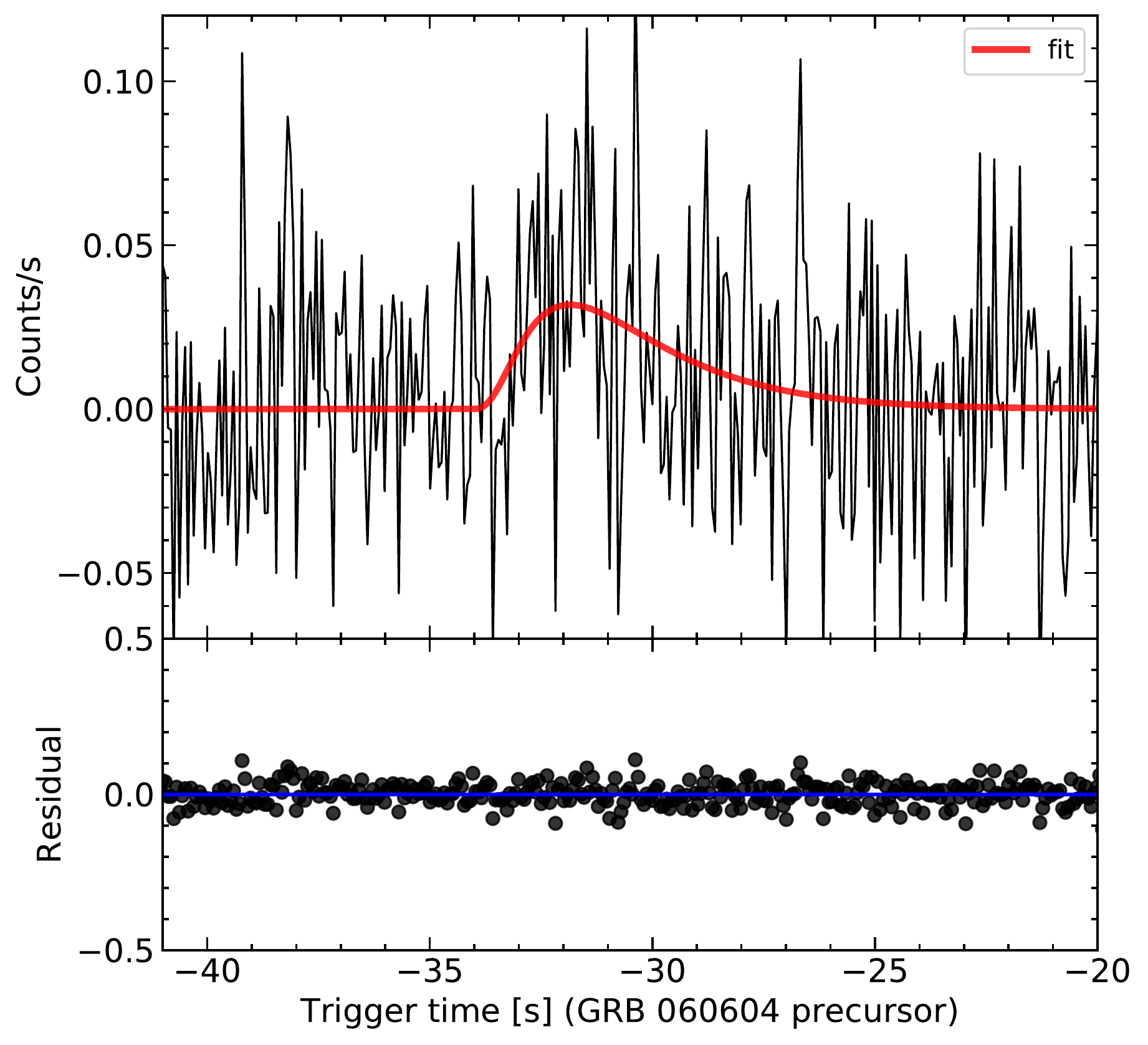}}
\subfigure{
\includegraphics[scale = 0.4]{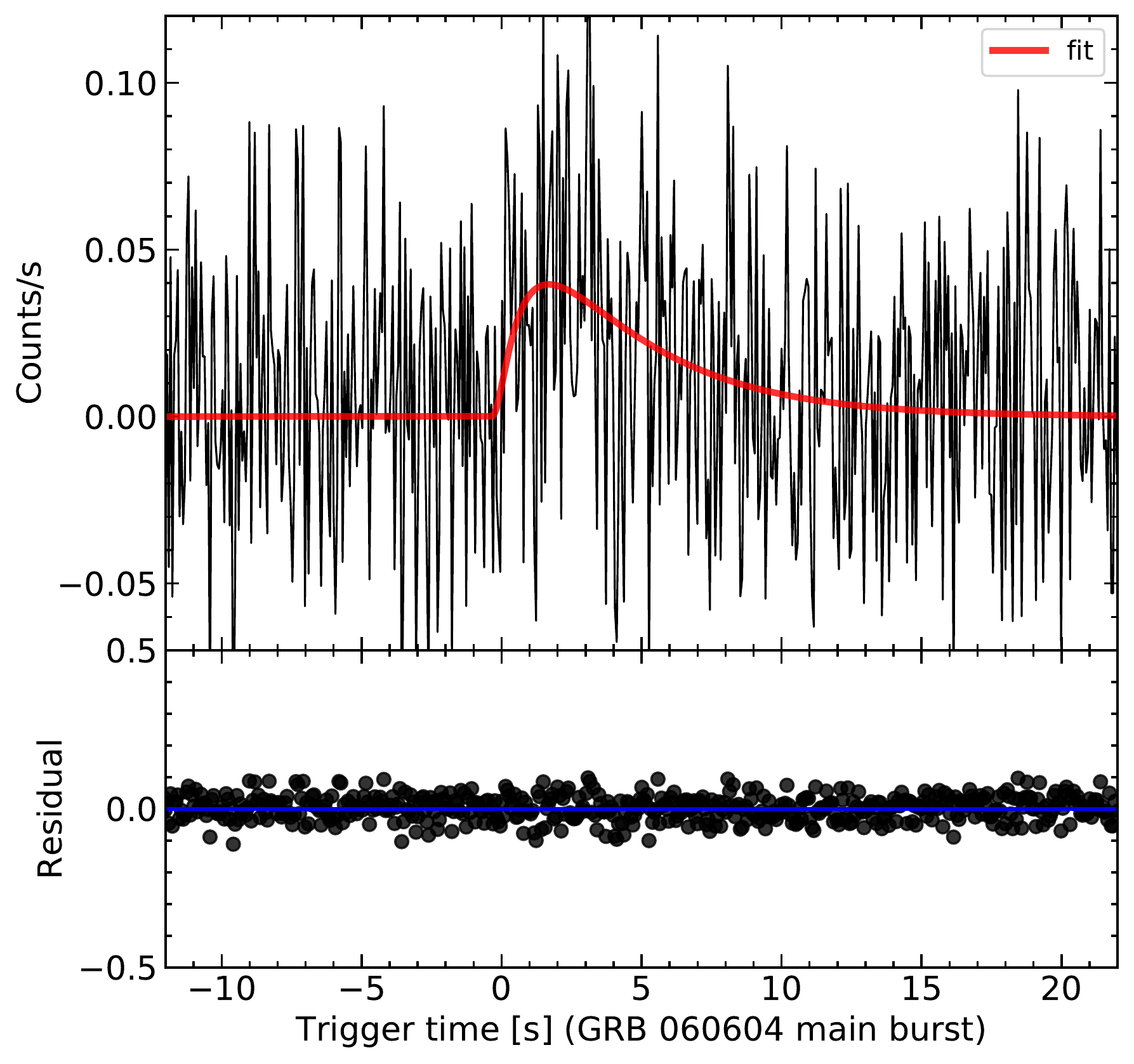}}
\caption{The lightcurve of GRB 060604
}
\end{figure}

\begin{figure}[!htp]
\centering
\subfigure{
\includegraphics[scale = 0.4]{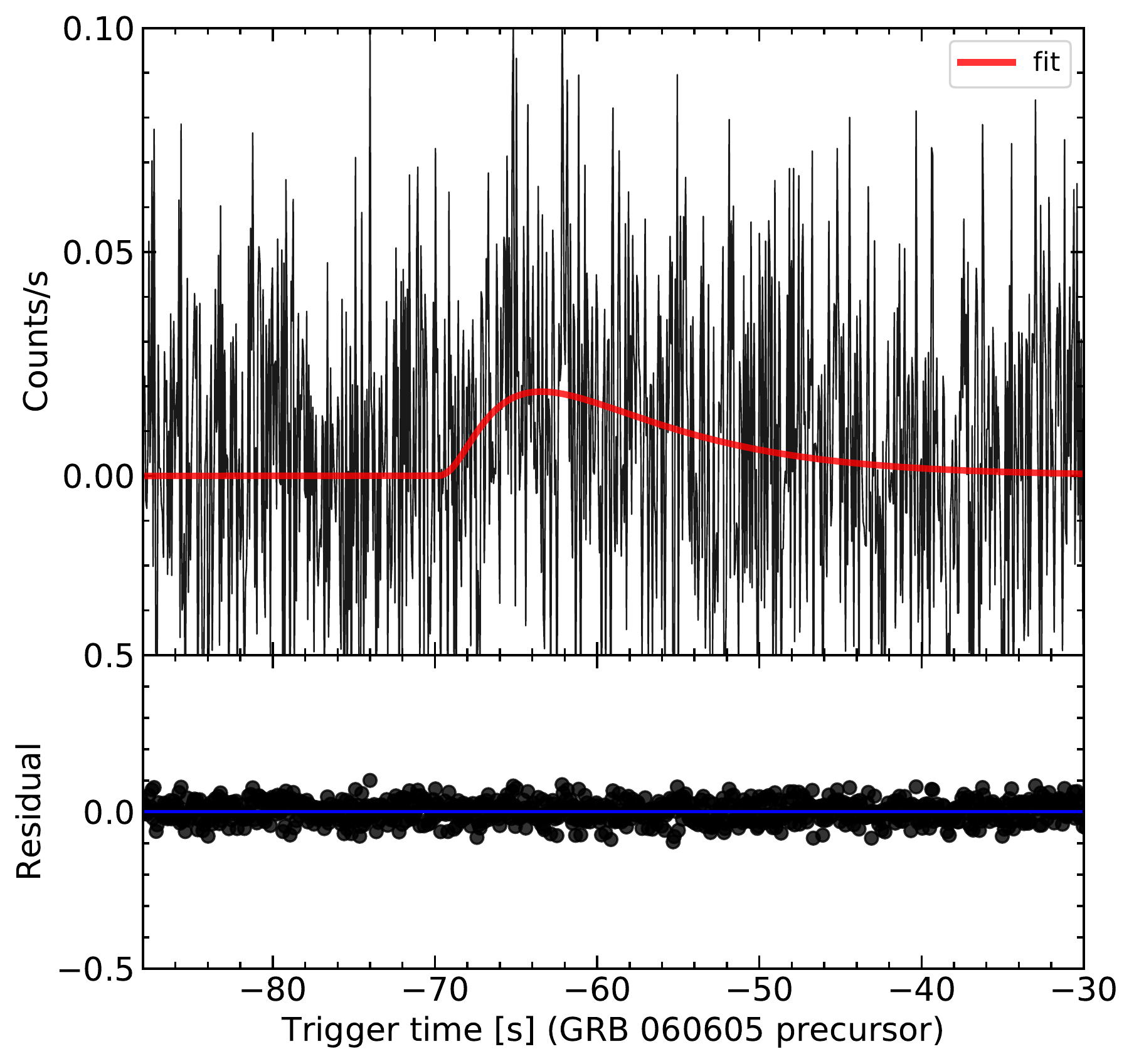}}
\subfigure{
\includegraphics[scale = 0.4]{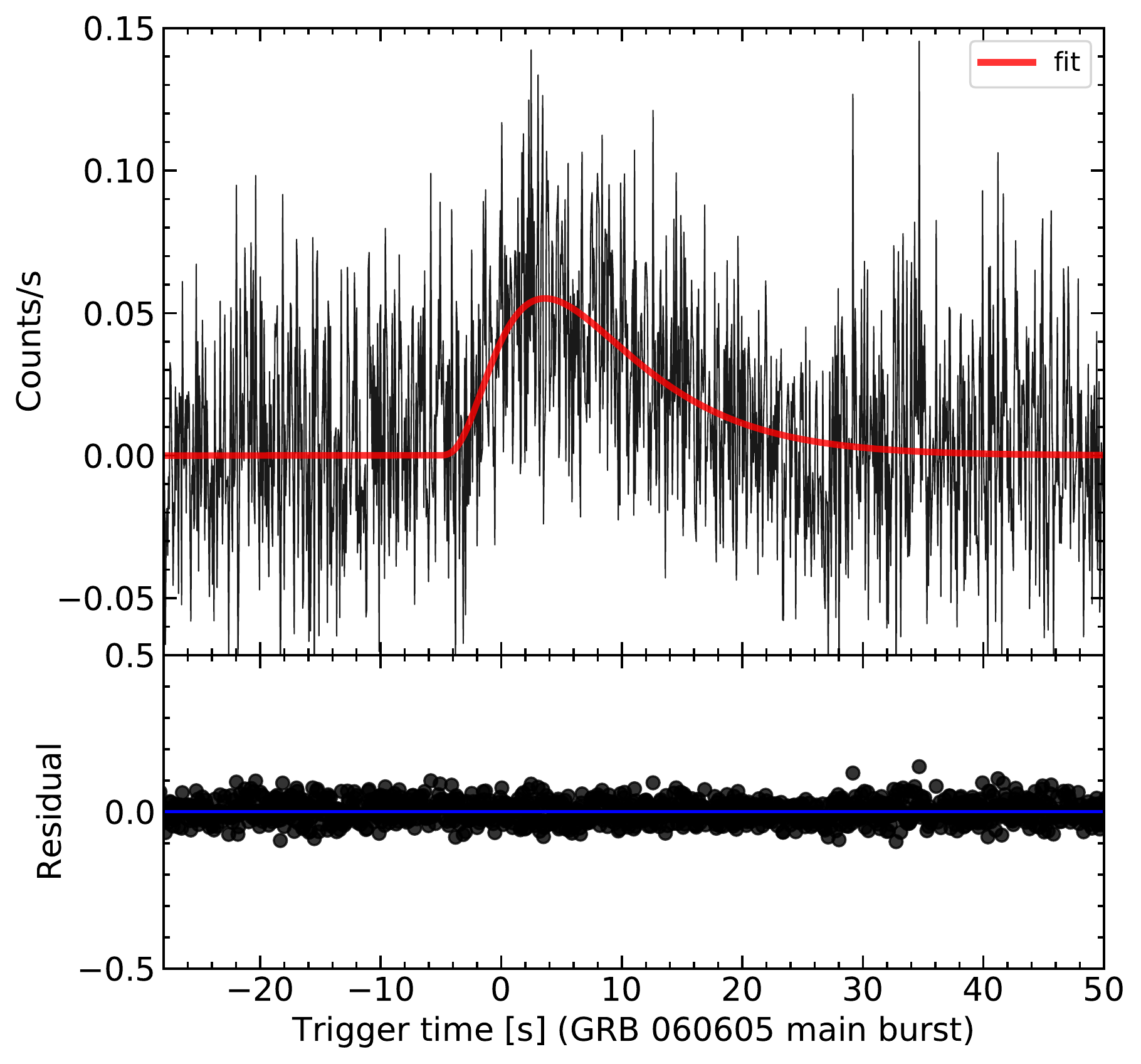}}
\caption{The lightcurve of GRB 060605
}
\end{figure}

\begin{figure}[!htp]
\centering
\subfigure{
\includegraphics[scale = 0.4]{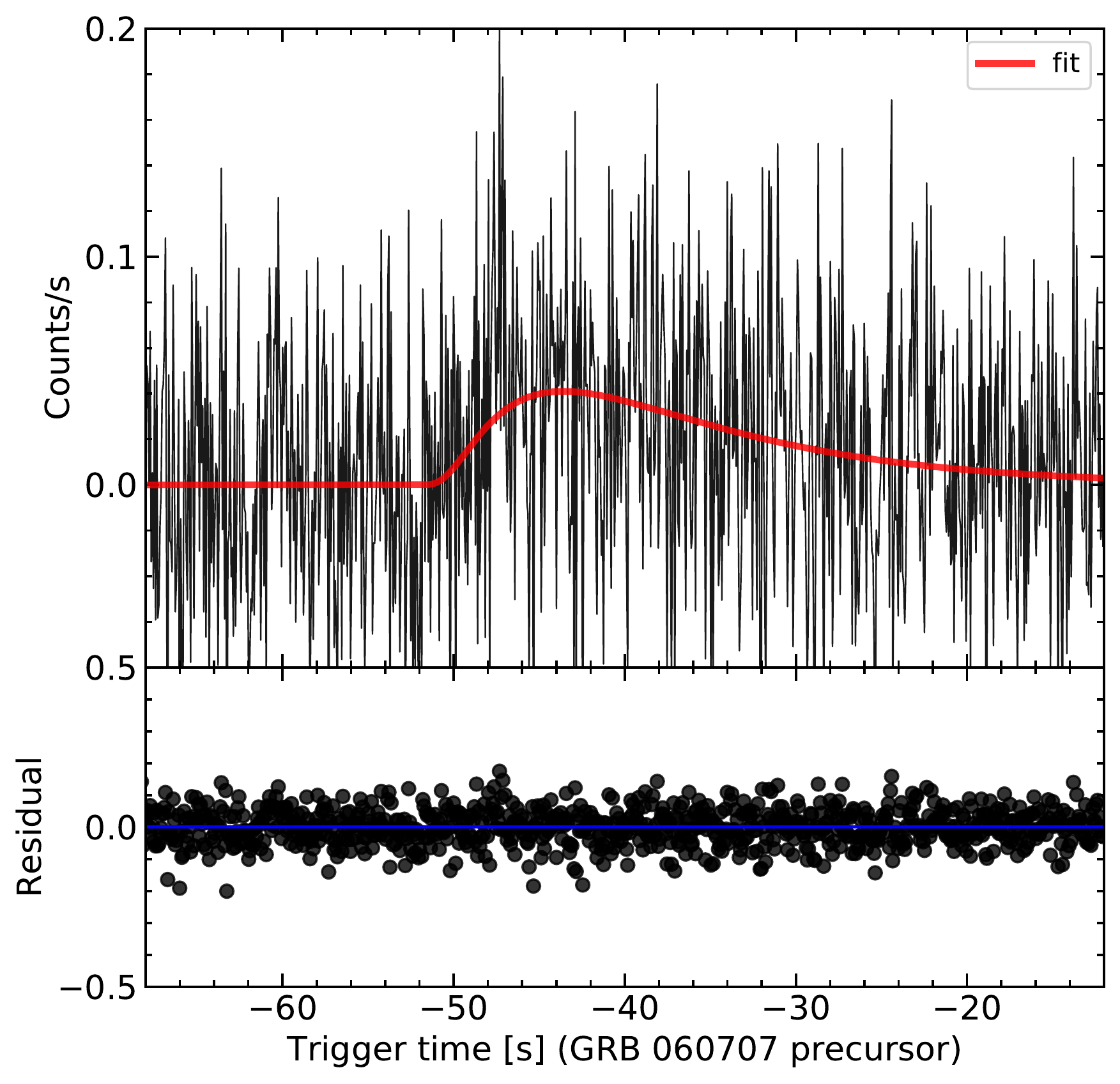}}
\subfigure{
\includegraphics[scale = 0.4]{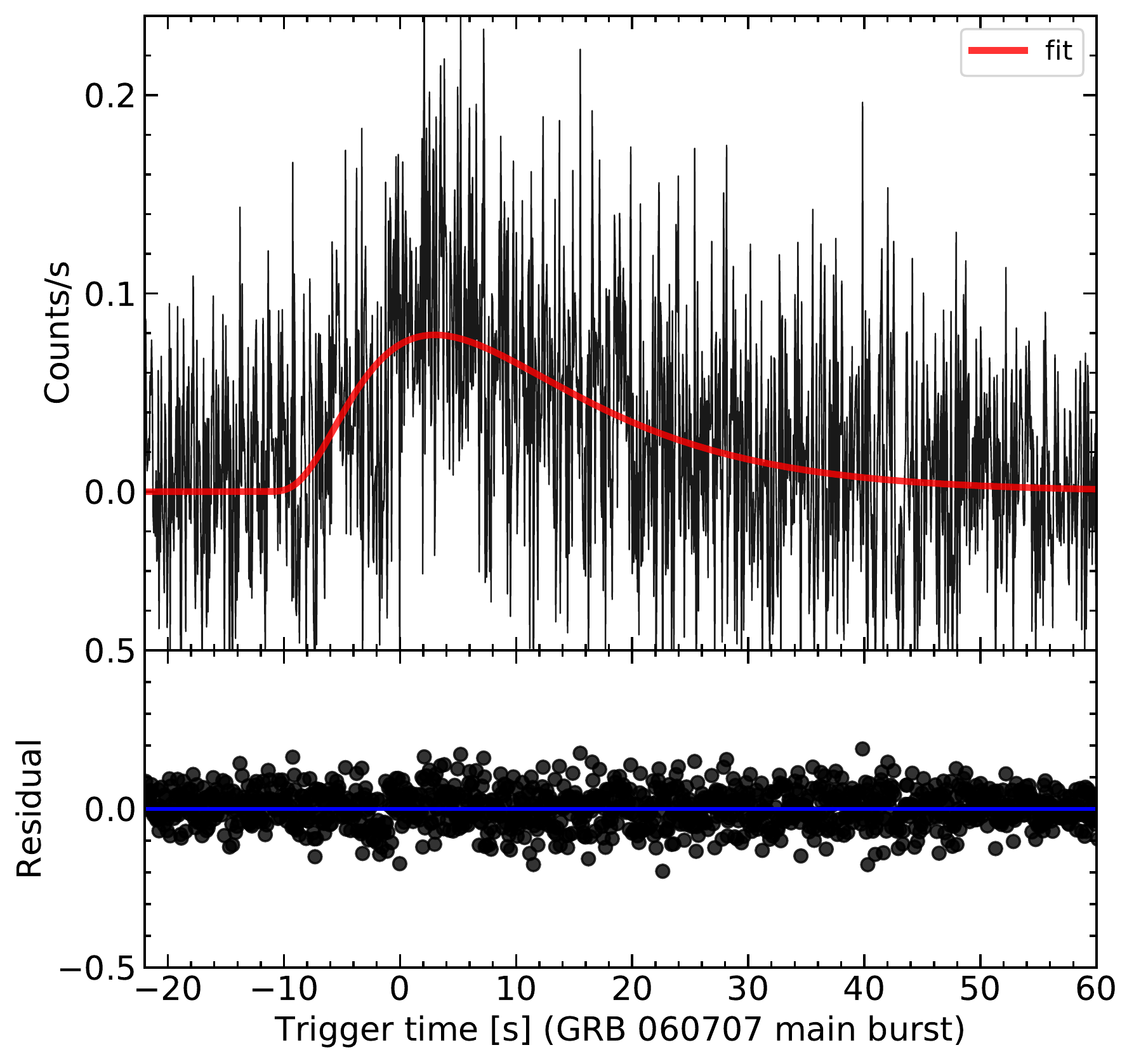}}
\caption{The lightcurve of GRB 060707
}
\end{figure}

\begin{figure}[!htp]
\centering
\subfigure{
\includegraphics[scale = 0.4]{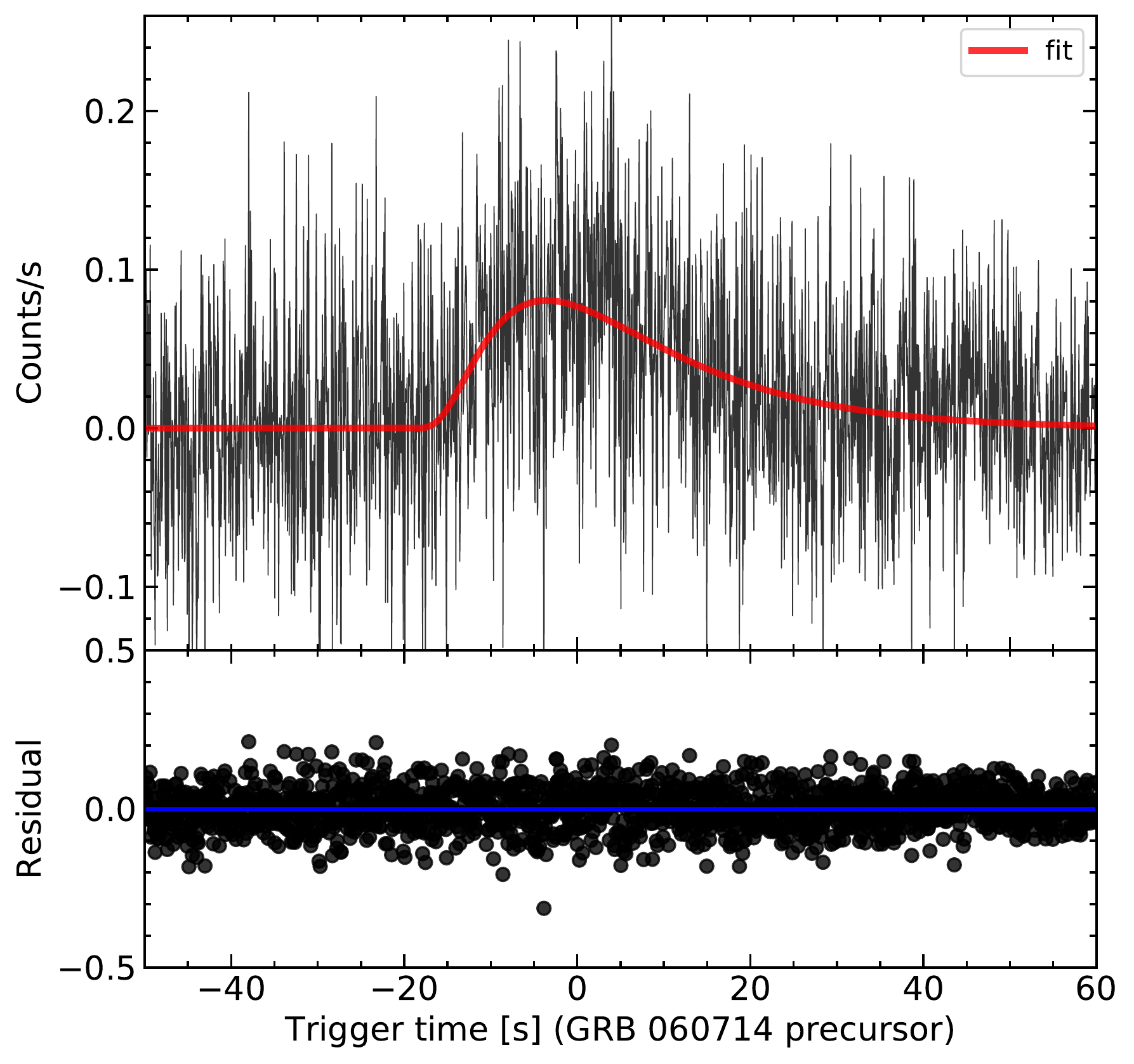}}
\subfigure{
\includegraphics[scale = 0.4]{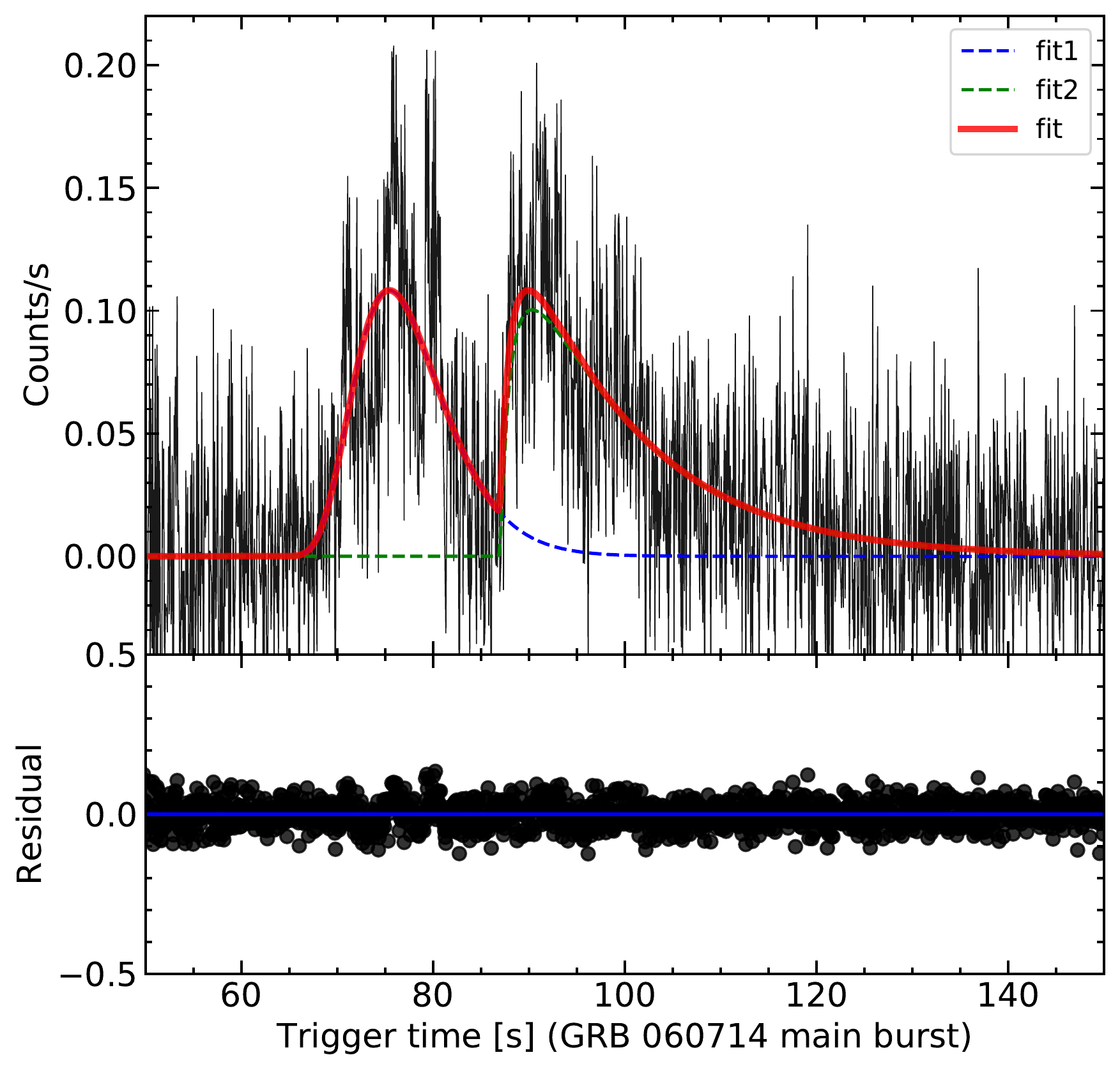}}
\caption{The lightcurve of GRB 060714
}
\end{figure}

\begin{figure}[!htp]
\centering
\subfigure{
\includegraphics[scale = 0.4]{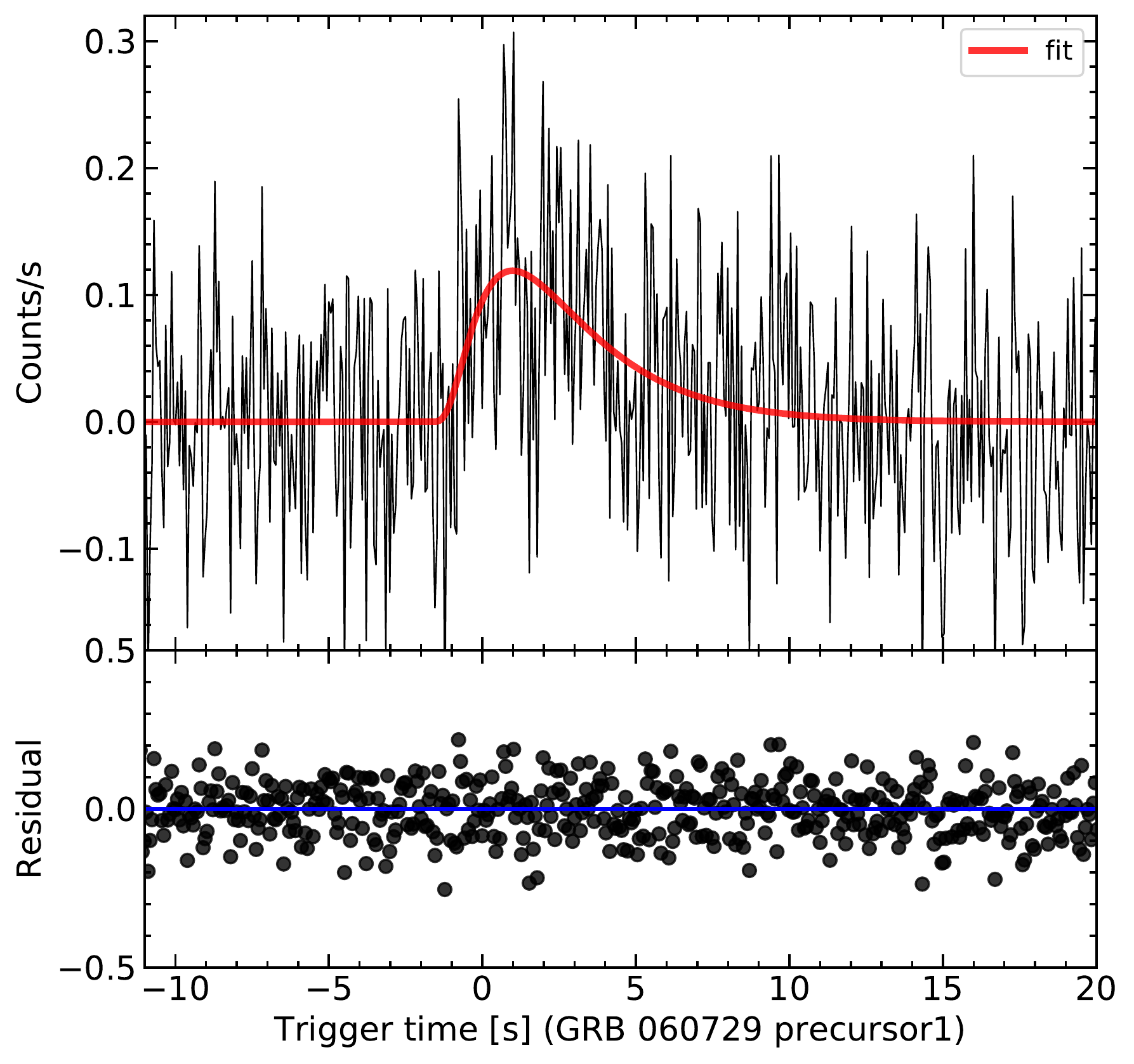}}
\subfigure{
\includegraphics[scale = 0.4]{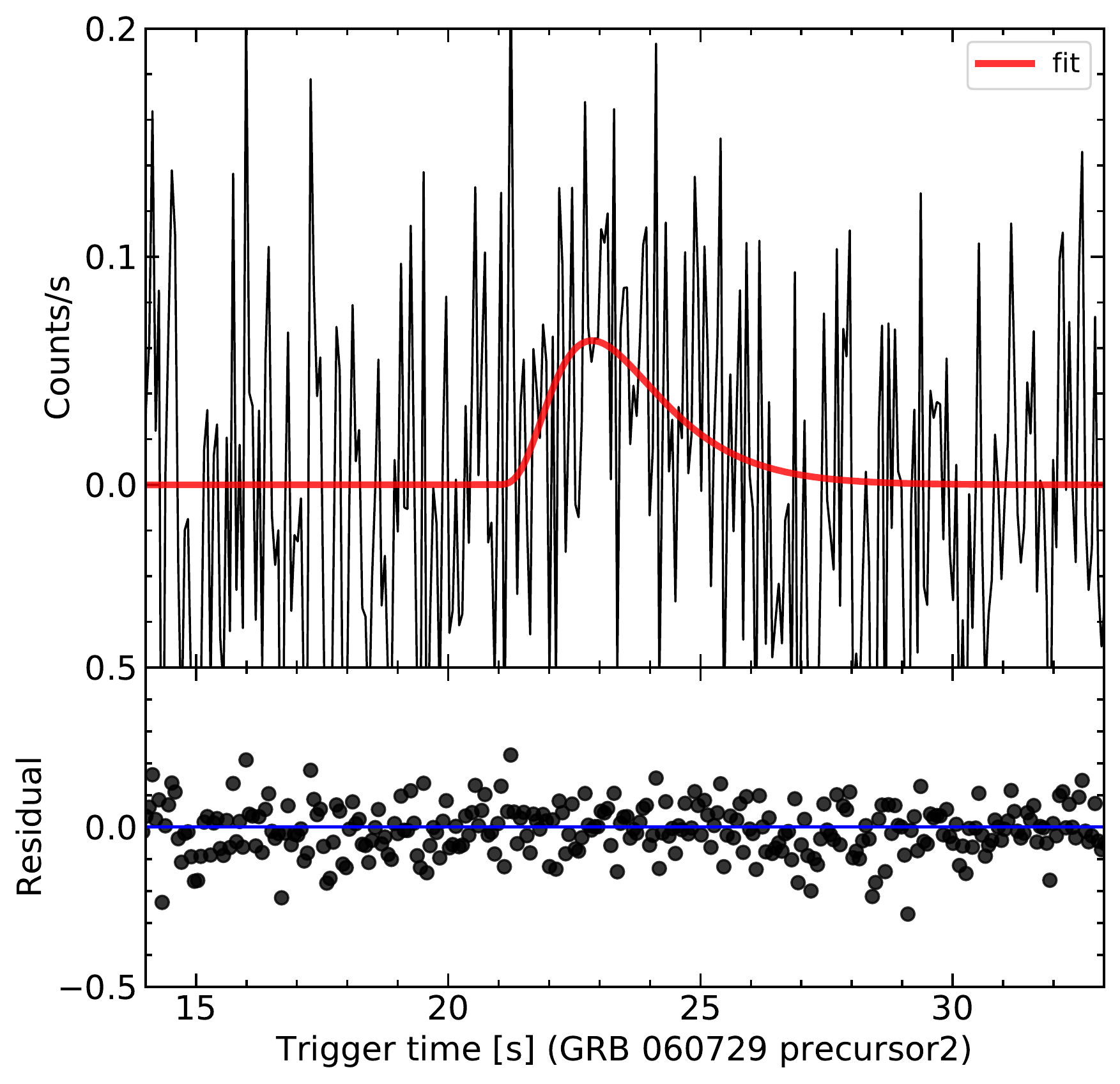}}
\subfigure{
\includegraphics[scale = 0.4]{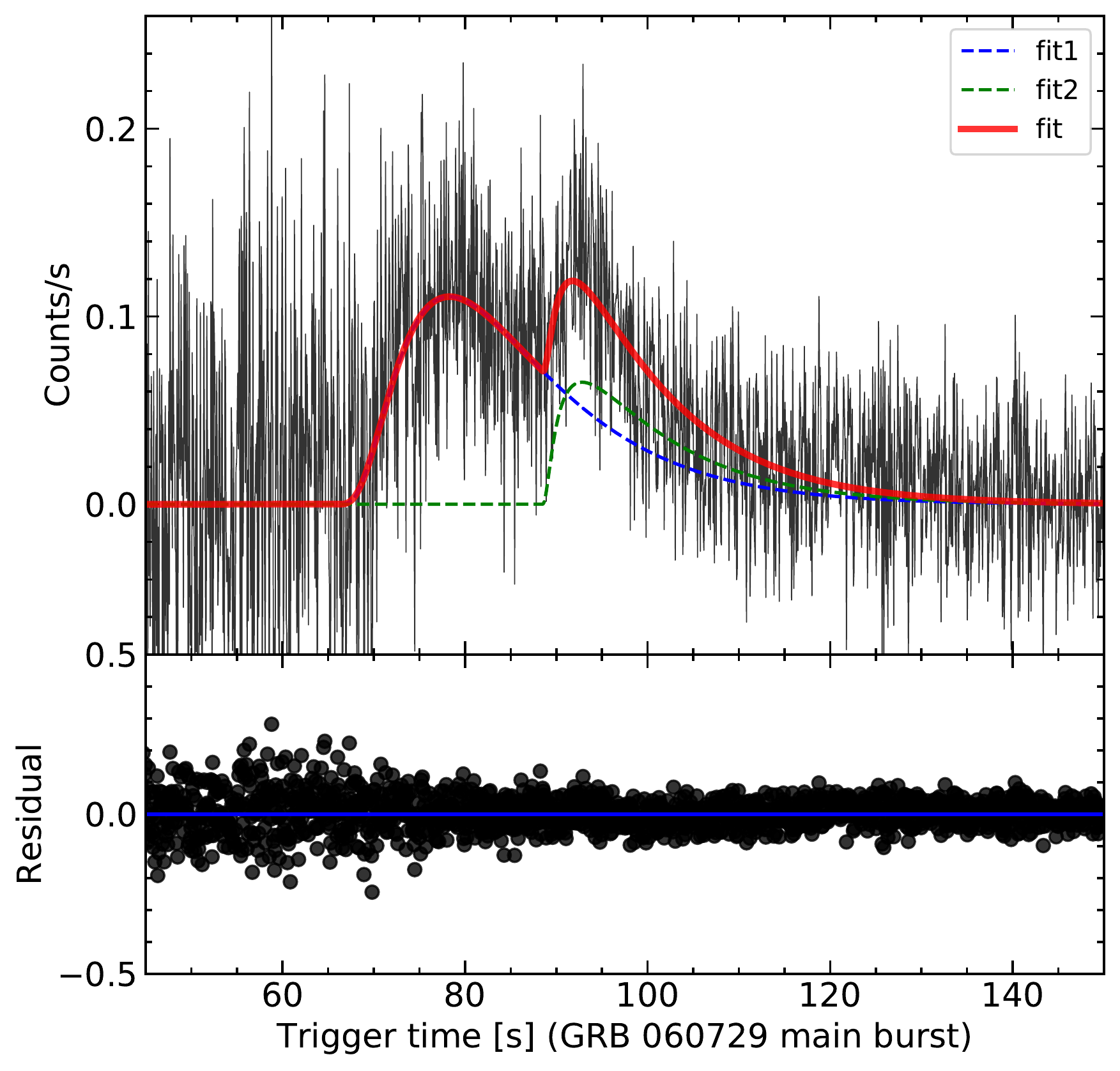}}
\caption{The lightcurve of GRB 060729
}
\end{figure}

\begin{figure}[!htp]
\centering
\subfigure{
\includegraphics[scale = 0.4]{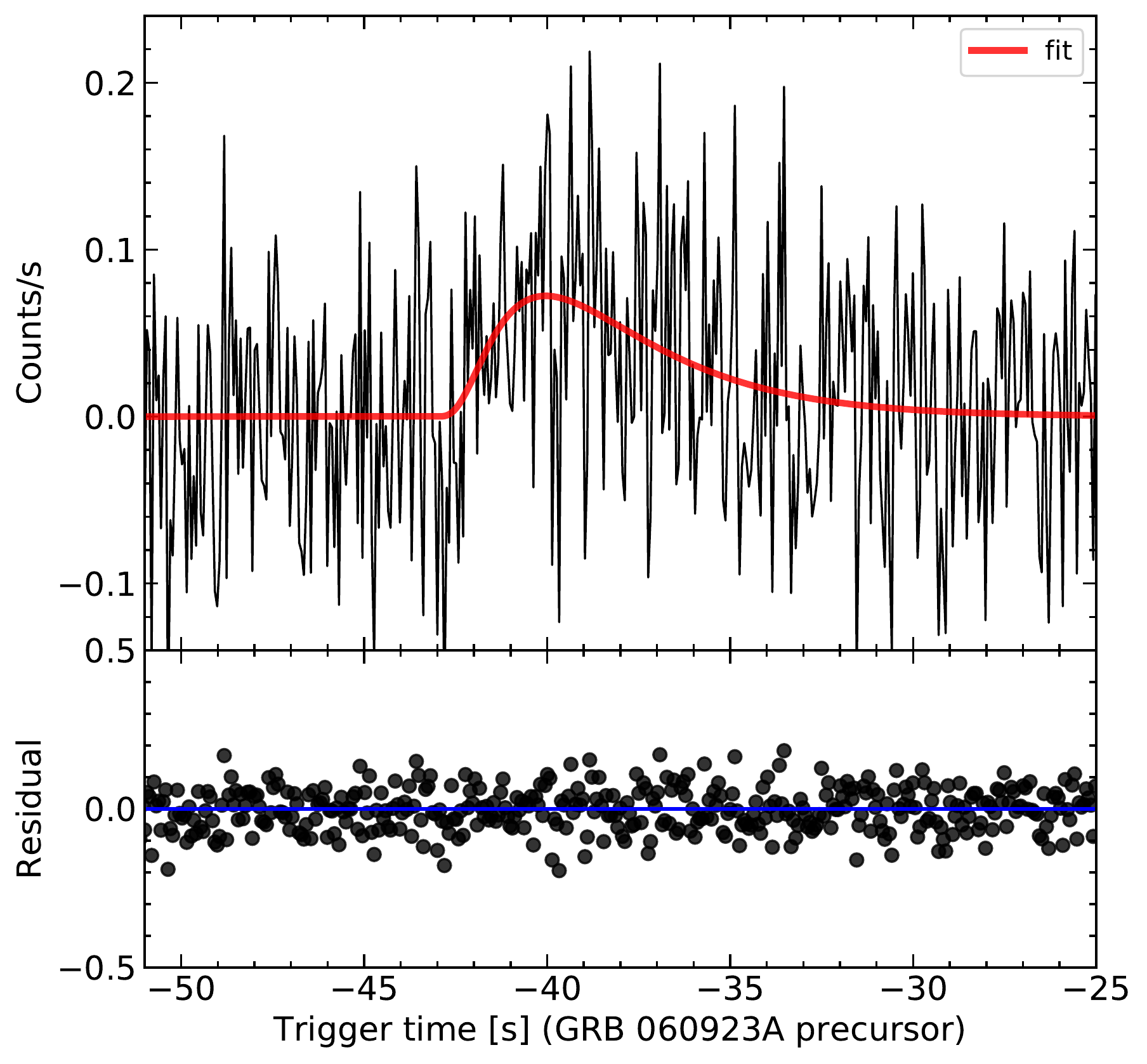}}
\subfigure{
\includegraphics[scale = 0.4]{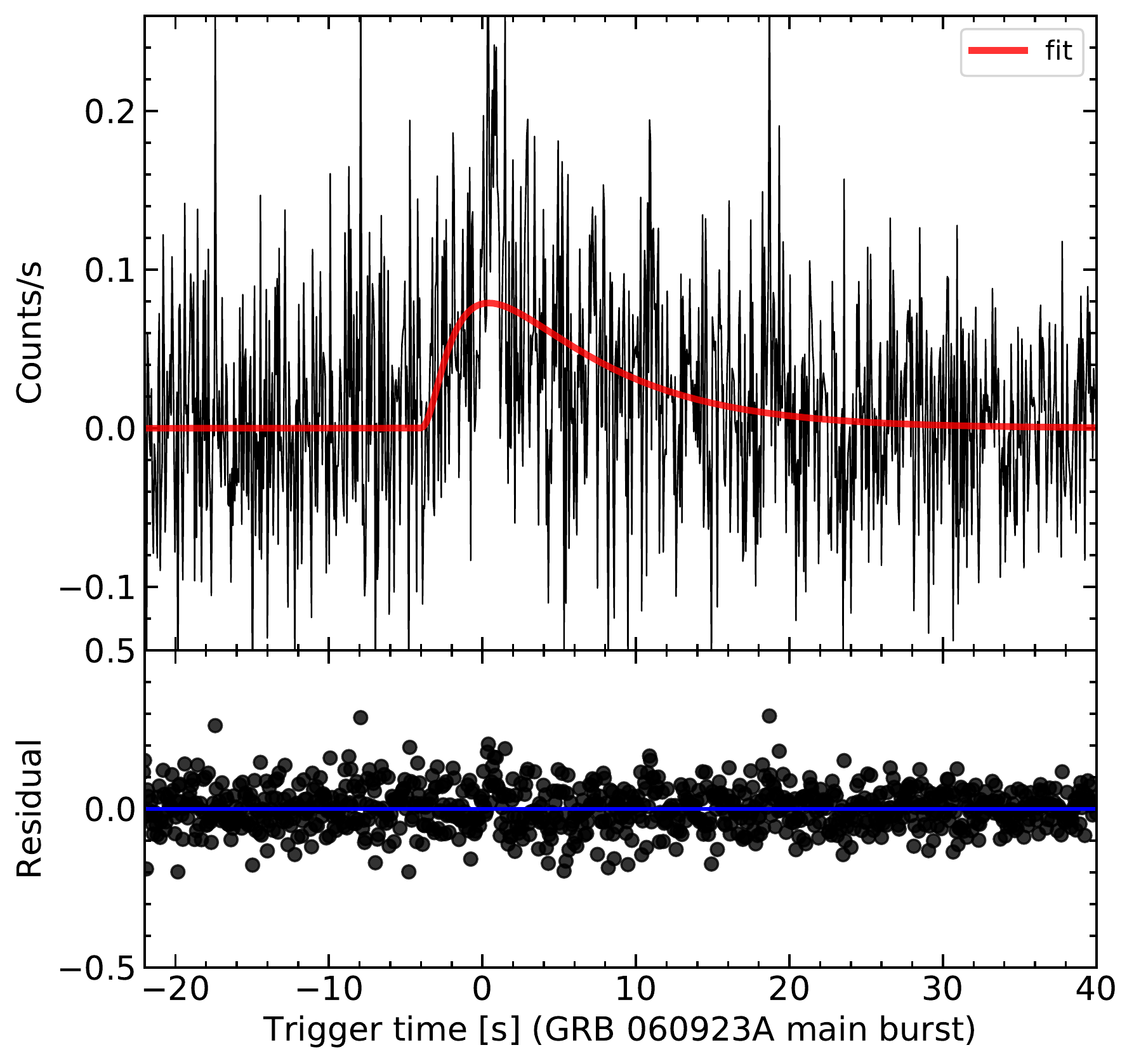}}
\caption{The lightcurve of GRB 060923A
}
\end{figure}

\begin{figure}[!htp]
\centering
\subfigure{
\includegraphics[scale = 0.4]{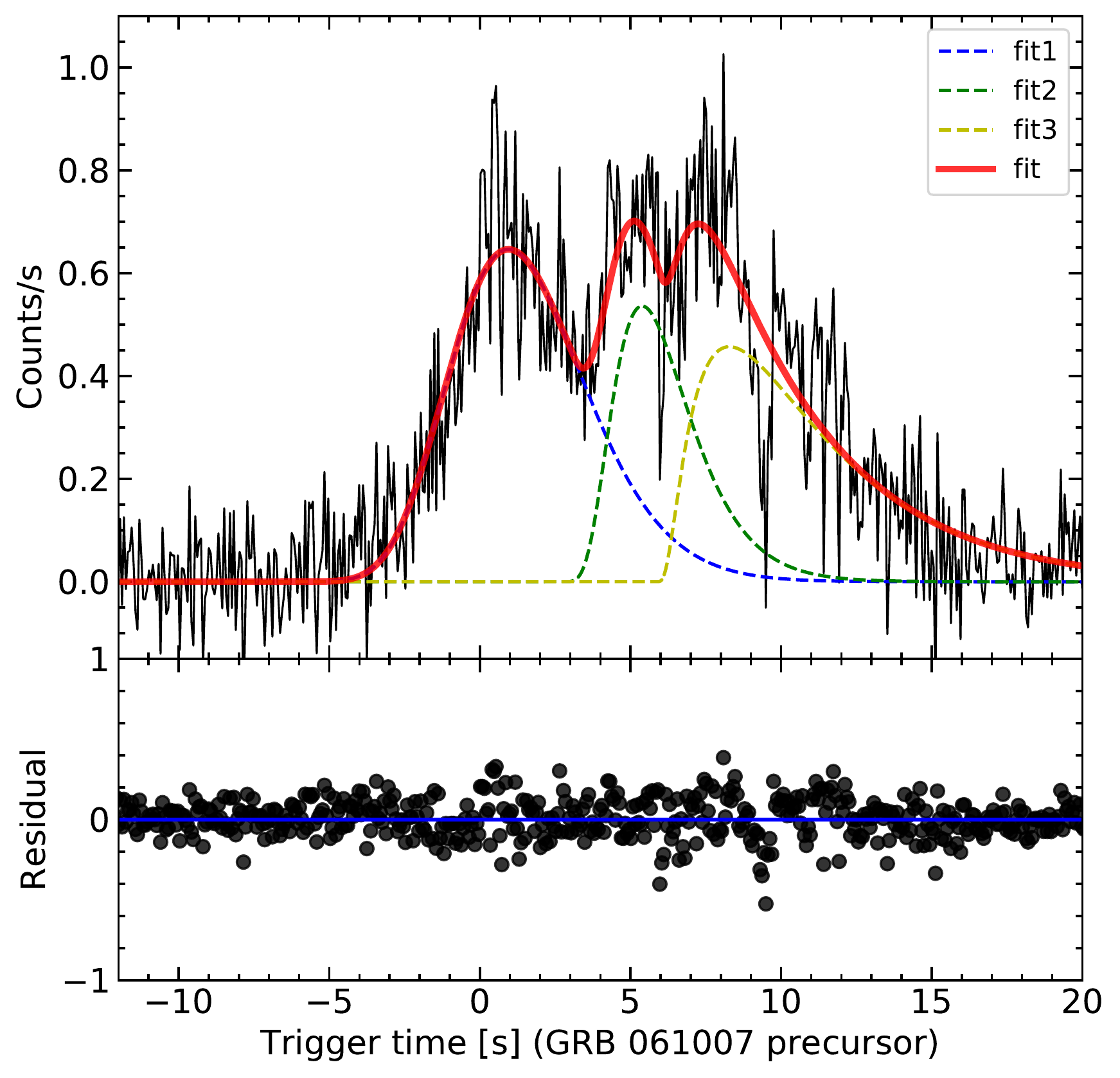}}
\subfigure{
\includegraphics[scale = 0.4]{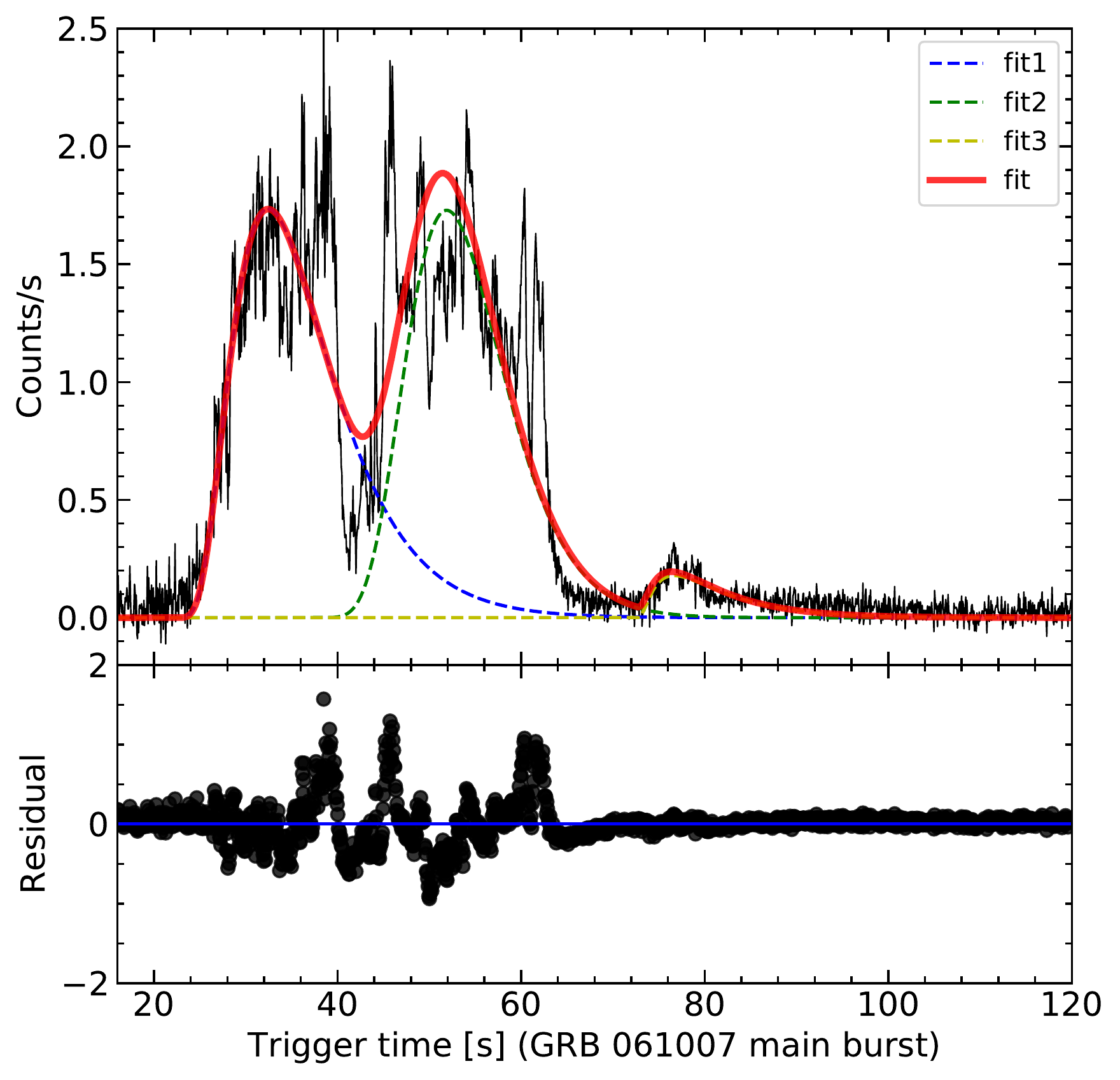}}
\caption{The lightcurve of GRB 061007
}
\end{figure}

\begin{figure}[!htp]
\centering
\subfigure{
\includegraphics[scale = 0.4]{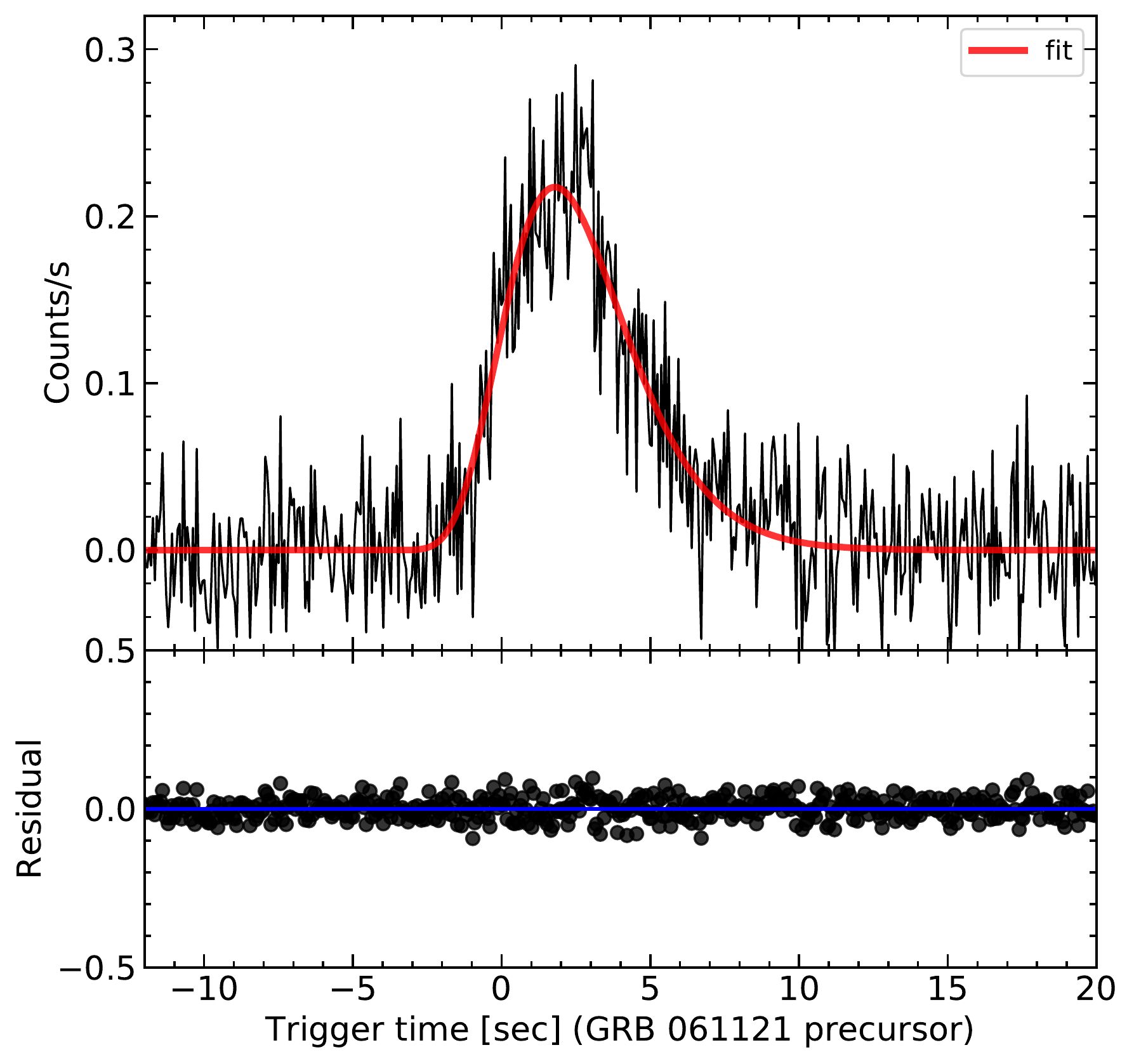}}
\subfigure{
\includegraphics[scale = 0.4]{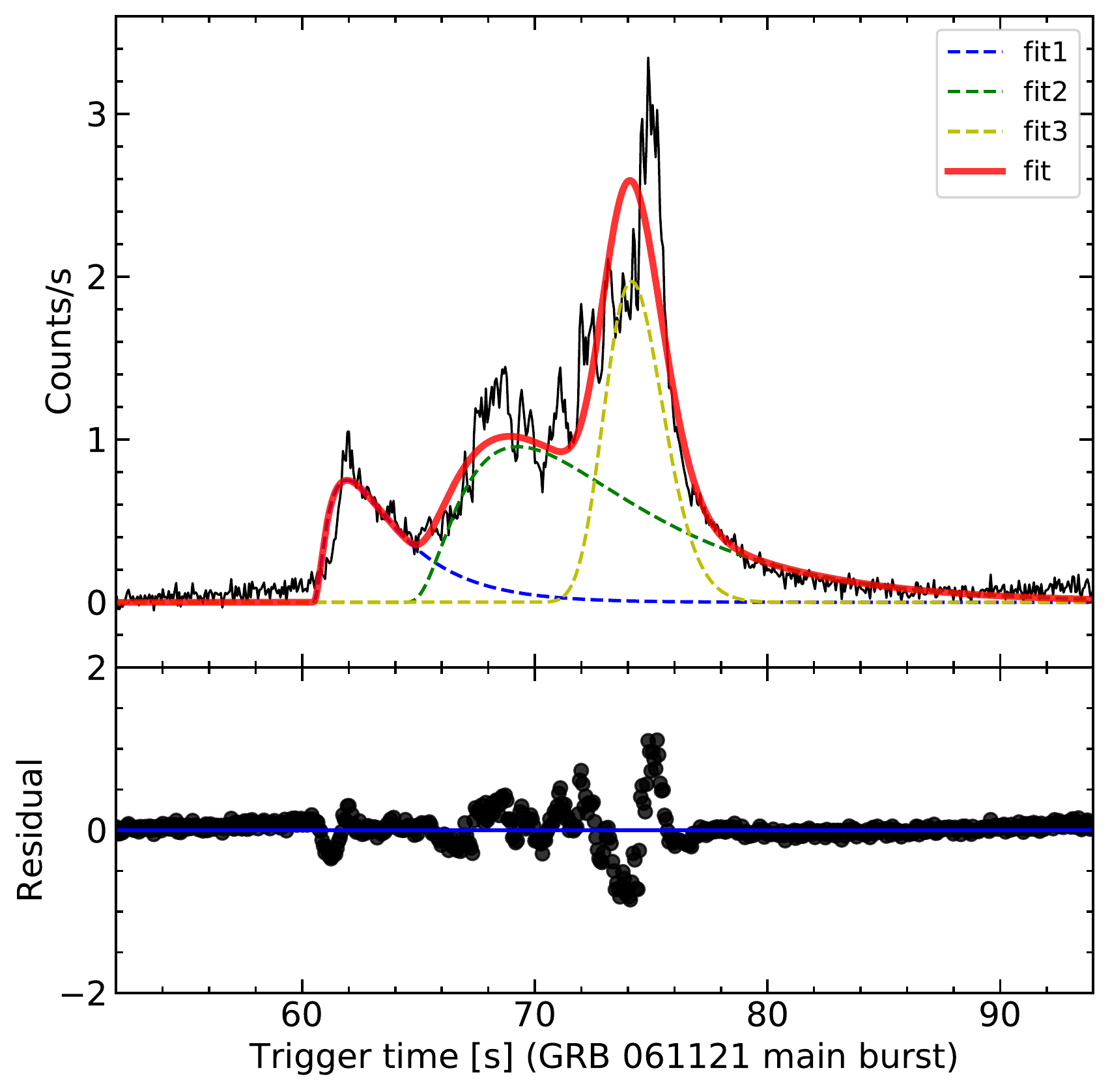}}
\caption{The lightcurve of GRB 061121
}
\end{figure}
\clearpage

\begin{figure}[!htp]
\centering
\subfigure{
\includegraphics[scale = 0.4]{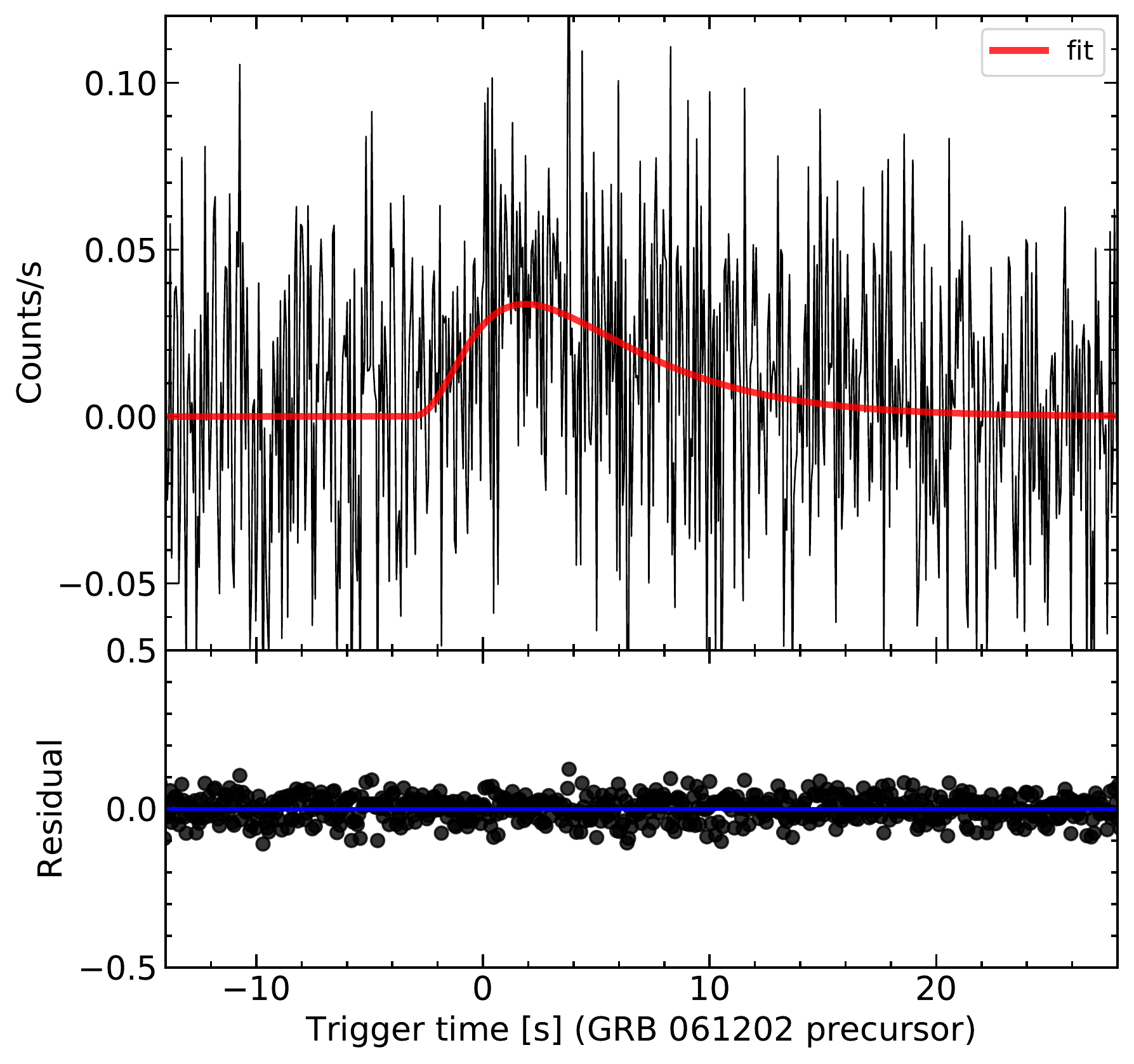}}
\subfigure{
\includegraphics[scale = 0.4]{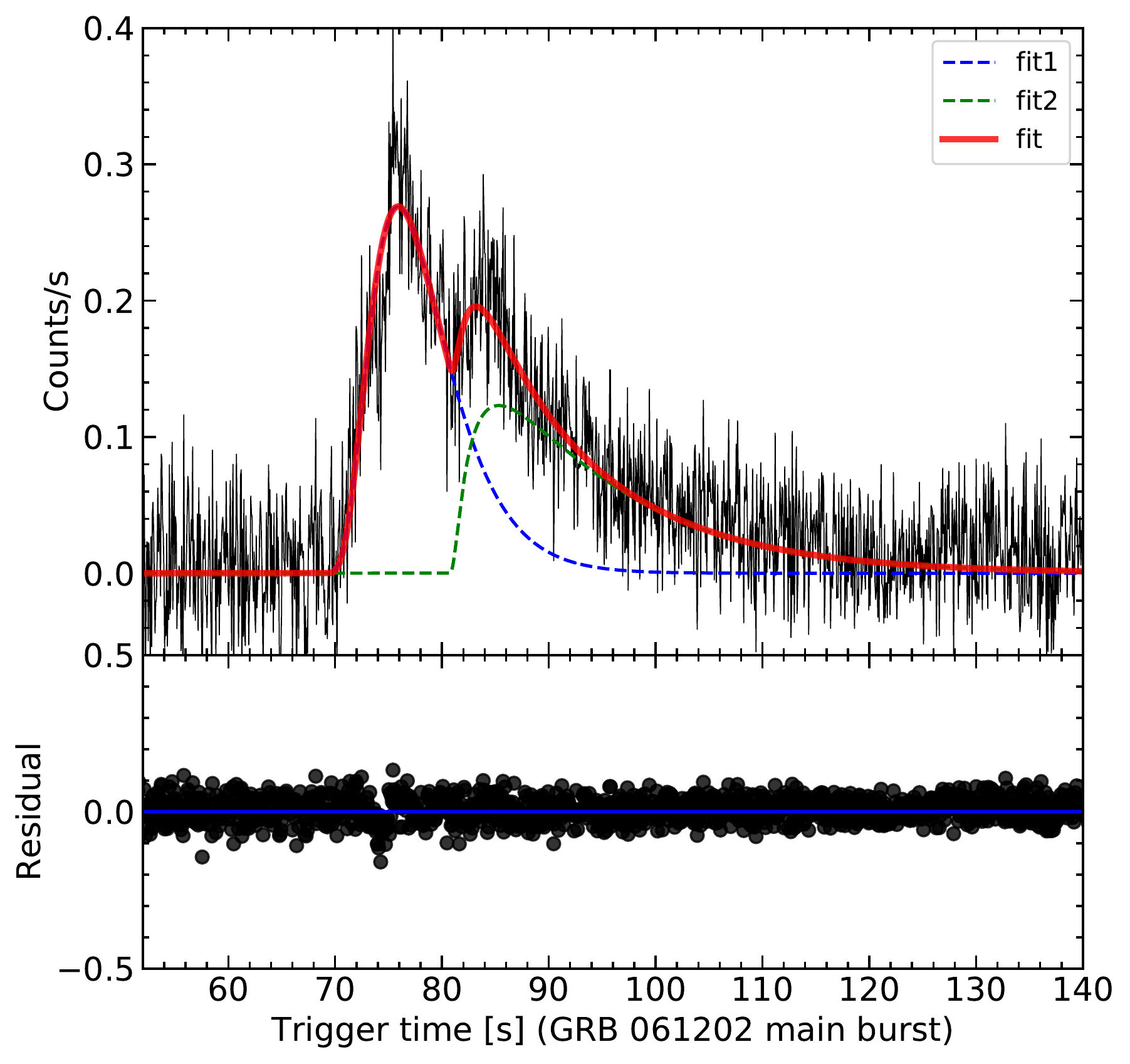}}
\caption{The lightcurve of GRB 061202
}
\end{figure}

\begin{figure}[!htp]
\centering
\subfigure{
\includegraphics[scale = 0.4]{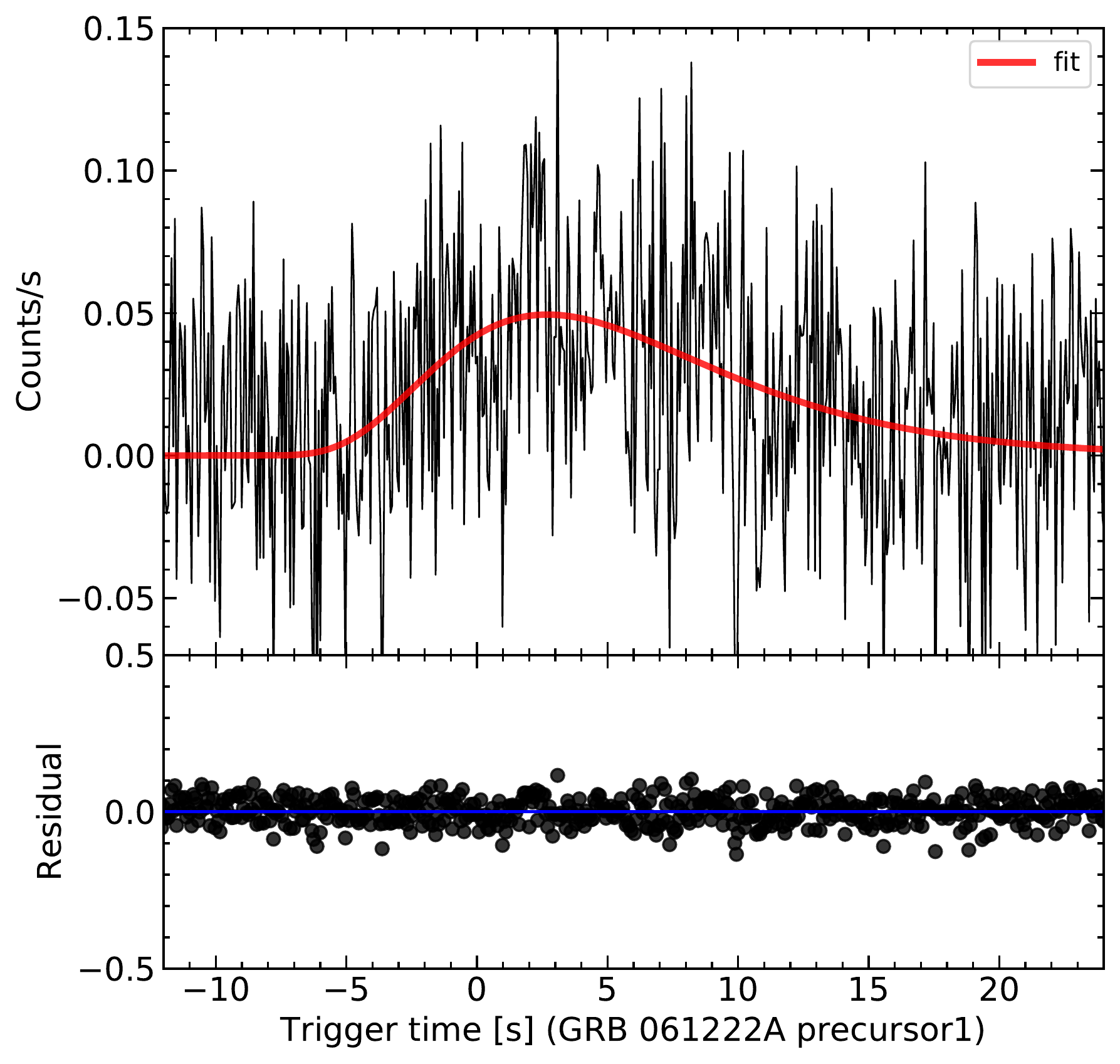}}
\subfigure{
\includegraphics[scale = 0.4]{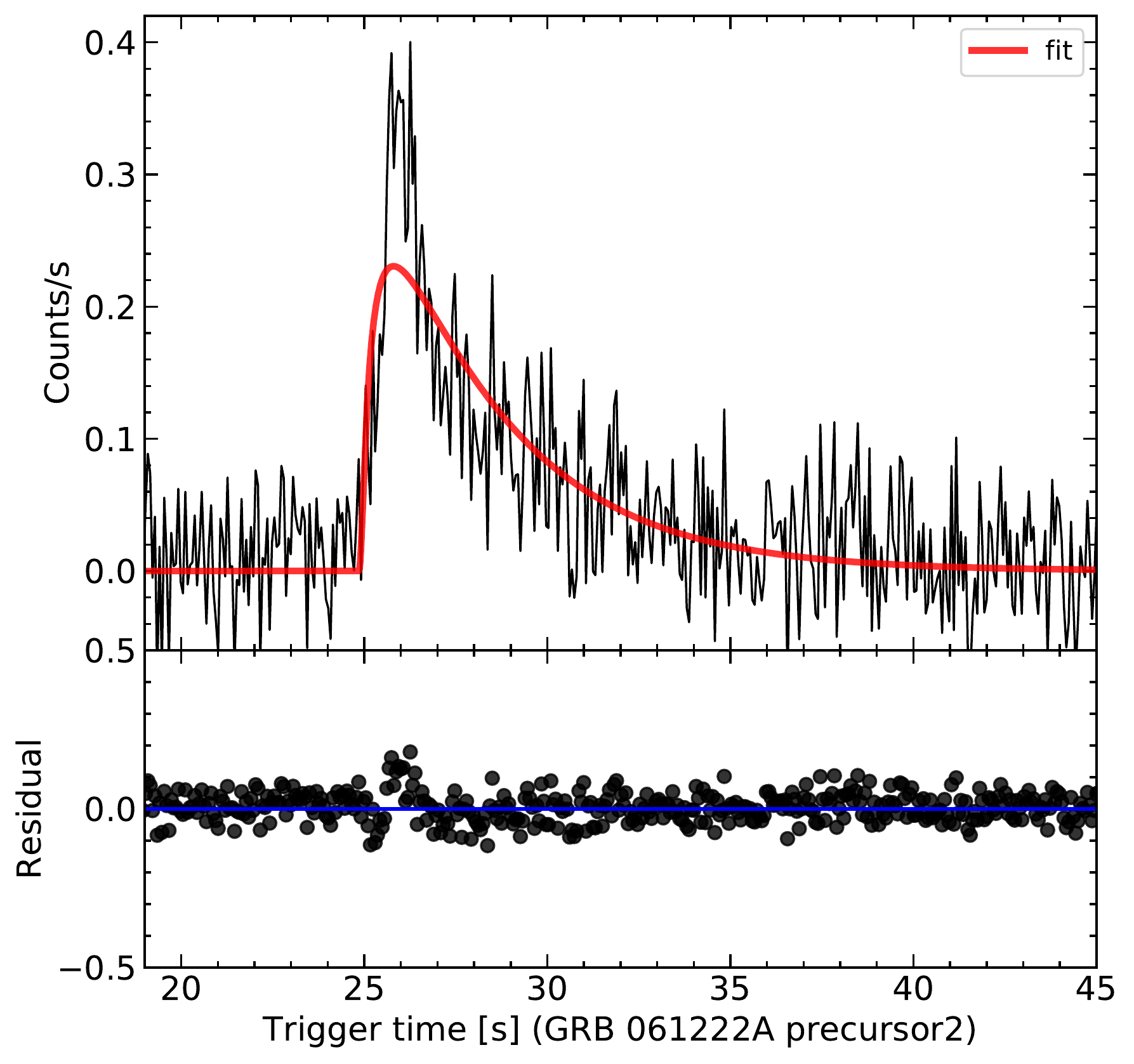}}
\subfigure{
\includegraphics[scale = 0.4]{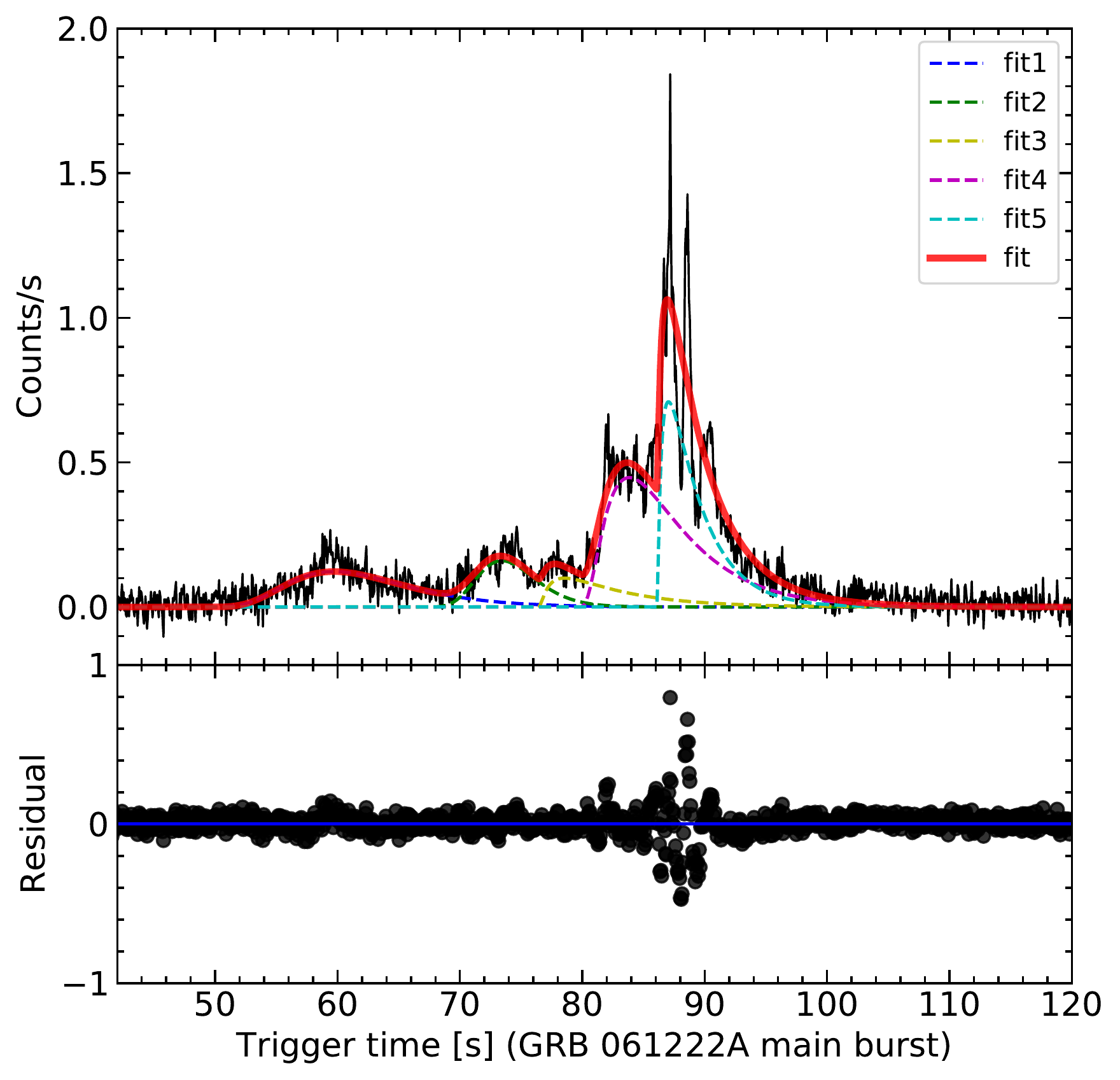}}
\caption{The lightcurve of GRB 061222A
}
\end{figure}

\begin{figure}[!htp]
\centering
\subfigure{
\includegraphics[scale = 0.4]{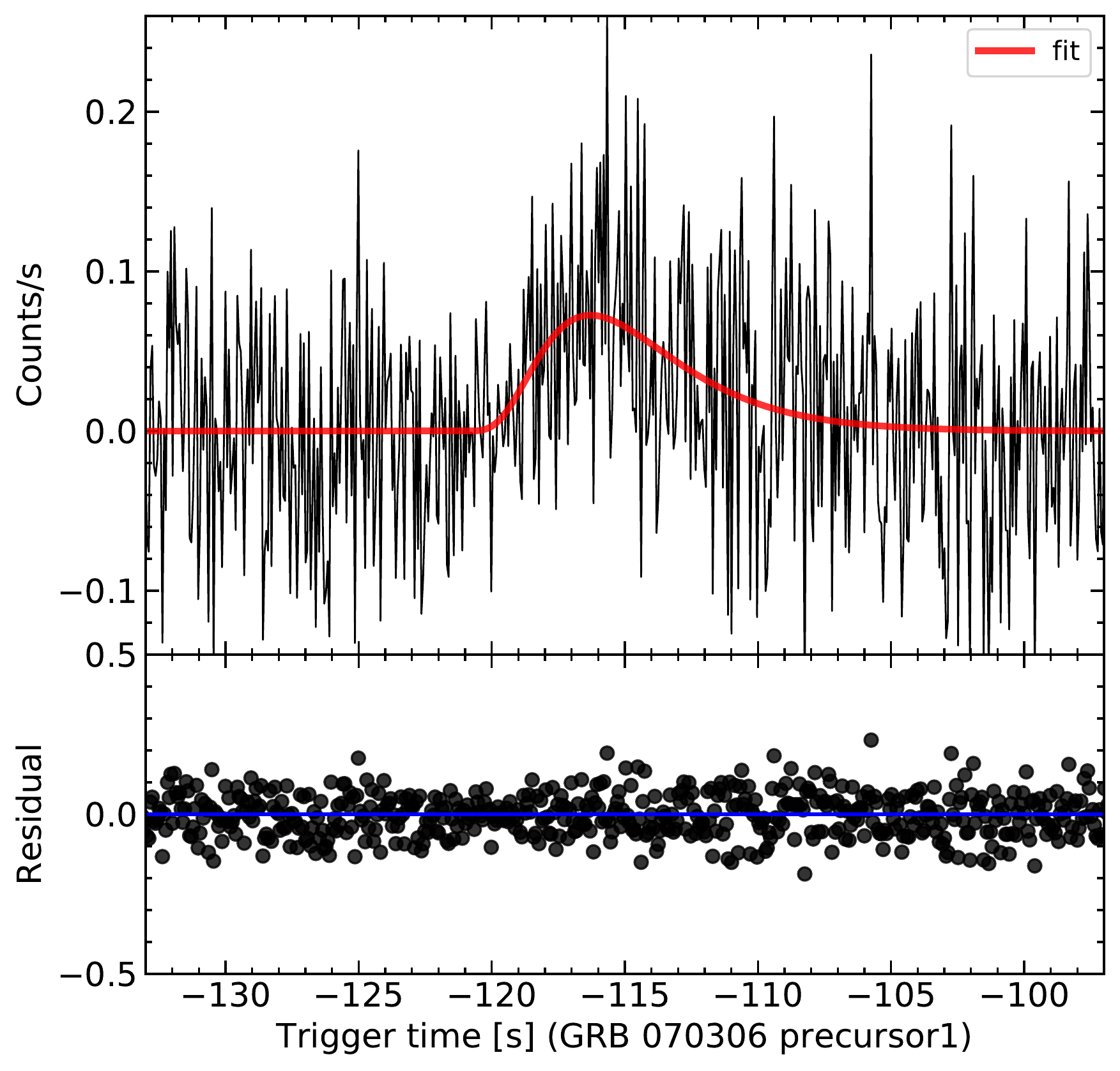}}
\subfigure{
\includegraphics[scale = 0.4]{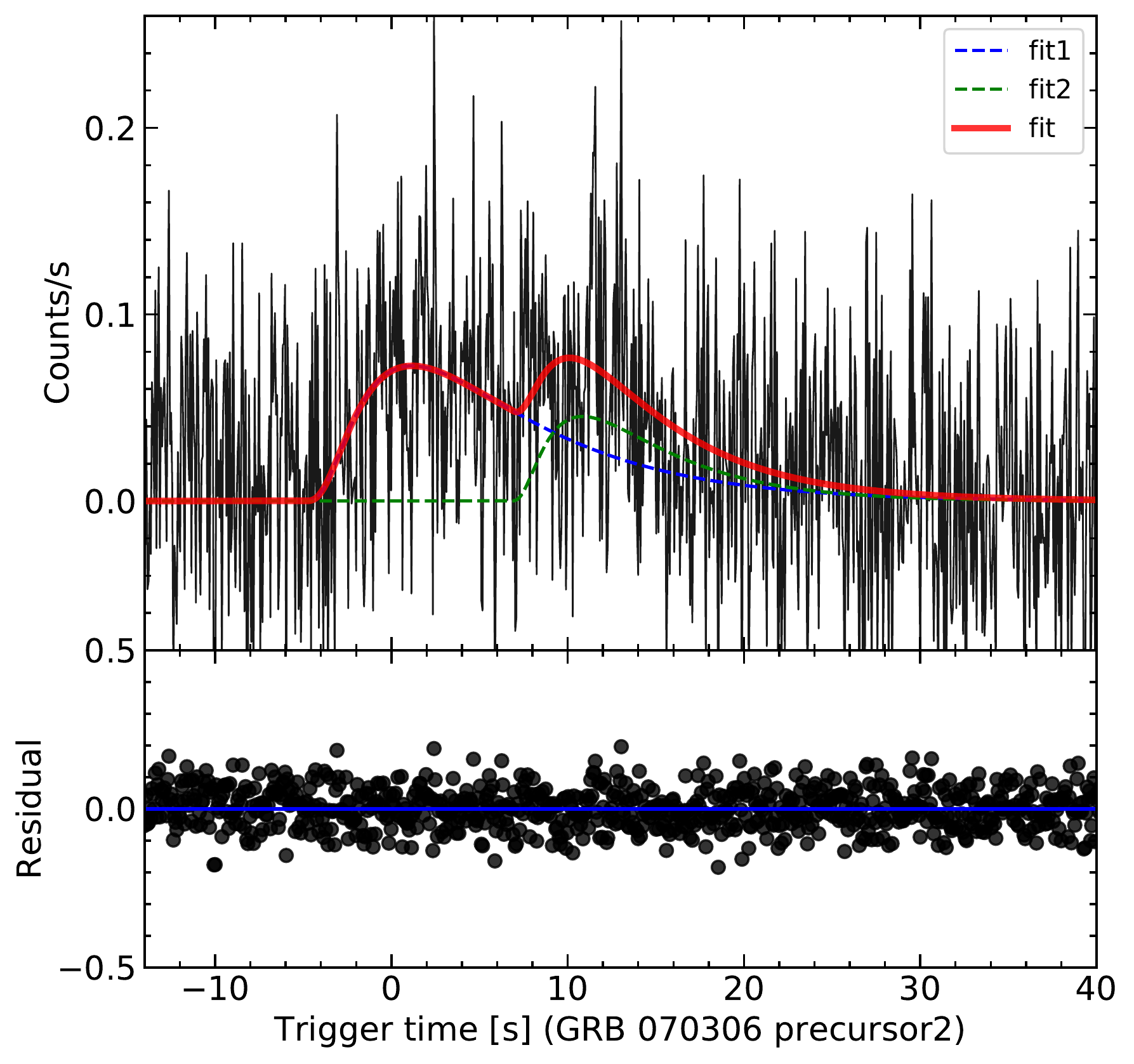}}
\subfigure{
\includegraphics[scale = 0.4]{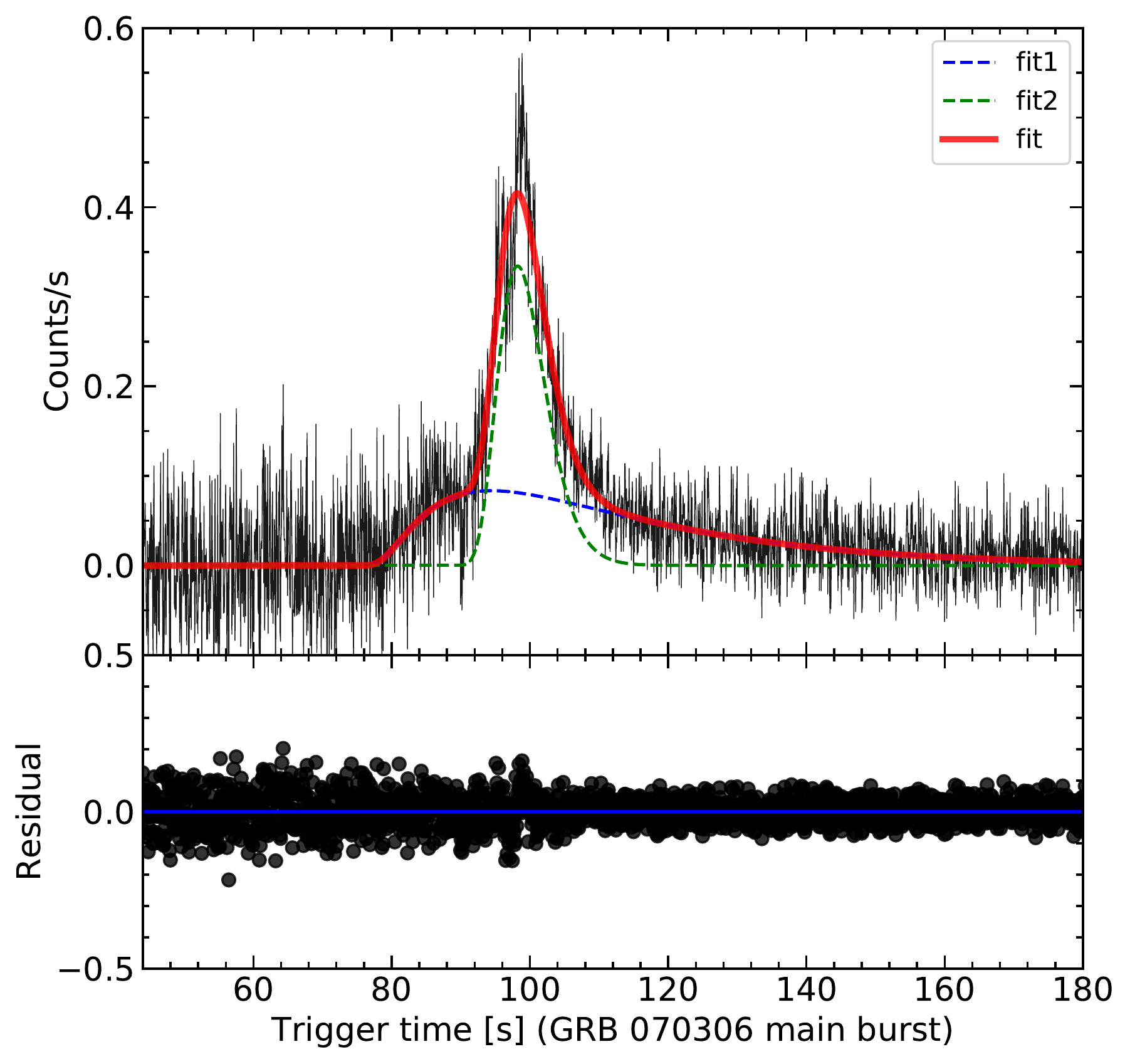}}
\caption{The lightcurve of GRB 070306
}
\end{figure}

\begin{figure}[!htp]
\centering
\subfigure{
\includegraphics[scale = 0.4]{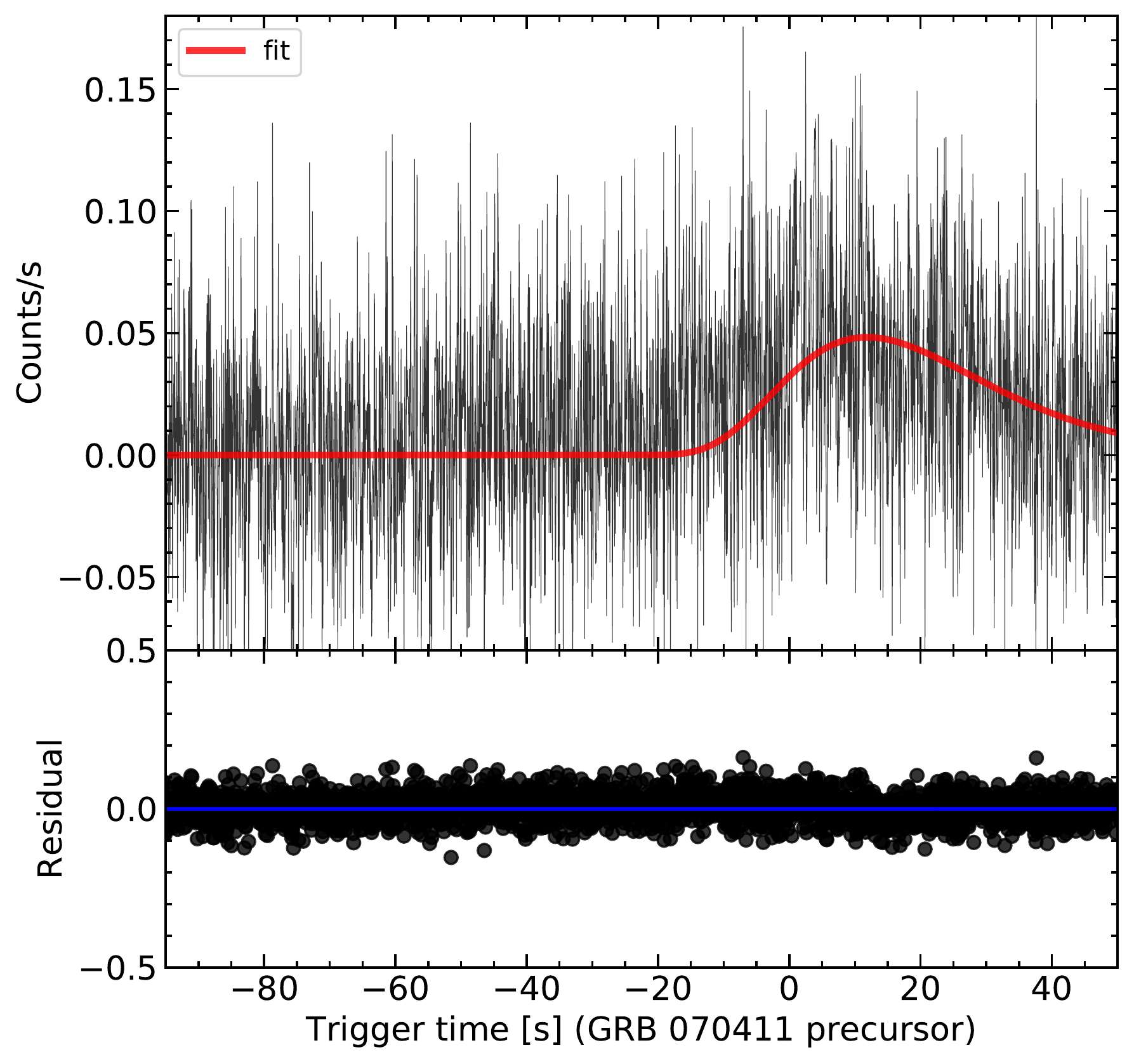}}
\subfigure{
\includegraphics[scale = 0.4]{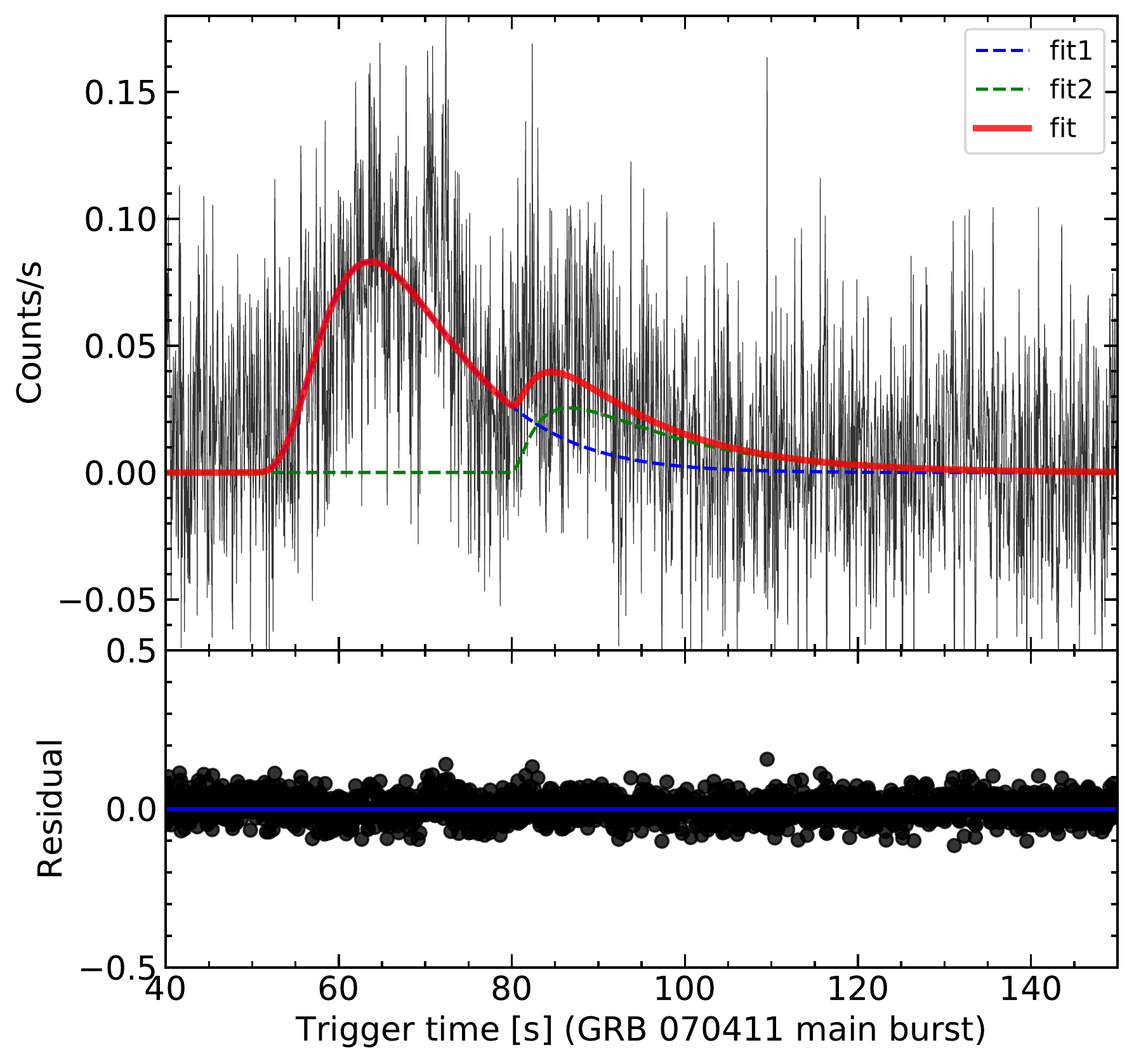}}
\caption{The lightcurve of GRB 070411
}
\end{figure}

\begin{figure}[!htp]
\centering
\subfigure{
\includegraphics[scale = 0.4]{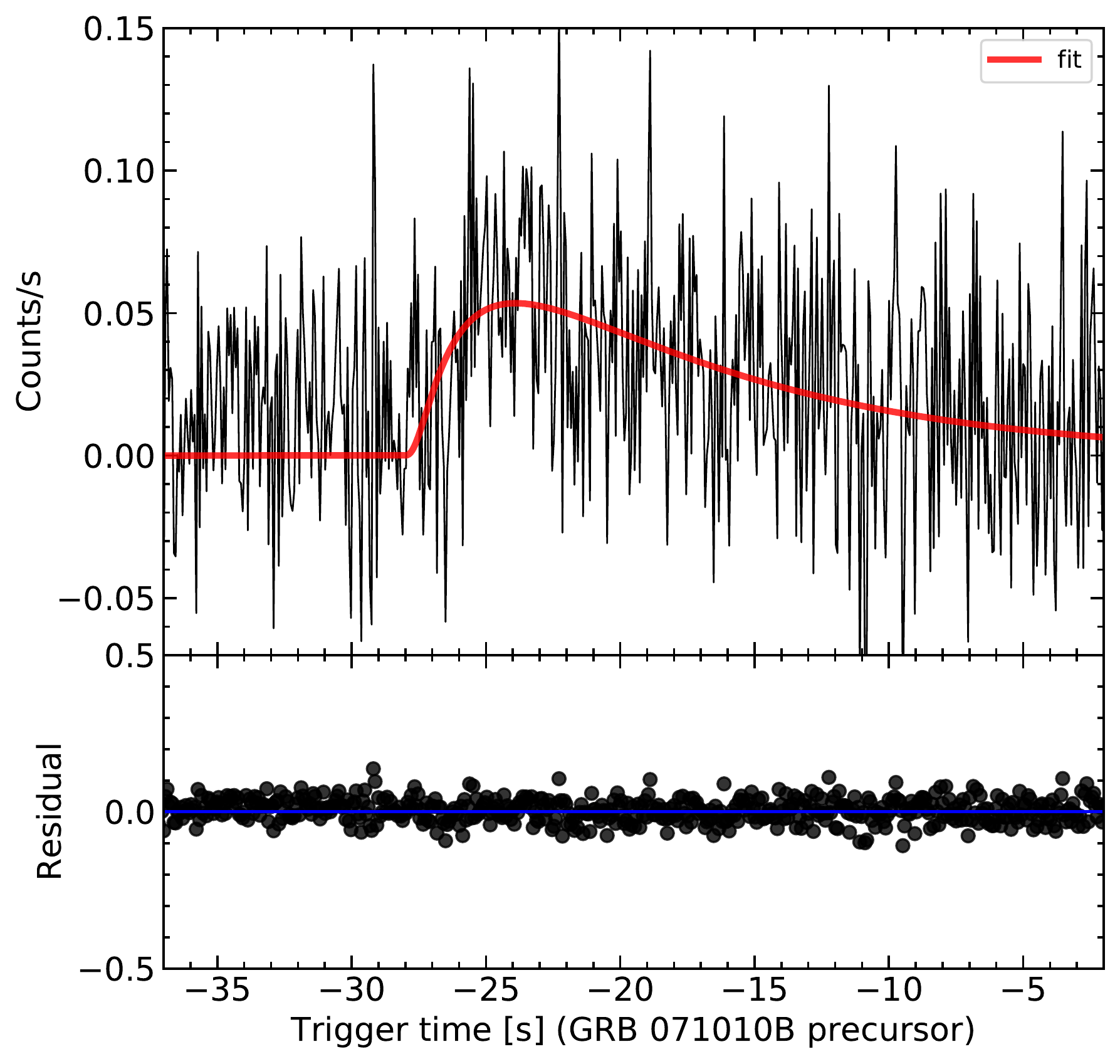}}
\subfigure{
\includegraphics[scale = 0.4]{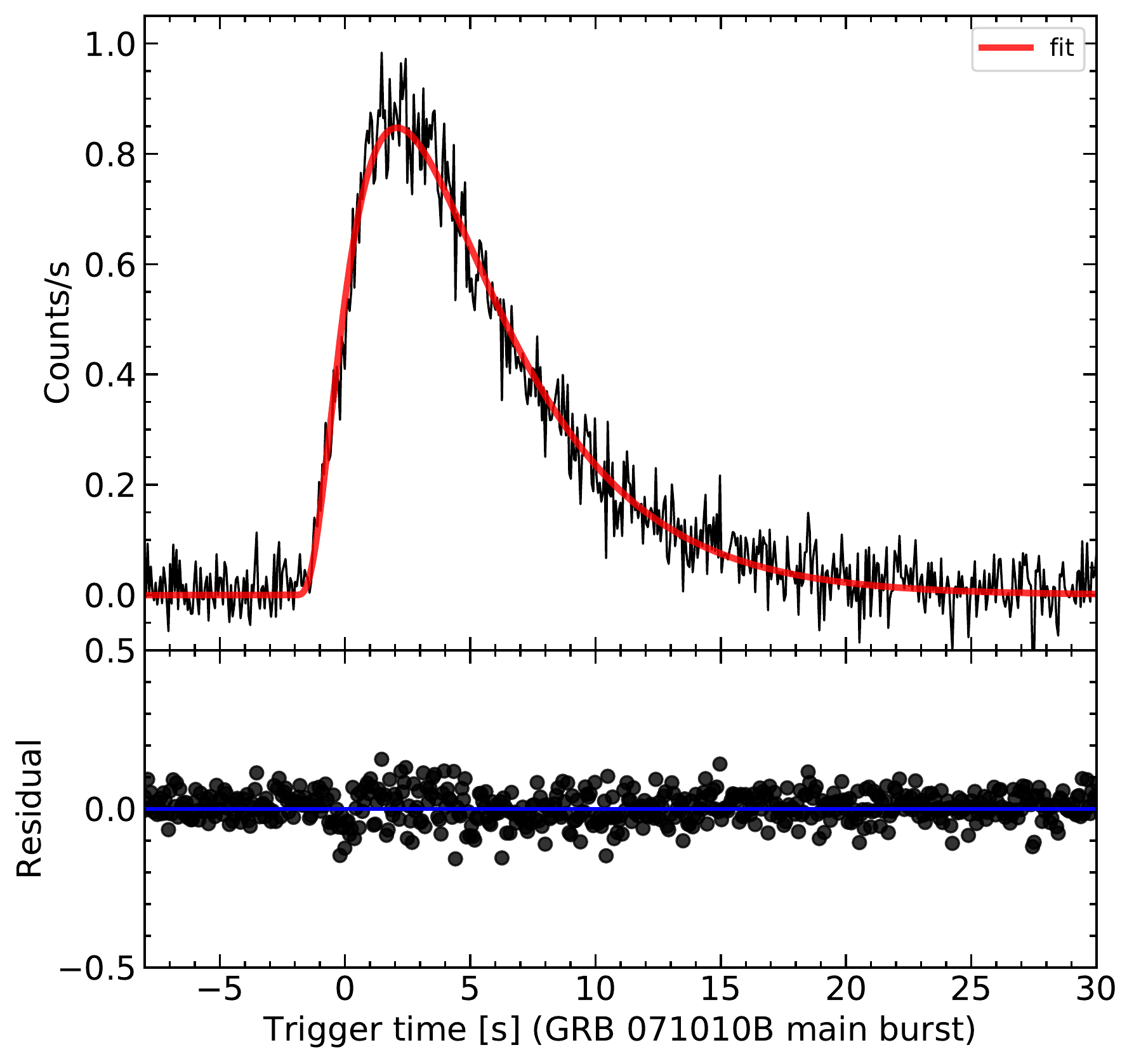}}
\caption{The lightcurve of GRB 071010B
}
\end{figure}

\begin{figure}[!htp]
\centering
\subfigure{
\includegraphics[scale = 0.4]{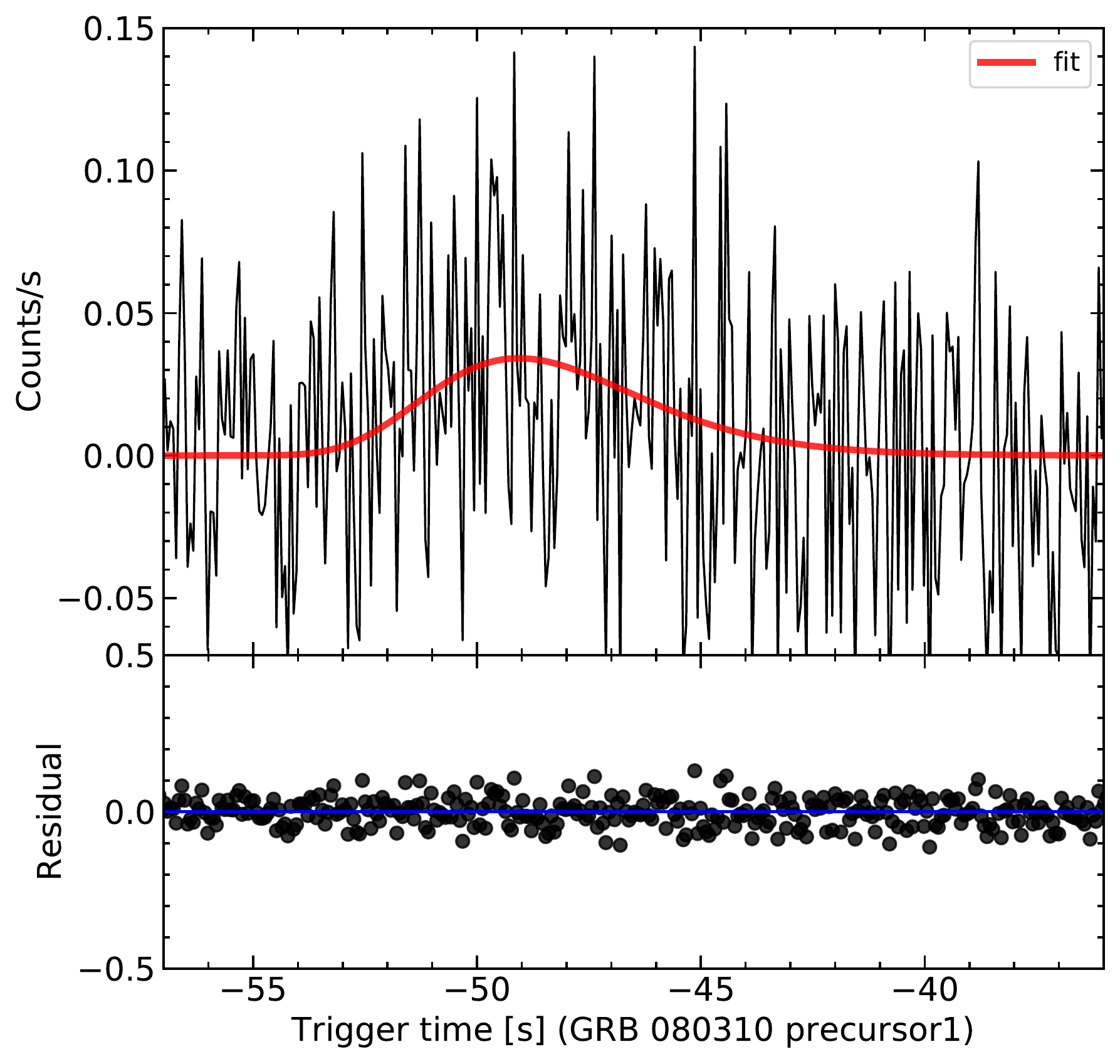}}
\subfigure{
\includegraphics[scale = 0.4]{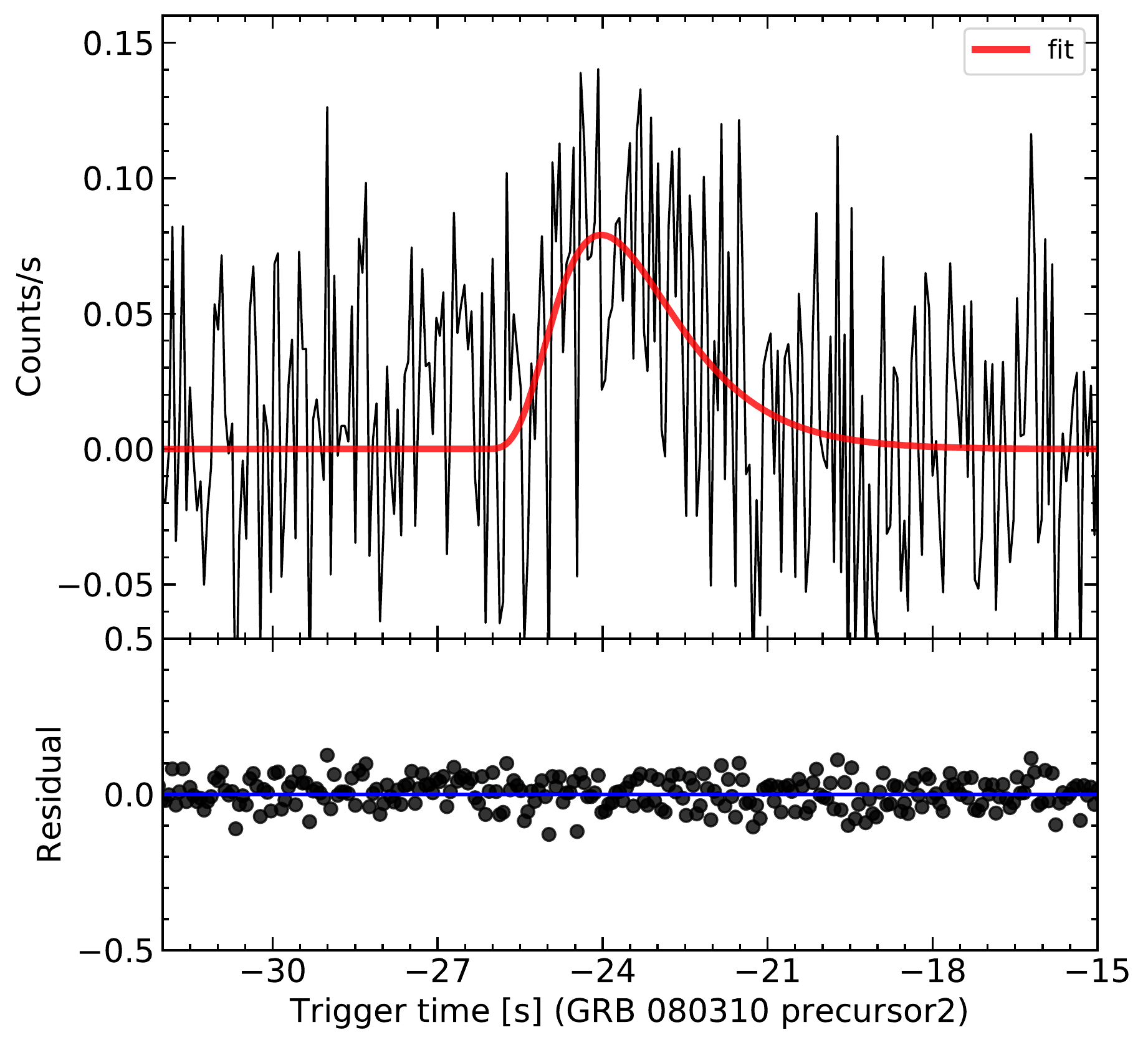}}
\subfigure{
\includegraphics[scale = 0.4]{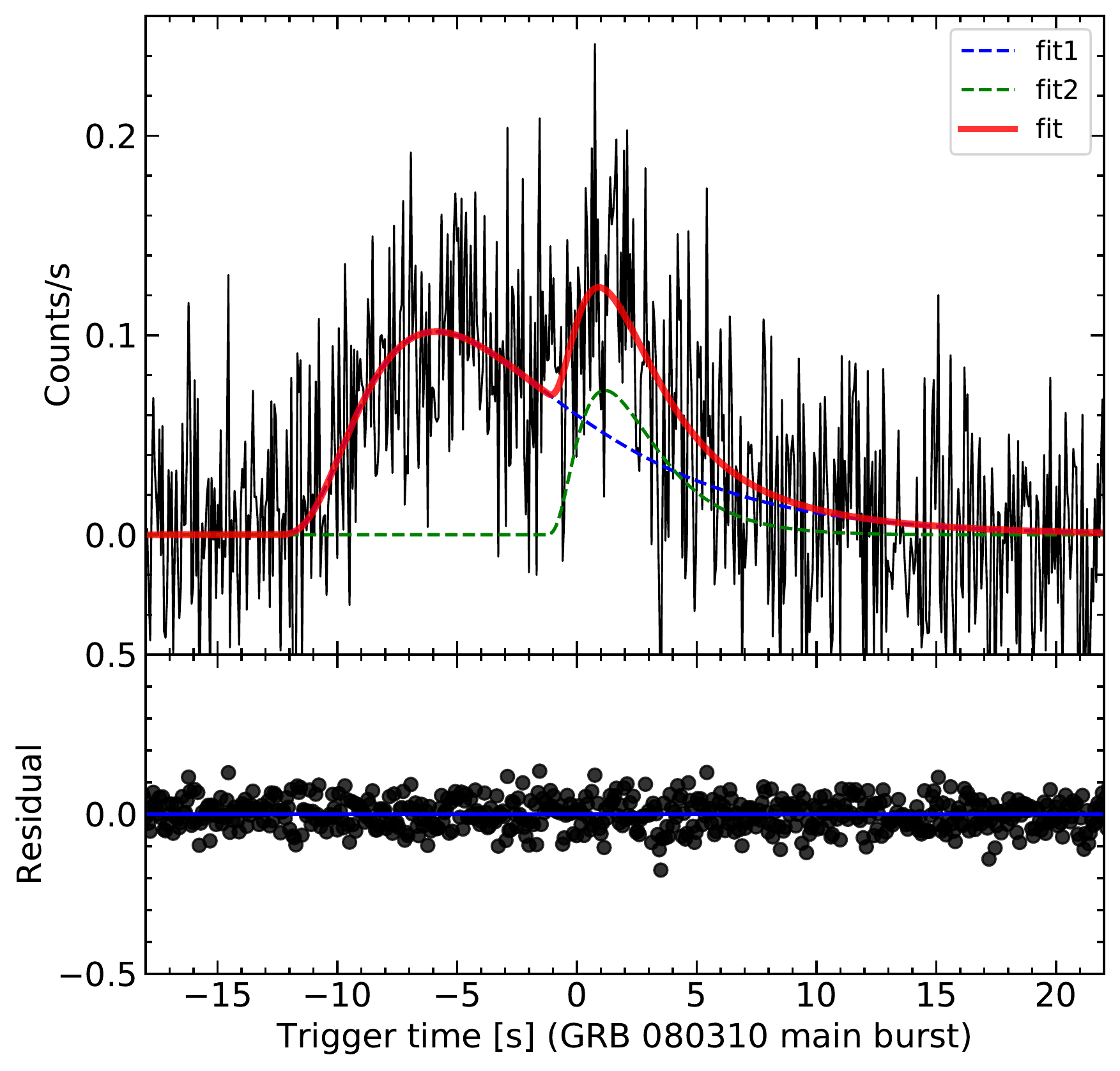}}
\caption{The lightcurve of GRB 080310
}
\end{figure}

\begin{figure}[!htp]
\centering
\subfigure{
\includegraphics[scale = 0.4]{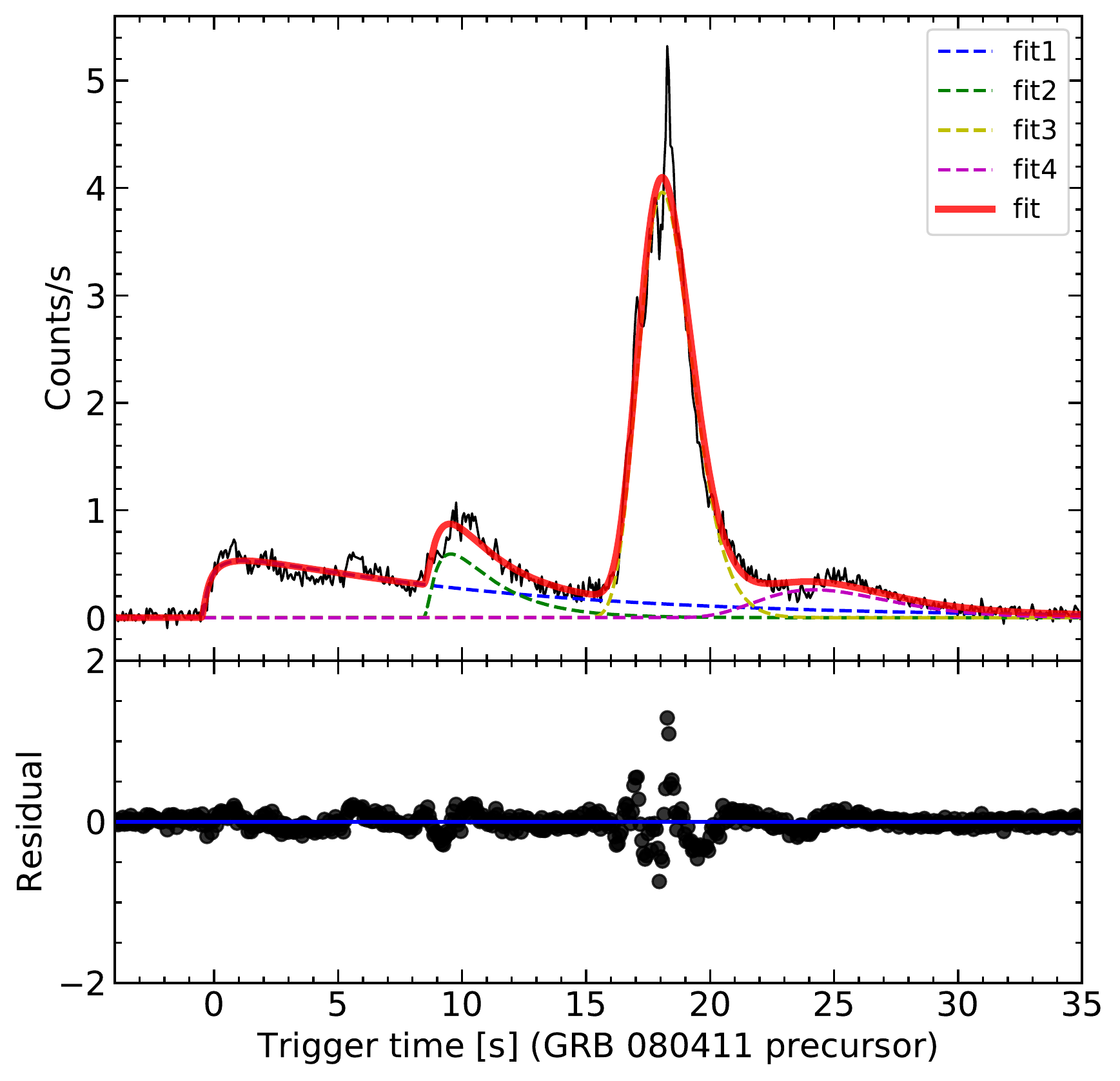}}
\subfigure{
\includegraphics[scale = 0.4]{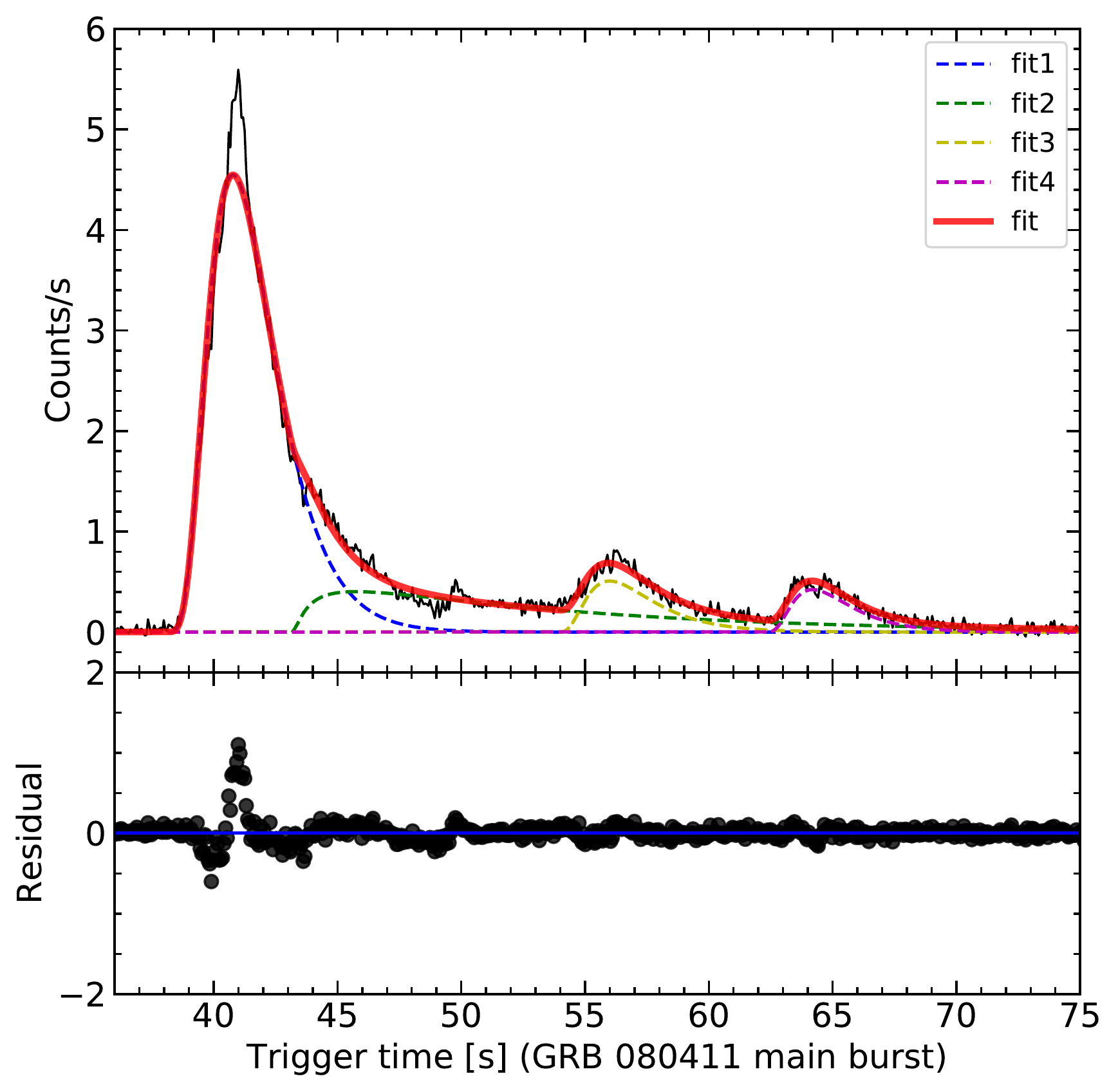}}
\caption{The lightcurve of GRB 080411
}
\end{figure}

\begin{figure}[!htp]
\centering
\subfigure{
\includegraphics[scale = 0.4]{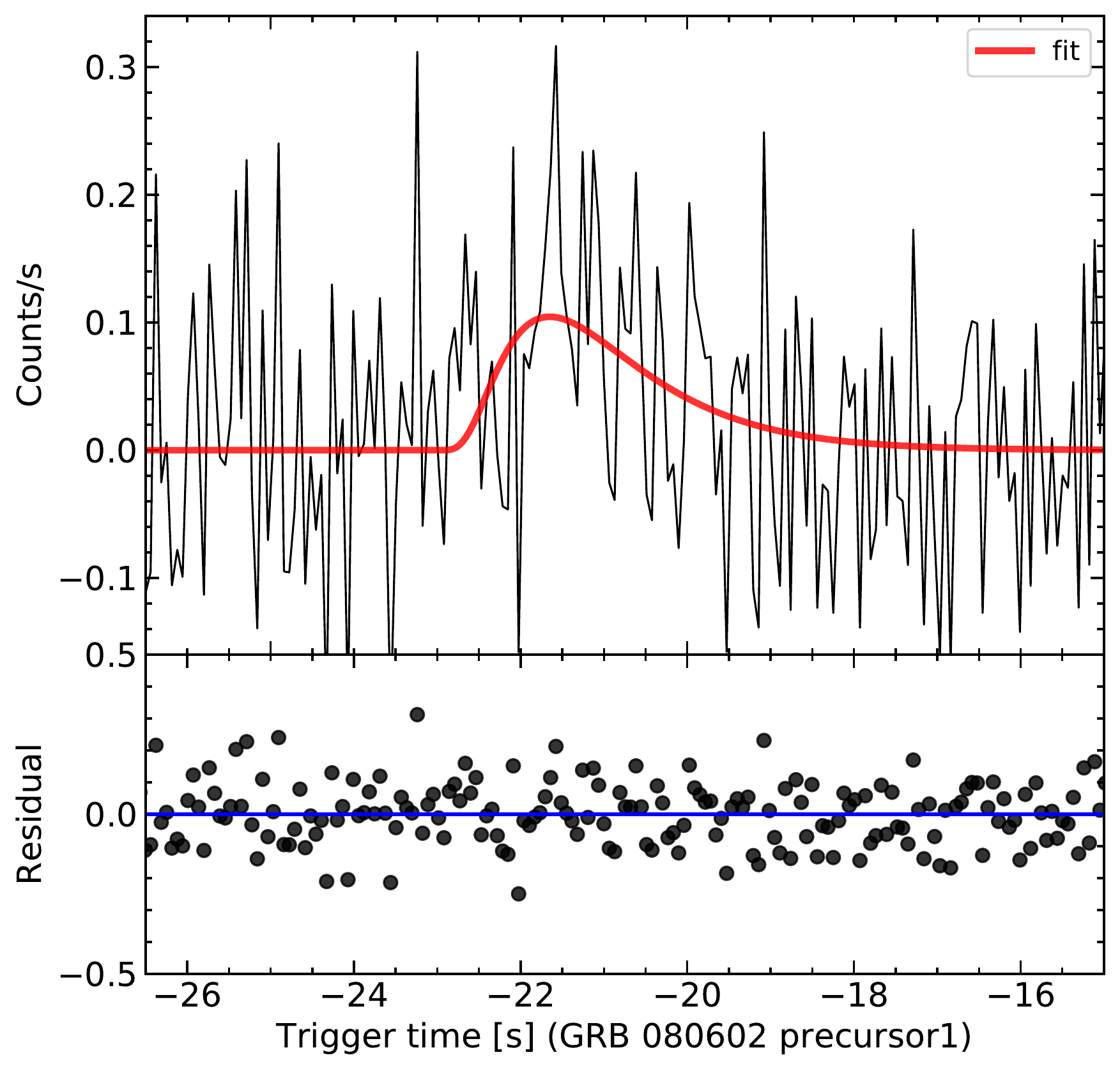}}
\subfigure{
\includegraphics[scale = 0.4]{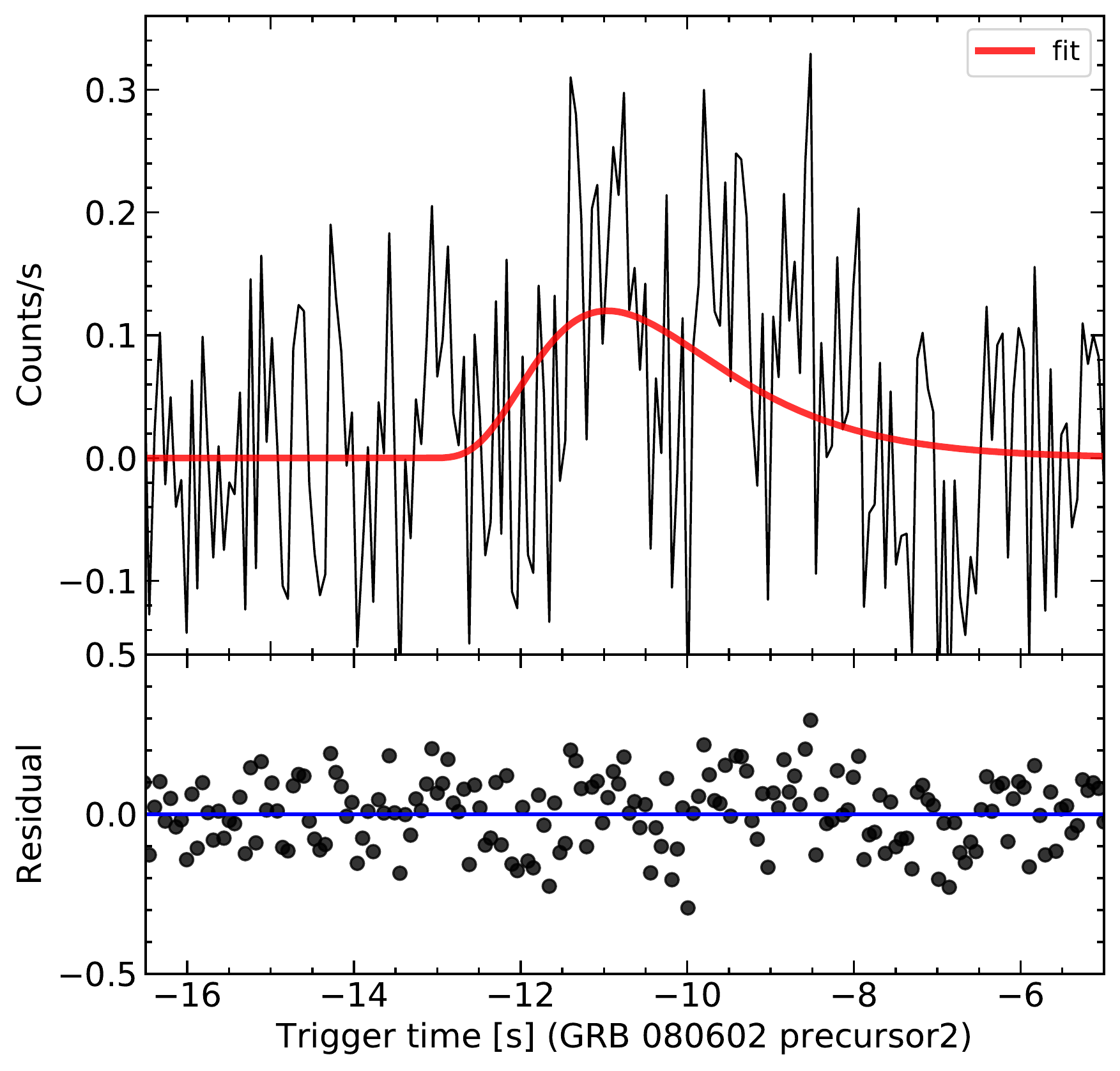}}
\subfigure{
\includegraphics[scale = 0.4]{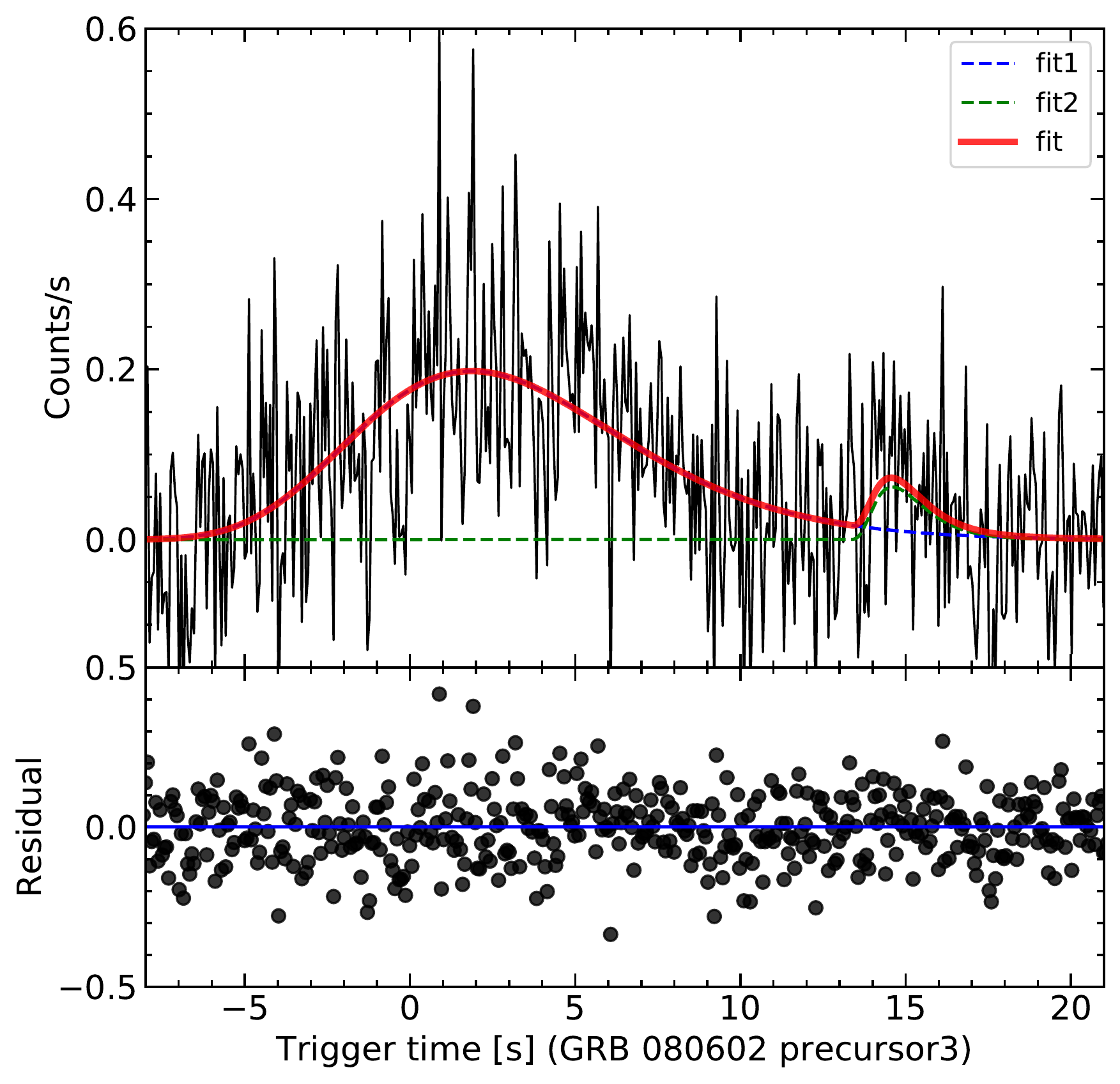}}
\subfigure{
\includegraphics[scale = 0.4]{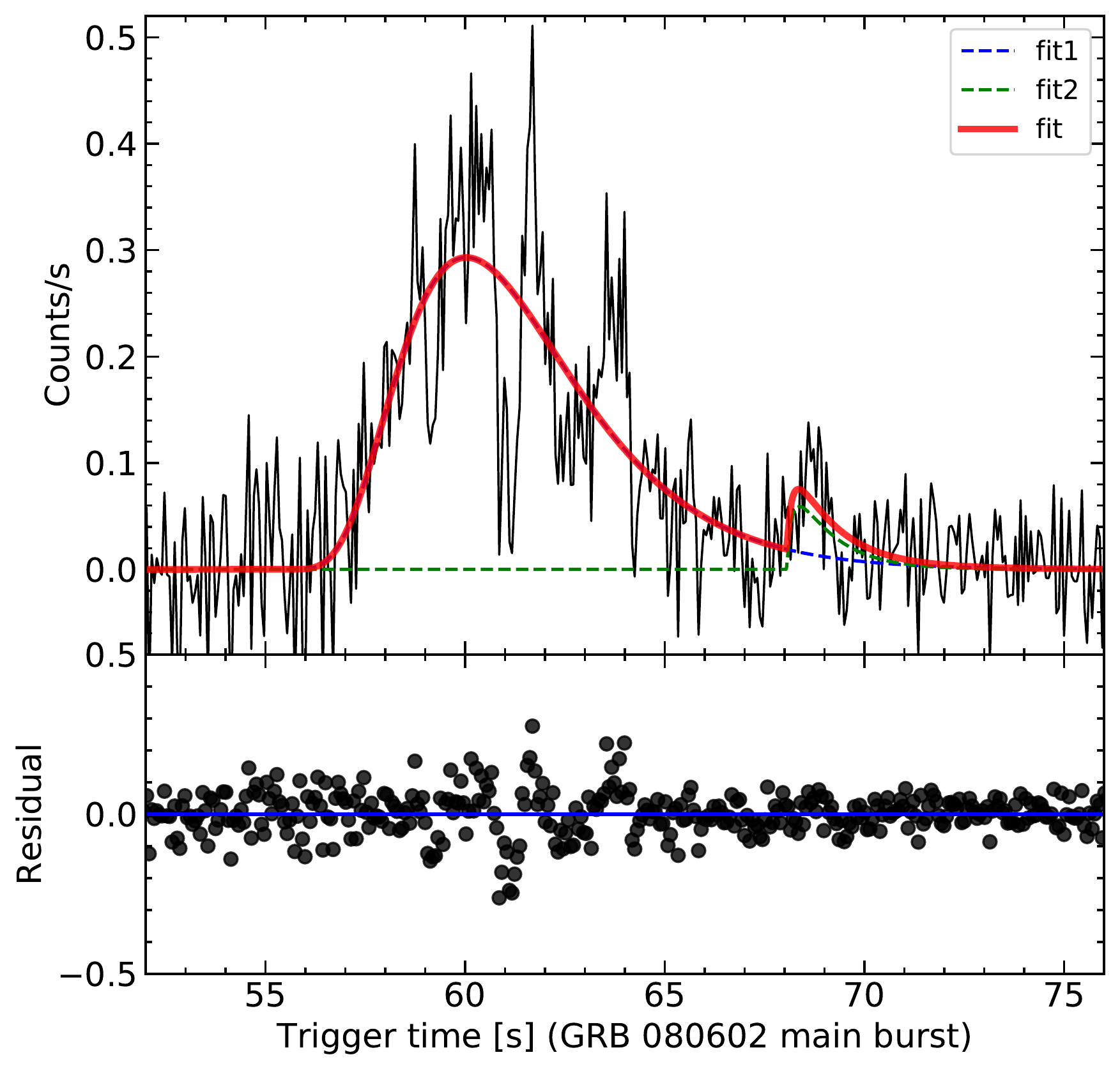}}
\caption{The lightcurve of GRB 080602
}
\end{figure}

\begin{figure}[!htp]
\centering
\subfigure{
\includegraphics[scale = 0.4]{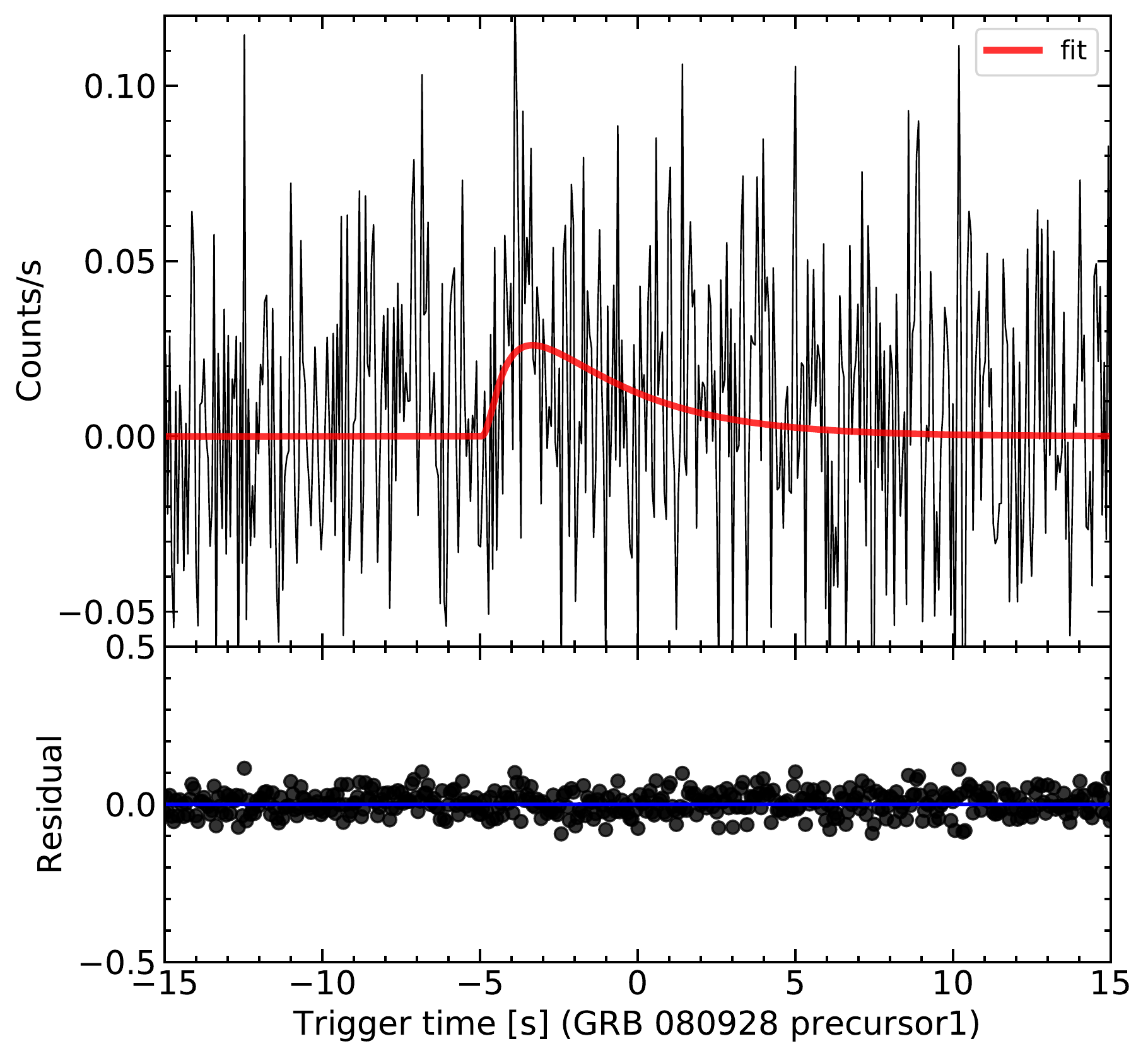}}
\subfigure{
\includegraphics[scale = 0.4]{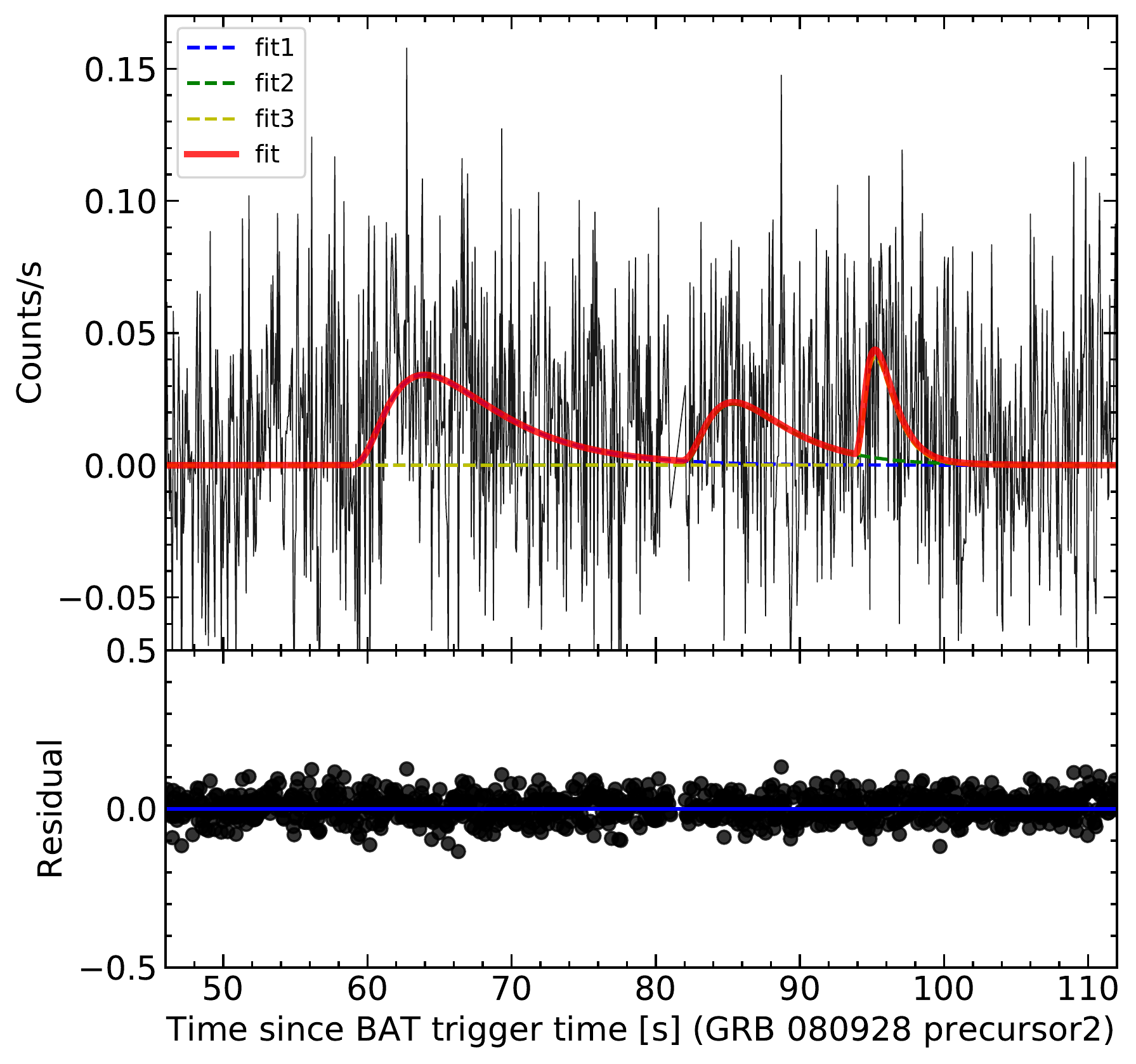}}
\subfigure{
\includegraphics[scale = 0.4]{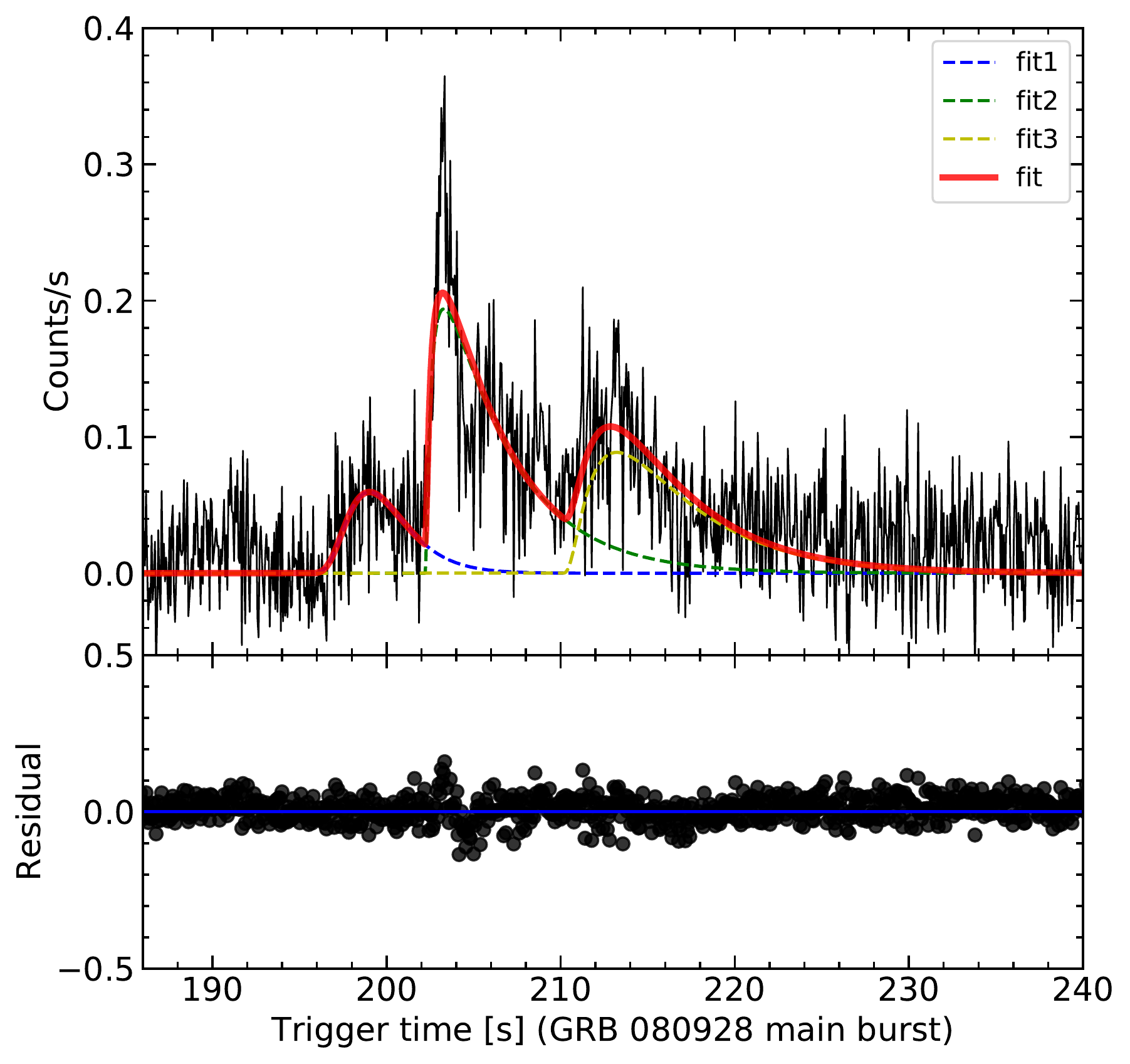}}
\caption{The lightcurve of GRB 080928
}
\end{figure}

\begin{figure}[!htp]
\centering
\subfigure{
\includegraphics[scale = 0.4]{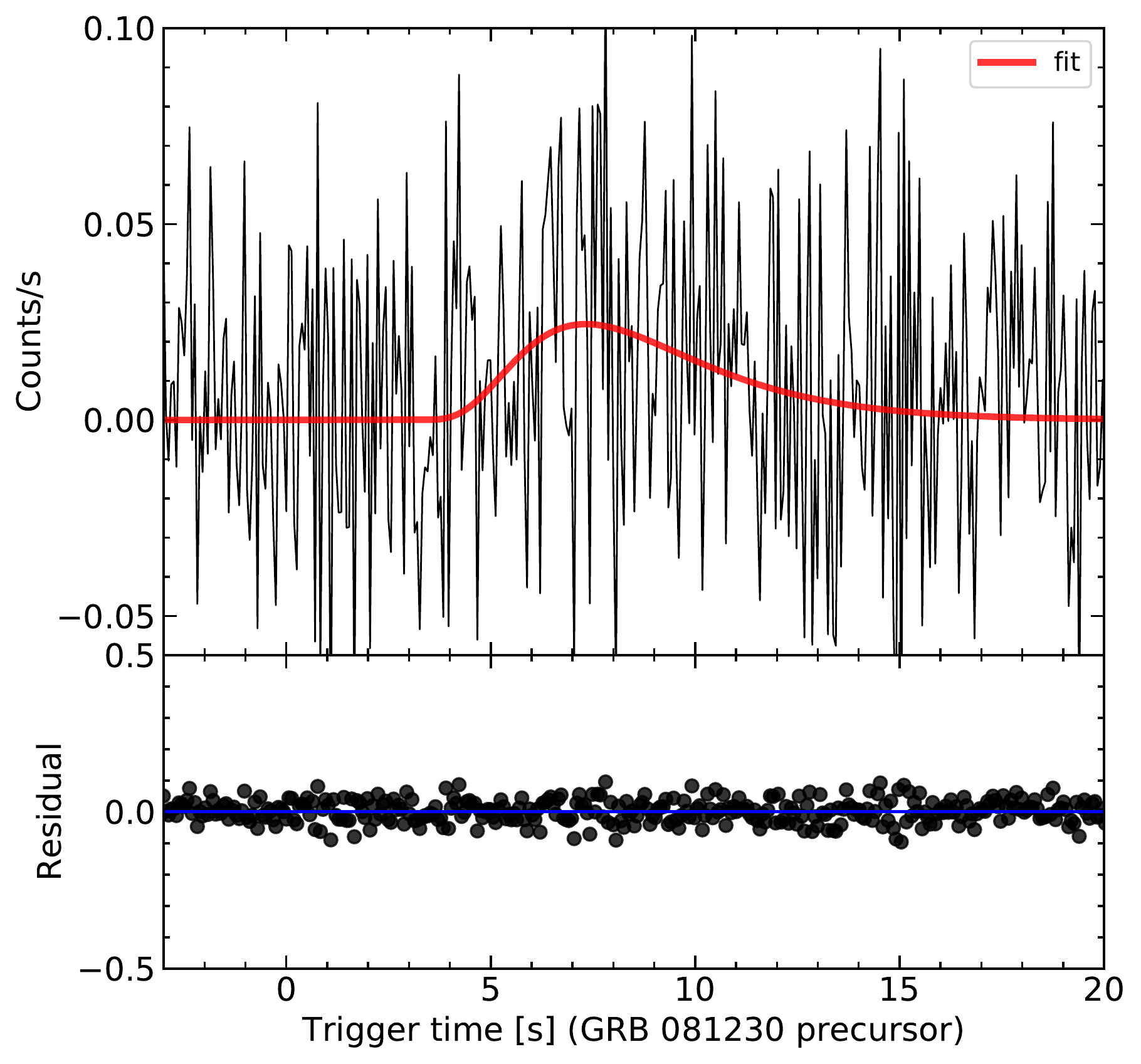}}
\subfigure{
\includegraphics[scale = 0.4]{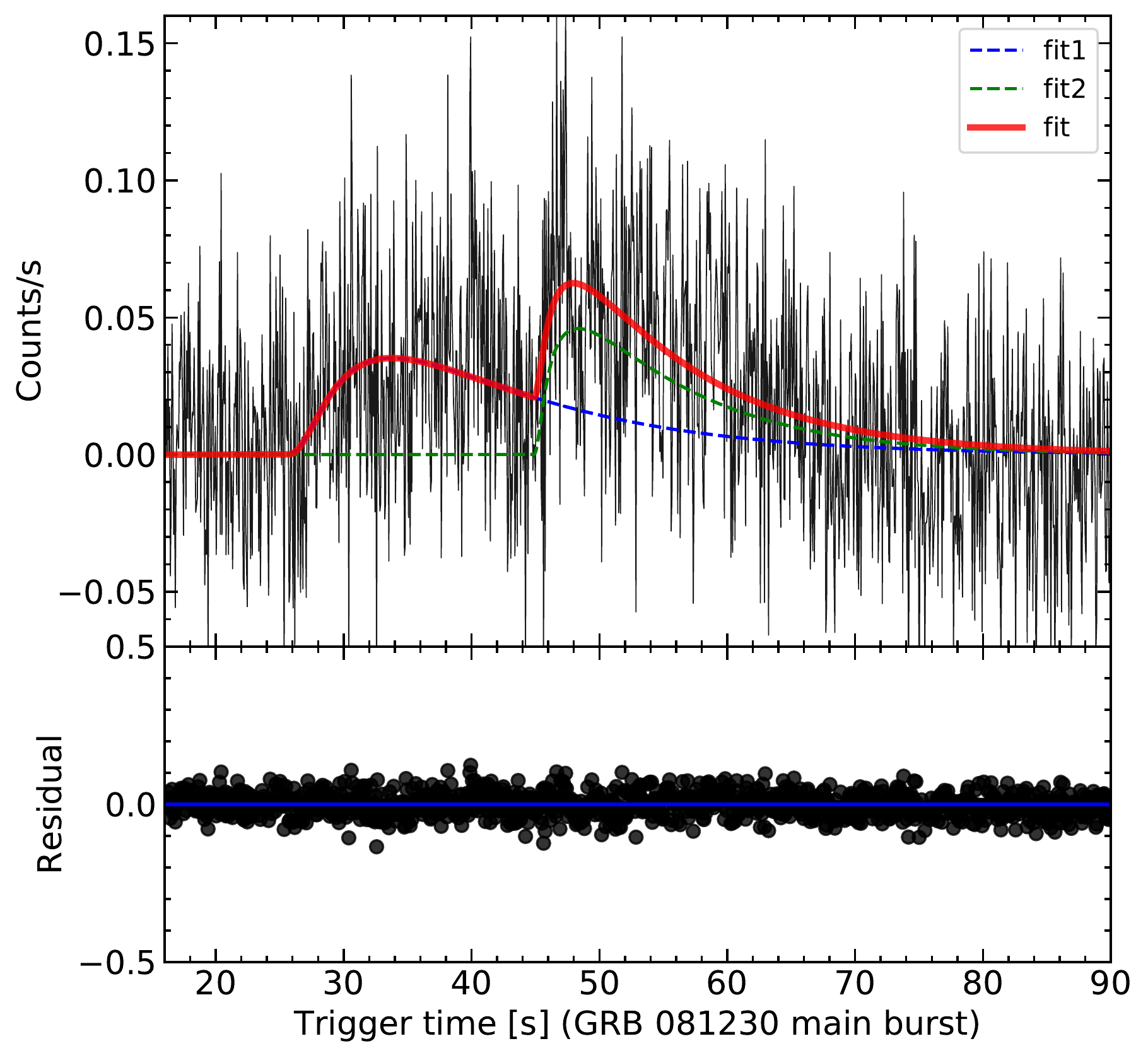}}
\caption{The lightcurve of GRB 081230
}
\end{figure}

\begin{figure}[!htp]
\centering
\subfigure{
\includegraphics[scale = 0.4]{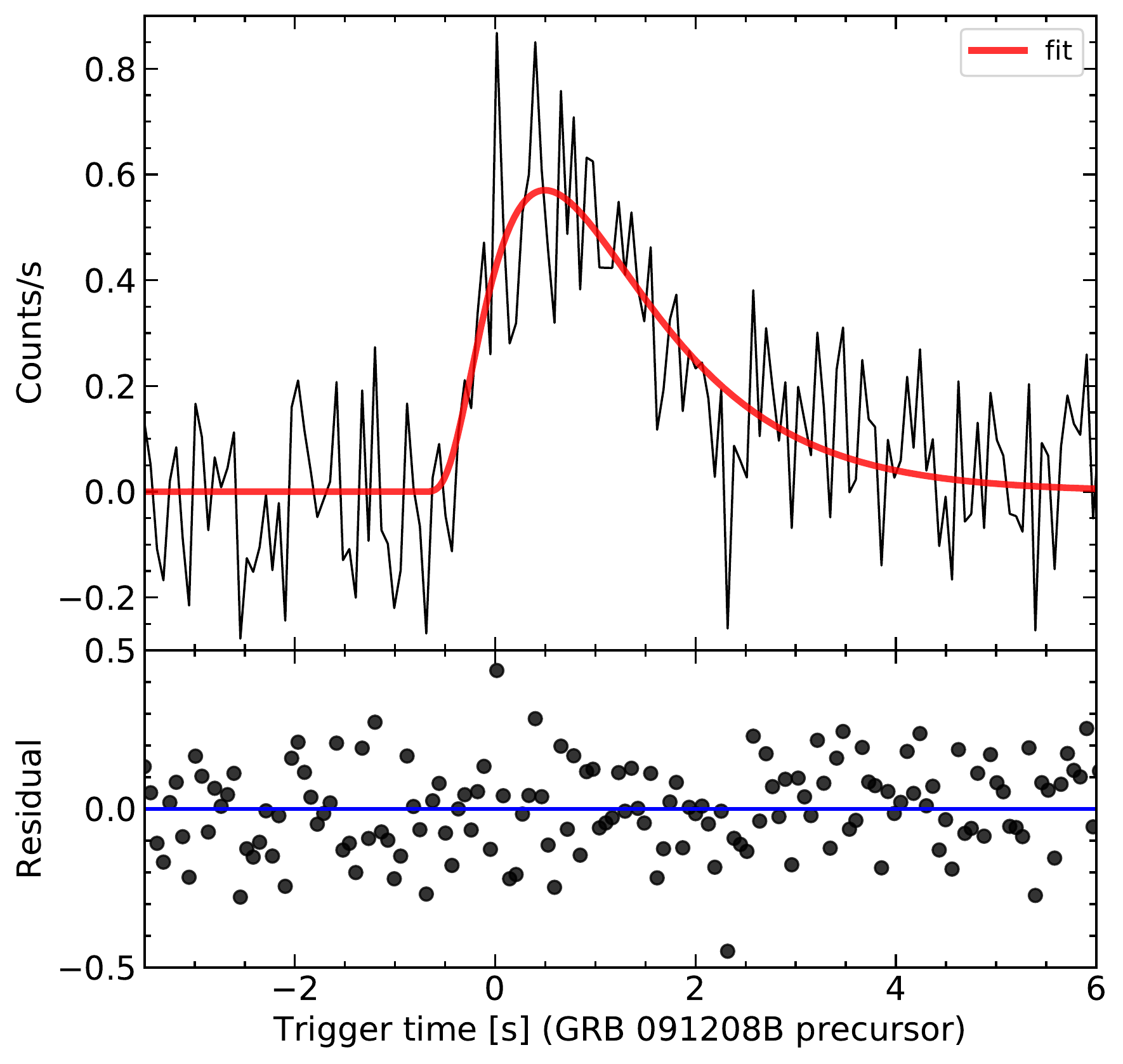}}
\subfigure{
\includegraphics[scale = 0.4]{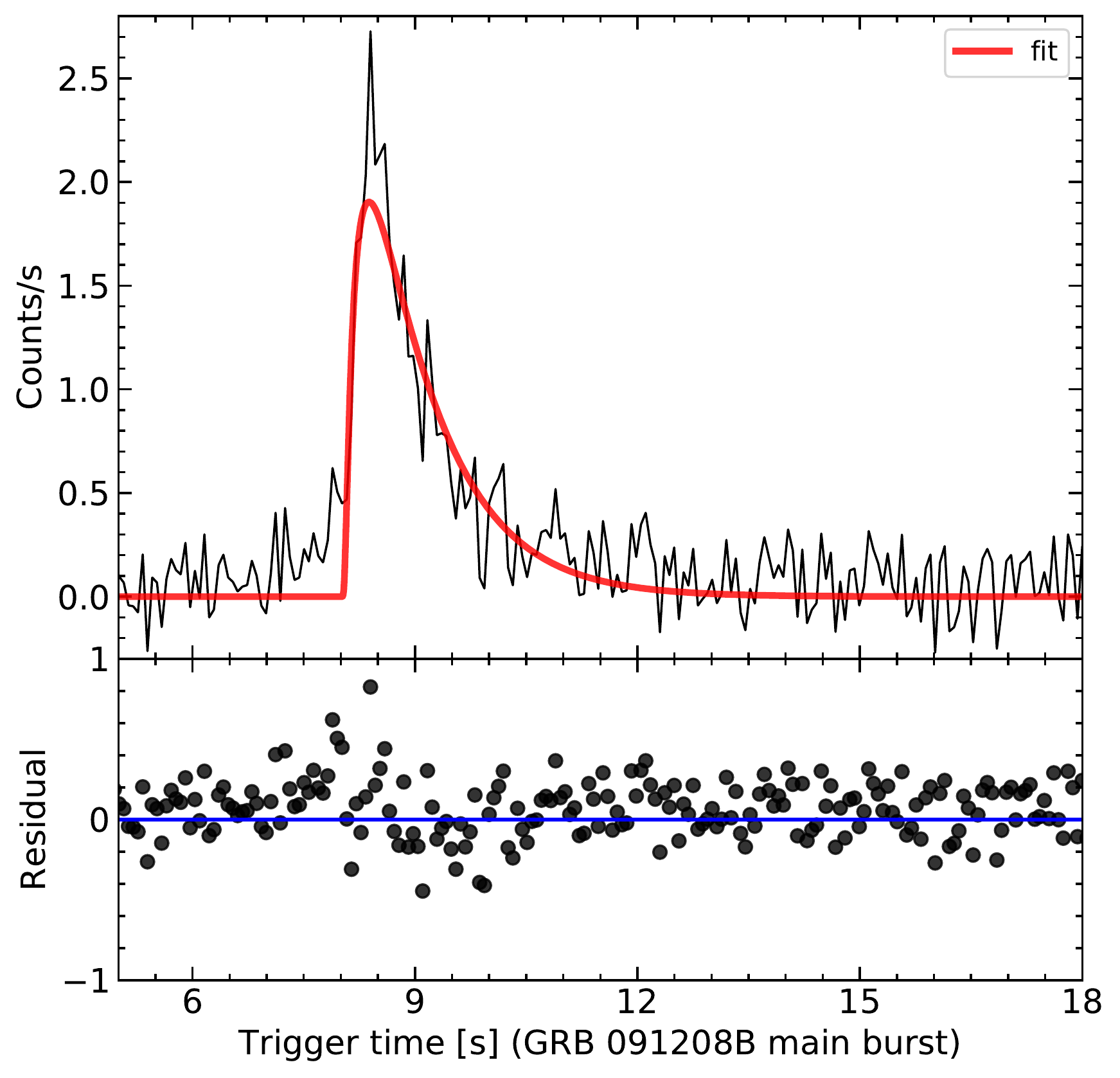}}
\caption{The lightcurve of GRB 091208B
}
\end{figure}

\begin{figure}[!htp]
\centering
\subfigure{
\includegraphics[scale = 0.4]{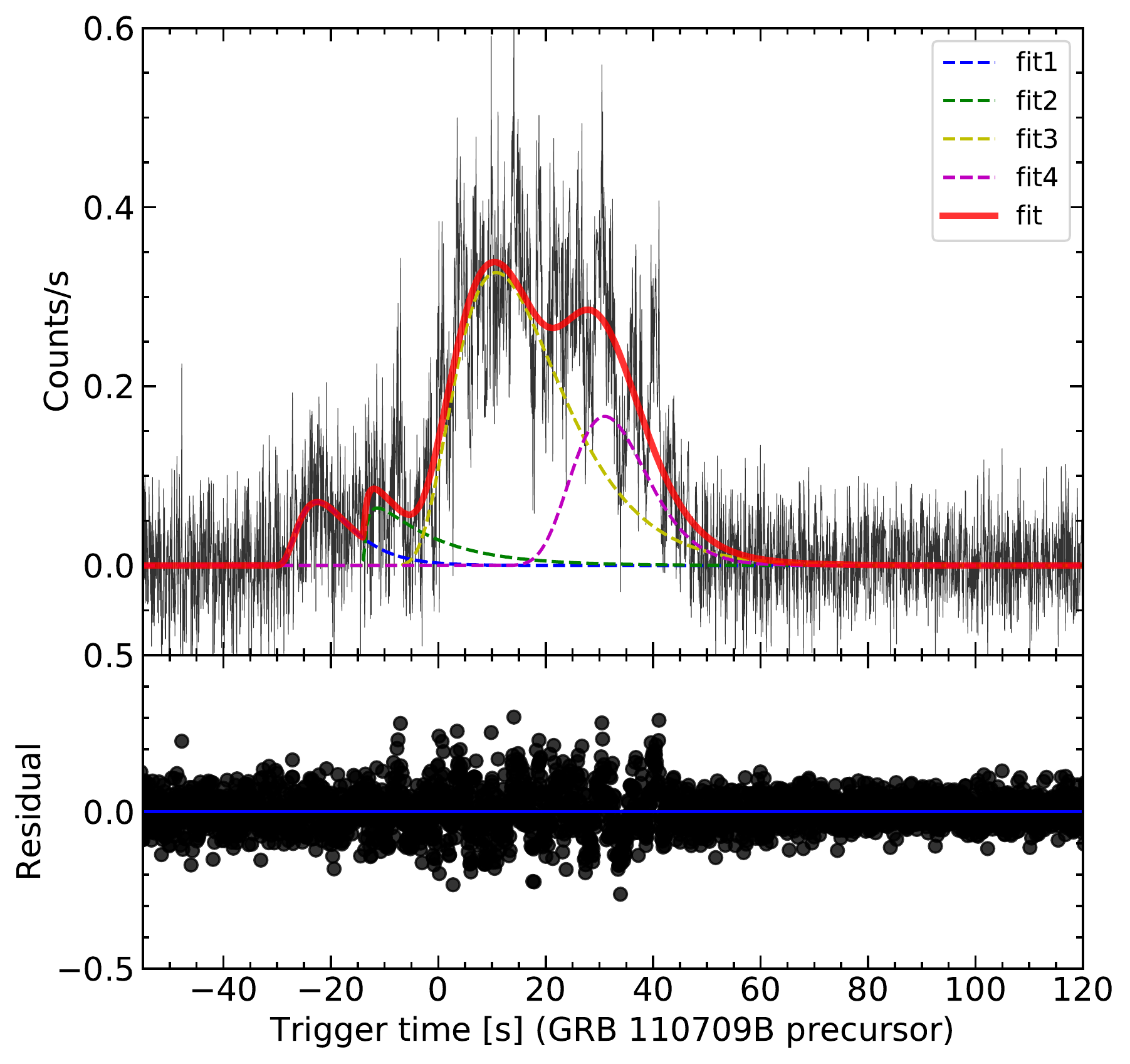}}
\subfigure{
\includegraphics[scale = 0.4]{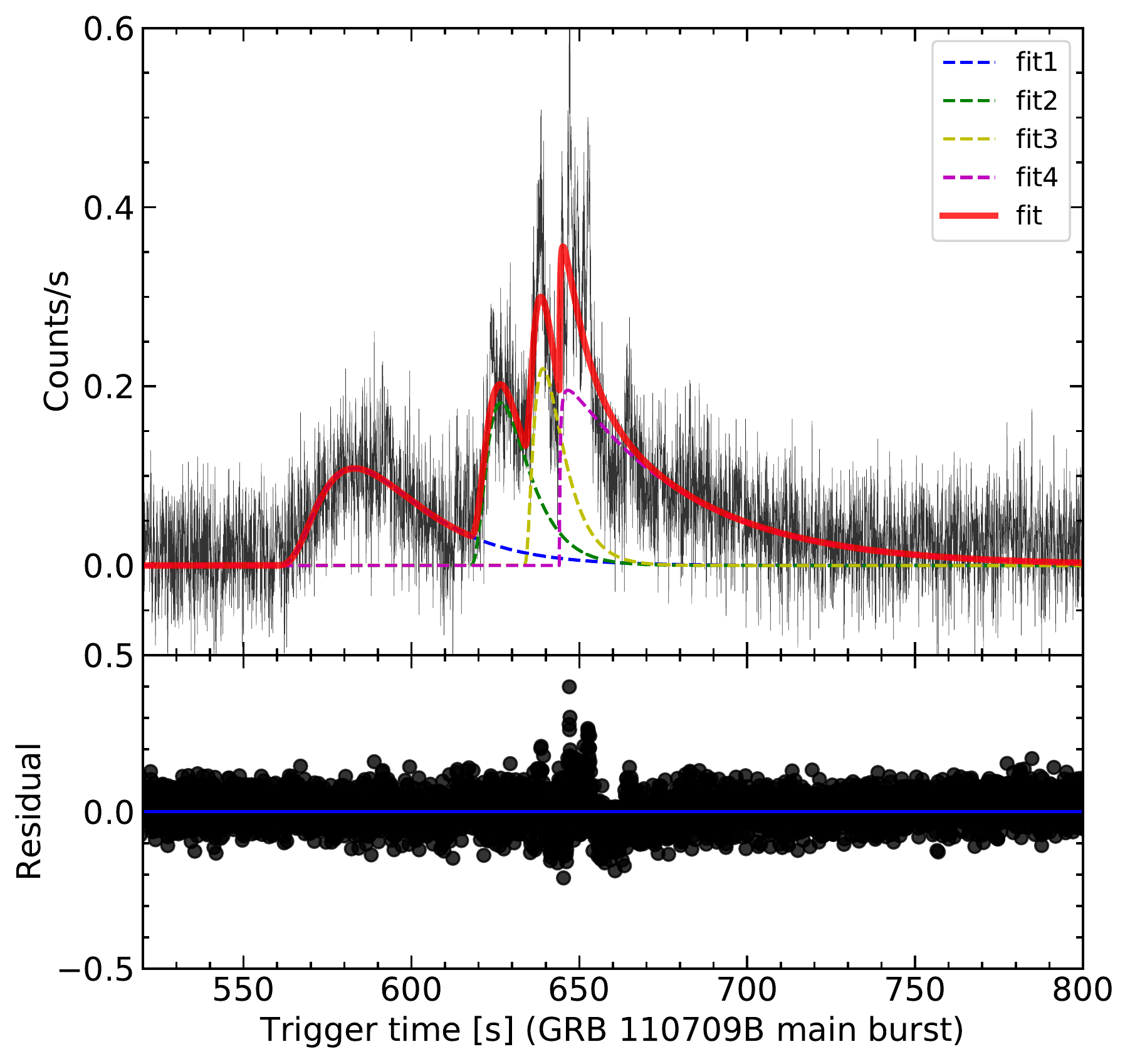}}
\caption{The lightcurve of GRB 110709B
}
\end{figure}

\begin{figure}[!htp]
\centering
\subfigure{
\includegraphics[scale = 0.4]{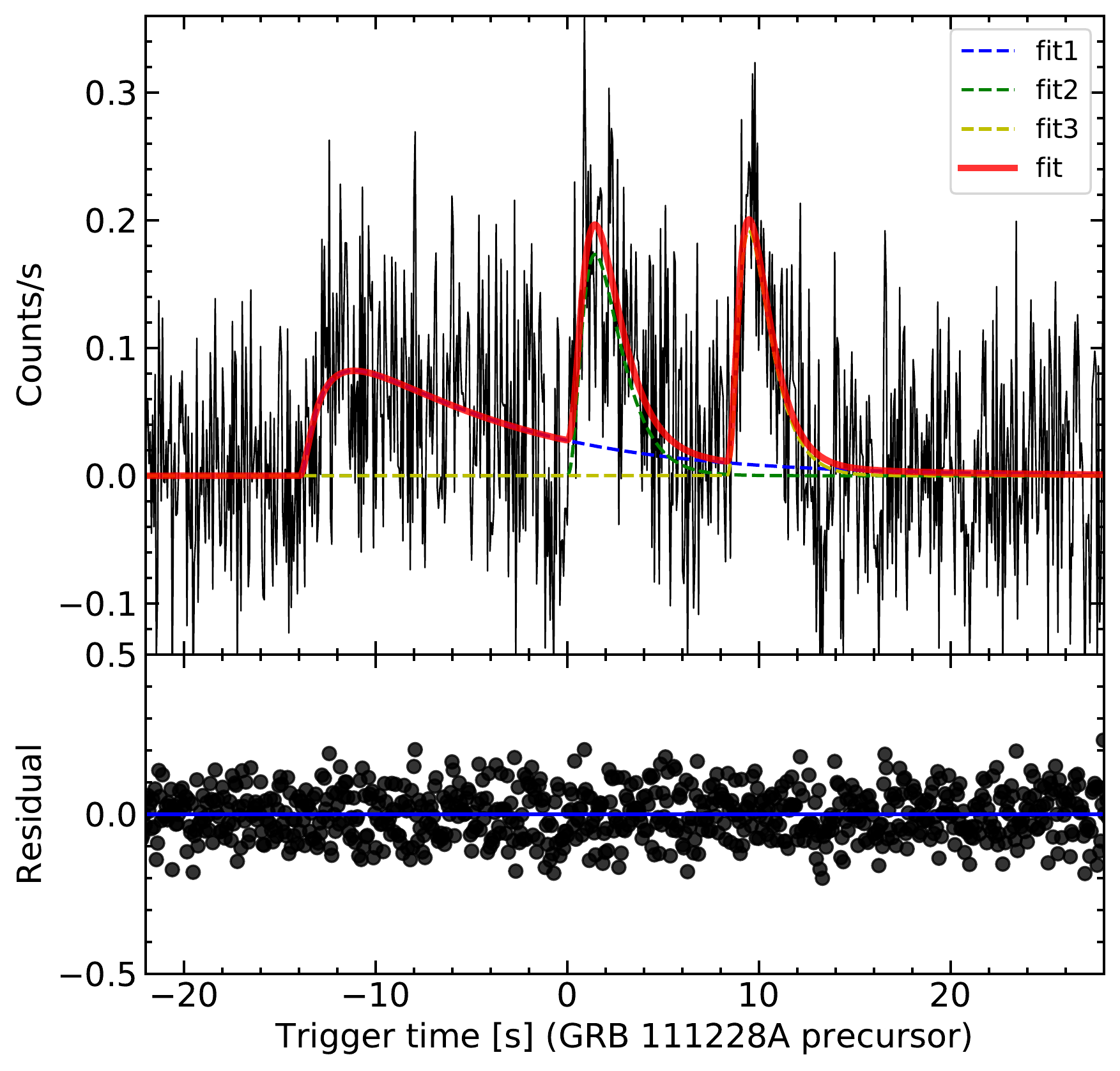}}
\subfigure{
\includegraphics[scale = 0.4]{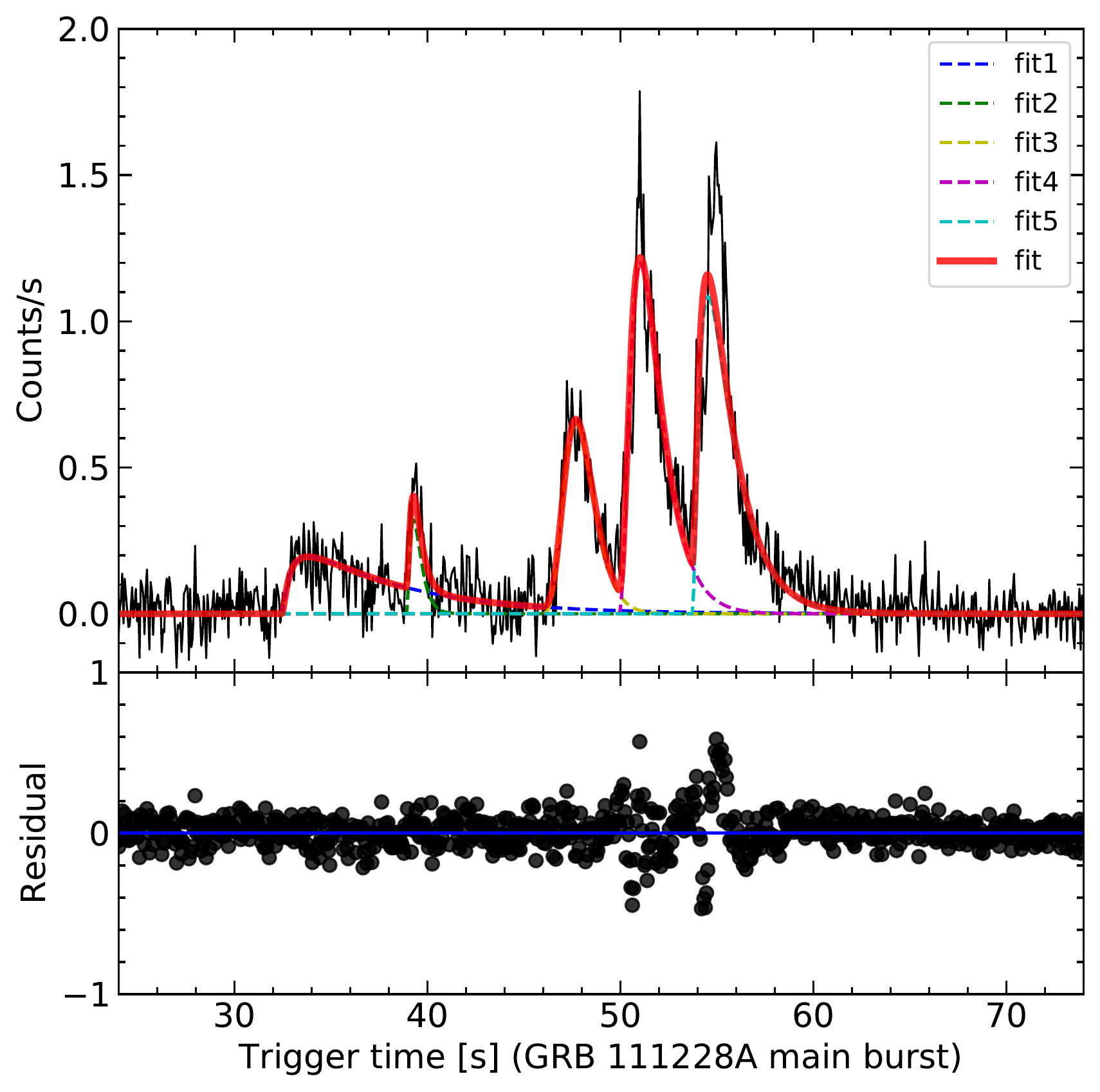}}
\caption{The lightcurve of GRB 111228A
}
\end{figure}

\begin{figure}[!htp]
\centering
\subfigure{
\includegraphics[scale = 0.4]{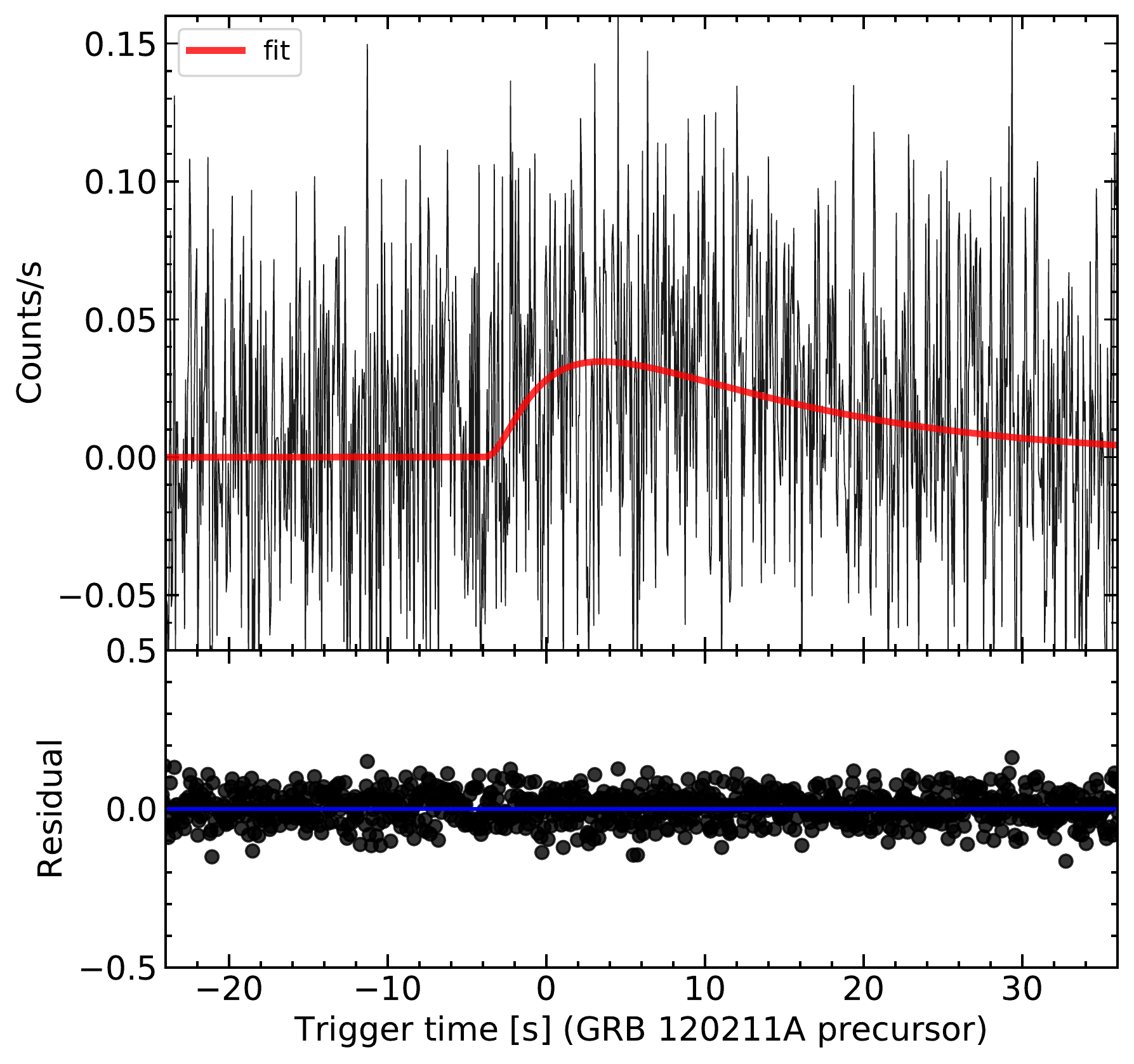}}
\subfigure{
\includegraphics[scale = 0.4]{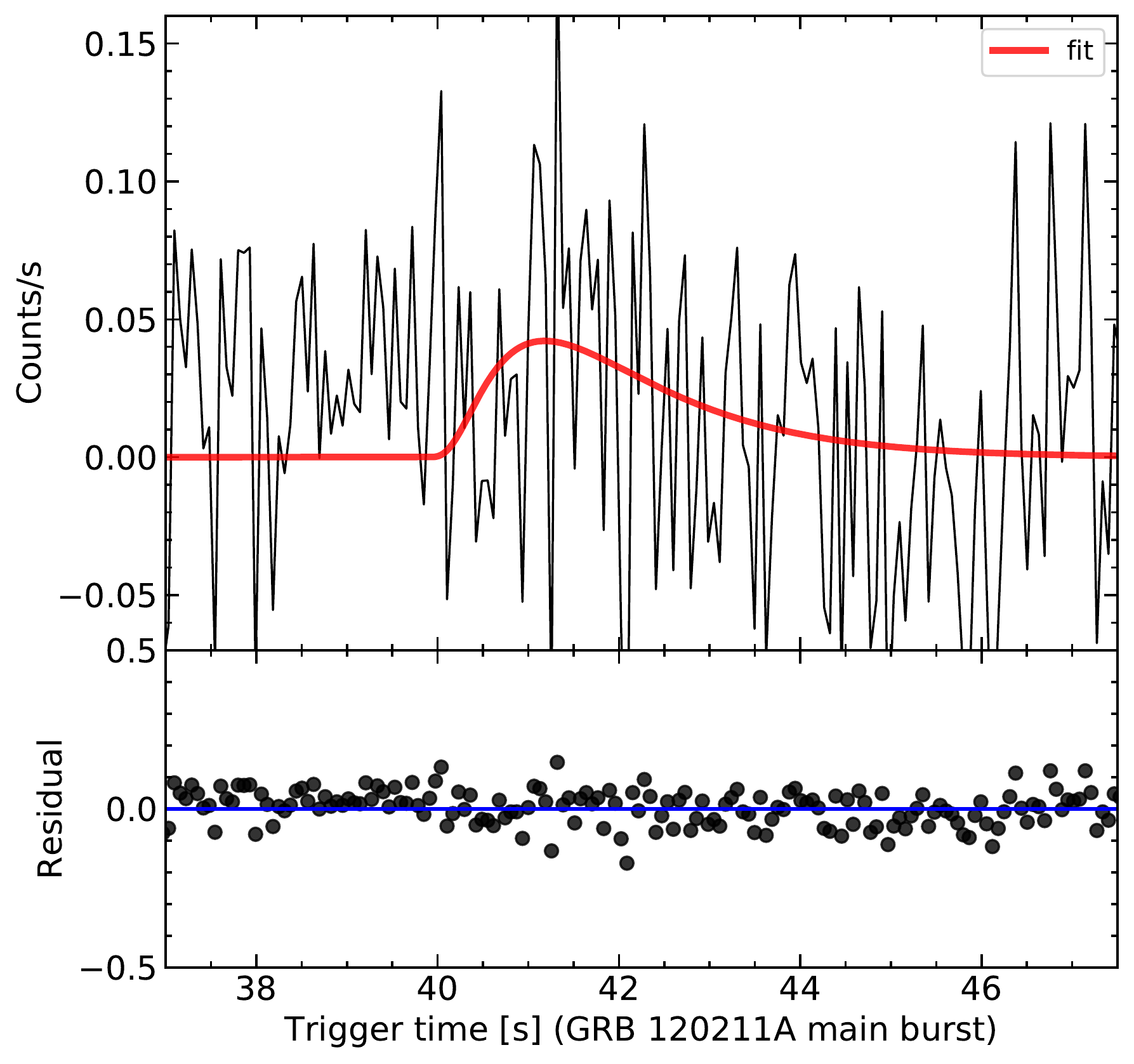}}
\caption{The lightcurve of GRB 120211A
}
\end{figure}
\clearpage

\begin{figure}[!htp]
\centering
\subfigure{
\includegraphics[scale = 0.4]{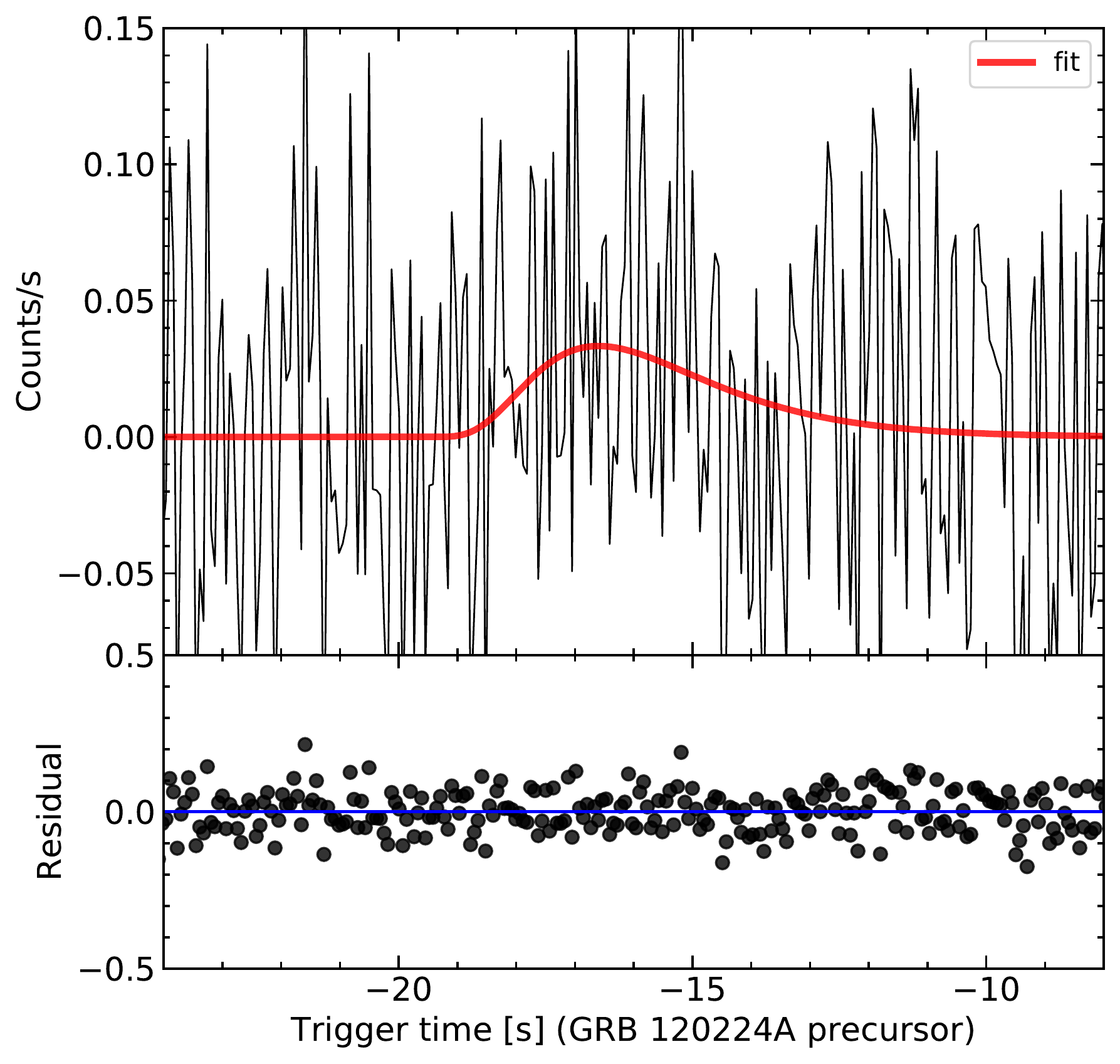}}
\subfigure{
\includegraphics[scale = 0.4]{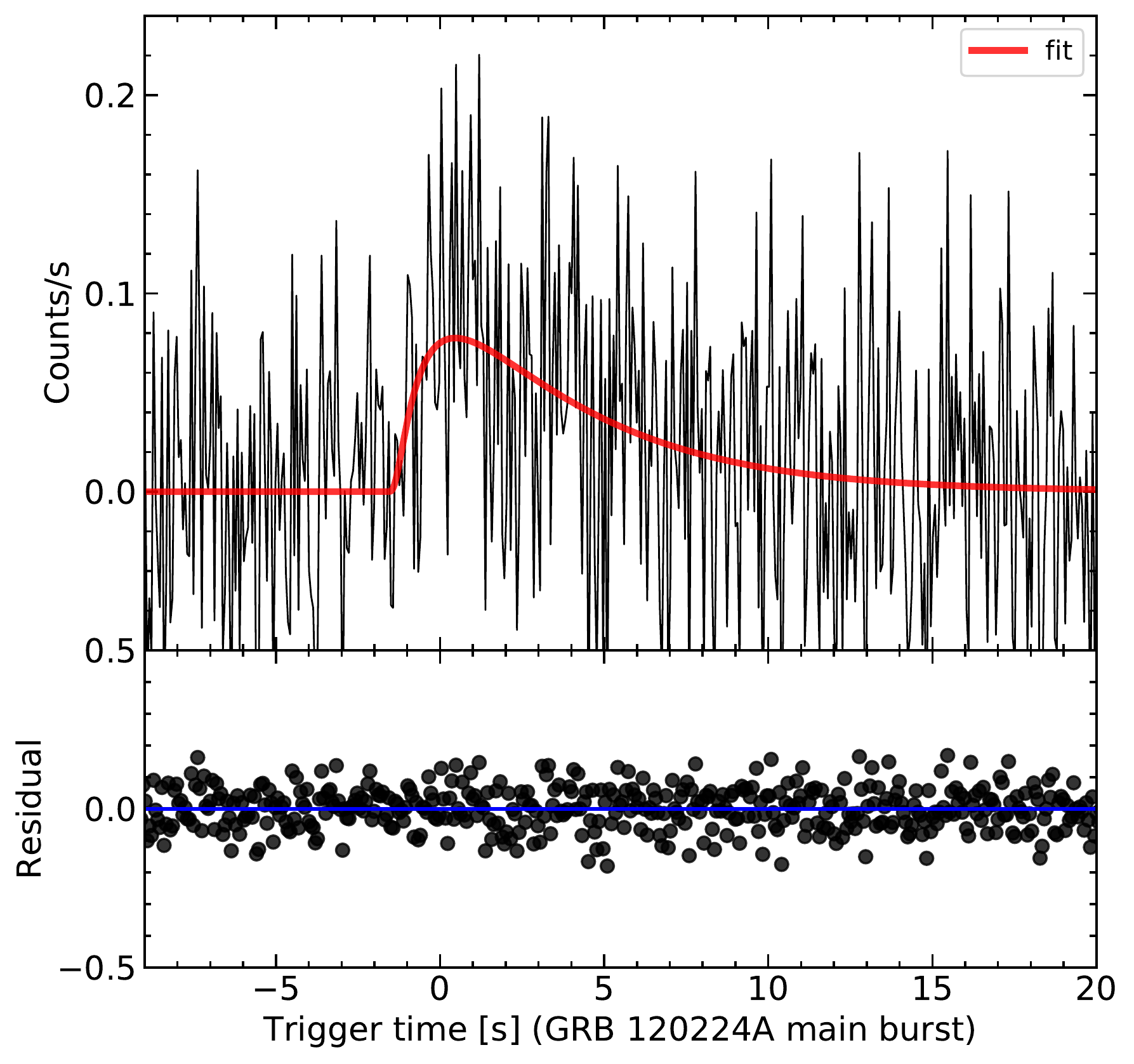}}
\caption{The lightcurve of GRB 120224A
}
\end{figure}

\begin{figure}[!htp]
\centering
\subfigure{
\includegraphics[scale = 0.4]{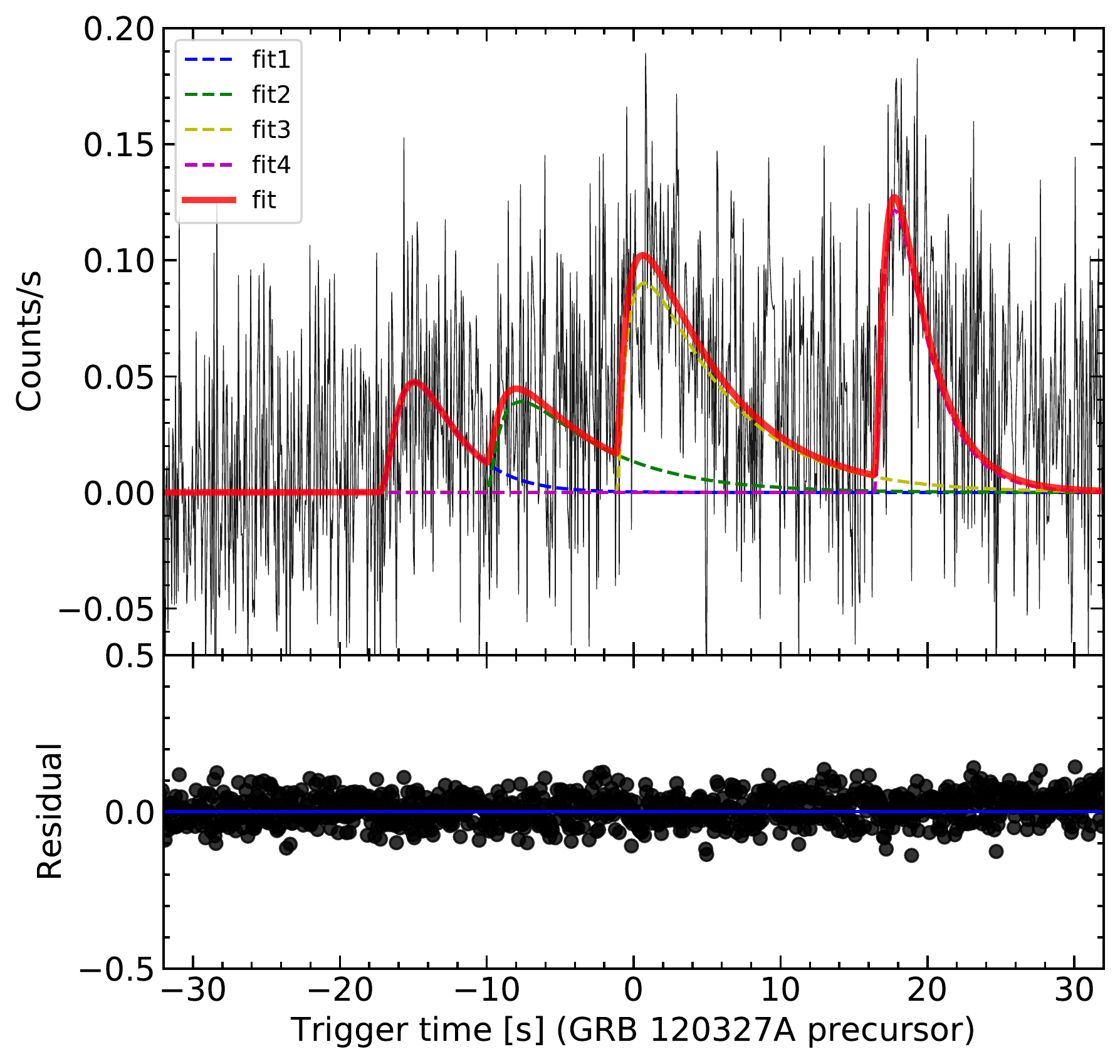}}
\subfigure{
\includegraphics[scale = 0.4]{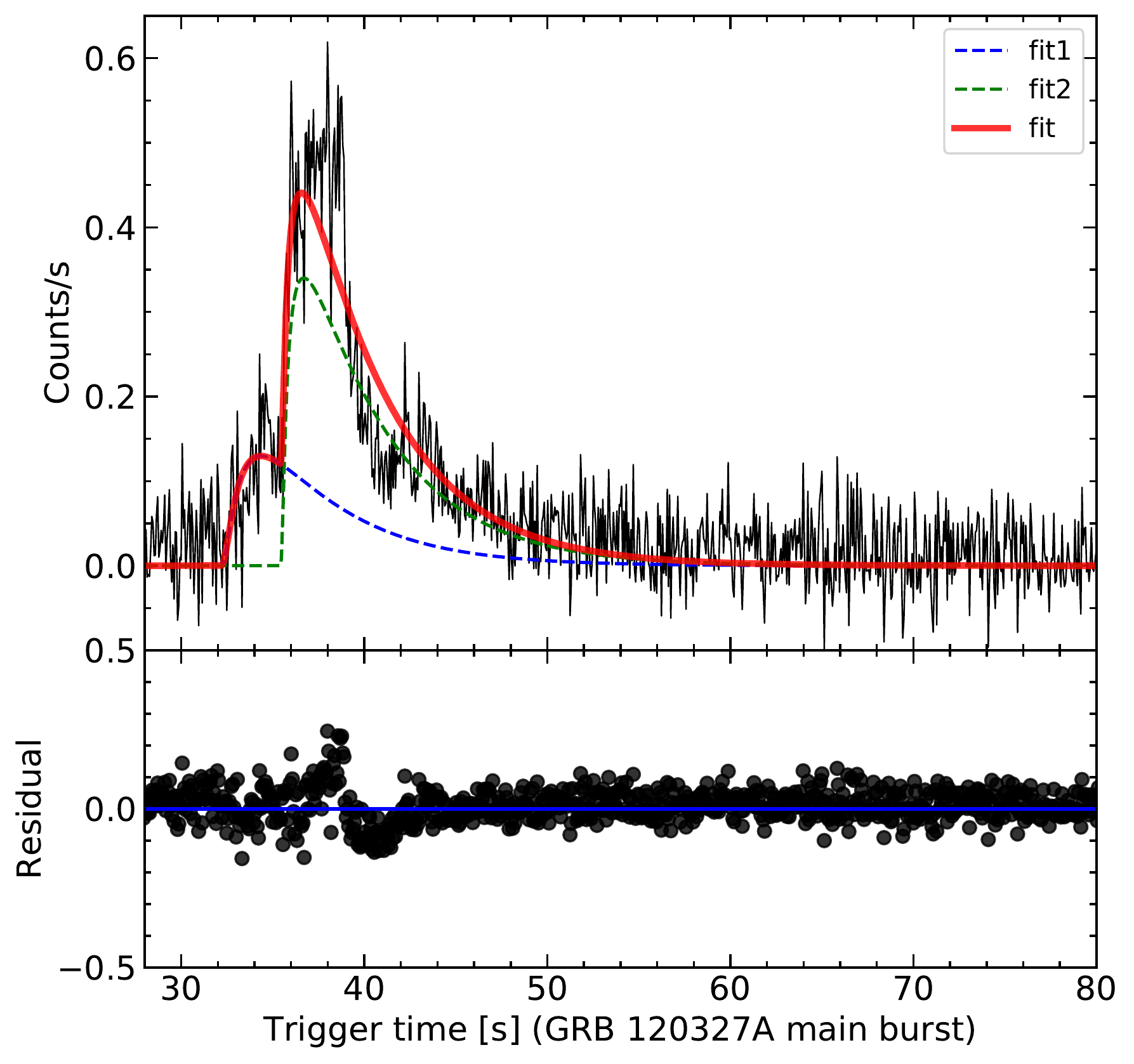}}
\caption{The lightcurve of GRB 120327A
}
\end{figure}

\begin{figure}[!htp]
\centering
\subfigure{
\includegraphics[scale = 0.4]{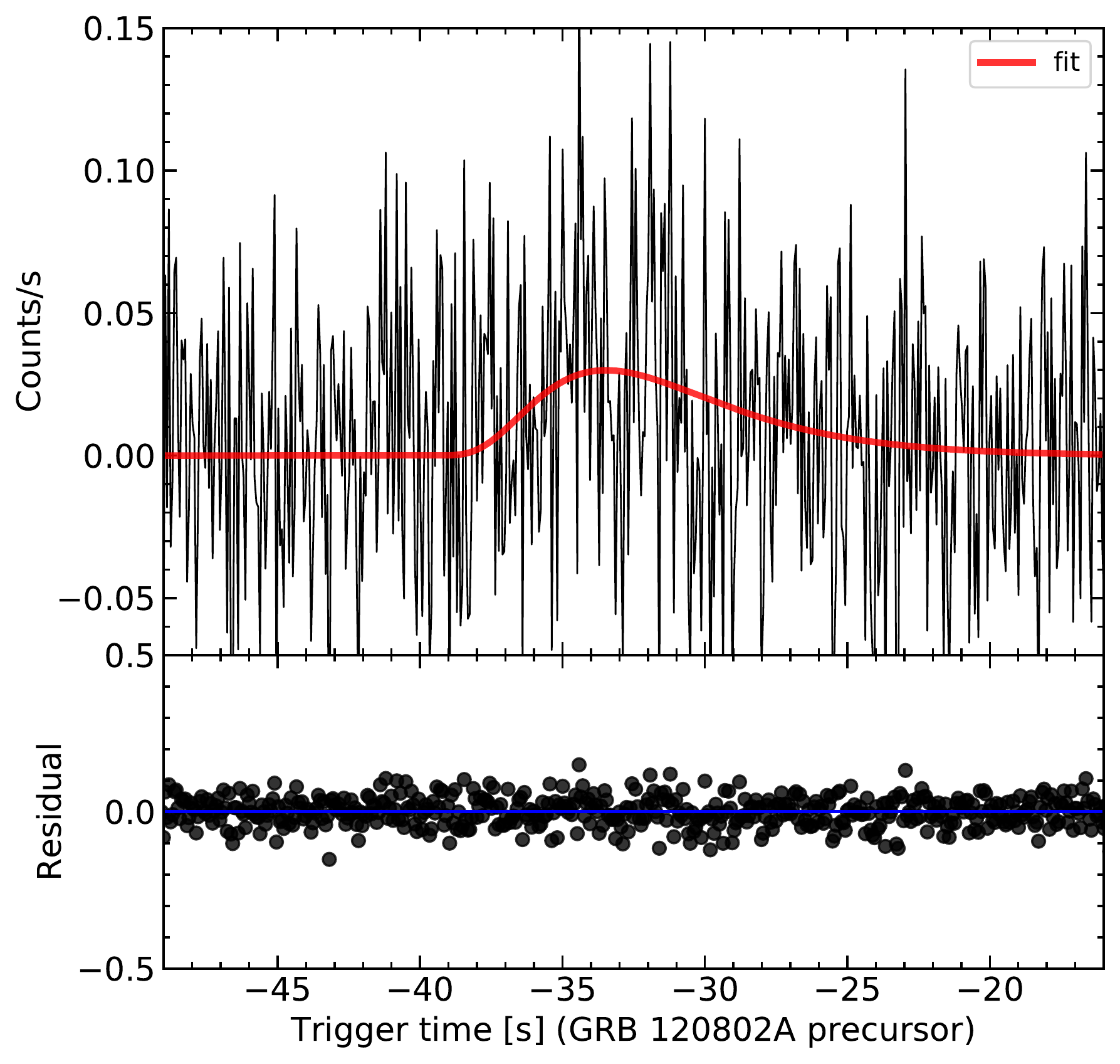}}
\subfigure{
\includegraphics[scale = 0.4]{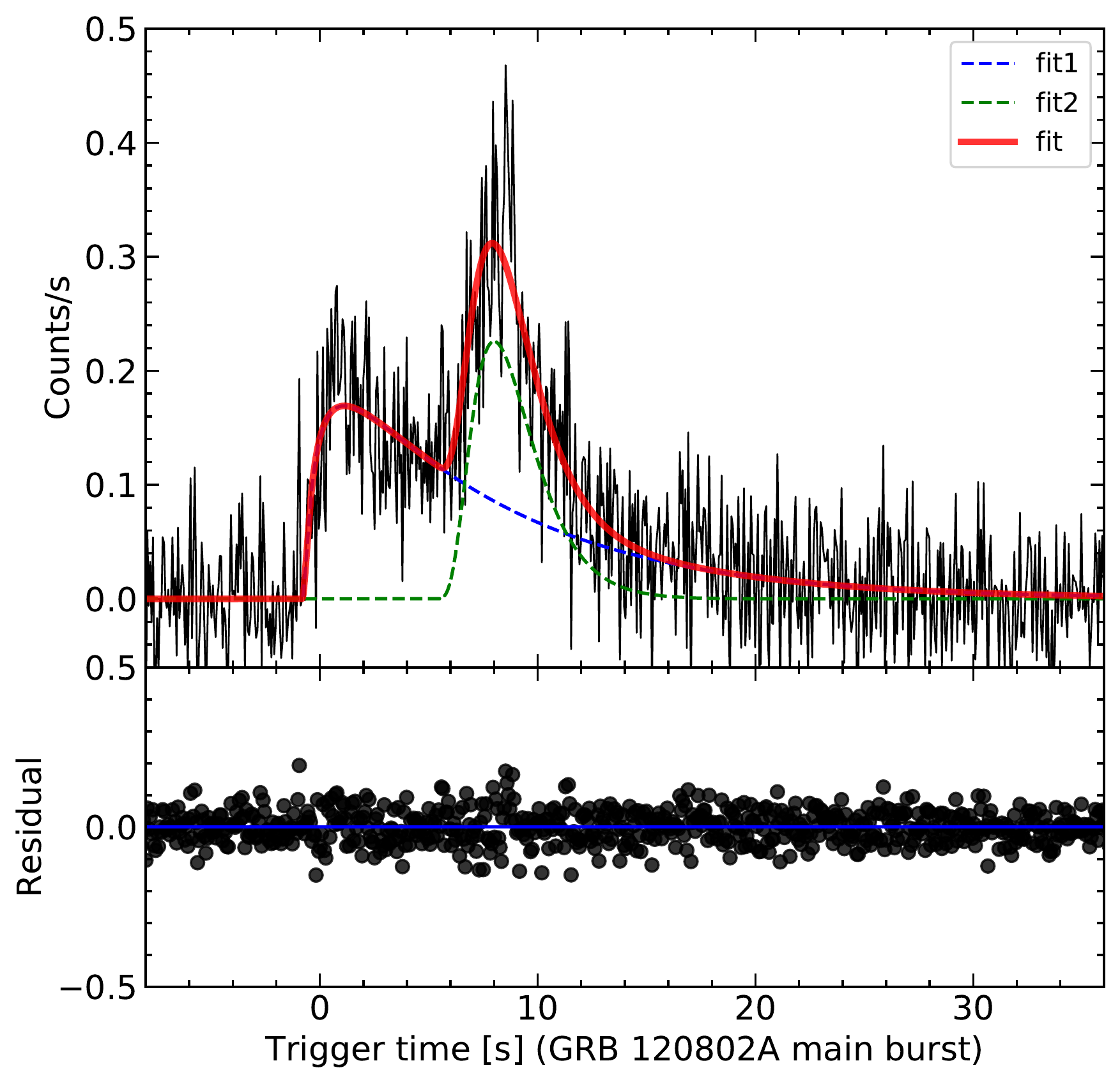}}
\caption{The lightcurve of GRB 120802A
}
\end{figure}

\begin{figure}[!htp]
\centering
\subfigure{
\includegraphics[scale = 0.4]{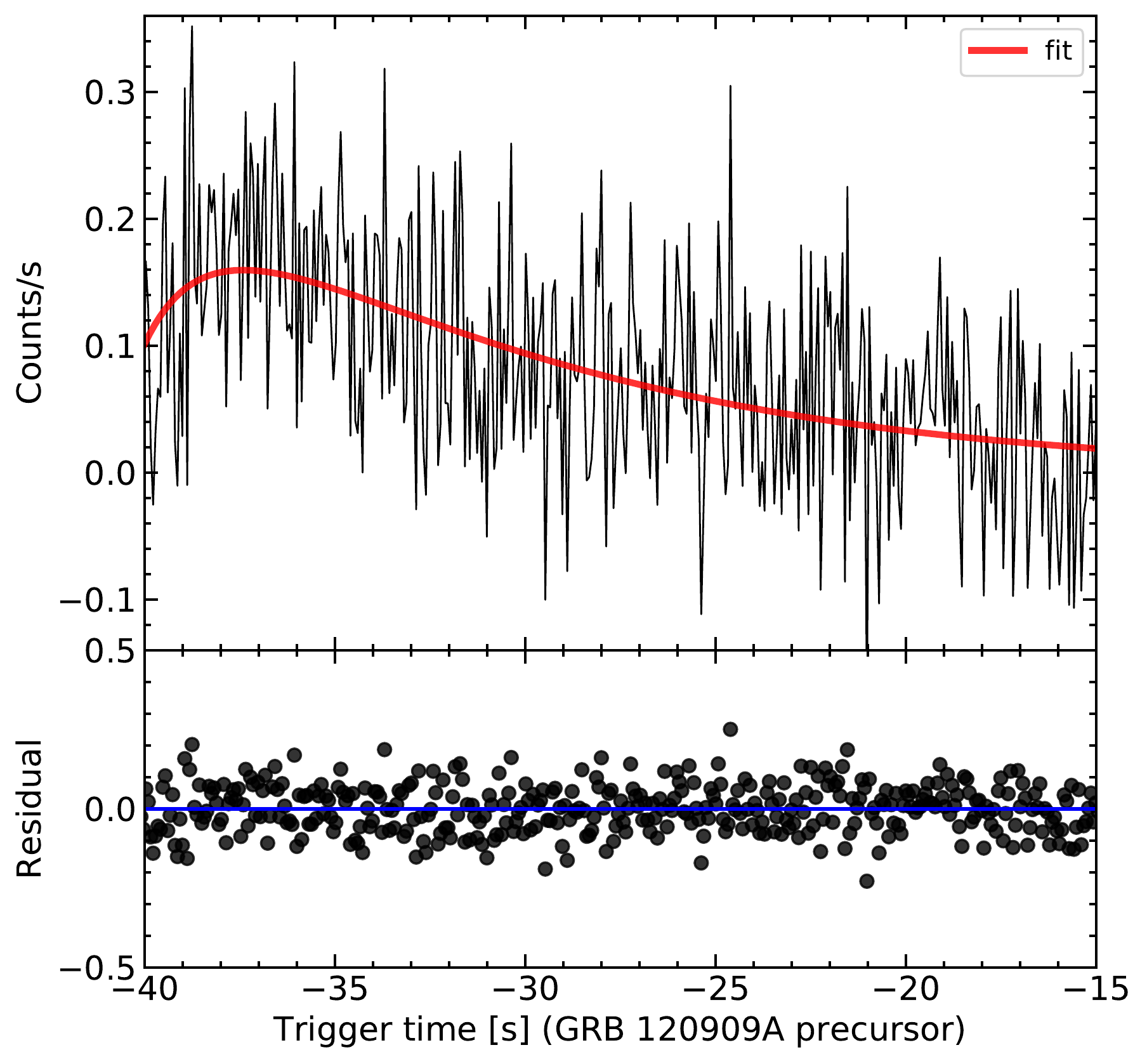}}
\subfigure{
\includegraphics[scale = 0.4]{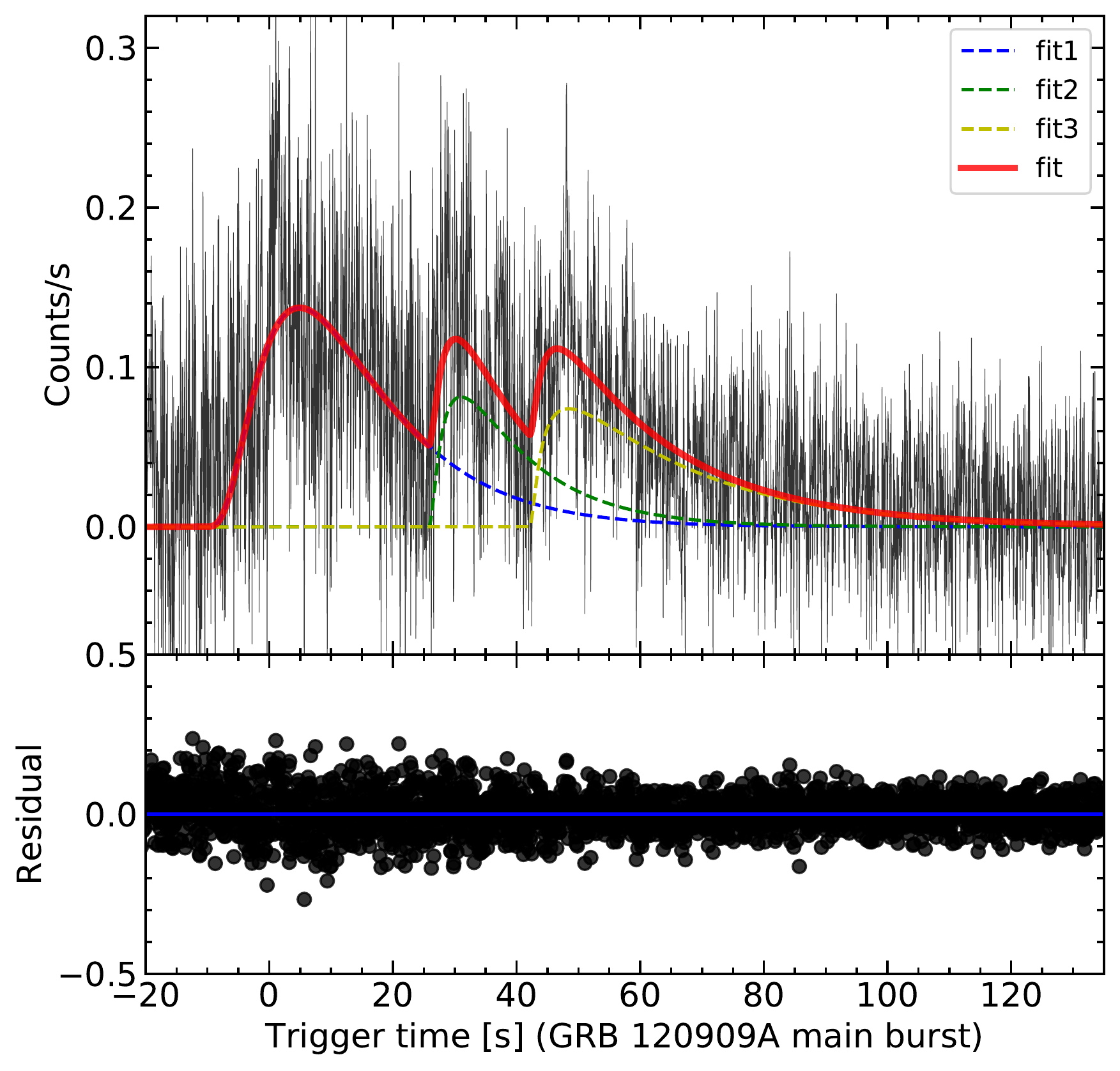}}
\caption{The lightcurve of GRB 120909A
}
\end{figure}

\begin{figure}[!htp]
\centering
\subfigure{
\includegraphics[scale = 0.4]{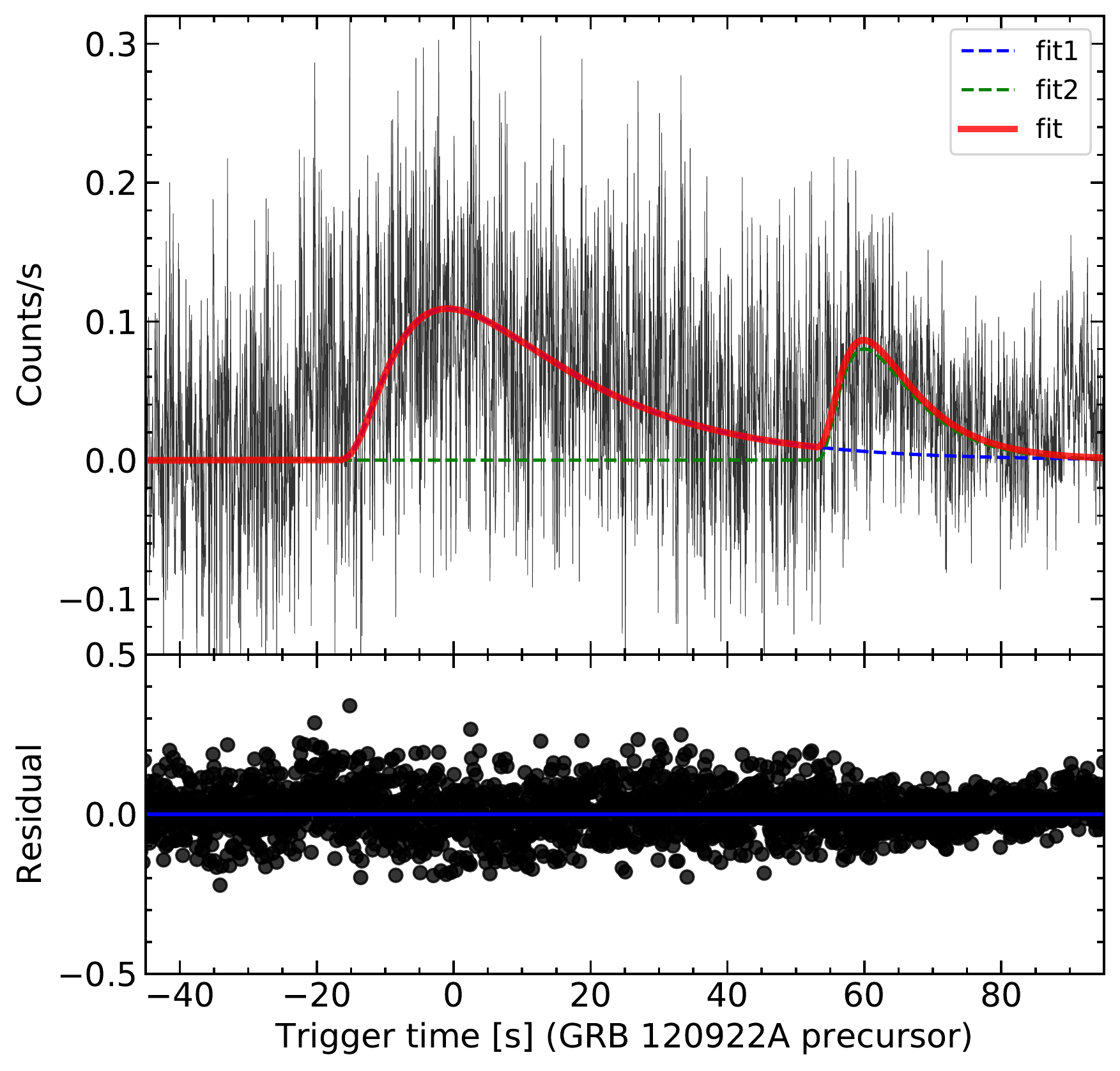}}
\subfigure{
\includegraphics[scale = 0.4]{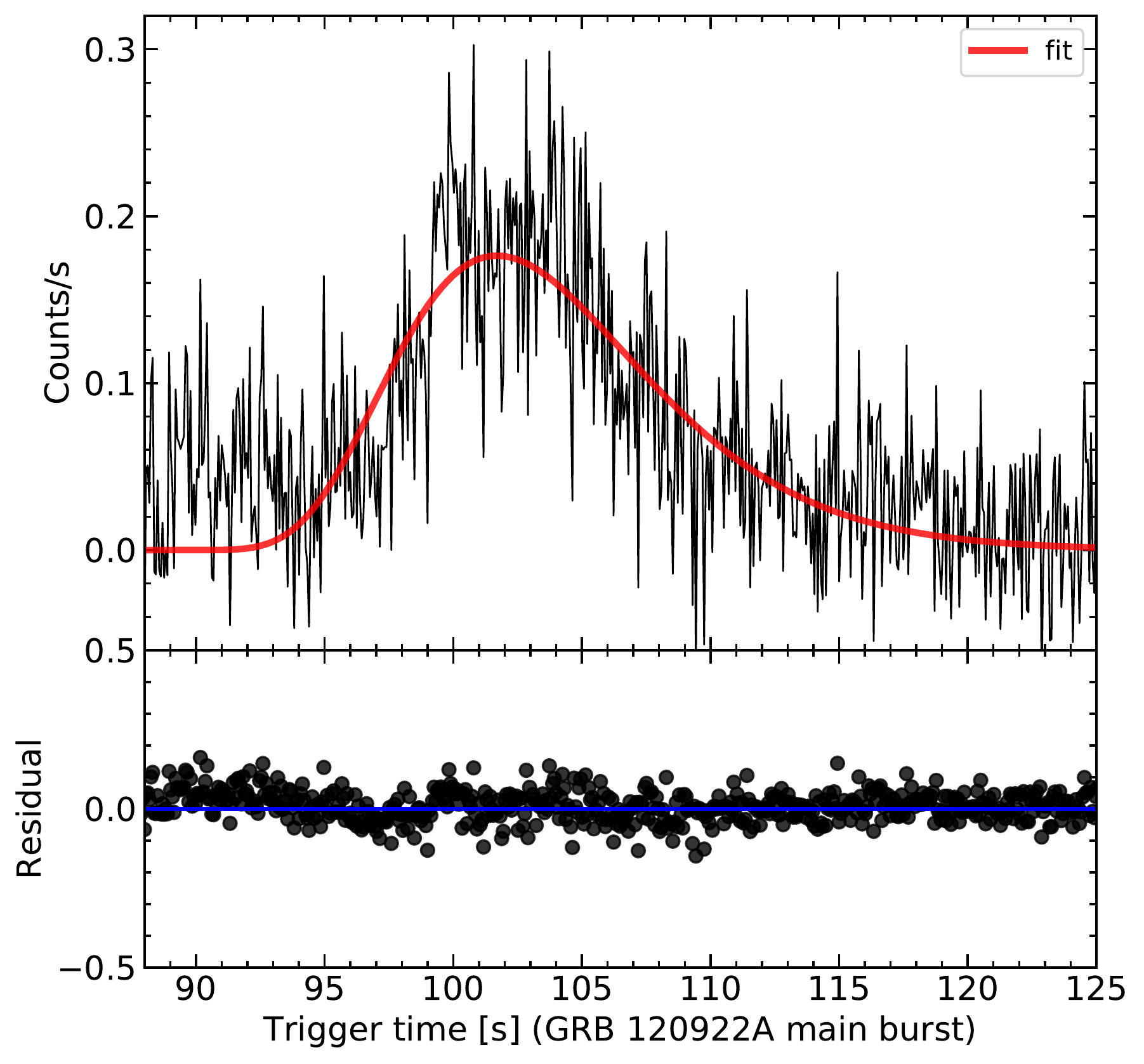}}
\caption{The lightcurve of GRB 120922A
}
\end{figure}
\clearpage

\begin{figure}[!htp]
\centering
\subfigure{
\includegraphics[scale = 0.4]{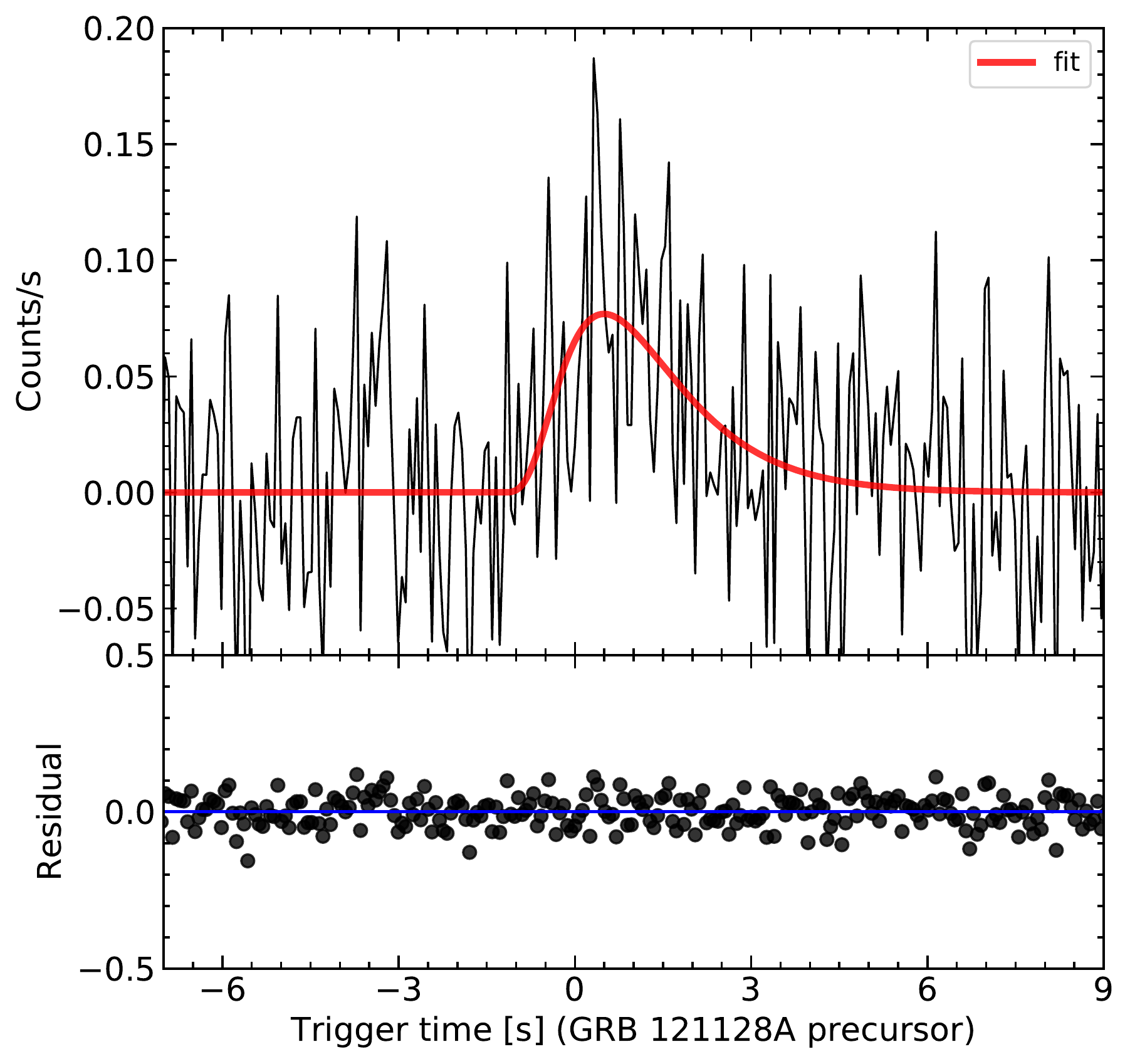}}
\subfigure{
\includegraphics[scale = 0.4]{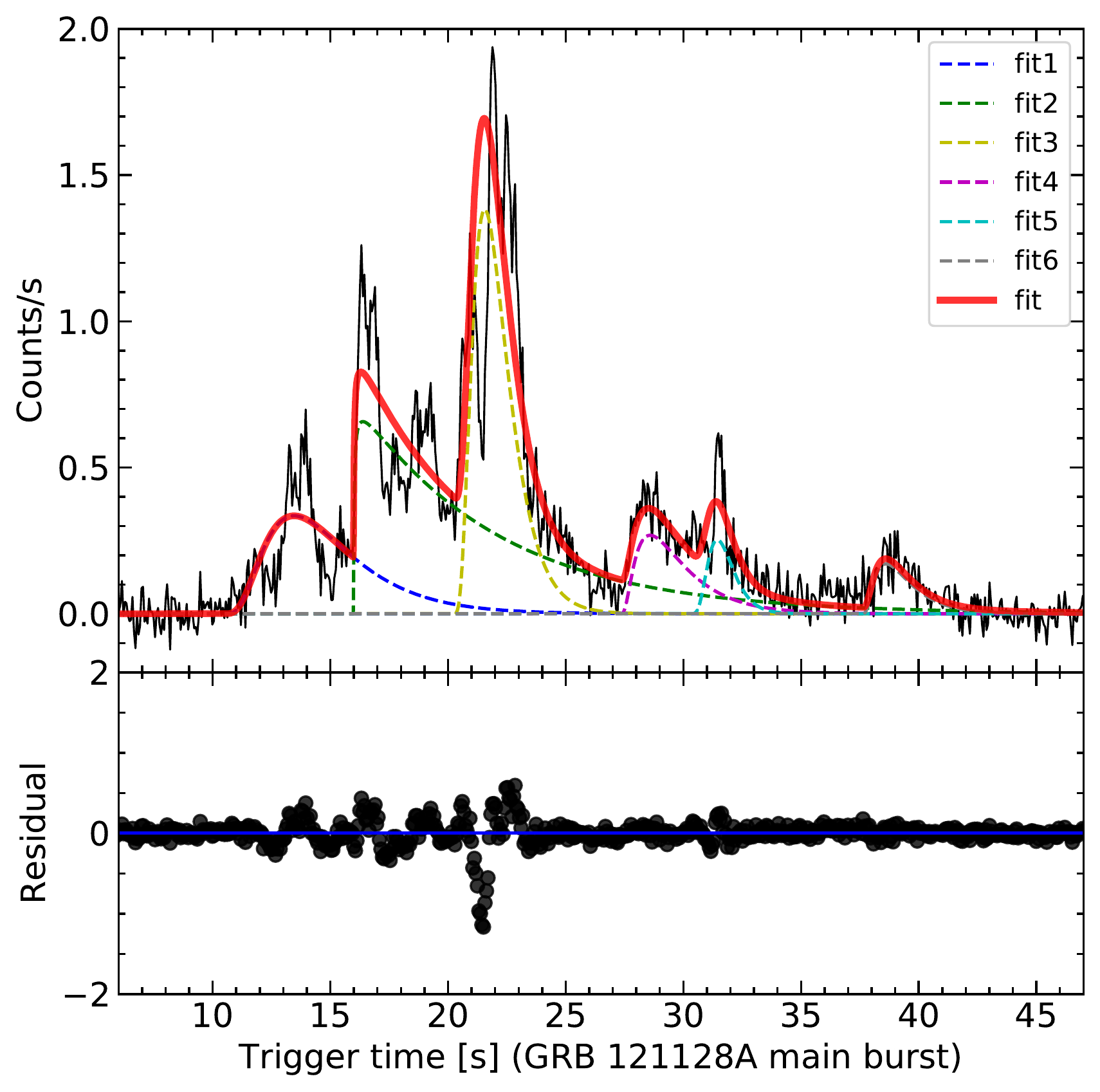}}
\caption{The lightcurve of GRB 121128A
}
\end{figure}

\begin{figure}[!htp]
\centering
\subfigure{
\includegraphics[scale = 0.4]{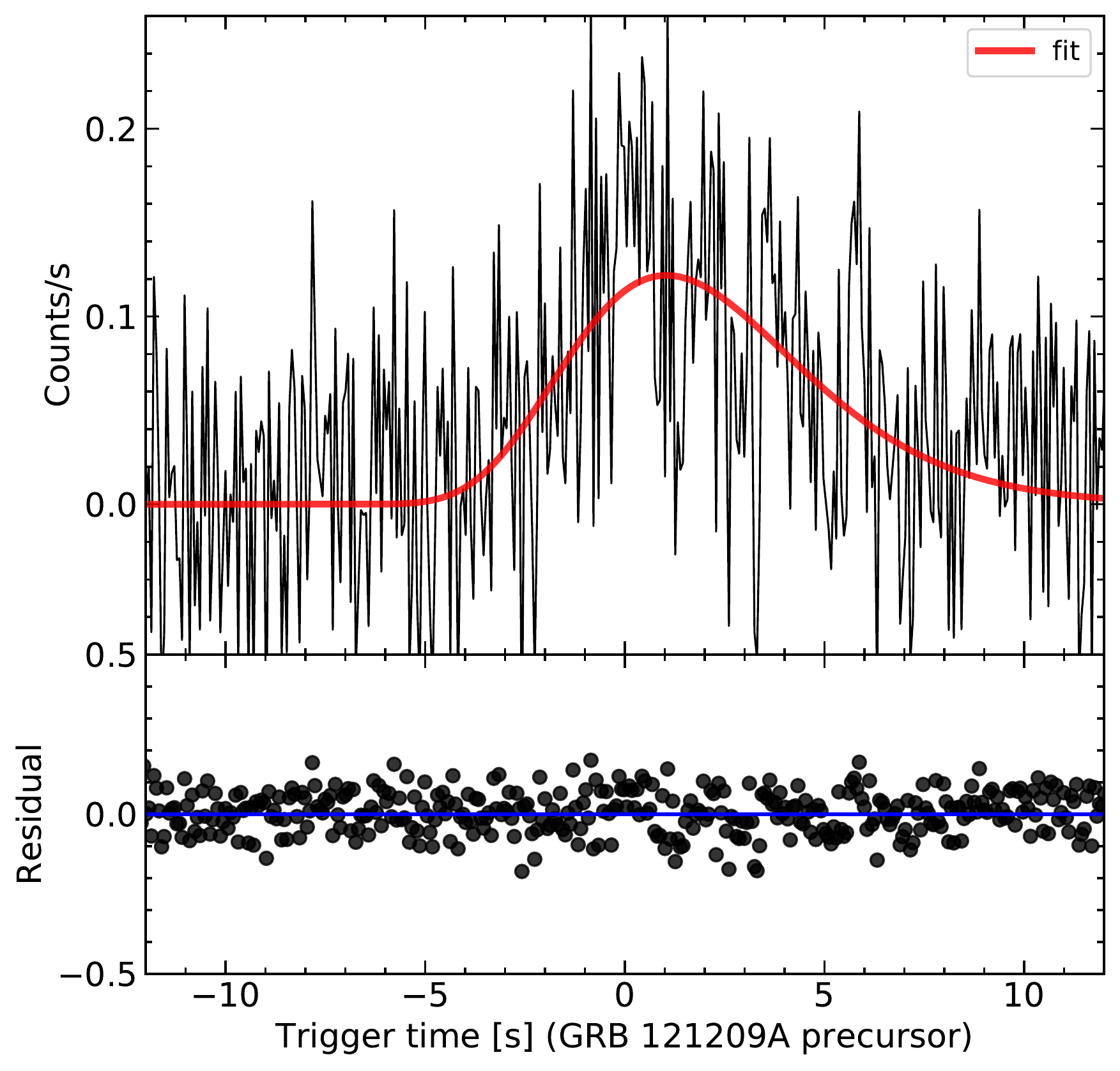}}
\subfigure{
\includegraphics[scale = 0.4]{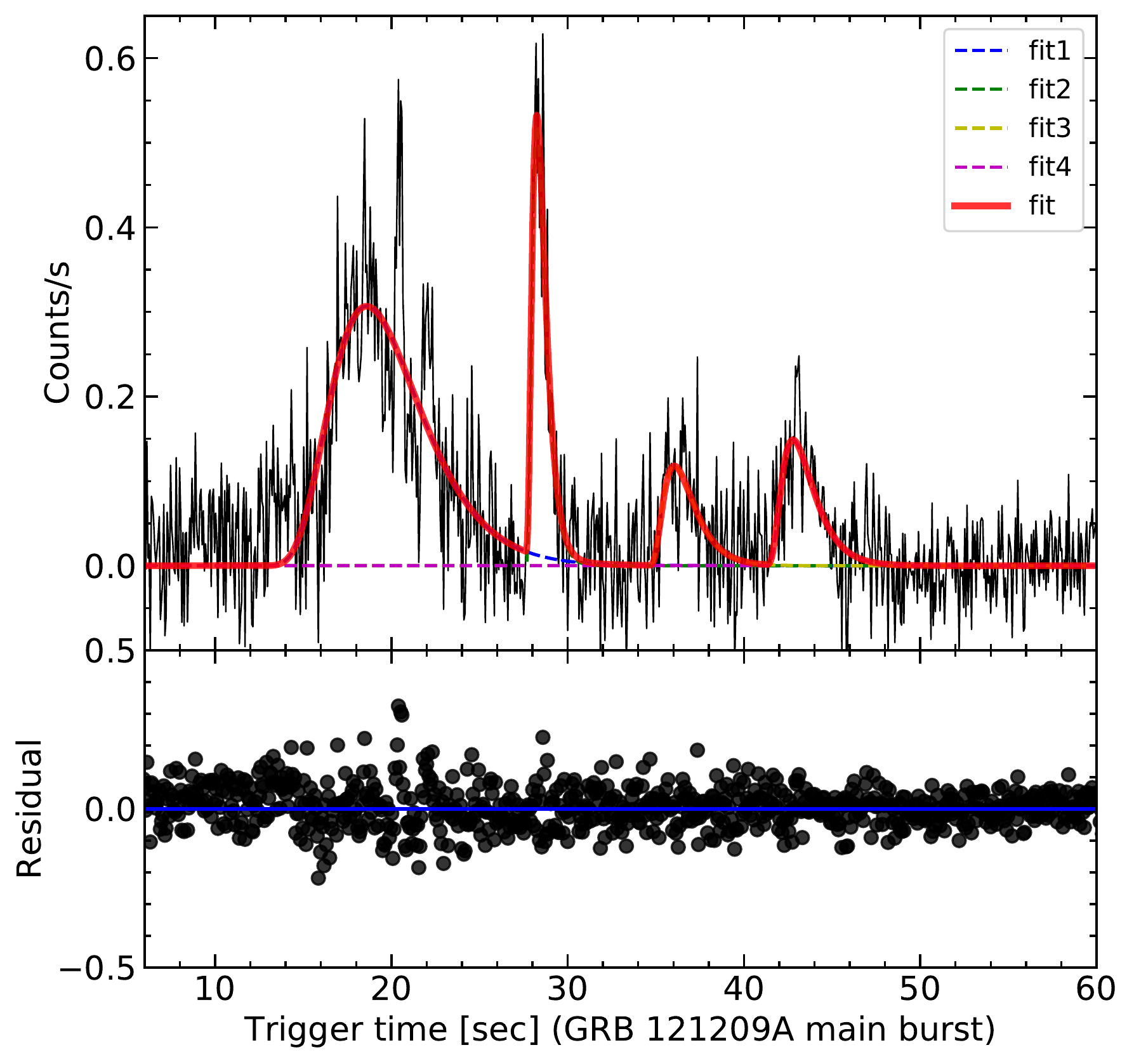}}
\caption{The lightcurve of GRB 121209A
}
\end{figure}

\begin{figure}[!htp]
\centering
\subfigure{
\includegraphics[scale = 0.4]{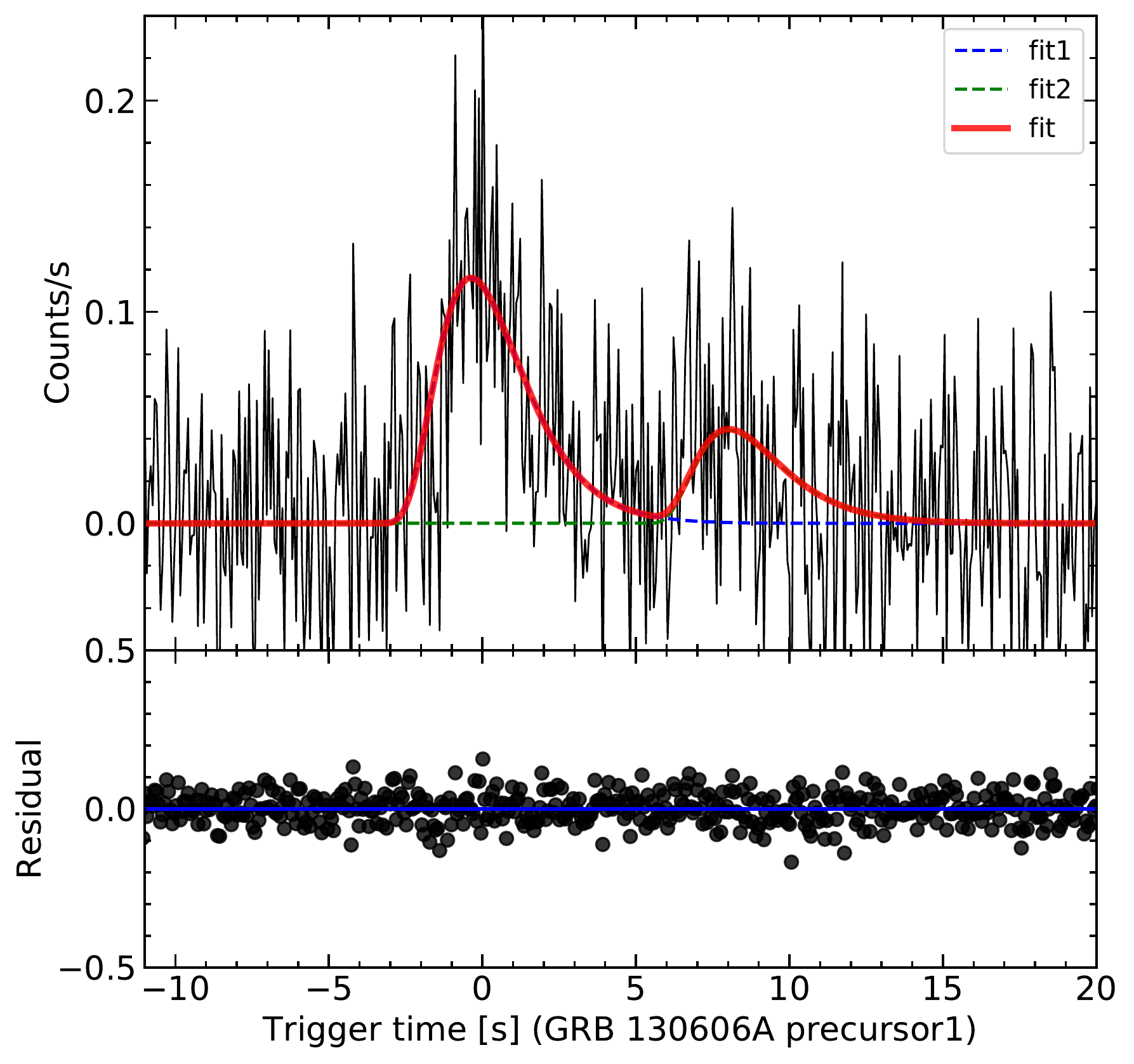}}
\subfigure{
\includegraphics[scale = 0.4]{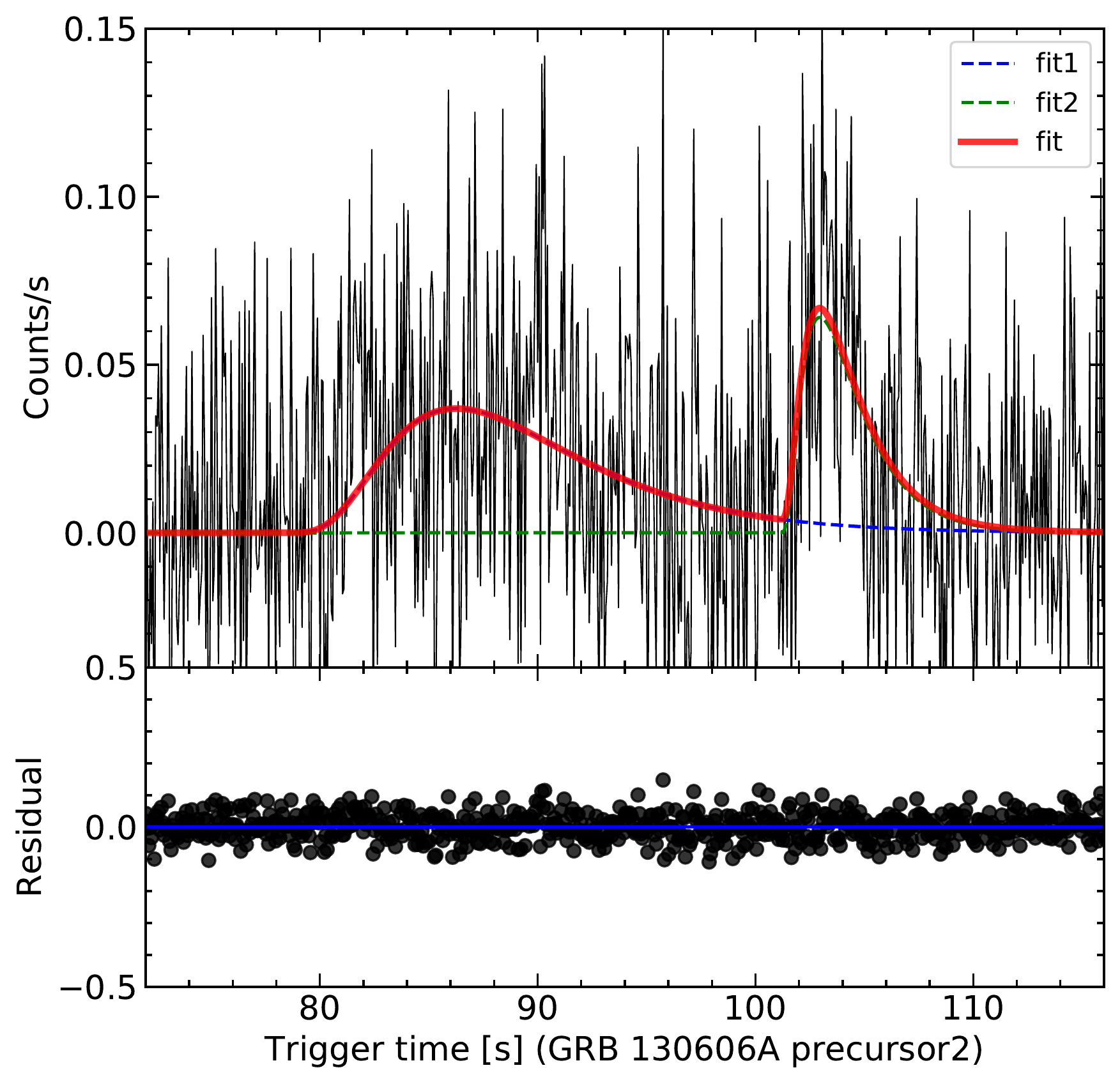}}
\subfigure{
\includegraphics[scale = 0.4]{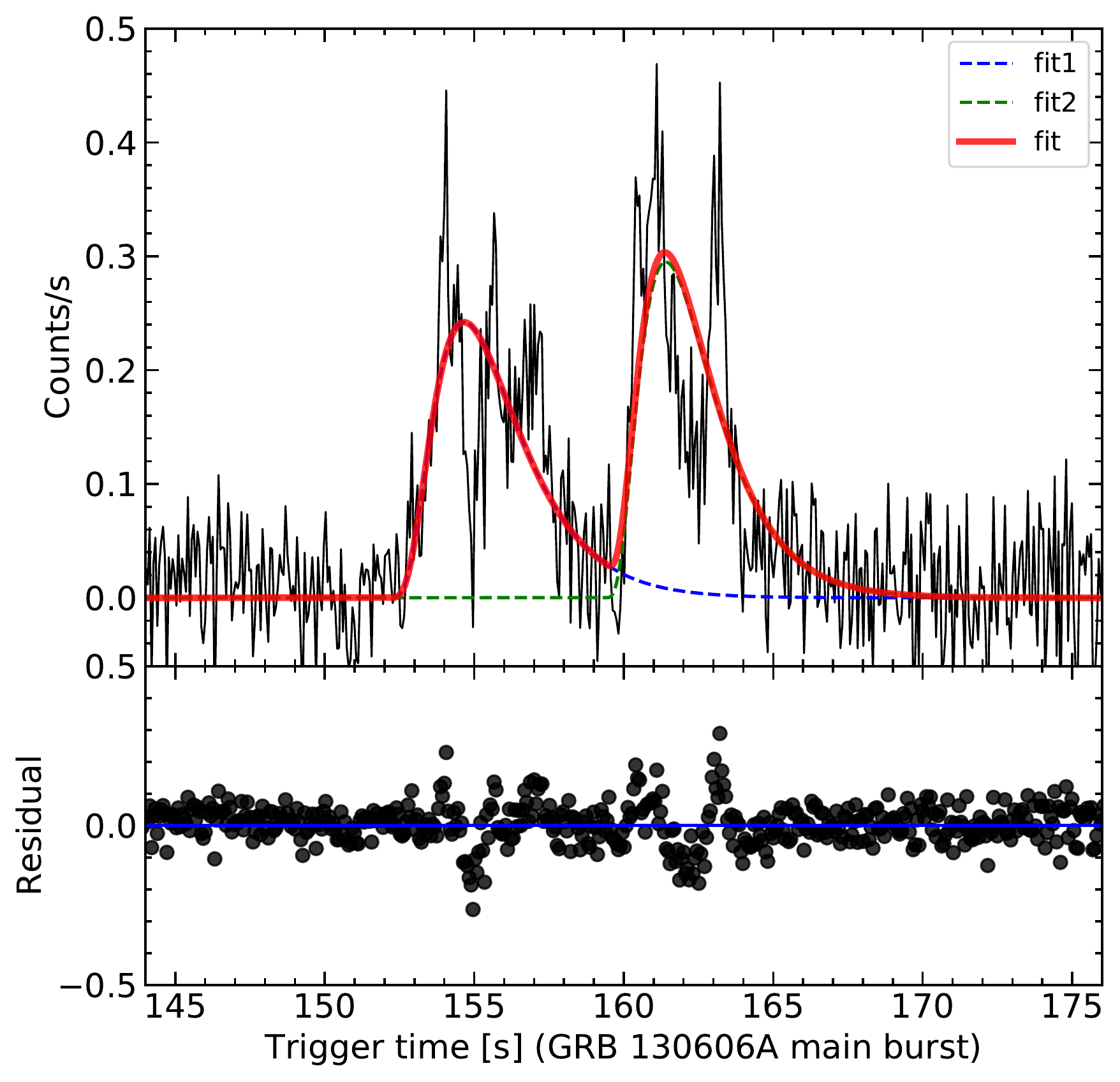}}
\caption{The lightcurve of GRB 130606A
}
\end{figure}

\begin{figure}[!htp]
\centering
\subfigure{
\includegraphics[scale = 0.4]{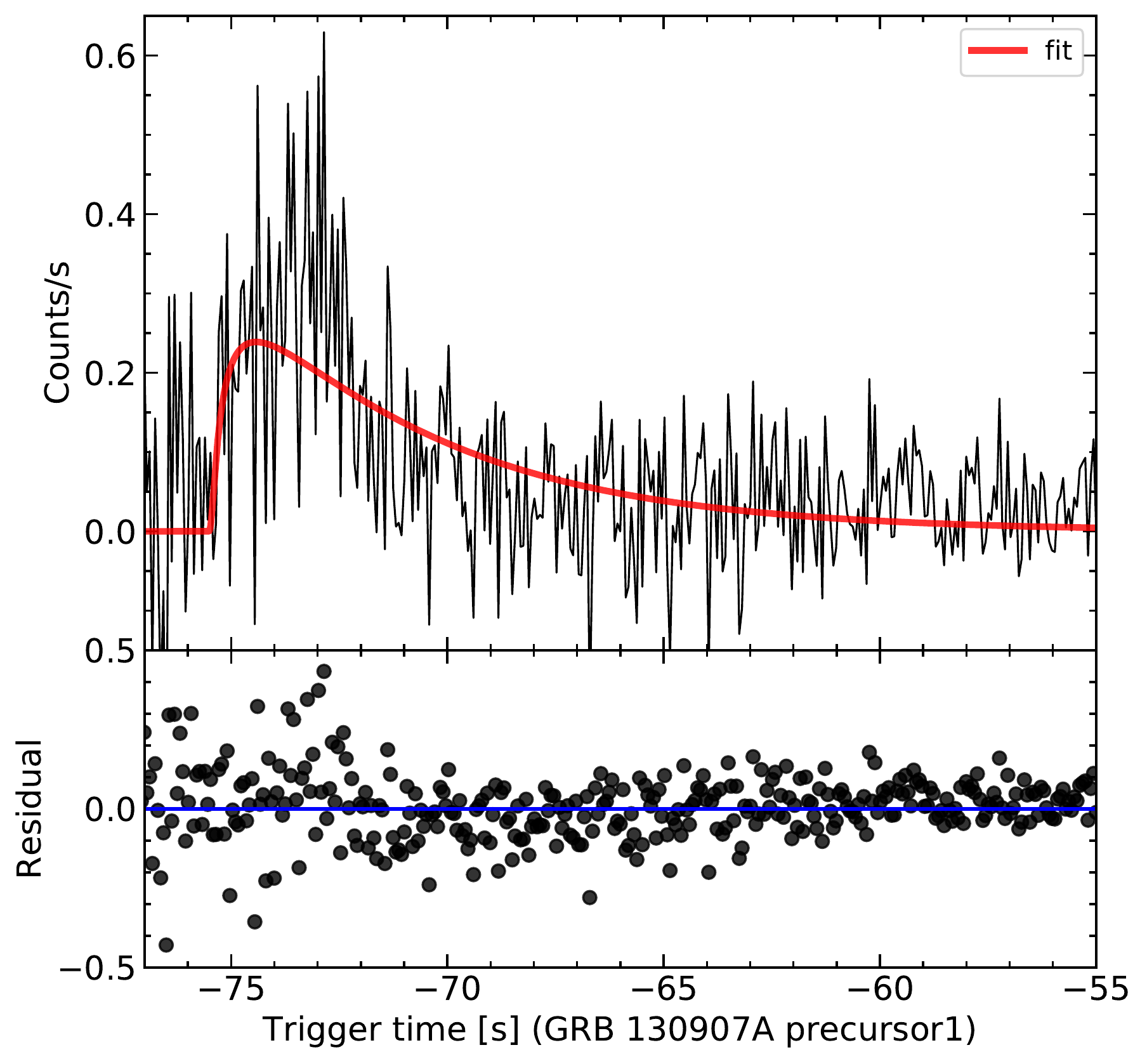}}
\subfigure{
\includegraphics[scale = 0.4]{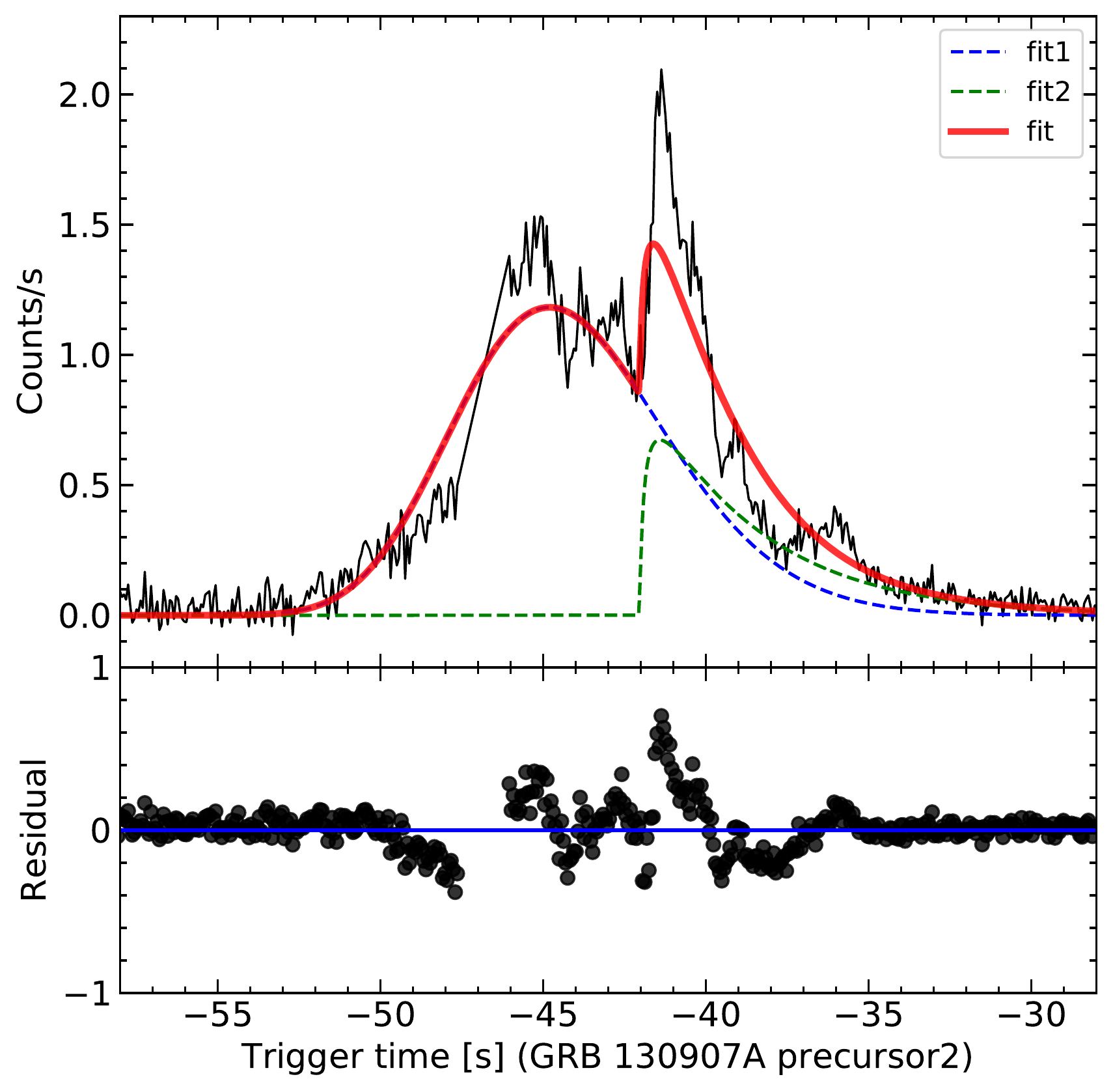}}
\subfigure{
\includegraphics[scale = 0.4]{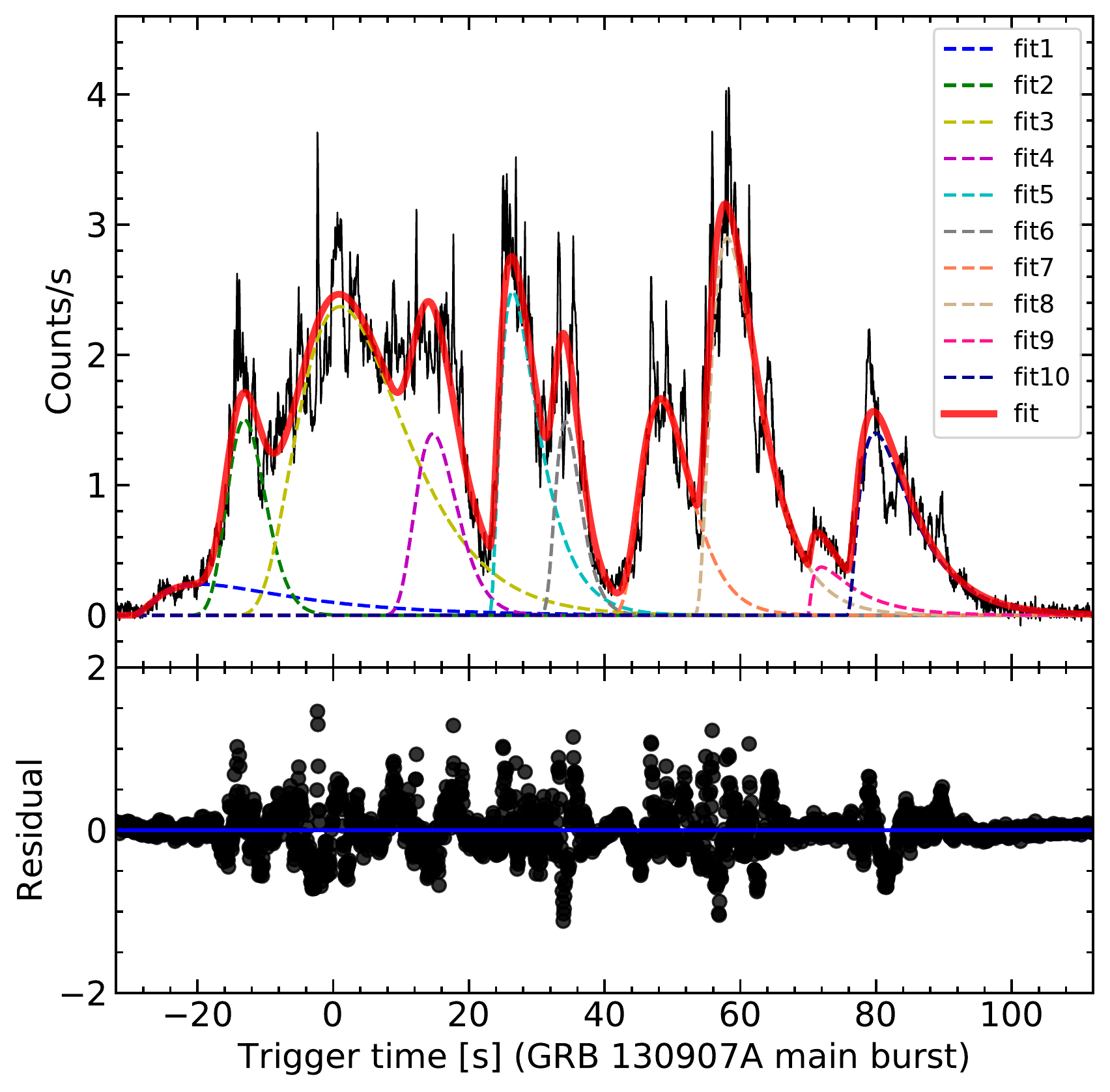}}
\caption{The lightcurve of GRB 130907A
}
\end{figure}

\begin{figure}[!htp]
\centering
\subfigure{
\includegraphics[scale = 0.4]{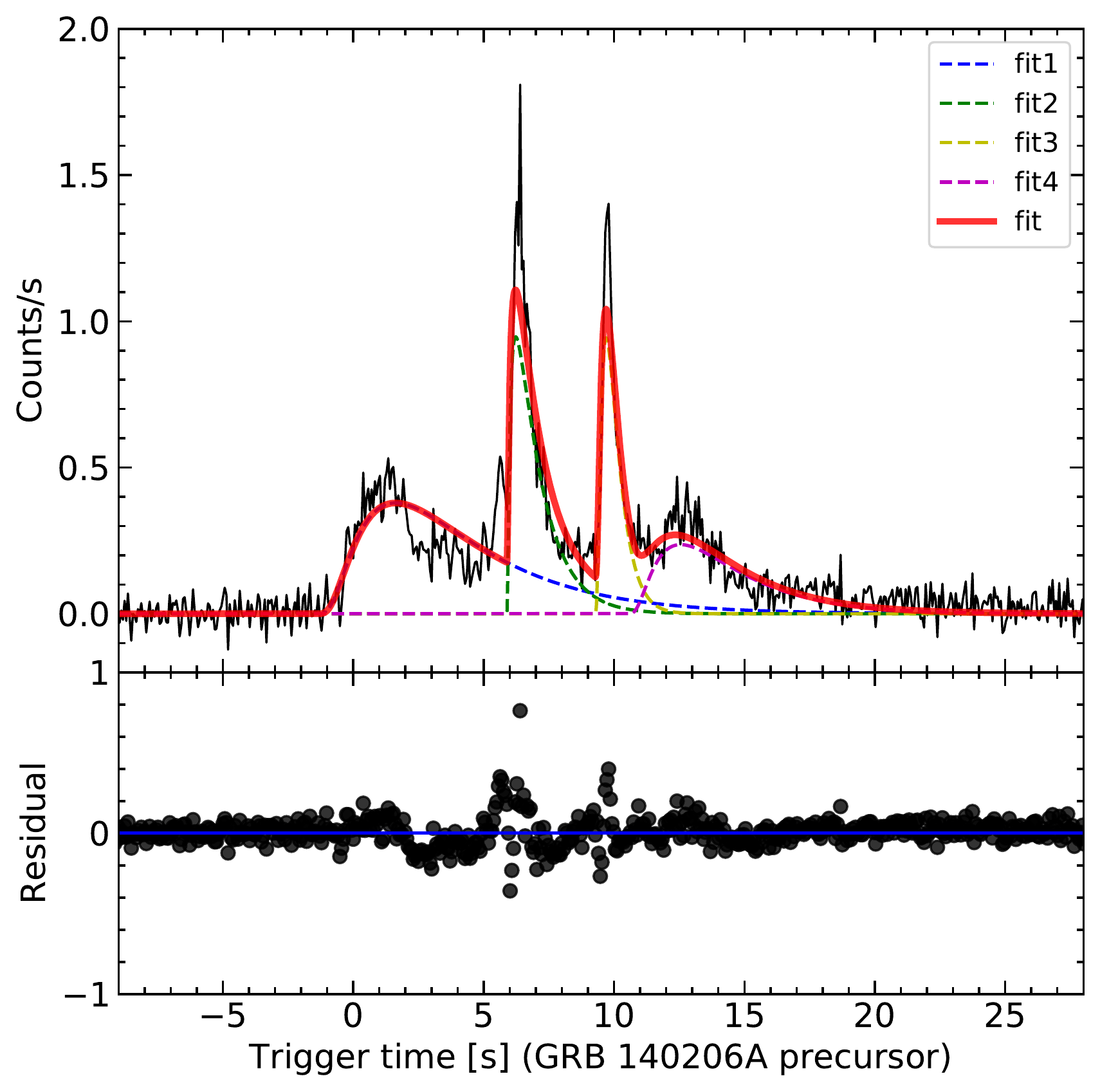}}
\subfigure{
\includegraphics[scale = 0.4]{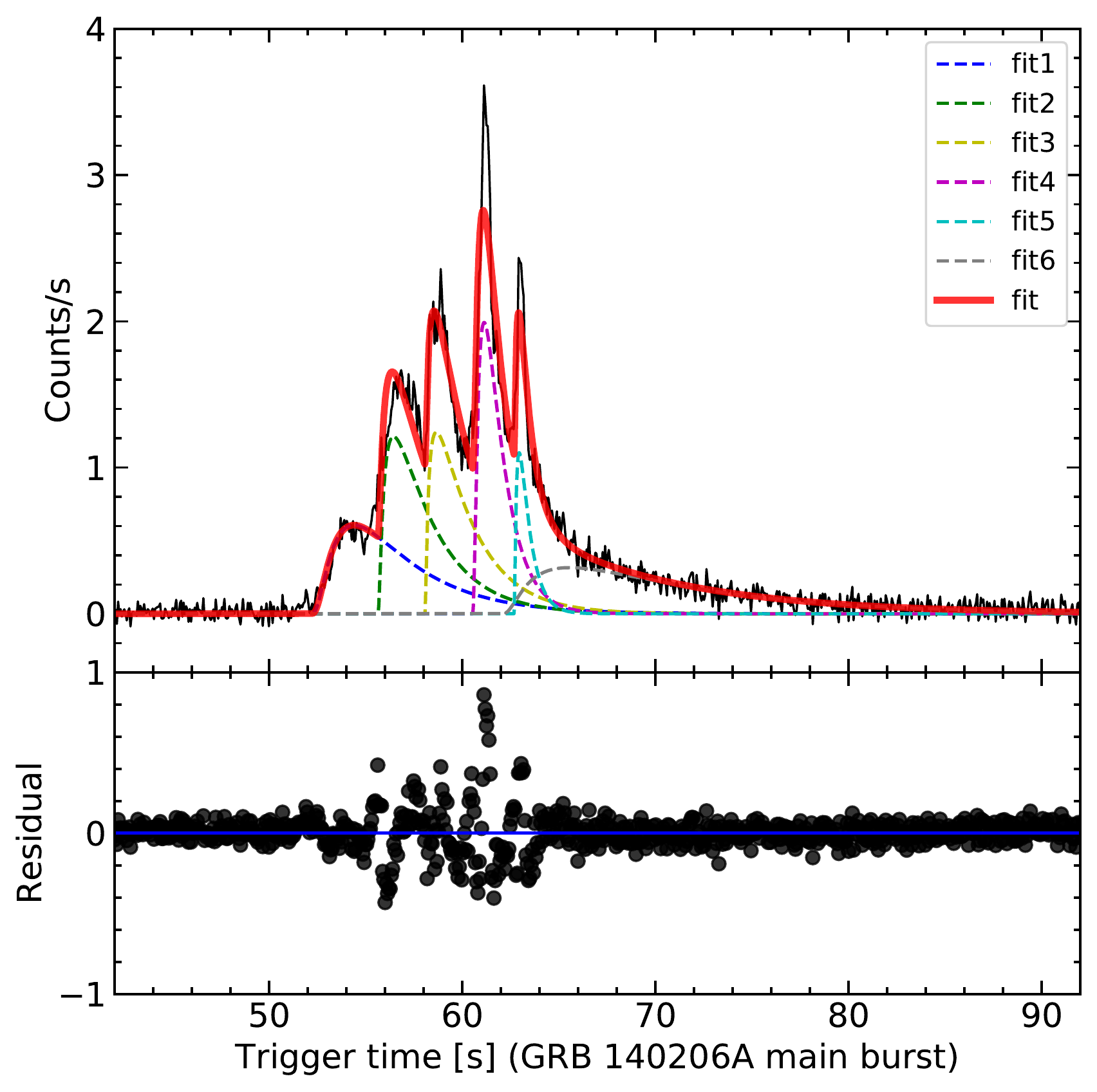}}
\caption{The lightcurve of GRB 140206A
}
\end{figure}

\begin{figure}[!htp]
\centering
\subfigure{
\includegraphics[scale = 0.4]{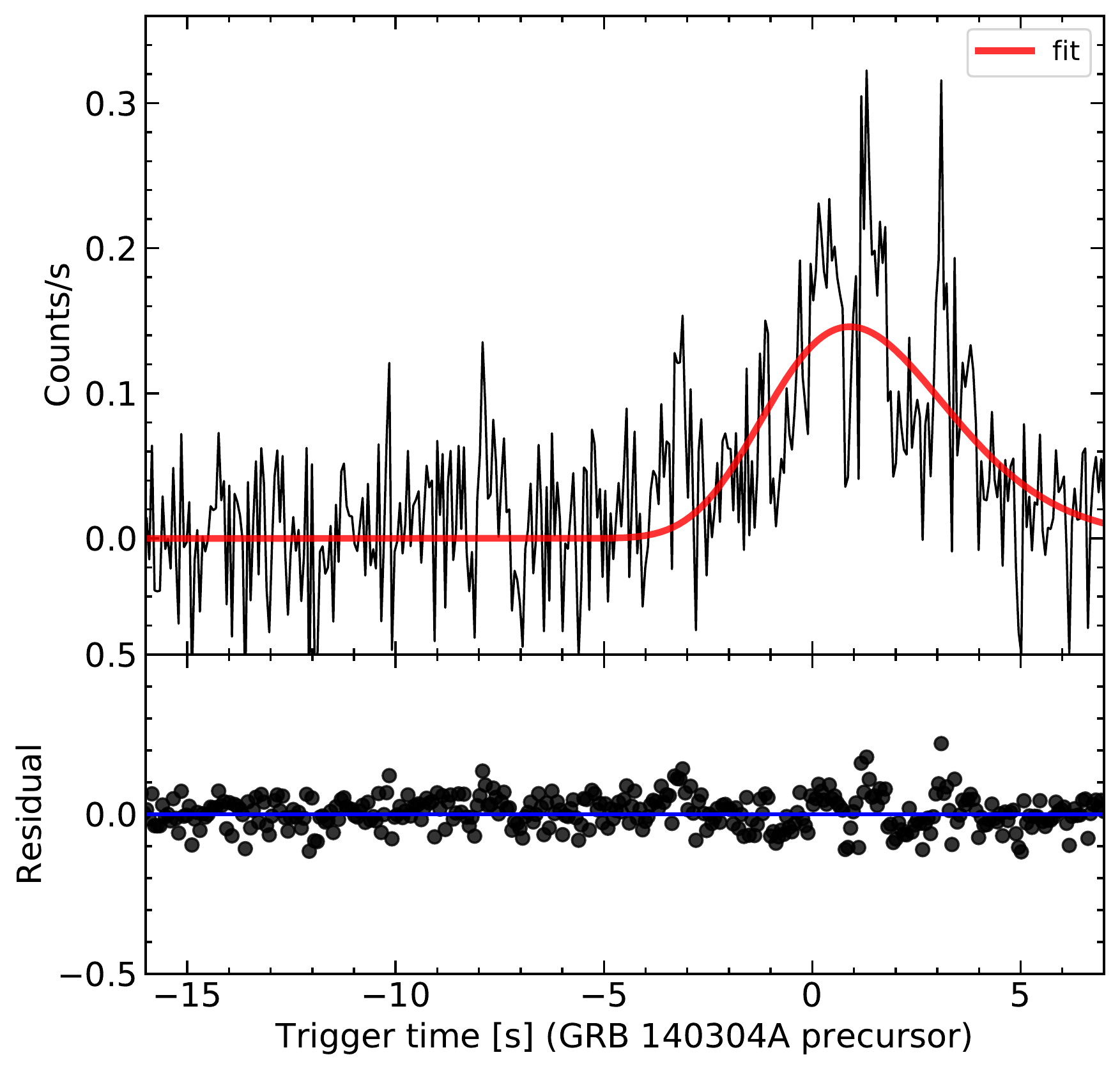}}
\subfigure{
\includegraphics[scale = 0.4]{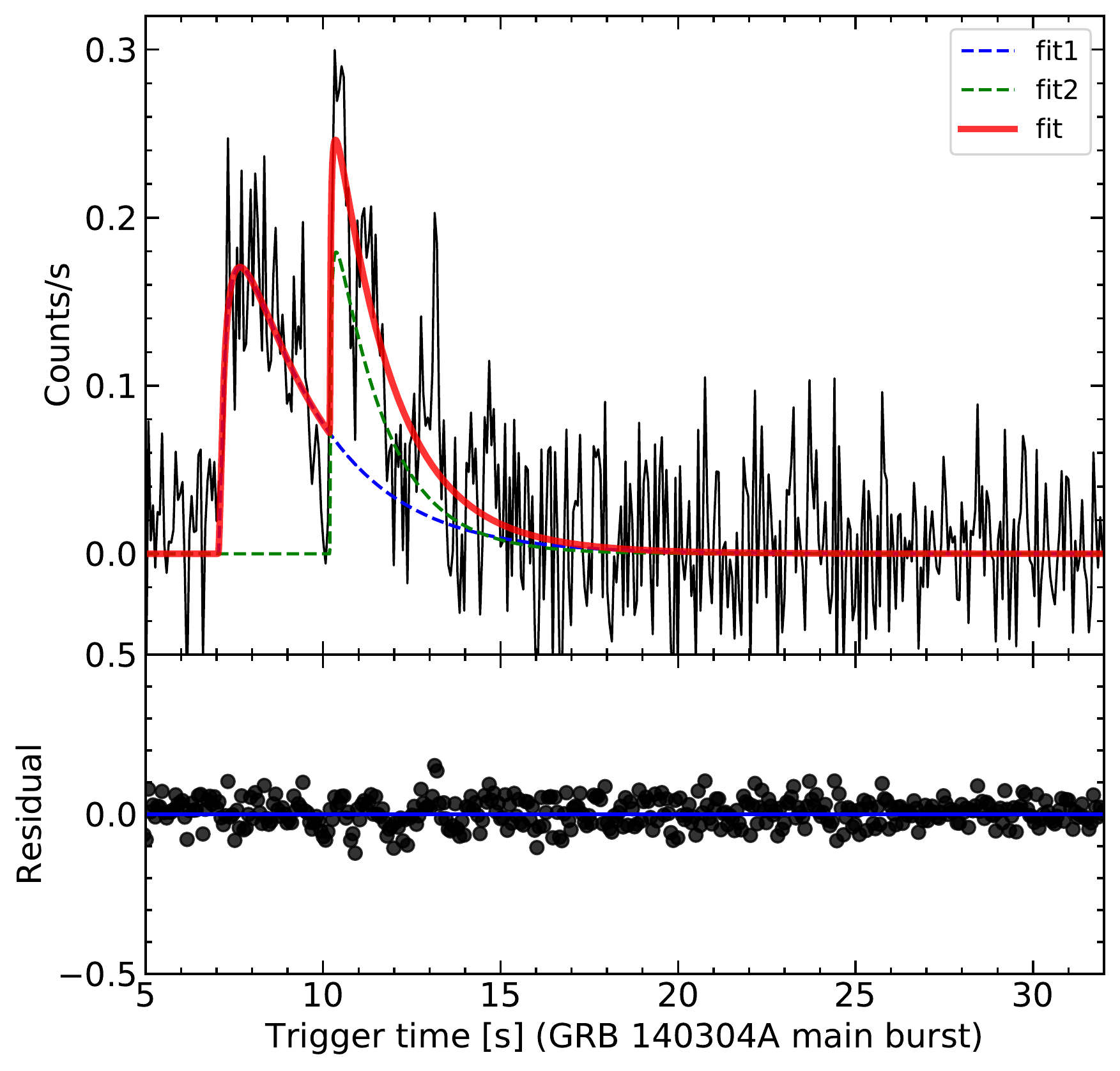}}
\caption{The lightcurve of GRB 140304A
}
\end{figure}

\begin{figure}[!htp]
\centering
\subfigure{
\includegraphics[scale = 0.4]{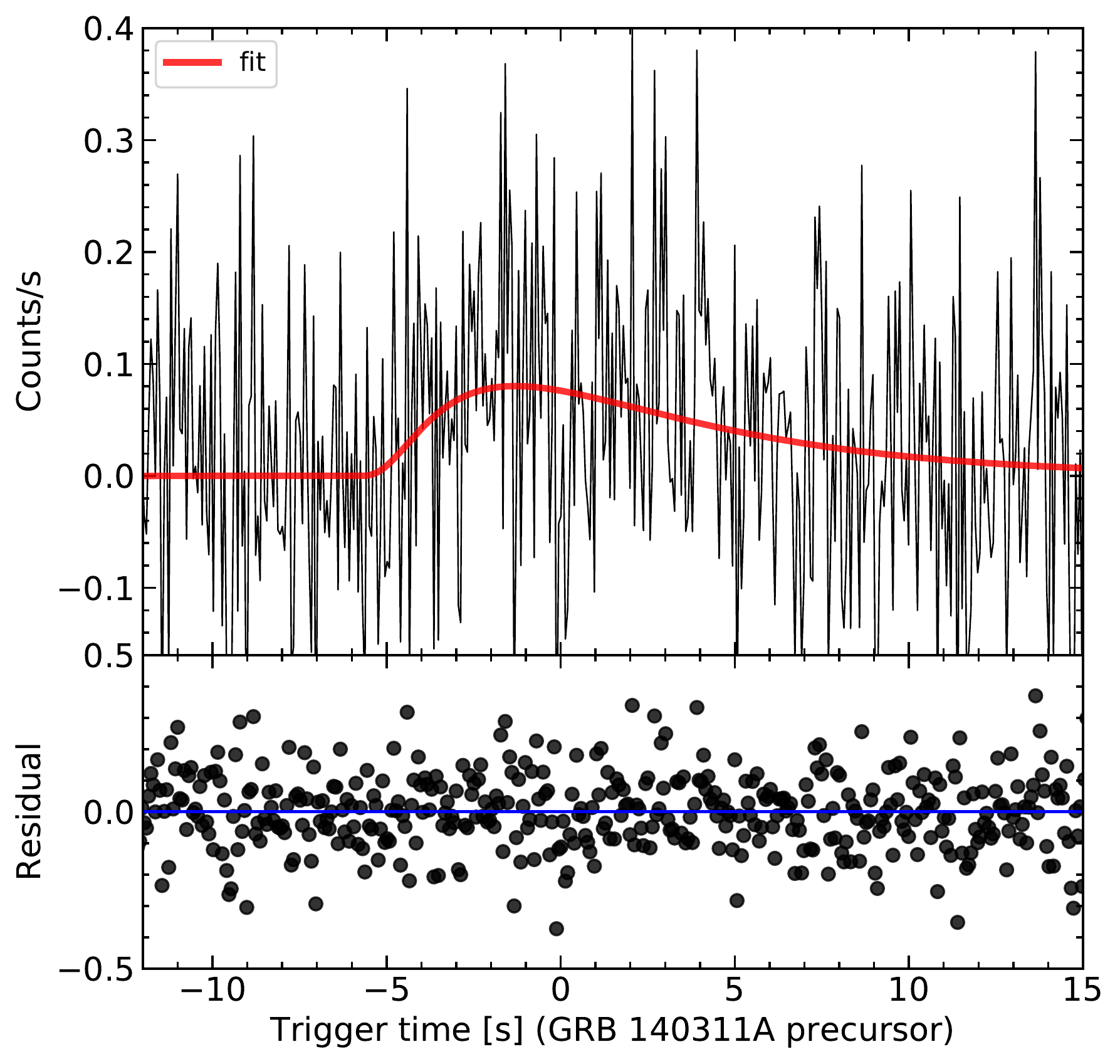}}
\subfigure{
\includegraphics[scale = 0.4]{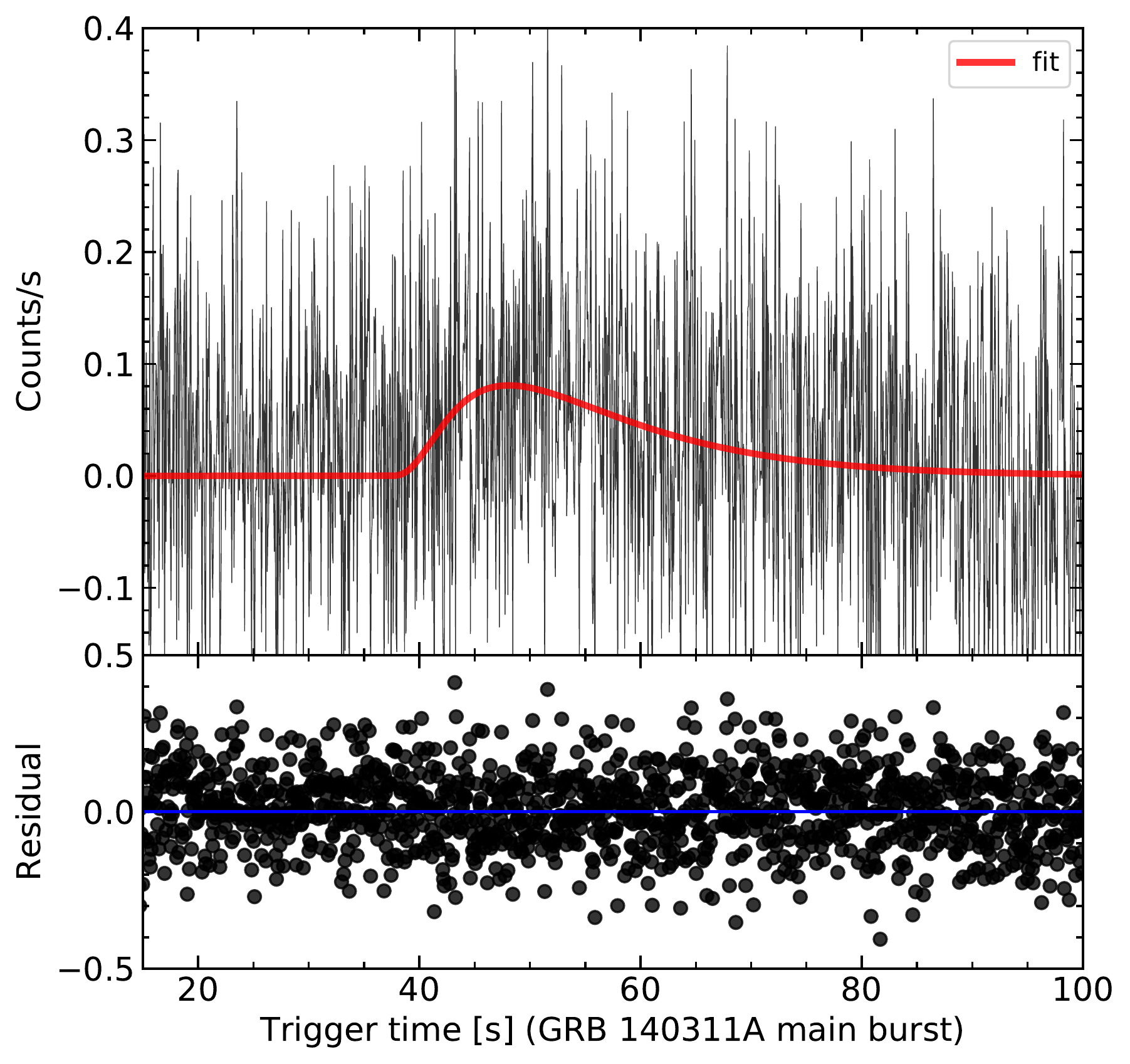}}
\caption{The lightcurve of GRB 140311A
}
\end{figure}

\begin{figure}[!htp]
\centering
\subfigure{
\includegraphics[scale = 0.4]{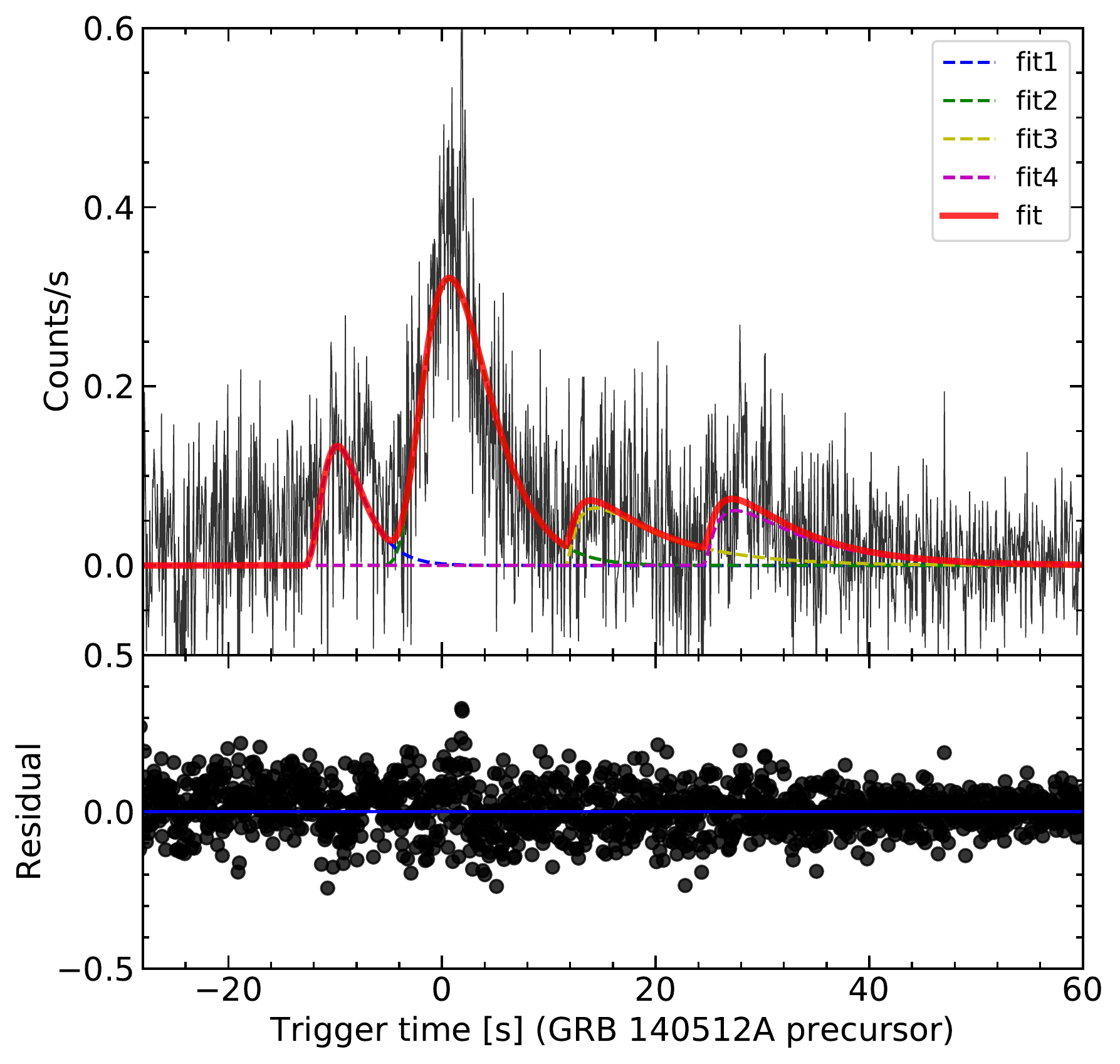}}
\subfigure{
\includegraphics[scale = 0.4]{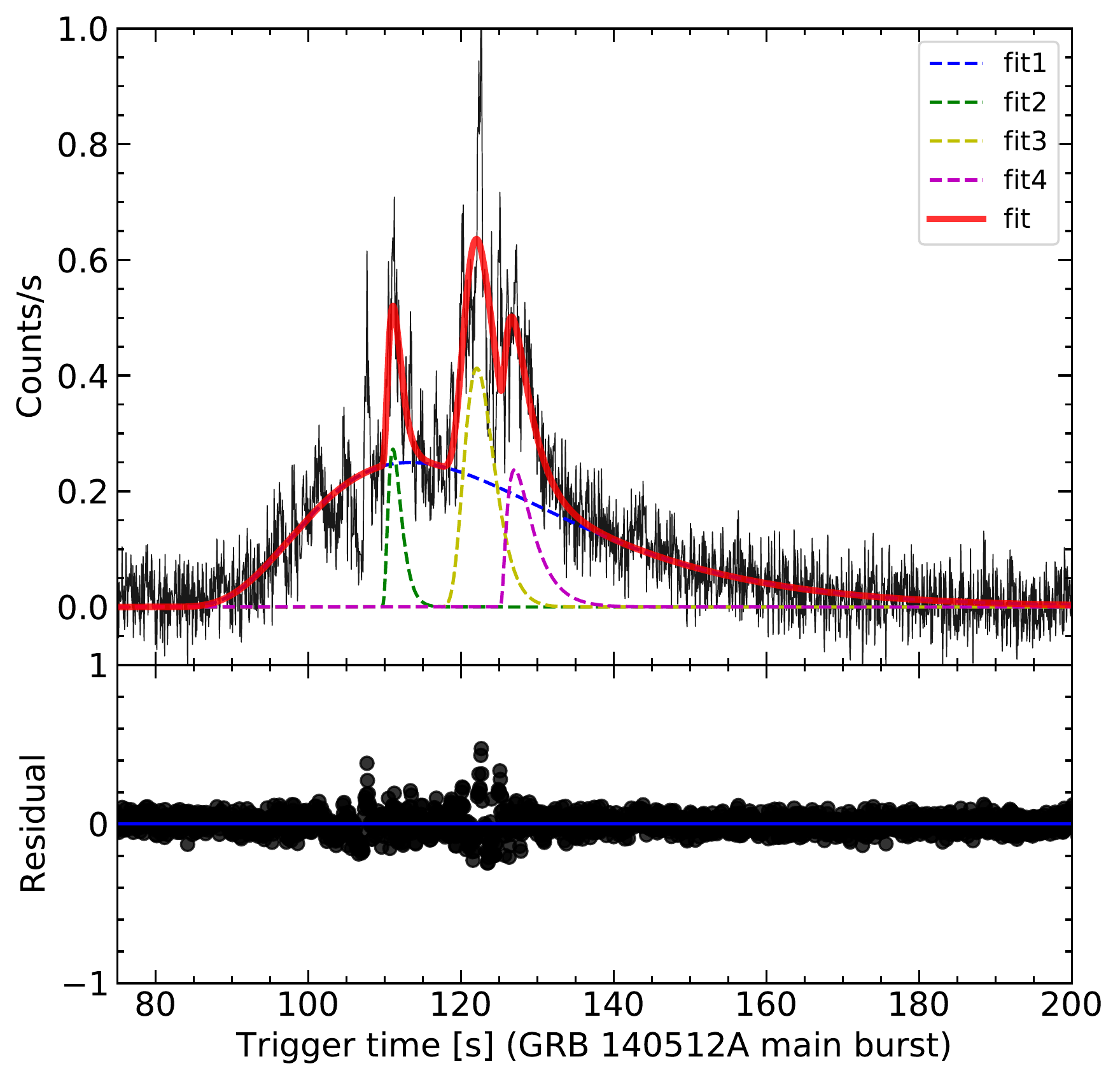}}
\caption{The lightcurve of GRB 140512A
}
\end{figure}

\begin{figure}[!htp]
\centering
\subfigure{
\includegraphics[scale = 0.4]{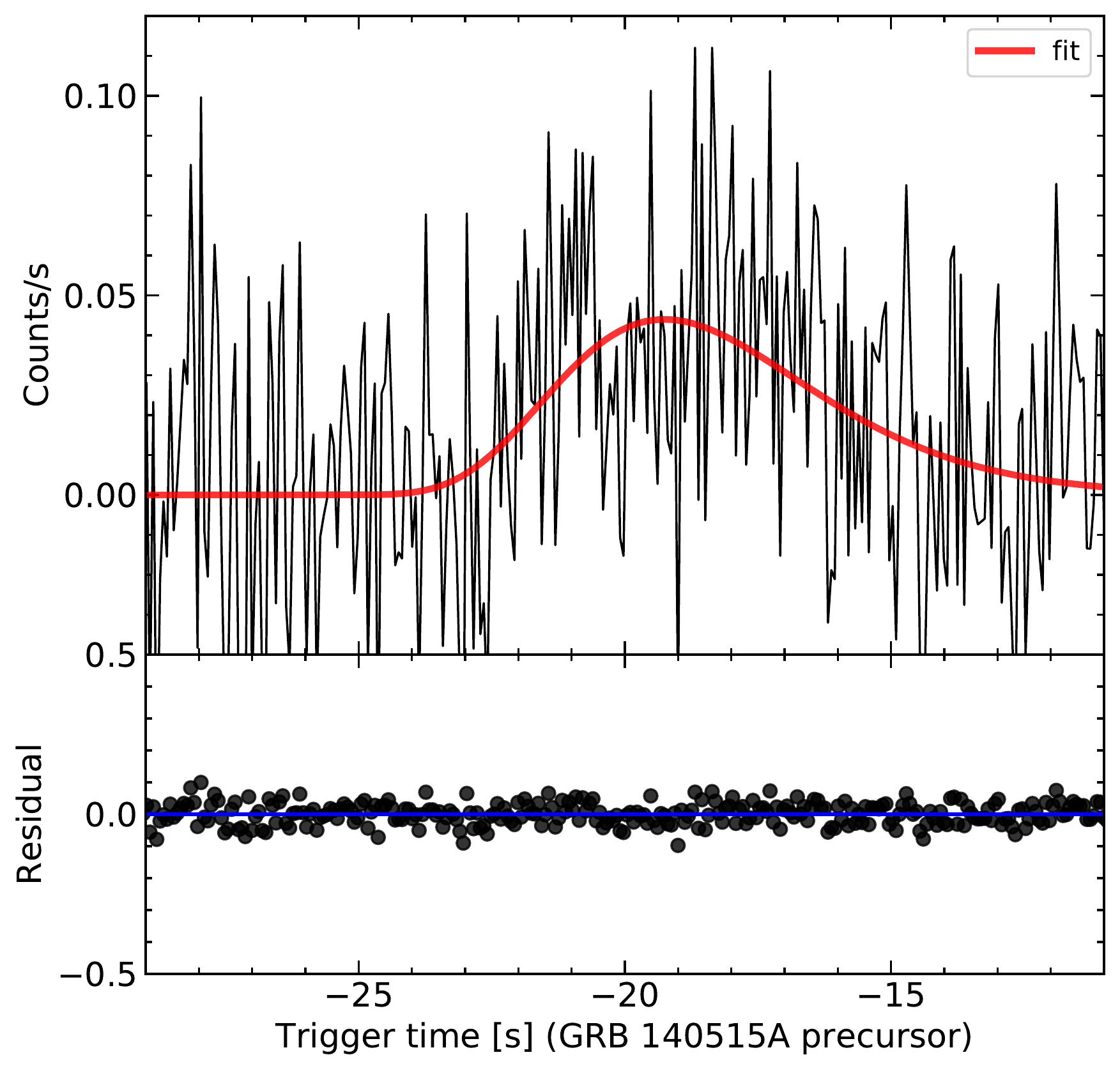}}
\subfigure{
\includegraphics[scale = 0.4]{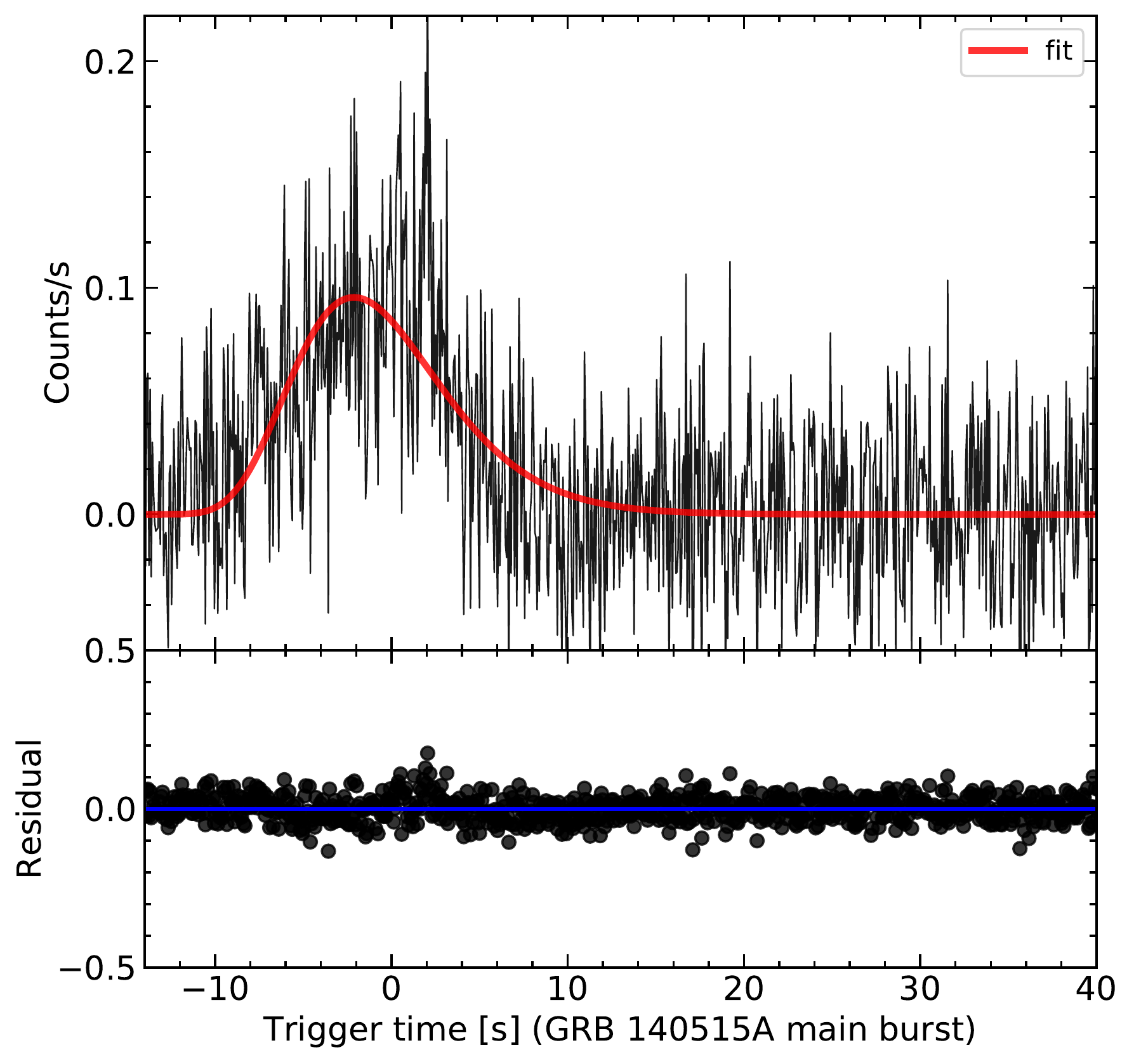}}
\caption{The lightcurve of GRB 140515A
}
\end{figure}

\begin{figure}[!htp]
\centering
\subfigure{
\includegraphics[scale = 0.4]{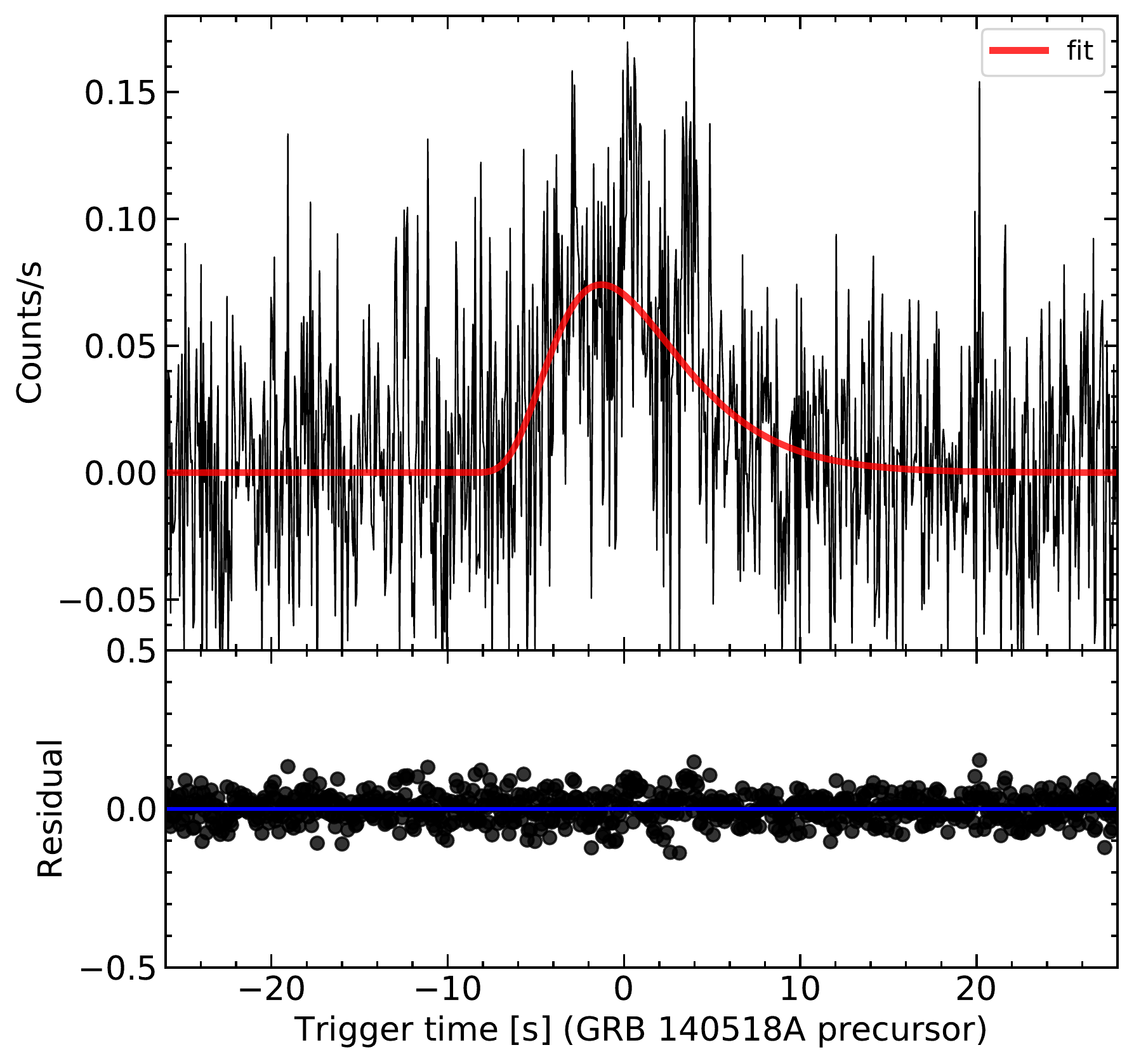}}
\subfigure{
\includegraphics[scale = 0.4]{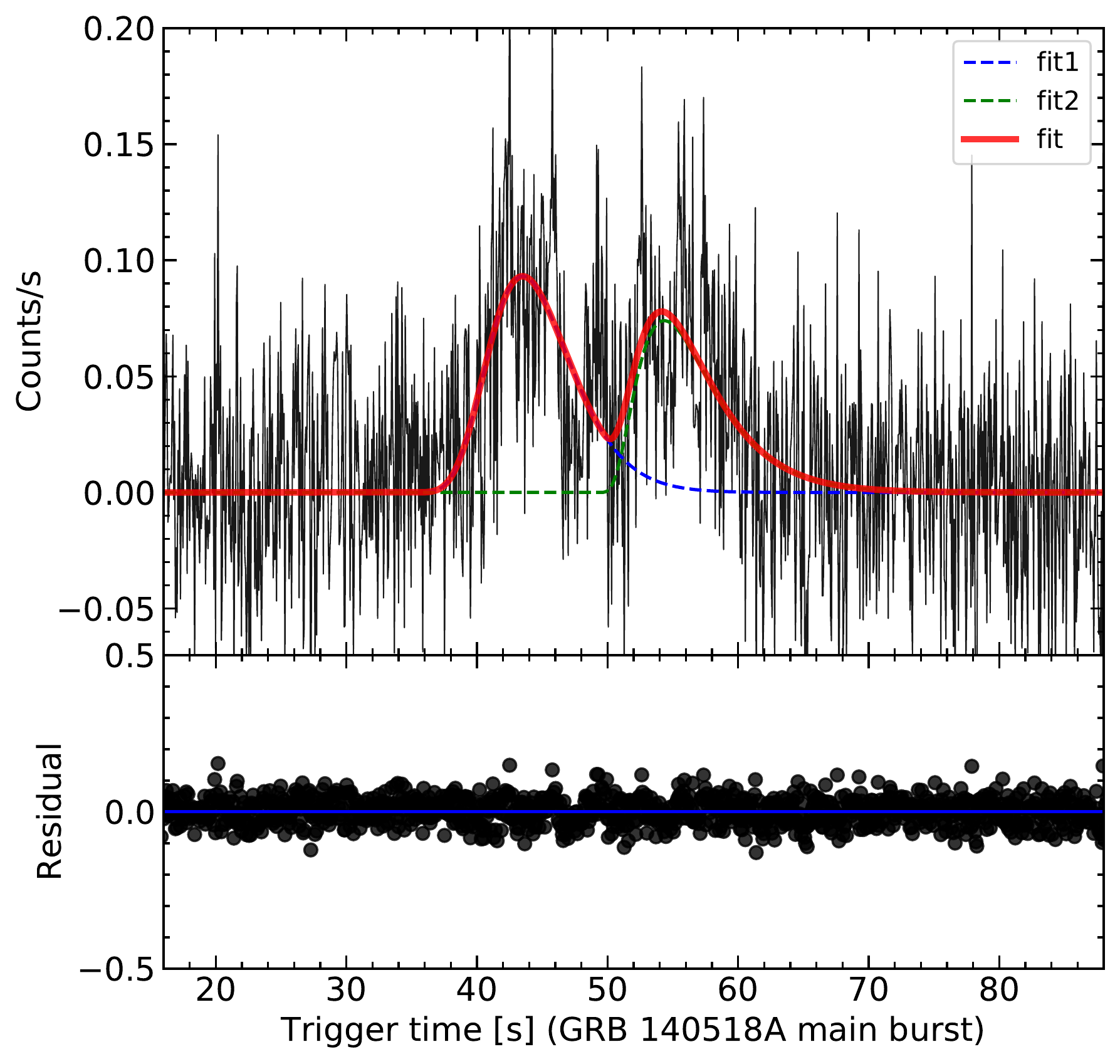}}
\caption{The lightcurve of GRB 140518A
}
\end{figure}

\begin{figure}[!htp]
\centering
\subfigure{
\includegraphics[scale = 0.4]{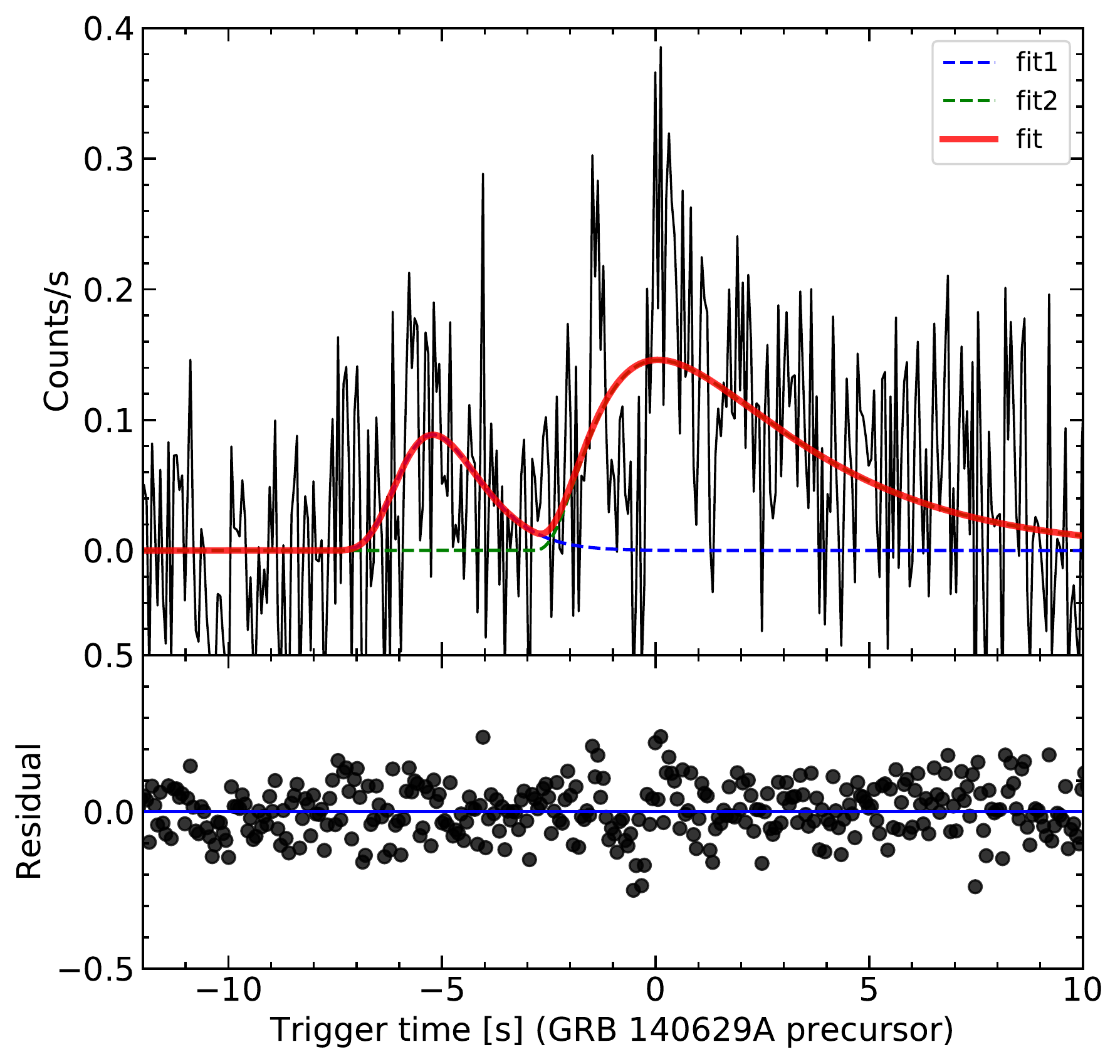}}
\subfigure{
\includegraphics[scale = 0.4]{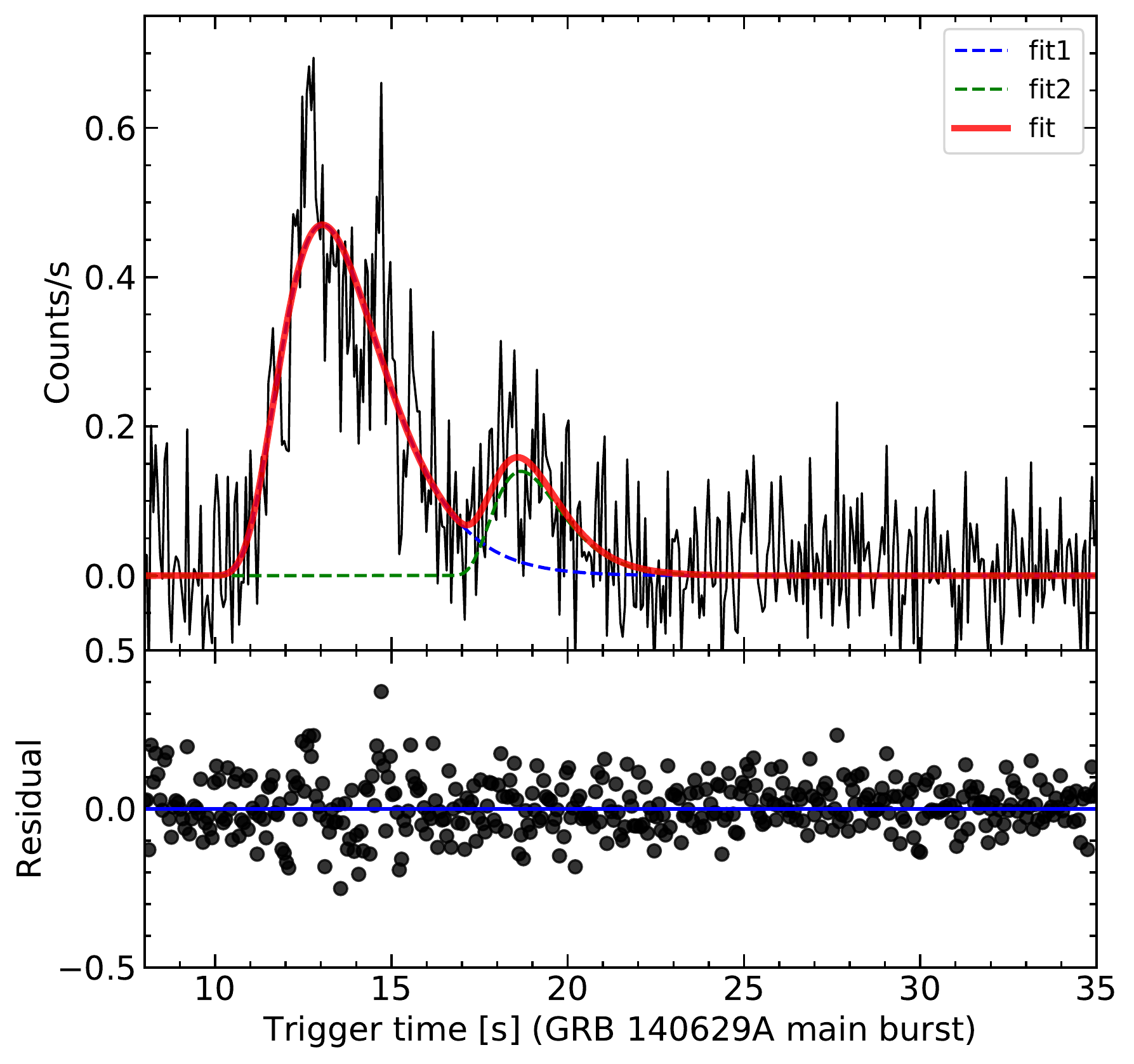}}
\caption{The lightcurve of GRB 140629A
}
\end{figure}

\begin{figure}[!htp]
\centering
\subfigure{
\includegraphics[scale = 0.4]{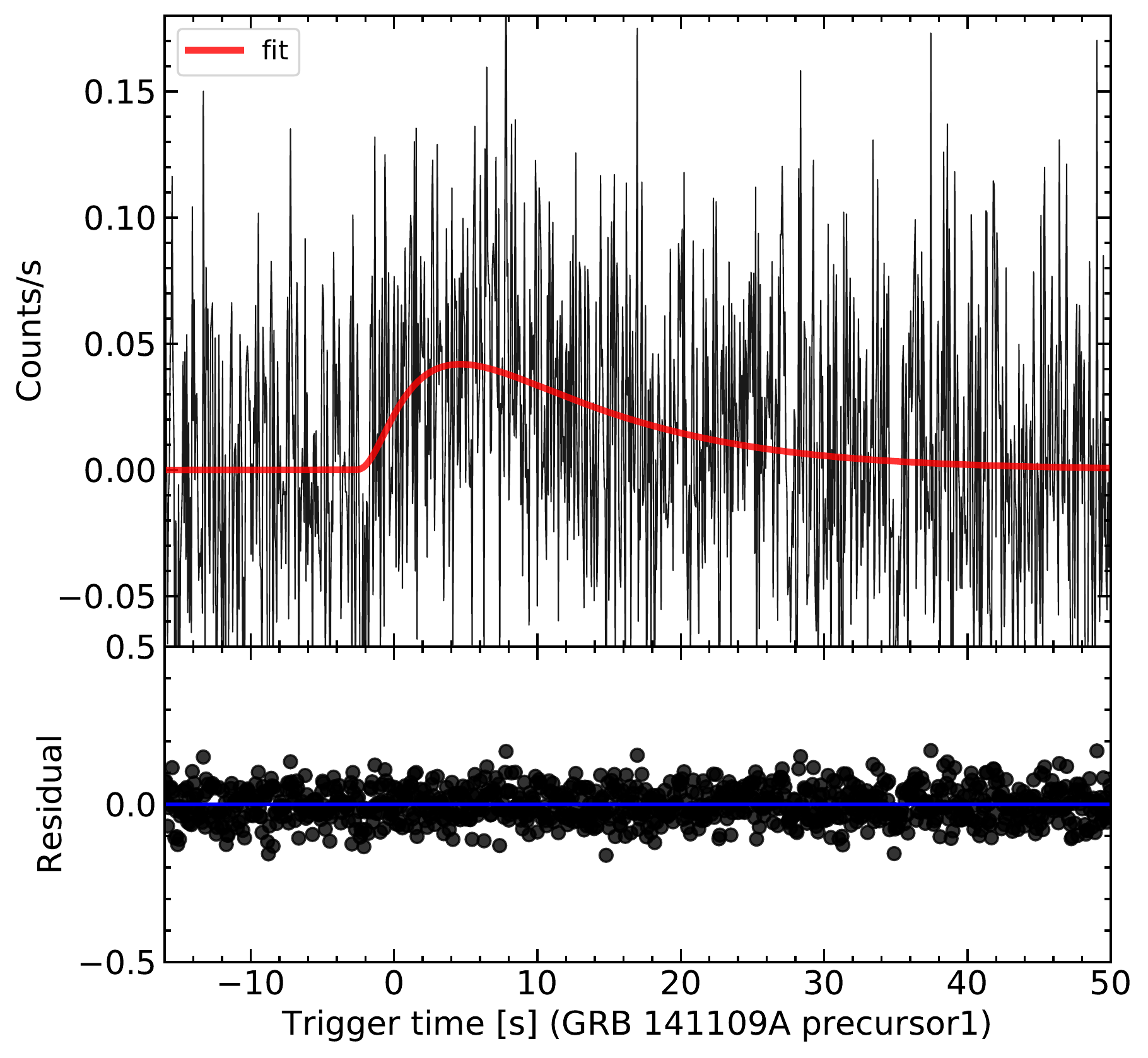}}
\subfigure{
\includegraphics[scale = 0.4]{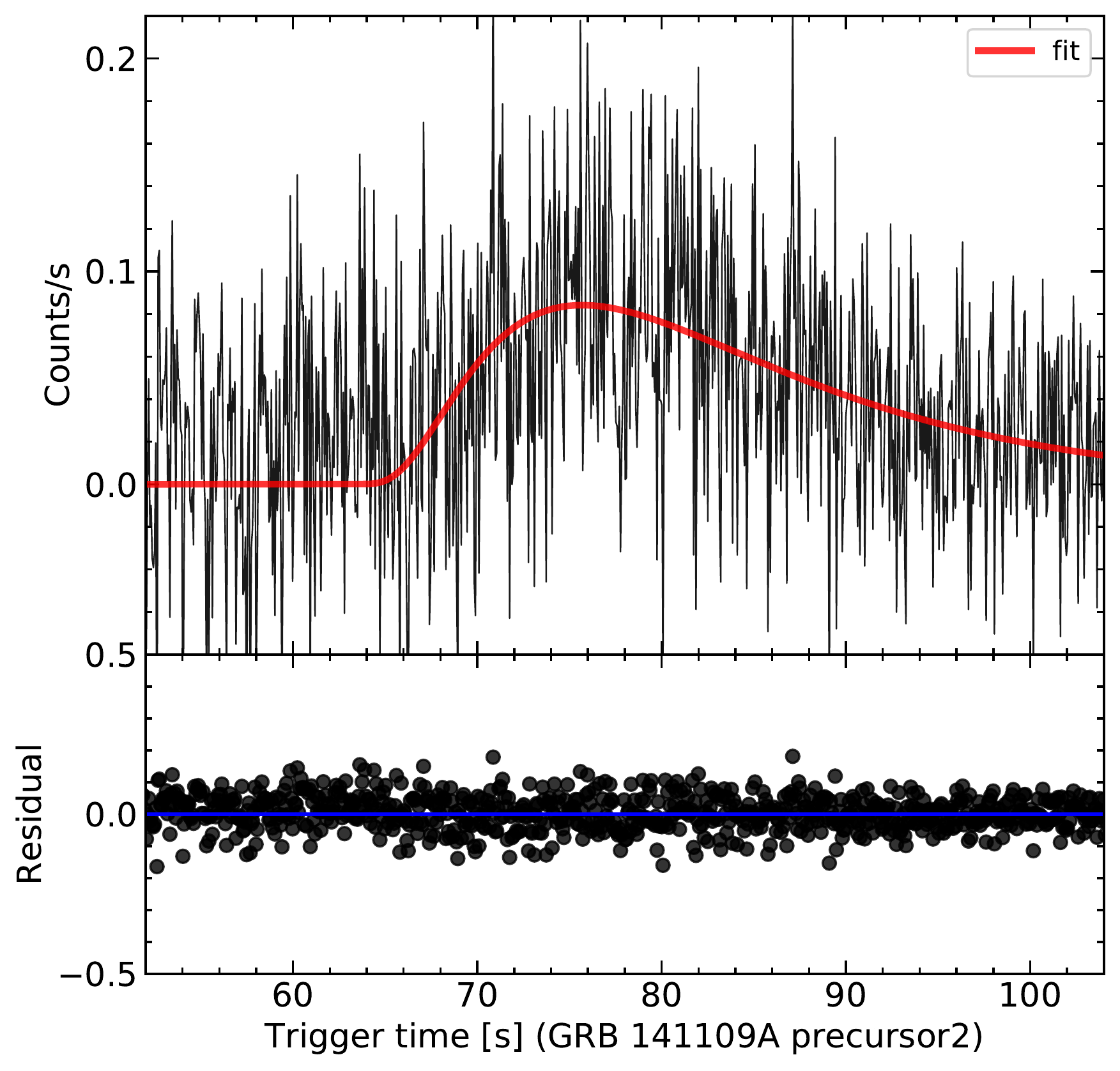}}
\subfigure{
\includegraphics[scale = 0.4]{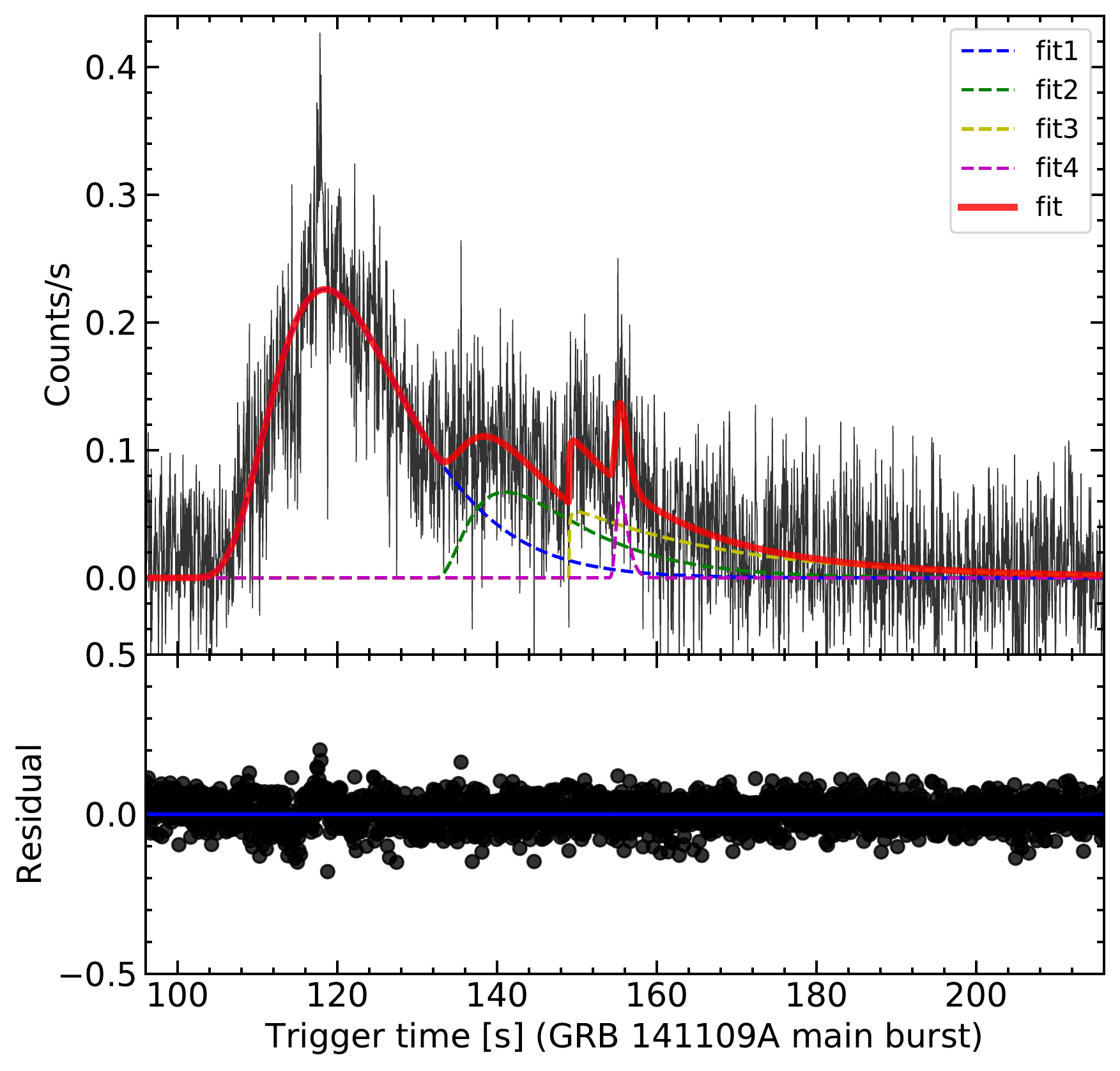}}
\caption{The lightcurve of GRB 141109A
}
\end{figure}

\begin{figure}[!htp]
\centering
\subfigure{
\includegraphics[scale = 0.4]{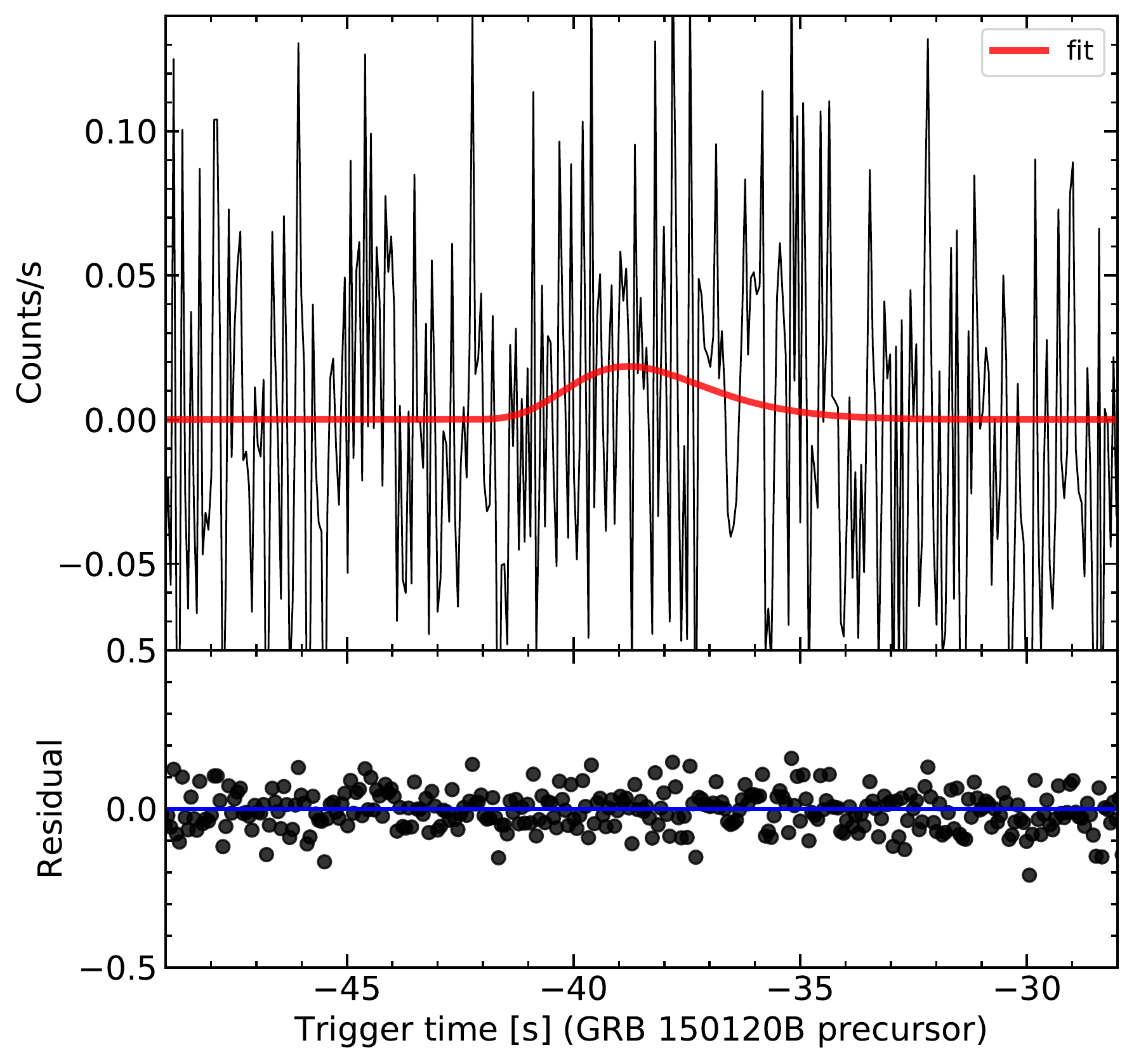}}
\subfigure{
\includegraphics[scale = 0.4]{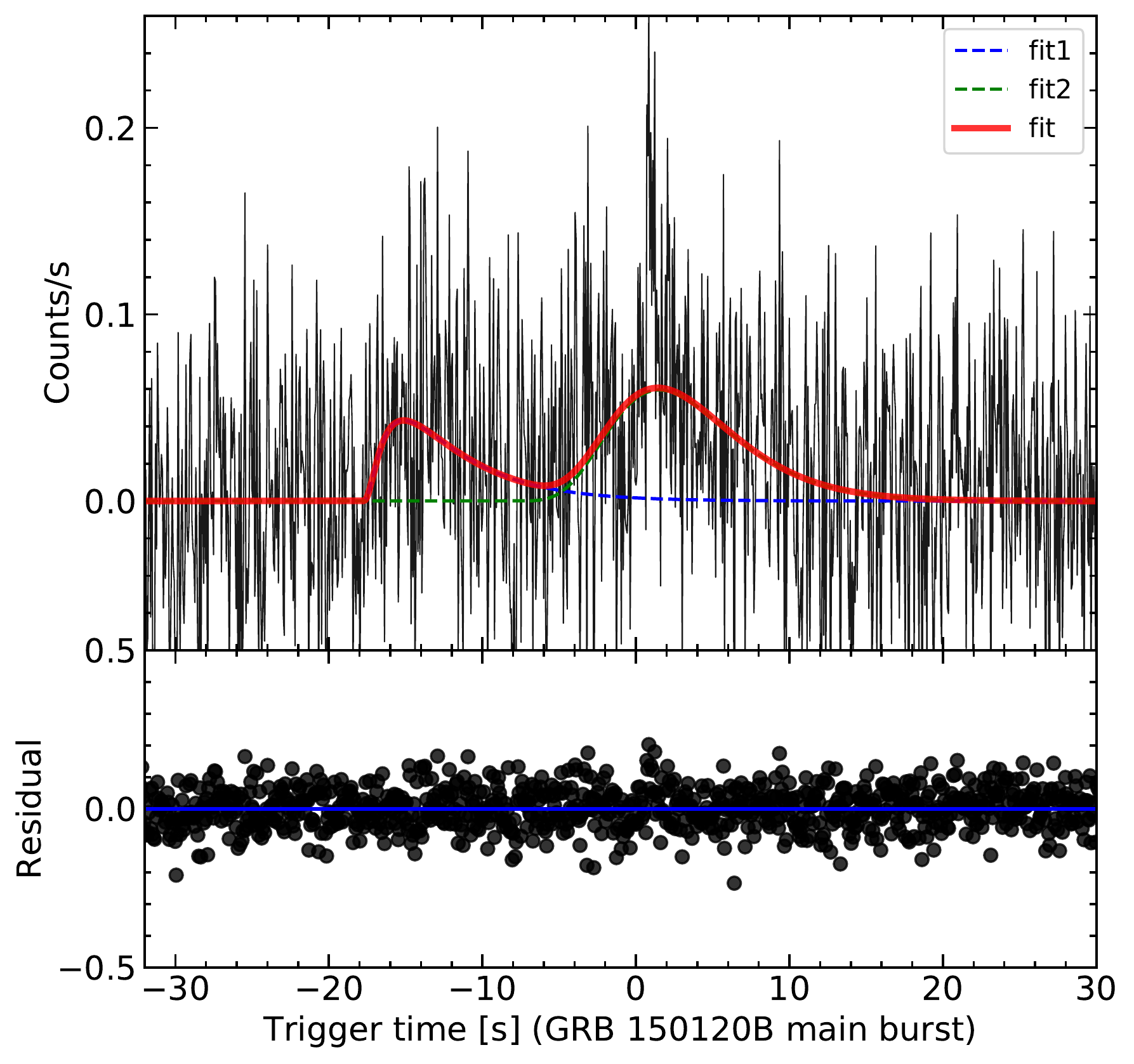}}
\caption{The lightcurve of GRB 150120B
}
\end{figure}

\begin{figure}[!htp]
\centering
\subfigure{
\includegraphics[scale = 0.4]{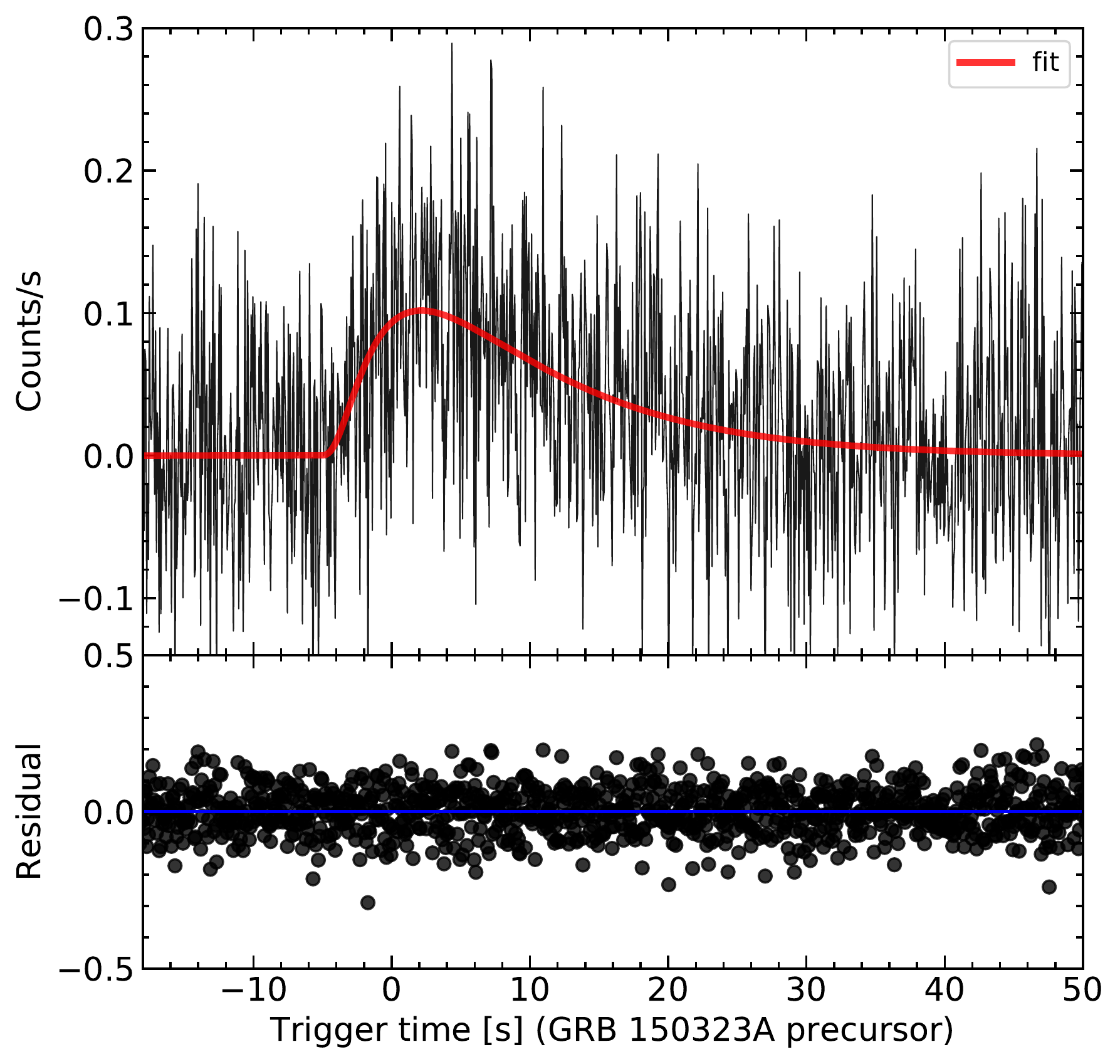}}
\subfigure{
\includegraphics[scale = 0.4]{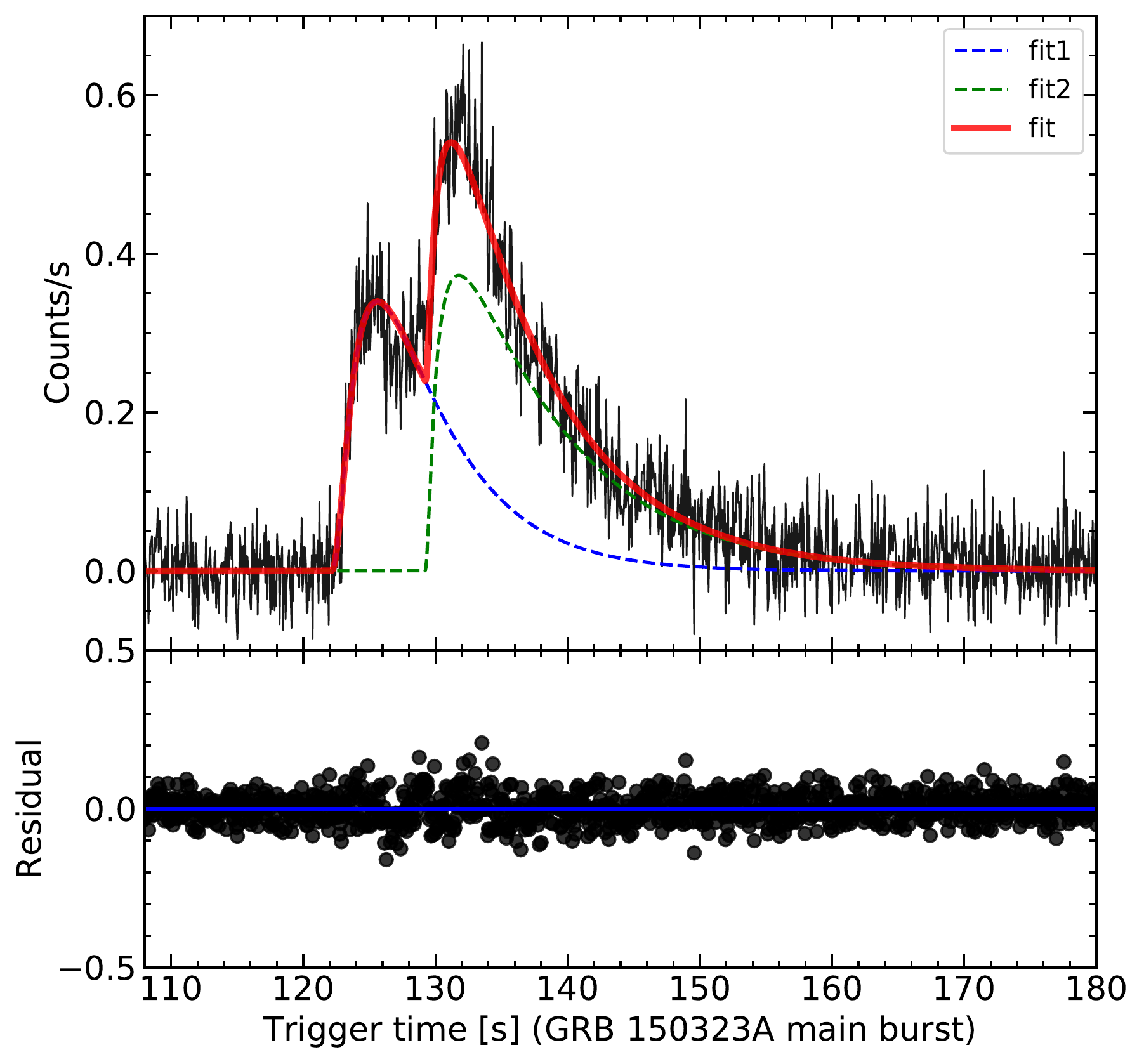}}
\caption{The lightcurve of GRB 150323A
}
\end{figure}

\begin{figure}[!htp]
\centering
\subfigure{
\includegraphics[scale = 0.4]{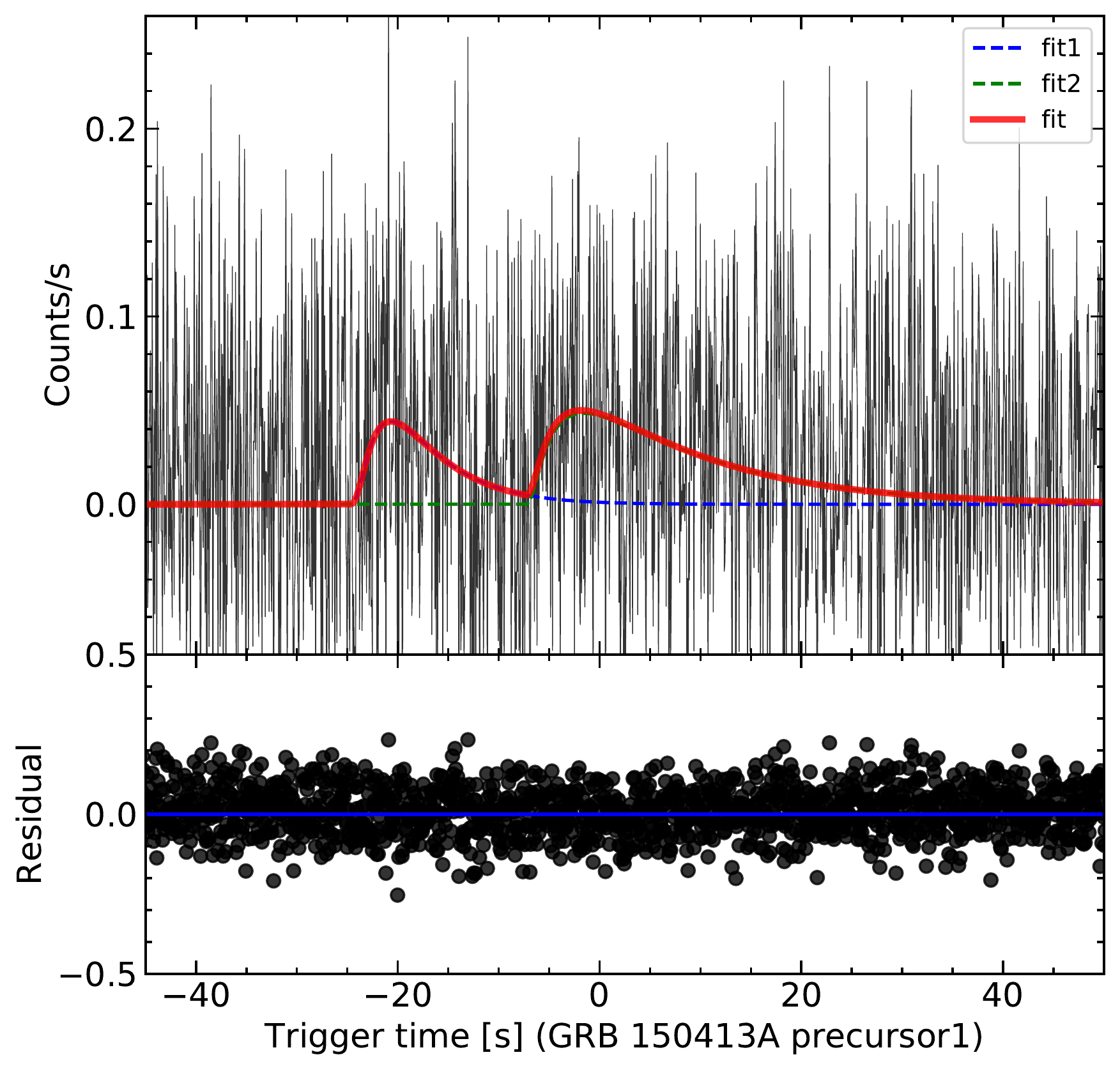}}
\subfigure{
\includegraphics[scale = 0.4]{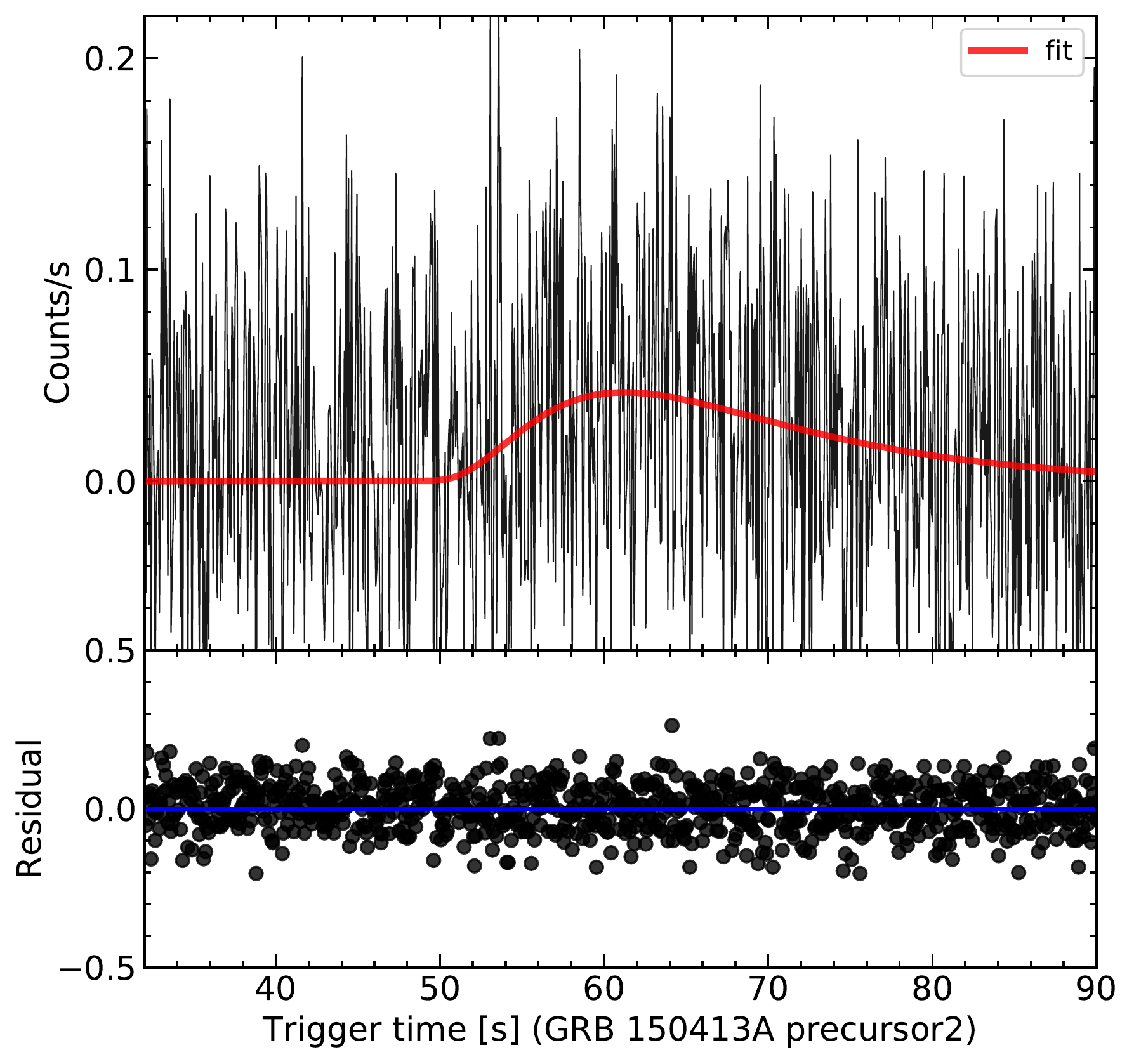}}
\subfigure{
\includegraphics[scale = 0.4]{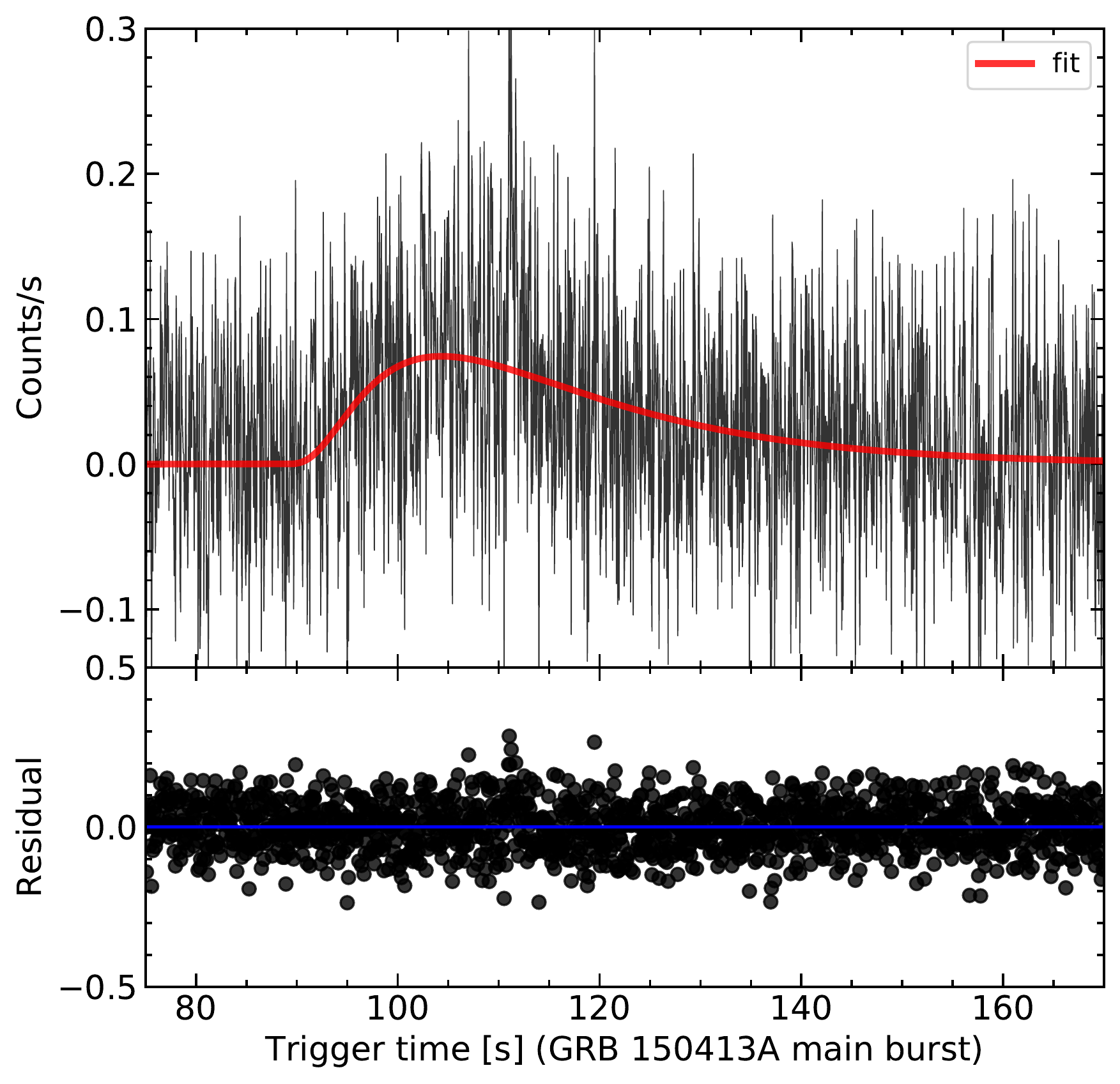}}
\caption{The lightcurve of GRB 150413A
}
\end{figure}


\begin{thebibliography}{}
\bibitem[Amati et al.(2008)]{2008MNRAS.391..577A} Amati, L., Guidorzi, C., Frontera, F., et al.\ 2008, \mnras, 391, 577
\bibitem[Bhatt \& Bhattacharyya(2012)]{bhatt12} Bhatt, N., \& Bhattacharyya, S. 2012, \mnras, 420, 1706
\bibitem[Bernardini et al.(2011)]{be11} Bernardini, M. G., Margutti, R., Chincarini, G., Guidorzi, C., \& Mao, J. 2011, \aap, 526, A27
\bibitem[Bernardini et al.(2013)]{be13} Bernardini, M. G., Campana, S., Ghisellini, G., et al. 2013, \apj, 775, 67
\bibitem[Bi et al.(2018)]{bi18} Bi, X., Mao, J., Liu, C., \& Bai, J.-M. 2018, \apj, 866, 97
\bibitem[Burlon et al.(2008)]{2008ApJ...685L..19B} Burlon, D., Ghirlanda, G., Ghisellini, G., et al.\ 2008, \apjl, 685, L19
\bibitem[Burlon et al.(2009)]{2009A&A...505..569B} Burlon, D., Ghirlanda, G., Ghisellini, G., et al.\ 2009, \aap, 505, 569
\bibitem[Charisi et al.(2015)]{2015MNRAS.448.2624C} Charisi, M., M{\'a}rka, S., \& Bartos, I.\ 2015, \mnras, 448, 2624
\bibitem[Cheng \& Dai(2001)]{2001APh....16...67C} Cheng, K.~S. \& Dai, Z.~G.\ 2001, Astroparticle Physics, 16, 67
\bibitem[Chincarini et al.(2010)]{chincarini10} Chincarini, G., Mao, J., Margutti, R., et al. 2010, \mnras, 406, 2113
\bibitem[Coppin et al.(2020)]{2020arXiv200403246C} Coppin, P., de Vries, K.~D., \& van Eijndhoven, N.\ 2020, Phys. Rev. D. 102, 103014
\bibitem[Daigne \& Mochkovitch(2002)]{2002MNRAS.336.1271D} Daigne, F., \& Mochkovitch, R.\ 2002, \mnras, 336, 1271
\bibitem[Fenimore et al.(1996)]{1996ApJ...473..998F} Fenimore, E.~E., Madras, C.~D., \& Nayakshin, S.\ 1996, \apj, 473, 998
\bibitem[Hu et al.(2014)]{2014ApJ...789..145H} Hu, Y.-D., Liang, E.-W., Xi, S.-Q., et al.\ 2014, \apj, 789, 145
\bibitem[Koshut et al.(1995)]{1995ApJ...452..145K} Koshut, T.~M., Kouveliotou, C., Paciesas, W.~S., et al.\ 1995, \apj, 452, 145
\bibitem[Kouveliotou et al.(1993)]{1993ApJ...413L.101K} Kouveliotou, C., Meegan, C.~A., Fishman, G.~J., et al.\ 1993, \apjl, 413, L101
\bibitem[Kumar \& Narayan(2009)]{kumar09} Kumar, P., \& Narayan, R. 2009, \mnras, 395, 472
\bibitem[Lan et al.(2018)]{2018ApJ...862..155L} Lan, L., L{\"u}, H.-J., Zhong, S.-Q., et al.\ 2018, \apj, 862, 155
\bibitem[Lazzati(2005)]{2005MNRAS.357..722L} Lazzati, D.\ 2005, \mnras, 357, 722
\bibitem[Lazzati \& Begelman(2005)]{2005ApJ...629..903L} Lazzati, D. \& Begelman, M.~C.\ 2005, \apj, 629, 903
\bibitem[Lee et al.(2000)]{2000ApJS..131....1L} Lee, A., Bloom, E.~D., \& Petrosian, V.\ 2000, \apjs, 131, 1
\bibitem[Li(2007)]{2007MNRAS.380..621L} Li, L.-X.\ 2007, \mnras, 380, 621
\bibitem[Li et al.(2020)]{li20} Li, X.-J., Zhang, Z.-B., Zhang, C.-T., Zhang, K., Zhang, Y., \& Dong, X.-F. 2020, \apj, 892, 113
\bibitem[Li et al.(2021)]{li21} Li, X. J., Zhang, Z. B., Zhang, X. L., \& Zhen, H. Y. 2021, ApJS, 252, 16
\bibitem[Lien et al.(2016)]{2016ApJ...829....7L} Lien, A., Sakamoto, T., Barthelmy, S.~D., et al.\ 2016, \apj, 829, 7
\bibitem[Liu \& Mao(2019)]{liu19} Liu, C.-X., \& Mao, J. 2019, \apj, 884, 59
\bibitem[Lipunova et al.(2009)]{2009MNRAS.397.1695L} Lipunova, G.~V., Gorbovskoy, E.~S., Bogomazov, A.~I., et al.\ 2009, \mnras, 397, 1695
\bibitem[Lyutikov \& Usov(2000)]{2000ApJ...543L.129L} Lyutikov, M. \& Usov, V.~V.\ 2000, \apjl, 543, L129
\bibitem[M{\'e}sz{\'a}ros \& Rees(2000)]{2000ApJ...530..292M} M{\'e}sz{\'a}ros, P., \& Rees, M.~J.\ 2000, \apj, 530, 292
\bibitem[Murakami et al.(1991)]{1991Natur.350..592M} Murakami, T., Inoue, H., Nishimura, J., et al.\ 1991, \nat, 350, 592
\bibitem[Narayan \& Kumar(2009)]{narayan09} Narayan, R., \& Kumar, P. 2009, \mnras, 394, L117 
\bibitem[Norris et al.(2005)]{2005ApJ...627..324N} Norris, J.~P., Bonnell, J.~T., Kazanas, D., et al.\ 2005, \apj, 627, 324
\bibitem[Norris et al.(1996)]{1996ApJ...459..393N} Norris, J.~P., Nemiroff, R.~J., Bonnell, J.~T., et al.\ 1996, \apj, 459, 393
\bibitem[Ramirez-Ruiz et al.(2002)]{2002MNRAS.331..197R} Ramirez-Ruiz, E., MacFadyen, A.~I., \& Lazzati, D.\ 2002, \mnras, 331, 197
\bibitem[Ramirez-Ruiz, \& Merloni(2001)]{2001MNRAS.320L..25R} Ramirez-Ruiz, E., \& Merloni, A.\ 2001, \mnras, 320, L25
\bibitem[Sakamoto et al.(2005)]{saka05} Sakamoto, T., Lamb, D. Q., Kawai, N., et al. 2005, \apj, 629, 311
\bibitem[Sakamoto et al.(2008)]{2008ApJS..175..179S} Sakamoto, T., Barthelmy, S.~D., Barbier, L., et al.\ 2008, \apjs, 175, 179
\bibitem[Sari \& Piran(1997)]{sari97} Sari, R., \& Piran, T. 1997, \apj, 485, 270
\bibitem[Troja et al.(2010)]{2010ApJ...723.1711T} Troja, E., Rosswog, S., \& Gehrels, N.\ 2010, \apj, 723, 1711
\bibitem[Wang, \& M{\'e}sz{\'a}ros(2007)]{2007ApJ...670.1247W} Wang, X.-Y., \& M{\'e}sz{\'a}ros, P.\ 2007, \apj, 670, 1247
\bibitem[Waxman \& M{\'e}sz{\'a}ros(2003)]{2003ApJ...584..390W} Waxman, E. \& M{\'e}sz{\'a}ros, P.\ 2003, \apj, 584, 390
\bibitem[Zhong et al.(2019)]{2019ApJ...884...25Z} Zhong, S.-Q., Dai, Z. G., Cheng, J.-G., et al.\ 2019, \apj, 884, 25

\end{thebibliography}
\end{document}